\documentclass[a4]{aa}

\usepackage[dvipsnames]{xcolor}
\usepackage{xspace}
\usepackage{graphicx}
\usepackage{txfonts}
\usepackage[pdftex,pagebackref,hyperindex=true,colorlinks=true,bookmarks=true,bookmarksopen=true,bookmarksnumbered=true,citecolor=OliveGreen]{hyperref}
\usepackage{siunitx}
\usepackage[normalem]{ulem}
\usepackage{float}
\usepackage{caption}
\usepackage{subcaption}
\usepackage{longtable}
\usepackage{xtab}
\usepackage{pdflscape}
\usepackage{adjustbox}
\usepackage{booktabs}

\DeclareSIUnit\parsec{pc}
\DeclareSIUnit\Year{y}
\newlength\defaultparindent

\usepackage{natbib}
\bibpunct{(}{)}{;}{a}{,}{,}

\newcommand{\doe}{\textsc{Doe}\xspace}
\newcommand{\nacre}{\textsc{Nacre}\xspace}

\newcommand{\ie}{\emph{i.e.}\xspace}
\newcommand{\eg}{\emph{e.g.}\xspace}
\newcommand{\Gaia}{\emph{Gaia}\xspace}
\newcommand{\sbnine}{SB$_{9}$\xspace}

\newcommand{\abratio}[2]{\ensuremath{[\mathrm{#1}/\mathrm{#2}]}\xspace}
\newcommand{\snr}{\ensuremath{\mathrm{S}/\mathrm{N}}\xspace}

\DeclareSIUnit\angstrom{\text {Å}}
\DeclareSIUnit\mag{\text {mag}}

\begin{document} 

\title{The \Gaia-ESO Survey: new spectroscopic binaries in the Milky Way}

\author{M. Van der Swaelmen
  \inst{\ref{iaa},\ref{oaa}}
  \and
  T. Merle
  \inst{\ref{iaa}}
  \and
  S. Van Eck
  \inst{\ref{iaa}}
  \and
  A. Jorissen
  \inst{\ref{iaa}}
  \and
  L. Magrini
  \inst{\ref{oaa}}
  \and
  S. Randich
  \inst{\ref{oaa}}
  \and
  A. Vallenari
  \inst{\ref{oapd}}
  \and
  T. Zwitter
  \inst{\ref{ulslovenia}}
  \and
  G. Traven
  \inst{\ref{ulslovenia}}
  \and
  C. Viscasillas V{\'a}zquez\inst{\ref{vilnius}}
  \and
  A. Bragaglia\inst{\ref{oab}}
  \and
  A. Casey
  \inst{\ref{astro3d},\ref{monash}}
  \and
  A. Frasca
  \inst{\ref{oac}}
  \and
  F. Jim\'enez-Esteban
  \inst{\ref{damadrid}}
  \and
  E. Pancino
  \inst{\ref{oaa}}
  \and
  C. C. Worley
  \inst{\ref{cambridge},\ref{canterburynz}}
  \and
  S. Zaggia
  \inst{\ref{oapd}}
}

\institute{
  Institut d'Astronomie et d'Astrophysique, Universit\'e Libre de Bruxelles, CP. 226, Boulevard du Triomphe, 1050 Brussels, Belgium\label{iaa}
  \and
  INAF -- Osservatorio Astrofisico di Arcetri, Largo E. Fermi 5, 50125, Firenze, Italy\\ \email{mathieu.van+der+swaelmen@inaf.it}\label{oaa}
  \and
  INAF -- Osservatorio Astronomico di Padova, Vicolo dell’Osservatorio 5, I-35122 Padova, Italy\label{oapd}
  \and
  Faculty of Mathematics and Physics, University of Ljubljana, Jadranska 19, 1000 Ljubljana, Slovenia\label{ulslovenia}
  \and
  Institute of Theoretical Physics and Astronomy, Vilnius University, Sauletekio av. 3, 10257 Vilnius, Lithuania\label{vilnius}
  \and
  INAF -- Osservatorio di Astrofisica e Scienza dello Spazio, via P. Gobetti 93/3, 40129 Bologna, Italy\label{oab}
  \and
  Center of Excellence for Astrophysics in Three Dimensions (ASTRO-3D), Australia\label{astro3d}
  \and
  School of Physics \& Astronomy, Monash University, Wellington Road, Clayton 3800, Victoria, Australia\label{monash}
  \and
  INAF -- Osservatorio Astrofisico di Catania, Via S.Sofia 78, I-95123, Catania, Italy\label{oac}
  \and
  Departamento de Astrof\'{\i}sica, Centro de Astrobiolog\'{\i}a (CSIC-INTA), ESAC Campus, Camino Bajo del Castillo s/n, E-28692 Villanueva de la Ca\~nada, Madrid, Spain\label{damadrid}
  \and
  Institute of Astronomy, University of Cambridge, Madingley Road, Cambridge CB3 0HA, United Kingdom\label{cambridge}
  \and
  School of Physical and Chemical Sciences -- Te Kura Mat\={u}, University of Canterbury, Private Bag 4800, Christchurch 8140, New Zealand\label{canterburynz}
}

\date{Submitted: 06/06/2023; Accepted: 05/10/2023}

\abstract
    {The \Gaia-ESO Survey (GES) is a large public spectroscopic survey {which acquired spectra for more than \num{100000} stars} across all major components of the Milky Way. In addition to atmospheric parameters and stellar abundances that have been derived in previous papers of this series, the GES spectra allow us to detect spectroscopic binaries with one (SB1), two (SB2) or more (SB$n \ge 3$) components.}
    {The present paper {discusses} the statistics of {GES SB$n \ge 2$ after analysing \num{160727} GIRAFFE HR10 and HR21 spectra, amounting to \num{37565} unique Milky Way field targets.}}
    {Cross-correlation functions (CCFs) have been re-computed thanks to a dozen spectral masks probing a range of effective temperatures ($\SI{3900}{\kelvin} < T_{\mathrm{eff}} < \SI{8000}{\kelvin}$), surface gravities ($1.0 < \log g < 4.7$) and  metallicities ($-2.6 < \abratio{Fe}{H} < 0.3$). By optimising the mask choice for a given spectrum, the new computed so-called \nacre (Narrow cross-correlation experiment) CCFs are narrower and allow to unblend more stellar components than standard masks. The \doe (Detection of Extrema) extremum-finding code then selects the individual components and provides their radial velocities.}
    {From the {sample of HR10 and HR21 spectra corresponding to \num{37565} objects}, the present study leads to the detection of \num{322} SB2, ten (three of them being tentative) SB3, and two tentative SB4. {In particular, compared to our previous study, the \nacre CCFs allow us to multiply the number of SB2 candidates by $\approx 1.5$.} The colour-magnitude diagram reveals, as expected, the shifted location of the SB2 main sequence. A comparison between the SB identified in \Gaia DR3 and the ones detected in the present work is performed and the complementarity of the two censuses is discussed. An application to mass-ratio determination is presented, and the mass-ratio distribution of the GES SB2 is discussed. When accounting for the SB2 detection rate, an SB2 frequency of $\approx \SI{1.4}{\percent}$ is derived within the present stellar sample of mainly FGK-type stars.}
    {As primary outliers identified within the GES data, SB$n$ spectra produce a wealth of information and useful constraints for the binary population synthesis studies.}


\maketitle

\section{Introduction}

{The single-star fraction of solar-type stars (masses in the range $[0.8, 1.2] \mathrm{M}_{\odot}$) is robustly estimated at $\num{60}\pm\SI{4}{\percent}$. Speaking in terms of stellar systems, the binary-star fraction with mass ratios larger than \num{0.1} and periods $P$ in the range $0.2 < \log P (\mathrm{d}) < 8.0$, represent $\num{30}\pm \SI{4}{\percent}$ of all late-type systems \citep{2017ApJS..230...15M}. The orbital periods lower than $\log{P}\sim4$ are} efficiently sampled by spectroscopic binaries (SB), which are detected by measuring radial velocity (RV) variations. One advantage of SB over astrometric binaries (AB) is that they can probe more easily fainter objects: the power of cross-correlation technique combining the flux information of hundreds or thousands of pixels allows us to use spectra with signal-to-noise ratio as low as \num{5} to measure radial velocities (\eg \citealt{2017A&A...608A..95M}). {Moreover, anticipating a comment we will develop later in this article, SB can be detected at larger distances than AB.}

With the advent of spectroscopic surveys, {a very large number} of observed targets {with RV variability, likely due to their SB nature, are being discovered}. Among SB, we distinguish those showing only one component in their spectra (SB1) from those showing more than one component\footnote{{In the entire article, SB$n$ should be understood as SB$n \ge 2$ except when it is specified otherwise.}} (SB$n$, with $n \ge 2$). One way to detect such SB$n$ is by counting the number of components in the cross-correlation functions (CCFs) of observed spectra against spectral templates {(also called masks) of single stars}. This procedure can be performed automatically, by using the three first derivatives of the cross-correlation functions, as done for example by the \doe (Detection of Extrema) code \citep{2017A&A...608A..95M}. This technique was applied to a previous release of the \Gaia-ESO survey \citep[GES;][]{gilmore22, randich22} as well as to the GALactic Archeology with HERMES \citep[GALAH;][]{desilva15} one \citep{2017A&A...608A..95M,2020A&A...638A.145T}. Among other techniques used in various surveys, we can mention the one adopted by \citet{2010AJ....140..184M} in the RAVE survey, by \citet{2017PASP..129h4201F} and \citet{2021AJ....162..184K} in the infrared APOGEE survey, by \citet{2022ApJS..258...26Z} in the LAMOST medium resolution survey. New methods are also developed to detect and characterise unresolved SB$n$, \ie composite spectra with radial velocities differences between the components below the detection threshold of the spectrograph (see the pioneering work of \citealt{2018MNRAS.473.5043E,2018MNRAS.476..528E}) and a similar method by \citet{2022MNRAS.510.1515K} also fitting the projected rotational velocity. Moreover, in the hunt of detached stellar mass black holes in binaries, the method of spectral disentangling \citep{1994A&A...281..286S} has turned several SB1 claimed to harbour quiet black holes (\eg \citealt{2021MNRAS.504.2577J}) into SB2 (\eg with a hot massive rapidly-rotating companion, \citealt{2022MNRAS.512.5620E}). It is also worth mentioning that machine learning and Bayesian inference are now also employed to detect SB$n$ as in \citet{2020A&A...638A.145T}.

One limitation of the cross-correlation method originates from the intrinsic width of the CCF: for a given RV separation, if the CCF is too wide, two components might blend and an SB$2$ will not be detected as such. This paper presents an innovative method to compute the cross-correlation functions of stellar spectra with carefully-designed spectral templates so that the resulting CCF is significantly narrower than those obtained by standard methods. The sample data, consisting in GIRAFFE HR$10$ and HR$21$ spectra (\Gaia-ESO observations + ESO archives), is presented in Sect.~\ref{Sect:data}. Section~\ref{Sec:Methods} {describes the construction and use of the new \nacre (Narrow cross-correlation experiment) masks, compares them to the \Gaia-ESO CCFs and estimates the bias and uncertainties affecting the velocities derived from the \nacre CCF}. Finally, the census of multiple stars, the comparison with previous \Gaia-ESO data releases and some physical properties of the uncovered binaries are discussed in Sect.~\ref{Sec:Results}.

\section{Data}
\label{Sect:data}
\subsection{The \Gaia-ESO survey}

The \Gaia-ESO Survey collaboration obtained spectra with the mid- and high-resolution multi-fibre spectrograph FLAMES \citep{2000SPIE.4008..129P}, mounted on the Nasmyth focus of the \SI{8}{\meter} Kueyen telescope (UT2) at the VLT/ESO, in Chile. The FLAMES \SI{25}{\arcmin} field of view offers \num{132} fibres (MEDUSA mode) feeding the mid-resolution spectrograph GIRAFFE and \num{8} fibres feeding the red arm of the high-resolution spectrograph UVES. Two UVES setups have been used, namely U520 $[\num{4200}, \num{6200}] \si{\angstrom}$, and U580 $[\num{4800},\num{6800}] \si{\angstrom}$, with a resolution $R \sim \num{47000}$. On the other hand, nine GIRAFFE setups\footnote{More precisely, the HR4 setup has been added for iDR6 and was not in iDR5.} have been used, namely HR3, HR4, HR5A, HR6, HR9B, HR10, HR14A, HR15N and HR21, with a resolution ranging from $R \sim \num{18000}$ for HR14A up to $R \sim \num{31500}$ for HR3 or HR9B\footnote{Most recent estimates for the GIRAFFE setup resolution available at \url{http://www.eso.org/sci/facilities/paranal/instruments/flames/inst/specs1.html}}. The various GIRAFFE setups were used for observations of different spectral types, either alone, \eg HR15N and HR9B, or in combination (HR3, HR4, HR5A, HR6, HR14A and HR10-HR21). A description of the use of the various setups can be found in \citet{randich22} and \citet{gilmore22}. 

Like other large surveys, the \Gaia-ESO Survey publishes its data through regular data releases. Each data release provides us with new spectra and a reprocessing of already released data such that old spectra benefit from improved data reduction pipelines. The present study uses the fifth internal data release (iDR5), which consists in all \Gaia-ESO observations between 01/01/2012 ($\mathrm{MJD} = \num{55927.01874}$) and 01/01/2016 ($\mathrm{MJD} = \num{57388.09855}$). A \Gaia-ESO data release also includes archival spectra from the ESO archive, {that have been extracted from the original images and analysed with the \Gaia-ESO pipelines, homogeneously with the new spectra.} The archival ESO spectra included in iDR5 are observations {obtained} with the same \Gaia-ESO setups ($\sim \SI{90}{\percent}$ of the archival spectra) as well as with other GIRAFFE/UVES setups not used by the \Gaia-ESO Survey. The archival data provide supplementary observations for $\sim \num{7000}$ targets but, among them, only $\sim \num{1000}$ targets have {(new)} \Gaia-ESO observations. In total, iDR5 provides spectra for $\sim \num{83000}$ Milky Way targets.

Recorded spectra are reduced by two different {teams} within \Gaia-ESO, one for GIRAFFE spectra \citep{gilmore22} and one for UVES spectra \citep{2014A&A...565A.113S}. These {teams} performed {standard} data processing tasks, among which: a) data reduction (dark correction, bias correction, flat-fielding, wavelength calibration, spectrum extraction) of all individual spectra; b) radial velocity determination; c) normalisation of each individual spectrum to produce what is called a {\emph{nightly} spectrum (or epoch spectrum)} in the \Gaia-ESO terminology; d) for a given setup, the spectrum co-addition of multi-epoch observations (after correction for the {heliocentric} velocity) to produce a final spectrum per object (and per setup), called {\emph{stacked} spectra} in the \Gaia-ESO terminology. Cross-correlation functions, corresponding to each {epoch spectrum}, are also released to the \Gaia-ESO analysis nodes, in addition to the \emph{stacked} and the \emph{nightly} spectra.

\subsection{The iDR5 sub-sample used in this study}
\label{Sect:iDR5subsample}

When considering all setups (\ie \Gaia-ESO and non-\Gaia-ESO setups), iDR5 consists in $\num{379093}$ spectra (among which $\sim\num{31600}$ are from the ESO archive), corresponding to $\num{82294}$ unique targets. If one excludes non-\Gaia-ESO setups from iDR5, the numbers are smaller: $\num{376122}$ \Gaia-ESO iDR5 spectra for $\num{82031}$ unique targets. Three GIRAFFE setups, HR10, HR21 and HR15N, produce together \SI{79}{\percent} of the \Gaia-ESO iDR5 spectra: \SI{25.7}{\percent} for HR10 ($\num{96648}$ out of \num{376122}), \SI{29}{\percent} for HR21 ($\num{109007}$) and \SI{24}{\percent} for HR15N ($\num{90420}$). The UVES setup U580 is the fourth most used setup (\SI{11}{\percent} of the \Gaia-ESO iDR5 spectra) and all other setups build up the remaining \SI{10}{\percent}. While HR15N is mostly dedicated to Milky Way open cluster stars, HR10 and HR21 are mostly dedicated to Milky Way field stars {but have also been used for clusters, the Bulge and Corot targets}. About $\num{42300}$ targets are observed with HR10 and $\sim \num{49700}$ with HR21; most of the targets ($\sim \num{42000}$) observed with HR10 have also been observed with HR21 with the exception of the targets observed in the bulge for which HR10 was not used. The GIRAFFE HR10 setup has a spectral resolution $R \sim \num{21500}$ and covers the spectral range $[\num{5334}, \num{5612}] \si{\angstrom}$ while the GIRAFFE HR21 setup has a spectral resolution $R \sim \num{18000}$ and covers the spectral range $[\num{8475}, \num{8983}] \si{\angstrom}$. The present study analyses {only} the Milky Way field stars observed with the HR10 and HR21 \Gaia-ESO setups, thus {\num{160727} spectra, corresponding to \num{37565} unique Milky Way field targets. This selection is motivated by the following points: a) as explained in Sect.~\ref{Sec:narrower_HR21_CCFs}, we are primarily interested in improving the SB$n$ detection rate in HR21 observations; b) most of the selected targets have at least two spectra obtained with a different setup, offering two independent checks of stellar multiplicity per object and so, HR10 can play the role of control-sample; c) as a first step, we test our workflow on field stars to avoid the potential additional issue of crowding in cluster fields that could lead to a higher number of false SB2 detections.}

\section{Methods}
\label{Sec:Methods}

\subsection{\doe}

In \cite{2017A&A...608A..95M}, the semi-automated tool Detection of Extrema (\doe) was applied to the fourth internal \Gaia-ESO data release (iDR4) in order to identify spectroscopic binaries. The main characteristics of \doe are briefly summarised below; more details can be found in \cite{2017A&A...608A..95M} and in particular, Fig.~7 of \cite{2017A&A...608A..95M} illustrates the result provided by \doe on the simulated spectrum of a binary system. \doe takes as input the cross-correlation function (CCF) of a given spectrum and returns the number and position of the detected CCF peaks, \ie the number and the radial velocity of the stellar components forming the system.

To this end, \doe computes the first, second and third derivatives of the CCF by convolving it with, respectively, the first, second and third derivative of a narrow Gaussian kernel. This technique avoids the numerical noise that may appear when one does discrete differentiation. From the successive derivatives, \doe searches for local CCF maxima and for inflexion points, in order to disentangle the spectral components. The CCF is then fitted by a multi-Gaussian model around the local maxima and/or inflexion points in order to obtain the velocity of the identified spectral components. In \cite{2017A&A...608A..95M}, we explained how we have adjusted three user-defined parameters to allow an efficient detection of the CCF components and to automatically obtain results matching those of a visual inspection of the CCFs.

We note that several physical phenomena, distinct from binarity, can produce multiple peaks in a CCF. For example, the Schwarzschild scenario explains double-peak CCFs due to shock wave propagation in Mira star atmospheres \citep[\eg][]{2000A&A...362..655A}. Cepheids can also display CCFs with multiple peaks \citep{2016MNRAS.463.1707A}. {However,} such stars are rare within the \Gaia-ESO Survey.

\subsection{Getting narrower HR21 CCFs}
\label{Sec:narrower_HR21_CCFs}

\begin{figure*}
  \centering
  \includegraphics[width=0.85\columnwidth]{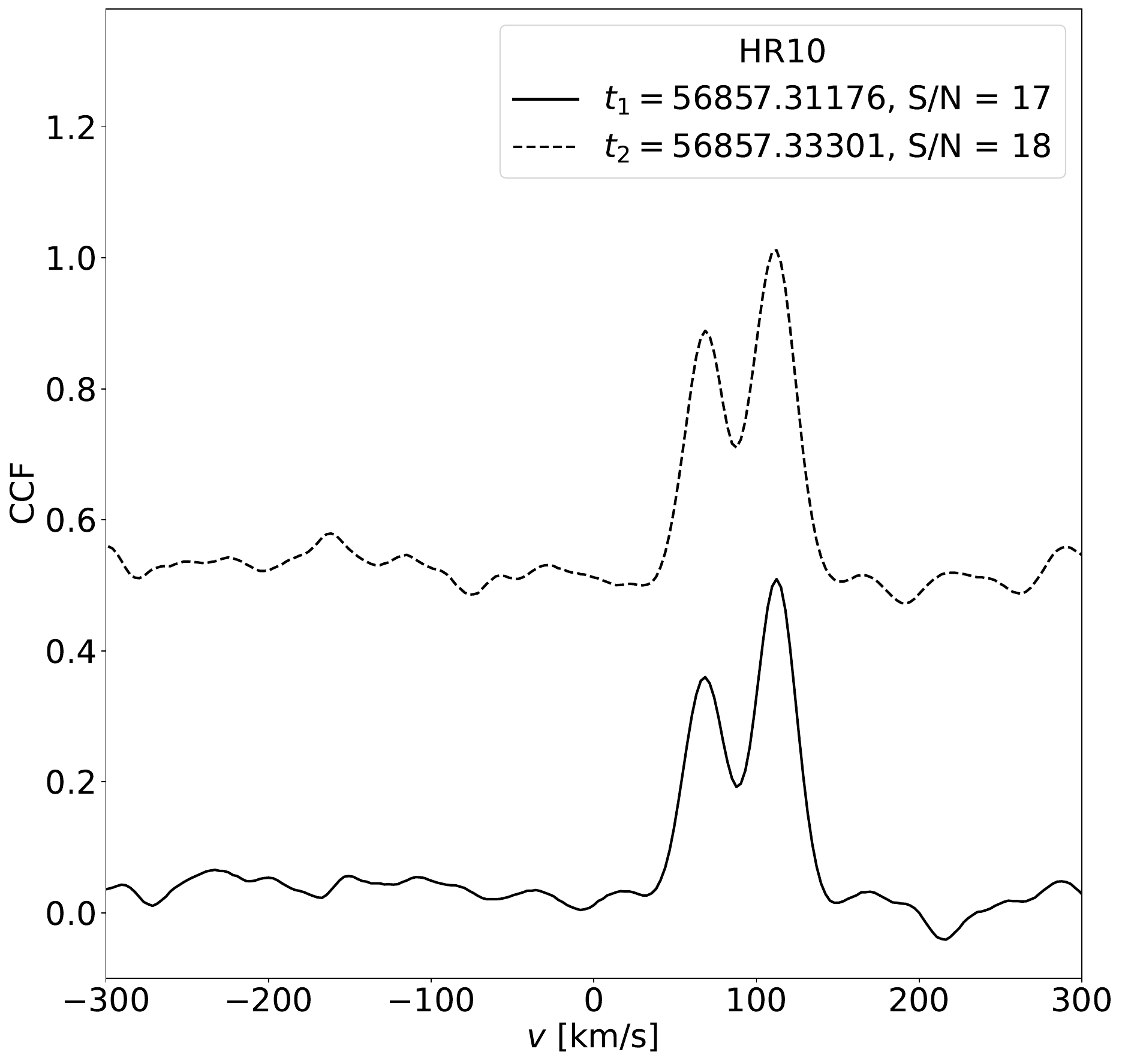}
  \includegraphics[width=0.85\columnwidth]{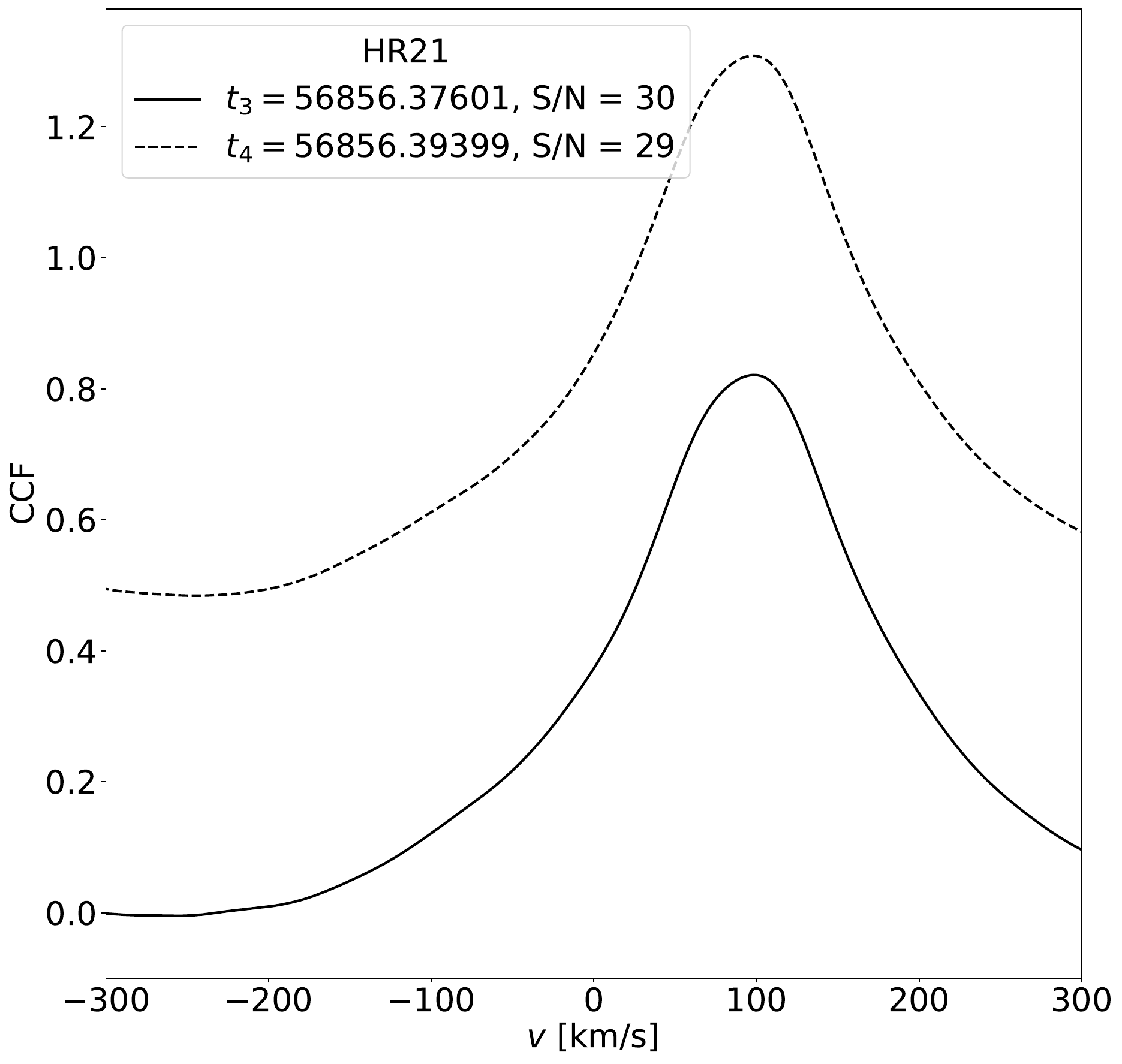}
  \caption{\label{Fig:HR10_vs_HR21_CASU_CCFs} Examples of HR10 (left) and HR21 (right) \Gaia-ESO CCFs for the star CNAME 23354061-4305405. Two consecutive observations have been made with each setup. The HR10 epochs $t_1 < t_2$ and the HR21 epochs $t_3 < t_4$ are separated by less than one day. The $\snr$ of each spectrum used to compute the \Gaia-ESO CCF is indicated as labelled. While the SB2 {nature} of the system is evident from the HR10 \Gaia-ESO CCFs, it is not the case from the HR21 CCFs. CCFs are shifted vertically by 0.5 for the sake of readability.}
\end{figure*}

When applying \doe to the iDR4 sample in \cite{2017A&A...608A..95M}, we realised that the efficiency of the SB$n \ge 2$ detection was somewhat {setup}-dependent as illustrated in Fig.~\ref{Fig:HR10_vs_HR21_CASU_CCFs}. Two pairs of observations of an SB2 are presented, namely two HR10 observations obtained during the same night (left panel) and two HR21 observations obtained the following night (right panel). The four spectra used to compute the CCFs all have $\snr > 15$. The puzzling fact is that two CCF peaks are clearly visible only in the HR10 observations and not in any of the HR21 observations while the two pairs of spectra are taken within \SI{24}{\hour}. Of course, this could be due to a rapid change of the radial velocities or to the slightly lower resolution of the HR21 settings (HR10: $R \sim \num{21500}$ vs. HR21: $R \sim \num{18000}$). However, this alone does not explain the boxy and wide peak with a full-width at half-maximum (FWHM) of about \SI{100}{\kilo\metre\per\second} of the HR21 CCFs. Actually, the HR21 setup hosts different spectral features {than HR10, some of which have broad wings (\eg \ion{Ca}{II} triplet lines) which can broaden the CCF and hide a double peak.} We {also} suspected the HR21 cross-correlating masks to be less adapted to the HR21 setting. Such investigations have led us to set up a procedure to {recompute CCFs} by using a set of optimised cross-correlating masks.

\paragraph{Choice of the template stars.}

We selected a dozen template stars among the \Gaia FGK benchmarks \citep{2015A&A...582A..81J}. These template stars were chosen such that they are representative of the stars observed in HR10 and HR21. Indeed, the stellar parameter space sampled by the HR10 and HR21 targets can be estimated from the recommended stellar parameters provided for iDR5: \SI{90}{\percent} of the \Gaia-ESO targets with a recommended temperature, surface gravity and metallicity span the ranges $[\num{4200}, \num{6700}]$~\si{\kelvin}, $[\num{2.2}, \num{4.7}]$ and $[\num{-1.1}, \num{0.3}]$ {dex} respectively. Table~\ref{Tab:Model_stars} lists our final choice of template stars and their stellar parameters. The stellar parameters of our template stars adequately match the \Gaia-ESO parameter ranges given above. The main selection criteria are the effective temperature and the surface gravity, such that our template sample contains {six main-sequence stars and six evolved stars with various metallicities:} hot (\SI{8000}{\kelvin}) and cold (\SI{4300}{\kelvin}) dwarfs ($\log g \approx 4$) and hot (\SI{5000}{\kelvin}) and cold (\SI{4000}{\kelvin}) giants ($\log g \approx 2$).

\begin{table*}
  \centering
  \caption{\label{Tab:Model_stars} List of template stars and their stellar parameters (taken from \citealt{2015A&A...582A..81J}). Columns are: effective temperature $T_{\mathrm{eff}}$; surface gravity $\log g$; metallicity $\abratio{Fe}{H}$; microturbulence velocity $\xi_{\mathrm{micro}}$; the last two columns indicate whether we used {that star to build an HR10 mask or an HR21 mask.}}
  \begin{tabular}{lS[table-format=5.0]S[table-format=4.2]S[table-format=5.2]S[table-format=4.2]cc}
    \toprule
    Star            & {$T_{\mathrm{eff}}$} & {$\log g$} & {$\abratio{Fe}{H}$} & {$\xi_{\mathrm{micro}}$} & HR10 & HR21\\
                    & {K}                  &            &                     & {\SI{}{\kilo\metre\per\second}} & & \\
    \midrule
    \midrule
    $\alpha$ Tau    &       3927    &       1.11    &       -0.37   &       1.63    & no  & yes\\
    $\beta$ Ara     &       4173    &       1.04    &       -0.05   &       1.88    & no  & yes\\
    Arcturus        &       4286    &       1.64    &       -0.52   &       1.58    & yes & yes\\
    $\mu$ Leo       &       4474    &       2.51    &        0.25   &       1.28    & no  & yes\\
    HD 122563        &       4587    &       1.61    &       -2.64   &       1.92    & yes & yes\\
    $\xi$ Hya       &       5044    &       2.87    &        0.16   &       1.40    & yes & yes\\
    \midrule
    61 Cyg A        &       4374    &       4.63    &       -0.33   &       1.07    & no  & yes\\
    $\alpha$ Cen A  &       5792    &       4.30    &        0.26   &       1.20    & yes & yes\\
    Sun             &       5777    &       4.44    &        0.03   &       1.06    & yes & yes\\
    HD 22879         &       5868    &       4.27    &       -0.86   &       1.05    & yes & yes\\
    Procyon         &       6554    &       3.99    &        0.01   &       1.66    & yes & yes\\
    Hot Dwarf 01\tablefootmark{a}      &       8000    &       4.50    &        0.00   &       1.50    & yes & yes\\
    \bottomrule
  \end{tabular}
  \tablefoot{
    \tablefoottext{a}{{This template star is not forged upon an existing star like the other template stars. It simply allows us to probe the late A dwarf stars.}}
  }
\end{table*}

\paragraph{Cross-correlating masks.}

\begin{figure*}
  \centering
  \includegraphics[width=\textwidth]{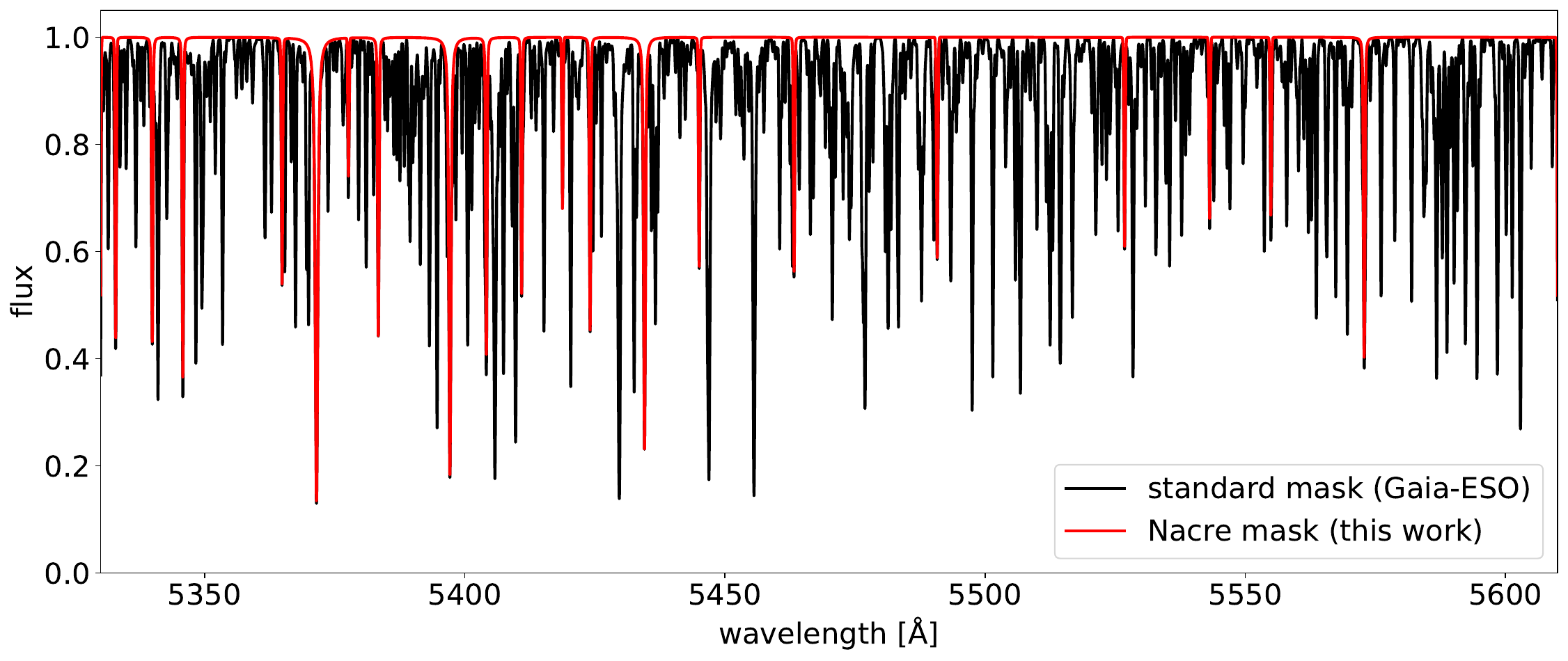}
  
  \includegraphics[width=\textwidth]{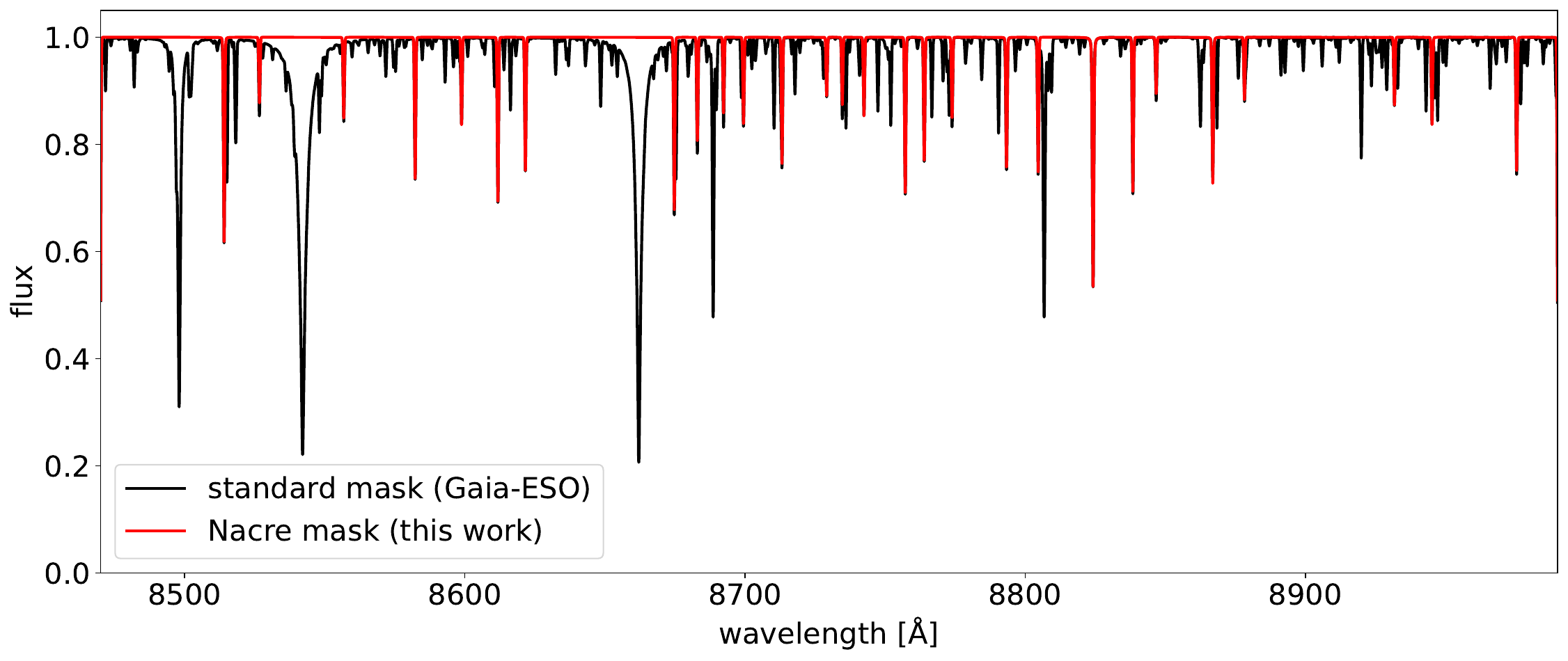}
  \caption{\label{Fig:HR10_HR21_std_vs_special_masks} Comparison for HR10 (top) and HR21 (bottom) of our cross-correlating mask (\nacre mask; red line) to a standard one (black line; \eg the one that would be used to compute the \Gaia-ESO CCFs) computed from a spectral synthesis with Arcturus stellar parameters. Our \nacre mask includes only weak and isolated lines. Despite our strong constraints, more than $20$ lines are kept in both HR10 and HR21 masks for Arcturus.}
\end{figure*}

One standard way to compute the cross-correlation function for a given observed spectrum is to convolve this observed spectrum with a template spectrum (\eg a synthetic spectrum or a reference observed spectrum). Usually, one tries to use a template spectrum that matches the target spectrum in order to avoid {spurious features in the CCF caused by} spectral mismatch and in order to get a very contrasted cross-correlation function (\ie its maximum or minimum, depending on how the cross-correlation function is computed, is significantly different from the typical level of the correlation noise). However, in order to obtain narrower HR21 cross-correlation functions, we noticed that a template spectrum built with only carefully selected absorption lines was a better choice, especially to ignore the very wide \ion{Ca}{II} triplet lines in the computation. Our criteria to keep a line for a given template star were as follows: a) the line should not be blended; b) the line should not be saturated; c) the normalised flux at the line centre should be higher than $\sim 0.3$. {To ease the selection of lines, we designed an automatic tool (Van der Swaelmen, private comm.) to scan atomic lines in a synthetic spectrum and quantify their blendedness and depths.} We limited our study to atomic lines of light, $\alpha$- and iron-peak elements: Na, Mg, Al, Si, Ca, Sc, Ti, V, Cr, Fe, Co. This choice of elements warrants a sufficient number of lines (\ie $> 10$) within the spectral range covered by both HR10 and HR21 setups. For a given template star of Table~\ref{Tab:Model_stars}, our automated line selection procedure is as follows: a) for a given wavelength range and for a given tested element, we compute two spectral syntheses for the star under consideration: one spectral synthesis $S_{\mathrm{total}}$ that includes all atomic and molecular transitions and one spectral synthesis $S_{\mathrm{partial}}$ that includes all atomic and molecular transitions except those for the targeted element; b) for each transition of the targeted element, we compute a blending score by comparing $S_{\mathrm{total}}$ and $S_{\mathrm{partial}}$ around the line under study; c) if the line is not blended or is weakly blended, then we check that the line is not too saturated by computing its equivalent width and we check that the intensity at the line centre is above our depth criterion. Figure~\ref{Fig:HR10_HR21_std_vs_special_masks} compares a standard cross-correlating mask (\ie a simple spectral synthesis {with all its absorption lines}, in black) to our fine-tuned cross-correlating mask (in red) when Arcturus is adopted as a benchmark and for both GIRAFFE setups. If, after applying our line selection criteria, it was not possible to keep at least ten lines for a given setup and for a given benchmark star, we discarded this star for the considered setup, {explaining the absence of HR10 masks for $\alpha$~Tau, $\beta$~Ara, $\mu$~Leo and 61~Cyg~A}. The last two columns of Table~\ref{Tab:Model_stars} list the template stars used for HR10 and HR21. In the remainder of the paper, these new masks are called \nacre masks and the resulting CCFs will be called \nacre CCFs. On the other hand, the CCFs provided by the \Gaia-ESO consortium will be designated \Gaia-ESO CCFs. {The library of synthetic spectra used for the selection of atomic lines, and the final cross-correlating masks were computed with the radiative-transfer code \emph{turbospectrum} \citep{2012ascl.soft05004P}, the grid of MARCS model atmospheres \citep{2008A&A...486..951G}, the VALD database of atomic transitions \citep{2015PhyS...90e4005R} and the molecular linelists maintained by B.~Plez\footnote{\url{https://nextcloud.lupm.in2p3.fr/s/r8pXijD39YLzw5T}}.}

\paragraph{Masking of spectral features in target spectra.}

\begin{figure*}
  \centering
  \includegraphics[width=\textwidth]{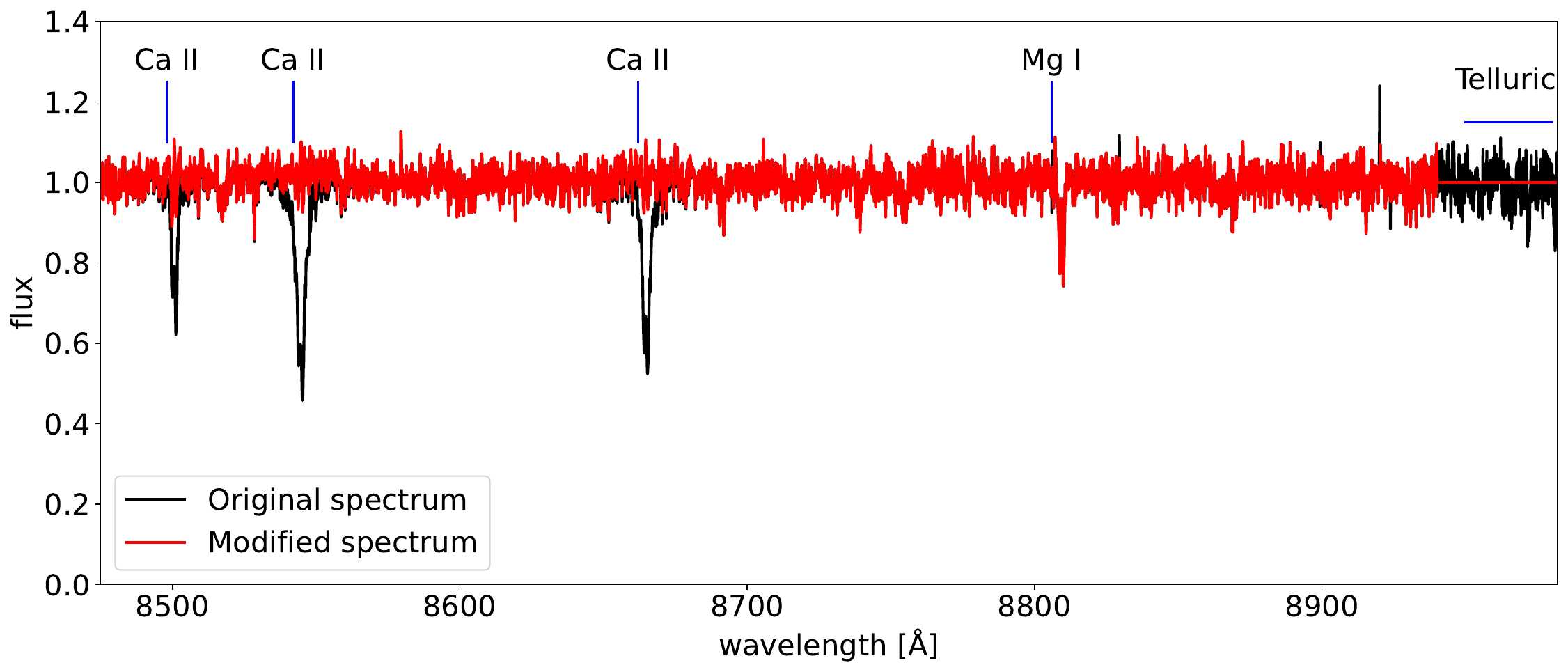}
  \caption{\label{Fig:HR21_spectral_feature_masking} Example of spectral feature removal and masking in a HR21 spectrum. The black line stands for the original HR21 spectrum of CNAME 23354061-4305405 ($\mathrm{MJD} = \num{56856.376009}$; median $\snr \approx 30$); the red line stands for the modified version of the spectrum, with the \ion{Ca}{II} triplet cancelled out and the telluric bands masked while the \ion{Mg}{I} line at \SI{8806}{\angstrom} is left untouched {because the criteria were not met (see text)}. There is a small shift between the vertical marks and the position of the absorption lines since the spectrum is not corrected for the stellar radial velocity (the spectrum is in the heliocentric frame); {our algorithm performs a small exploration around the laboratory wavelength to find the observed position of the calcium and magnesium lines to be removed}.}
\end{figure*}

By construction, our HR21 masks do not encompass the near-infrared \ion{Ca}{II} triplet at \SI{8498}{\angstrom}, \SI{8542}{\angstrom}, and \SI{8662}{\angstrom}, nor the H lines from the Paschen spectral series, nor the \ion{Mg}{I} line at \SI{8806}{\angstrom}. However, those lines {are generally} present in the target spectra and may contribute to broaden the HR21 CCF peak and/or create strong correlation noise for large velocity lags (with respect to the target star velocity). A special processing of the HR21 spectra was thus applied in order to mask such strong lines. To this end, we designed an automated procedure to locate the \ion{Ca}{II} triplet and the \ion{Mg}{I} lines in the HR21 spectrum, fit them with a Lorentzian profile and, if {the fit converges} and the (normalised) flux at the line centre is below an arbitrary threshold (fixed to \num{0.8}), we subtracted the fit from the observed spectrum, which resulted in an effective masking of these strong \ion{Ca}{II} and {\ion{Mg}{I} lines. Most of the Paschen lines, when they exist, are close to the \ion{Ca}{II} lines and we assume they are removed at the same time as the \ion{Ca}{II} lines are removed.} 

Telluric bands are also present at the red end of the HR21 spectral range. Since these telluric bands tend to decrease the contrast of the \nacre CCFs, we decided to mask the last $\approx \SI{40}{\angstrom}$ of the HR21 spectra {(we replace the fluxes by \num{1} since we compute the CCF for $1 - f(\lambda)$ where $f$ is the observed spectrum or the mask)}. The red spectrum in Fig.~\ref{Fig:HR21_spectral_feature_masking} is the result of the described masking procedure applied to one observation of the object with CNAME\footnote{‘CNAME' is a \Gaia-ESO survey specific stellar ID, formed by a concatenation of the right ascension and the declination of the target.} 23354061-4305405 while the black spectrum is the original observation. A spectral feature likely to be removed is indicated by a blue line above the spectrum. In the displayed example, the Lorentzian fit of the {\ion{Mg}{I}} line at \SI{8806}{\angstrom} failed {(line not deep enough with respect to our decision criterion)} and the spectral chunk is left untouched {at these wavelengths}. On the other hand, the \ion{Ca}{II} triplet is successfully subtracted and, over this range, the modified spectrum is similar to the spectrum in non-modified regions, \ie the subtraction procedure did not produce any spurious signal. {This exposure has a median \snr of \num{30}, which means that weak lines are still present in the spectrum after cancelling out the calcium lines.}

\paragraph{Analysis of \nacre CCFs.}
\label{Sec:CCF_analysis}

\begin{figure*}
  \centering
  \includegraphics[width=0.85\columnwidth]{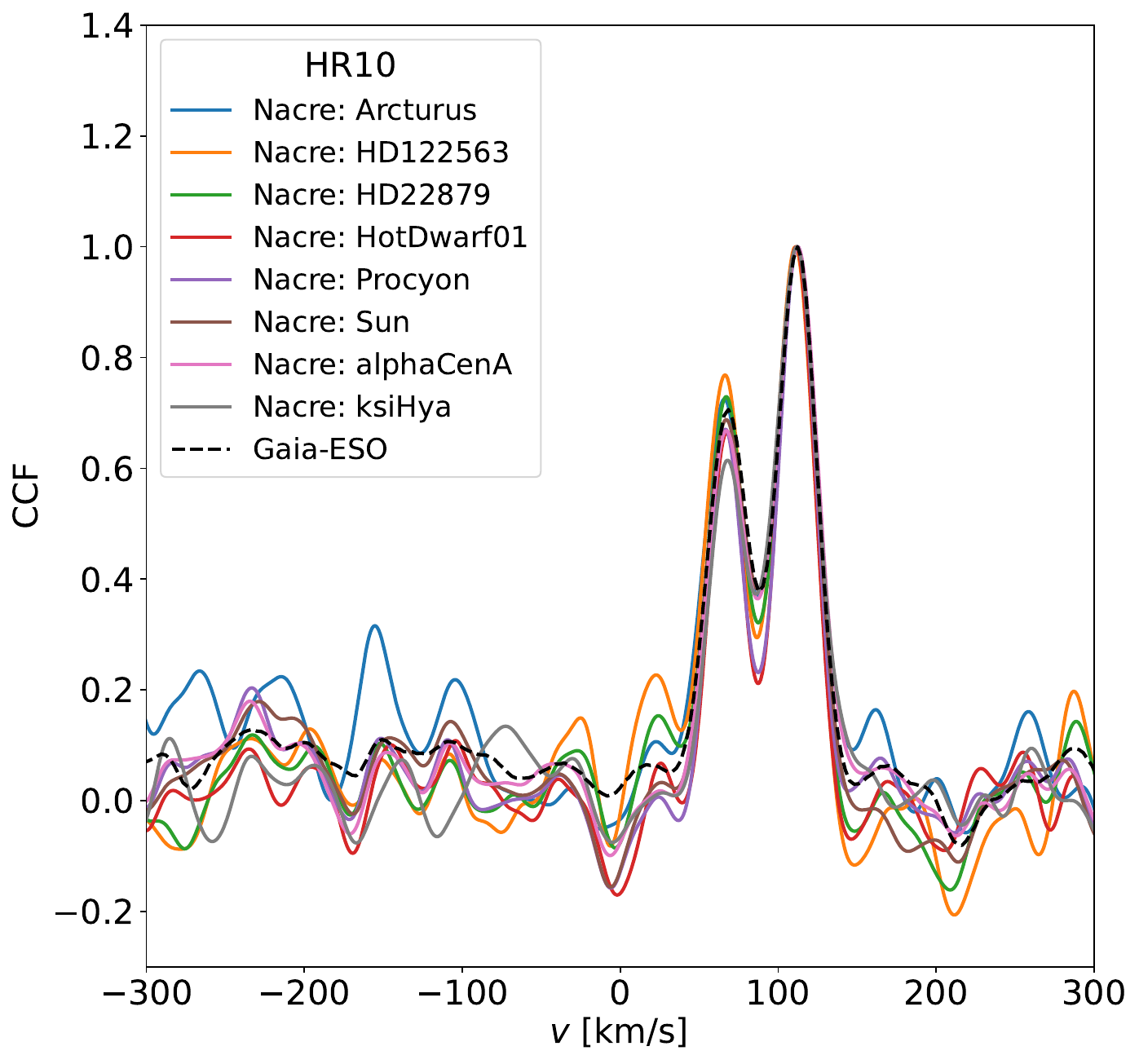}
  \includegraphics[width=0.85\columnwidth]{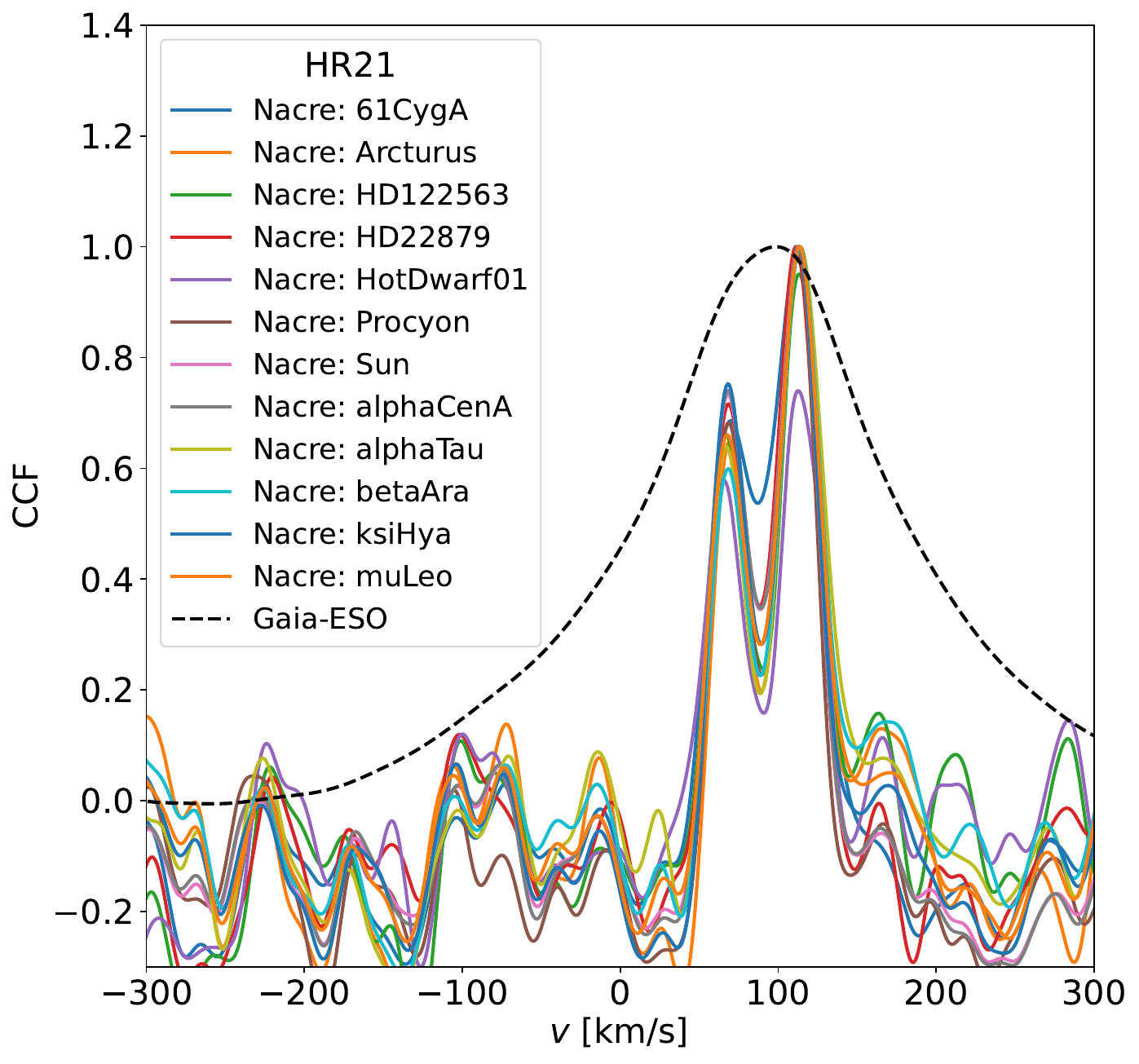}
  \caption{\label{Fig:GES_vs_Nacre_CCFs} Comparison of \Gaia-ESO (black dashed line) CCFs and \nacre (other coloured solid lines) CCFs for a given HR10 (left; $\mathrm{MJD} = \num{56857.311765}$) and the corresponding HR21 (right; $\mathrm{MJD} = \num{56856.376009}$) observation of CNAME 23354061-4305405. The mask names are provided in the legend boxes. For each observation, all \nacre CCFs agree very well: they have the same shape, they exhibit the same number of stellar components at nearly identical radial velocities. While the HR21 \Gaia-ESO CCF exhibits only one stellar component, our CCFs reveal the presence of two stars, in agreement with the HR10 observation. The CCFs are all normalised by their respective maximum for the sake of readability.}
\end{figure*}

With the \nacre masks and the cleaned observed spectrum at hand, the next step is to compute all CCFs for a given observation (obtained with a given setup) by using all our \nacre masks (prepared for a given setup). Figure~\ref{Fig:GES_vs_Nacre_CCFs} shows all HR10 and HR21 CCFs computed for the star with CNAME 23354061-4305405 and also displays the \Gaia-ESO CCF in black. 

One sees that despite the relatively small number of lines included in our masks, the HR10 \nacre CCFs overlap with the HR10 \Gaia-ESO CCF. On the other hand, {as sought for}, the HR21 \nacre CCFs have a very different shape from that of the \Gaia-ESO CCF: they are always narrower. For the {model-case} of the target with CNAME 23354061-4305405, the HR10 and HR21 \nacre CCFs now carry the \emph{same information} concerning the binary nature of the object.

At the end of the computation step, a dozen \nacre CCFs are available for each observation. In order to keep only one \nacre CCF per observation, we rank the series of CCFs by a quantity that estimates their contrast, \ie the {difference between} the maximum intensity of the CCF to the {typical intensity} of the cross-correlation noise measured in the side (or feet) of the CCF peak. The \nacre CCF with the highest grade is called the \emph{best \nacre CCF} and will serve as a reference during the analysis. The \nacre CCFs whose grade do not meet some conditions (adjusted empirically once and for all) are discarded since they are considered to be dominated by the correlation noise. Such cases become more frequent when the \snr of the spectrum decreases. The reader should note that all the CCFs displayed in this article have been normalised so that the highest CCF component always peaks at 1.

For each observation, \doe is run on the best \nacre CCF {as well as on the other grade-selected \nacre CCFs} and it returns {the radial velocity of each spectral component found in the different CCFs}, obtained by fitting the upper part of relevant CCF peaks by a multi-Gaussian function. For each CCF, we use an estimate of the correlation noise (measured in the left and right {feet} of the CCF peak) to adjust the intensity threshold needed by \doe {(one of its user-defined parameter)} to disentangle local maxima associated to a possible stellar component from local maxima due to the correlation noise. As for the results obtained for a given observation, different situations may happen, that we classified as follows:

\begin{itemize}
\item all the masks agree, \ie \doe finds the same number $n$ of stellar components in the different CCFs and they are at (nearly) identical velocities. The spectrum under study thus hosts $n$ stellar components and their respective velocities {are} read from the best-mask \nacre CCF (\ie with the highest rank);
\item the masks partially disagree, \ie $n$ stellar components are found in $i$ masks while $m$ stellar components are found in $j$ masks and the $n$ (respectively, $m$) radial velocities returned by the $i$ (respectively, $j$) masks agree within the error bars. Then, we read the results from the CCFs characterised by the largest number of components if it also includes our best \nacre CCF (if not, then the observation is flagged for visual check);
\item the masks strongly disagree, \ie it is not possible to reduce the results in only two classes of solutions as in the previous situation. Then, we are not able to make an automated decision and {the observation is flagged for visual check}.
\end{itemize}

This reasoning holds because, as explained in the first paragraph of this section, our \Gaia-ESO star sample spans a relatively small range of effective temperatures (FGK stars), surface gravities and metallicities, and therefore, the spectral mismatch between an observation at one end of the temperature (respectively, gravity, metallicity) range and a \nacre mask at the other end will not change drastically the shape of the \nacre CCF. In other words, when we are not working in the low \snr regime ($\snr < 5-10$), if a stellar component is at a given velocity in one \nacre CCF, it will also appear at a close velocity in another \nacre CCF. The small changes from one mask to the next in measured velocities can be explained by various factors such as the \snr of the spectrum, the set of lines used in the cross-correlating mask, the spectral mismatch, etc. and will be investigated in the next section. Figure~\ref{Fig:GES_vs_Nacre_CCFs} illustrates this fact for one HR10 and one HR21 observation. One sees that all our CCFs tend to {have a similar shape}: for both HR10 and HR21 \nacre CCFs, they exhibit two stellar components, located at similar radial velocities.

\subsection{Quality control}

\begin{figure*}
  \centering
  \includegraphics[width=0.85\columnwidth]{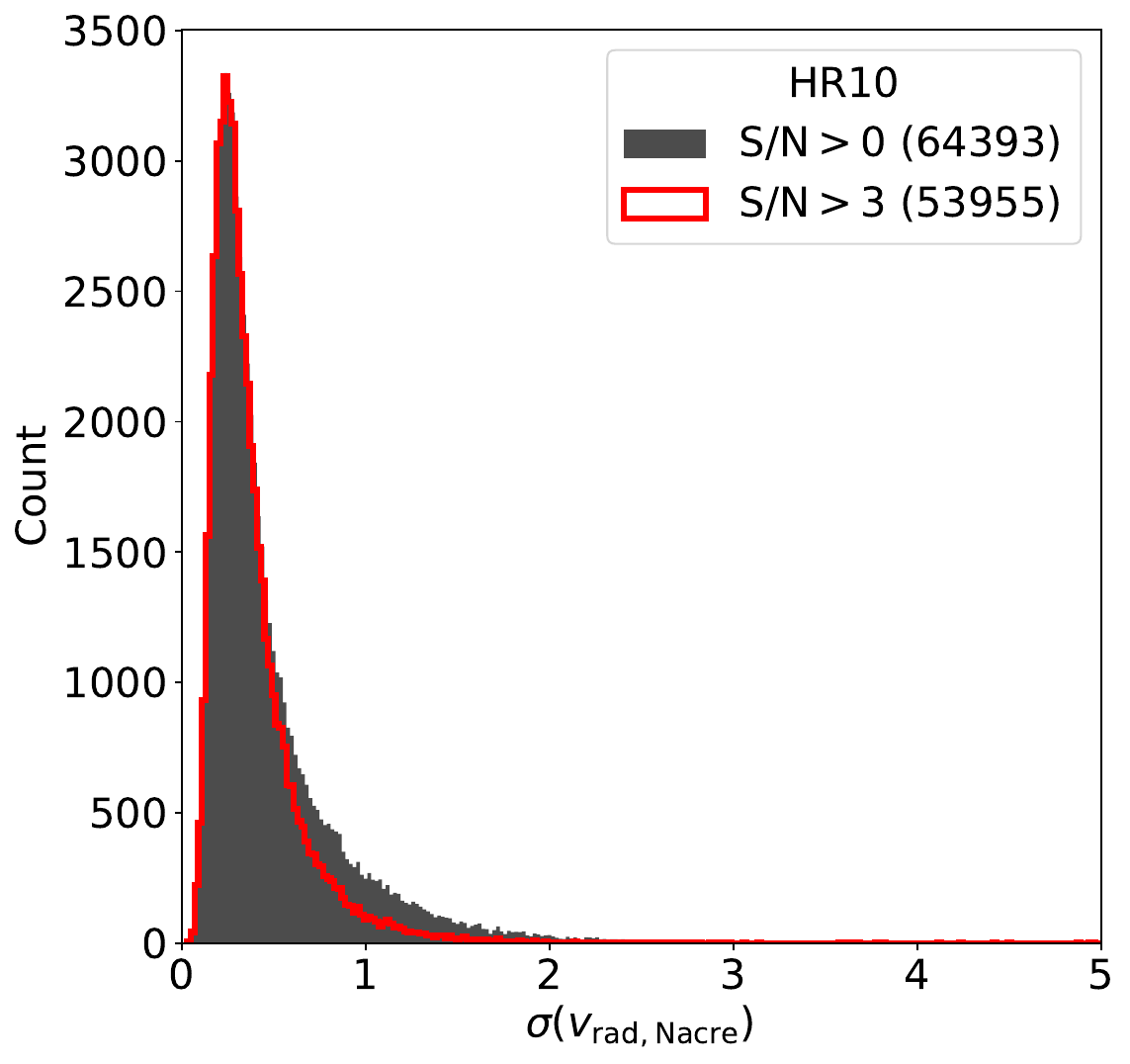}
  \includegraphics[width=0.85\columnwidth]{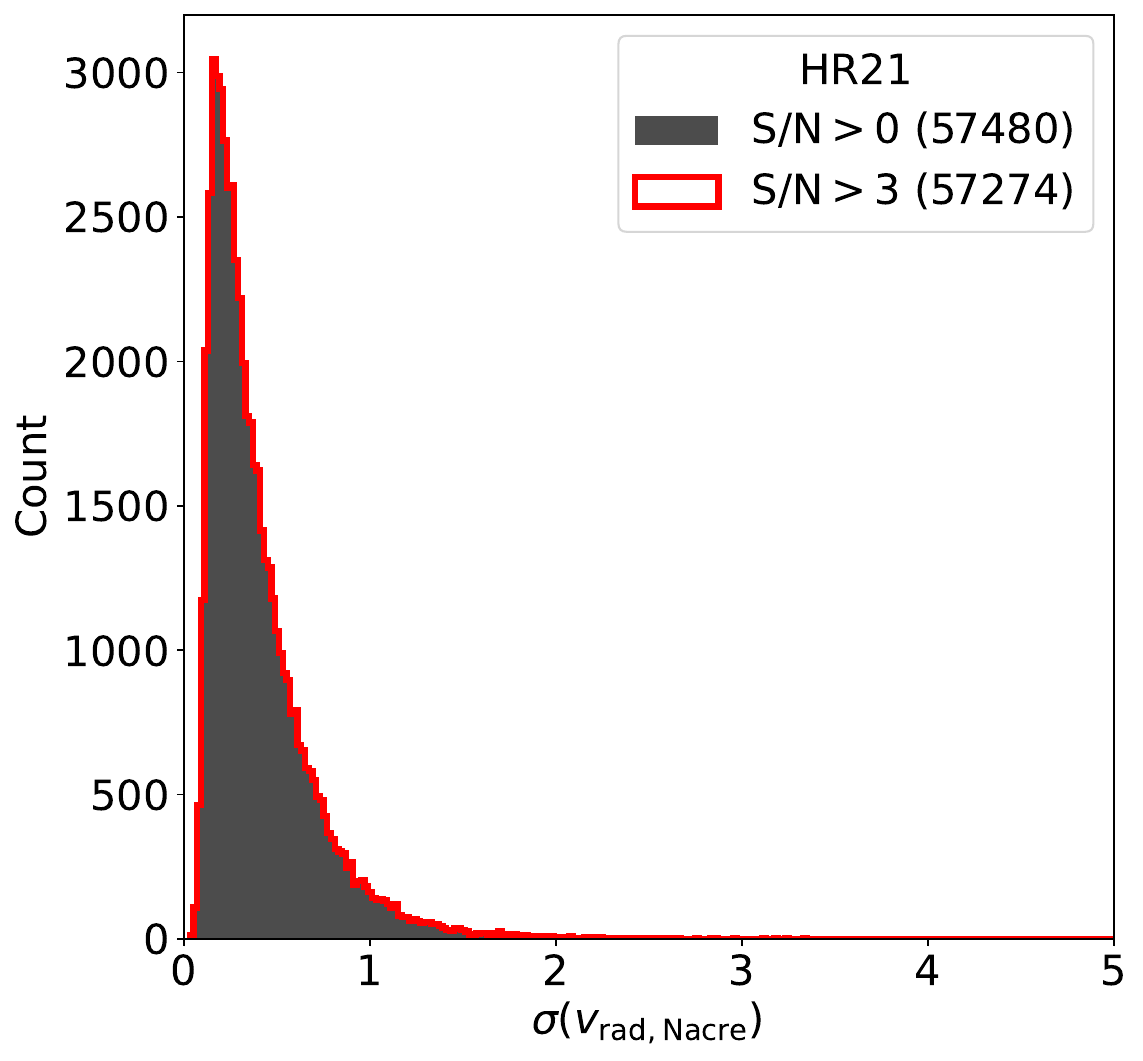}
  \caption{\label{Fig:iDR5_sigmaV_distribution} {Distribution of $\sigma[v_{\mathrm{rad,Nacre}}]$ with (red line) or without (filled black) a cut on the $\snr$ (binwidth = \SI{0.016}{\kilo\metre\per\second}). Left: HR10 setup; $\langle \sigma[v_{\mathrm{rad,Nacre}}] \rangle = \SI{0.39}{\kilo\metre\per\second}$ with $\snr \ge 3$. Right: HR21 setup; $\langle \sigma[v_{\mathrm{rad,Nacre}}] \rangle = \SI{0.40}{\kilo\metre\per\second}$ with $\snr \ge 3$.} With or without the $\snr$ cut, most of our sample is characterised by a standard deviation on the radial velocities measured from the different \nacre CCFs below \SI{1}{\kilo\metre\per\second}; the $\snr$ cut depopulates the right tail of the HR10 distribution.}
\end{figure*}

The nice agreement between the radial velocities provided by all \nacre CCFs is illustrated in Fig.~\ref{Fig:iDR5_sigmaV_distribution} plotting the distribution of the dispersion $\sigma[v_{\mathrm{rad,Nacre}}]$ of the \nacre radial velocities $v_{\mathrm{rad,Nacre}}$. In this figure, we consider only a) individual observations where \doe detects a single stellar component in the \nacre CCFs and b) spectra for which the \snr reported by the \Gaia-ESO exists and is either $> 0$ (in black) or $> 3$ ({in red}). {We note that this is an 'observation-wise' distribution, \ie it is based on individual observations and a given star may count more than once.} While both distributions peak at \SI{0.2}{\kilo\metre\per\second} for HR10 observations, the black distribution exhibits a more extended tail than the {red} distribution. The fact that this extended tail disappears {at higher} \snr clearly shows that the \snr is the main {responsible} of the velocity dispersion from one mask to the other. The {red} distribution shows that a $\snr$ as low as \num{3} is acceptable for our purposes since, in that case, most of the distribution is still below \SI{1}{\kilo\metre\per\second} and the largest dispersion is around \SI{3}{\kilo\metre\per\second}. On the other hand, for HR21, the cut on the $\snr$ has {no} effect on the shape distribution. This can be understood because a) HR21 spectra have on average a higher $\snr$ than HR10 spectra and b) the HR10 spectral range for FGK stars tends to contain more weak metallic lines than the HR21 spectral range. Therefore, at low $\snr$ ($< 3$), in HR10 spectra, the scatter in the flux due to the photon noise has a higher probability (compared to the HR21 case) to mimic a spectral feature that will cause a slight shift in the position of the CCF peak. Like for HR10, the HR21 {red} distribution peaks at \SI{0.2}{\kilo\metre\per\second} and most of the distribution is below \SI{1}{\kilo\metre\per\second}. We conclude that the velocities returned by the different masks for a specific observation are most of the time equivalent.

\begin{figure*}
  \centering
  \includegraphics[width=0.85\columnwidth]{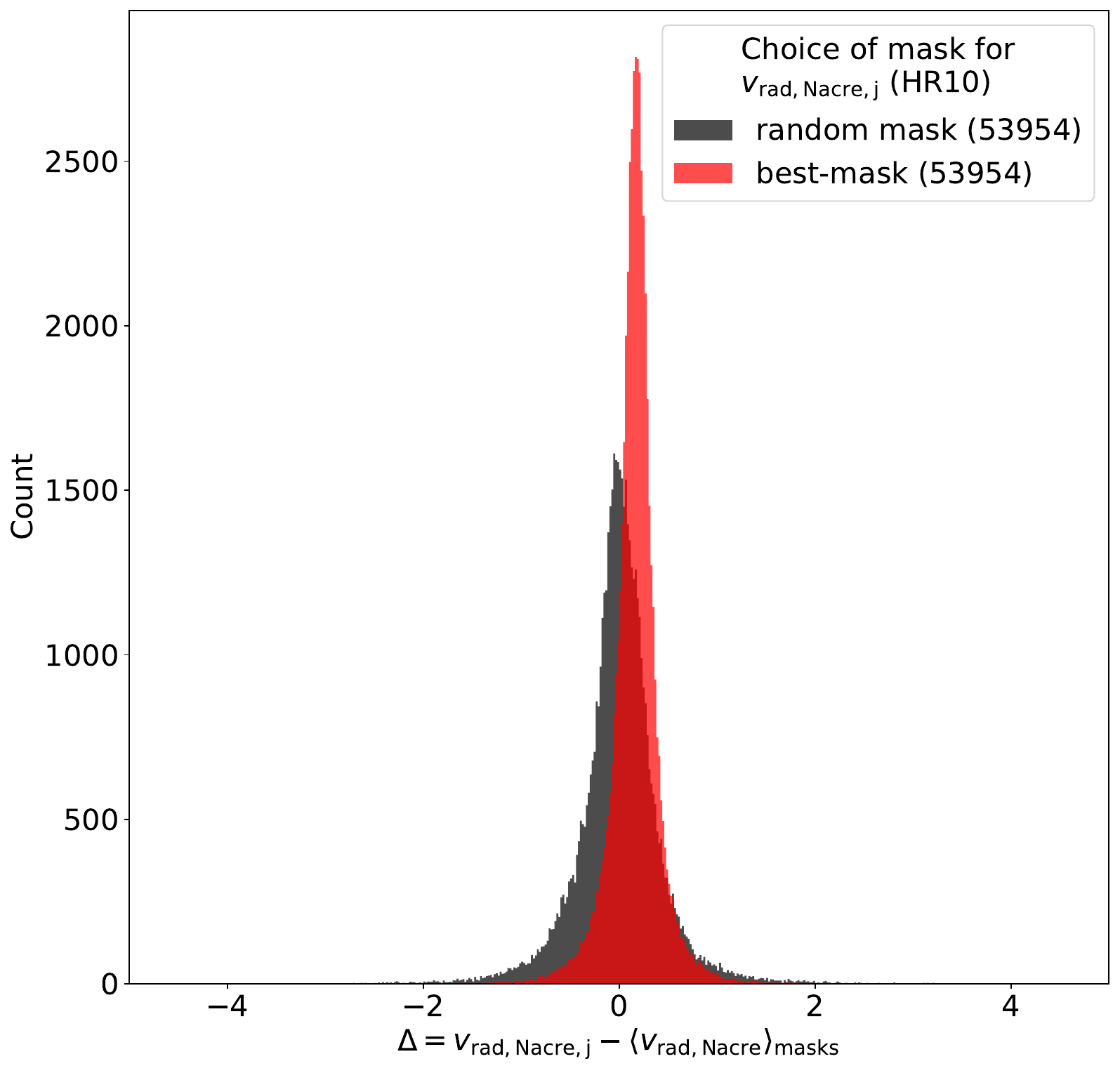}
  \includegraphics[width=0.85\columnwidth]{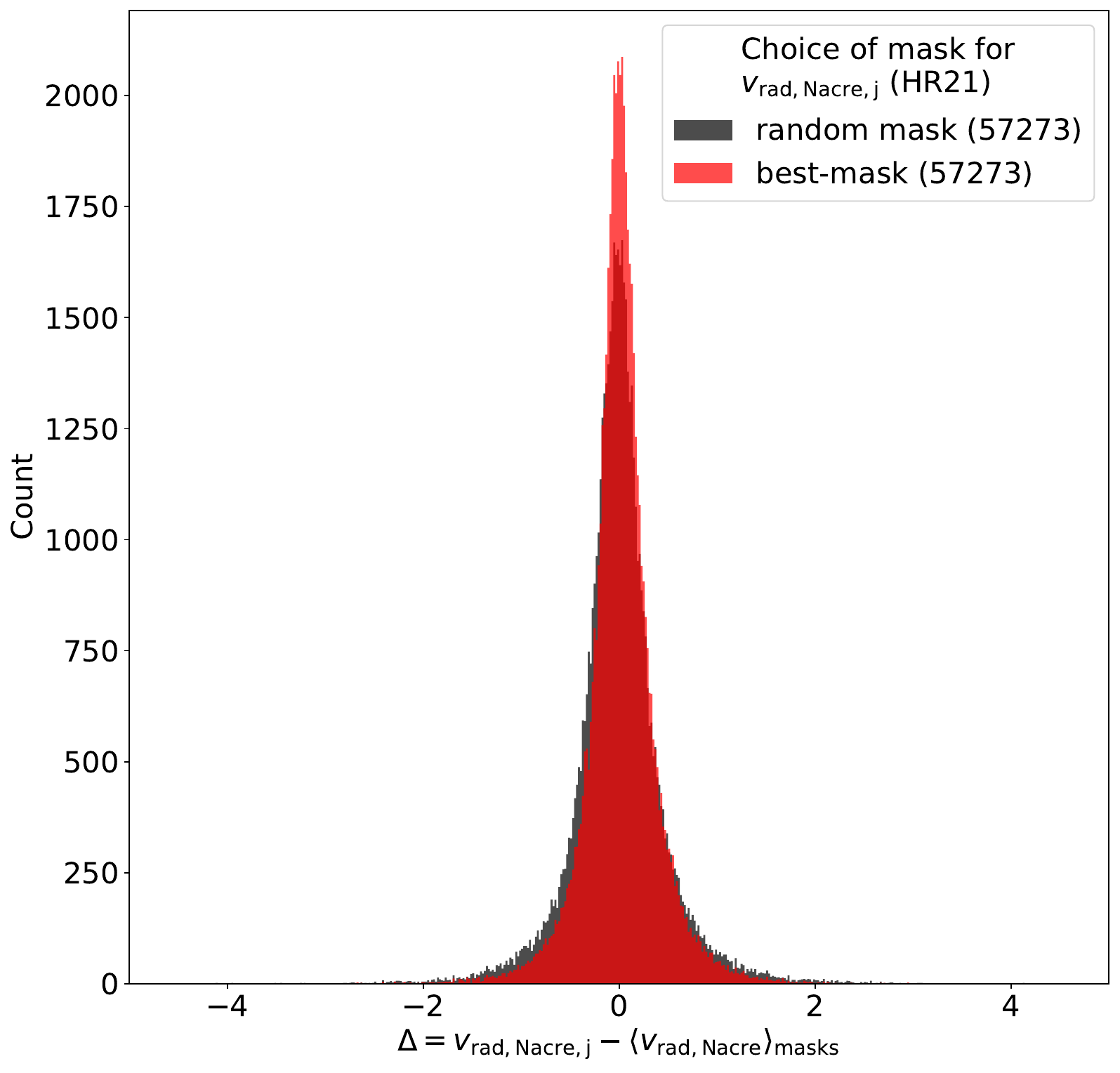}
  \caption{\label{Fig:iDR5_random_vs_best_Nacre_velocity} {Distribution of $\Delta = v_{\mathrm{rad,Nacre,ref}} - \langle v_{\mathrm{rad,Nacre}} \rangle_{\mathrm{masks}}$ (binwidth = \SI{0.02}{\kilo\metre\per\second}) where $v_{\mathrm{rad,Nacre,j}}$ is either the velocity $v_{\mathrm{rad,Nacre,best}}$ returned by the best-mask \nacre CCF (see text; red) or a randomly chosen \nacre velocity $v_{\mathrm{rad,Nacre,rand}}$ (black). Left: HR10 setup. Right: HR21. The mean, standard deviation and interquartile range of each distribution, expressed in \si{\kilo\metre\per\second}, are: $(0.17, 0.59, 0.22)$ for HR10 and 'best mask'; $(0.01, 0.77, 0.39)$ for HR10 and 'random mask'; $(0.02, 0.38, 0.33)$ for HR21 and 'best mask'; $(0., 0.50, 0.41)$ for HR21 and 'random mask'. The number of objects in a given selection is indicated in the legend box.} The distributions are narrower for $v_{\mathrm{rad,Nacre,best}}$ than for $v_{\mathrm{rad,Nacre,rand}}$.}
\end{figure*}

Figure~\ref{Fig:iDR5_random_vs_best_Nacre_velocity} shows the distribution (observation-wise) of the difference $v_{\mathrm{rad,Nacre,j}} - \langle v_{\mathrm{rad,Nacre}}\rangle_{\mathrm{masks}}$, where $v_{\mathrm{rad,Nacre,j}}$ is either the velocity $v_{\mathrm{rad,Nacre,best}}$ derived from the best-mask \nacre CCF (red histogram), or a velocity $v_{\mathrm{rad,Nacre,rand}}$ randomly picked up among the series of CCF velocities (black histogram). $\langle v_{\mathrm{rad,Nacre}} \rangle_{\mathrm{masks}}$ is the mean velocity over {all of the selected} \nacre CCFs. The black distribution is centred around \SI{0}{\kilo\metre\per\second} while the {red} distribution exhibits a small offset (\SI{0.2}{\kilo\metre\per\second}) for HR10 and shows no offset for HR21. However, the {red} distribution is narrower (standard deviations: \SI{0.59}{\kilo\metre\per\second} for HR10; \SI{0.41}{\kilo\metre\per\second} for HR21) than the black one (standard deviations: \SI{0.69}{\kilo\metre\per\second} for HR10; \SI{1.5}{\kilo\metre\per\second} for HR21), indicating that our ranking criterion for the CCFs is more efficient than a random selection among the measured velocities.

\begin{figure*}
  \centering
  \includegraphics[width=0.85\columnwidth]{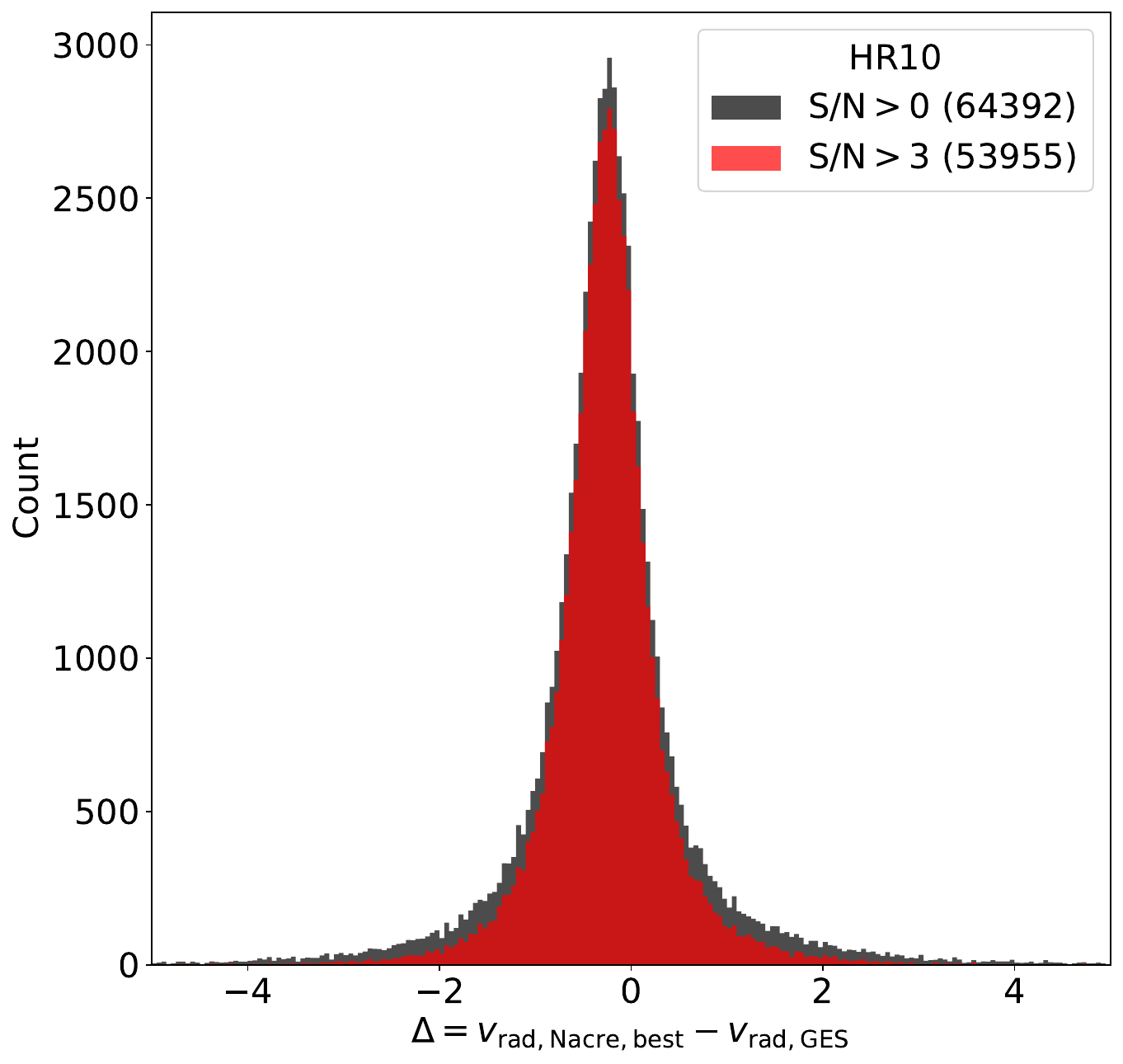}
  \includegraphics[width=0.85\columnwidth]{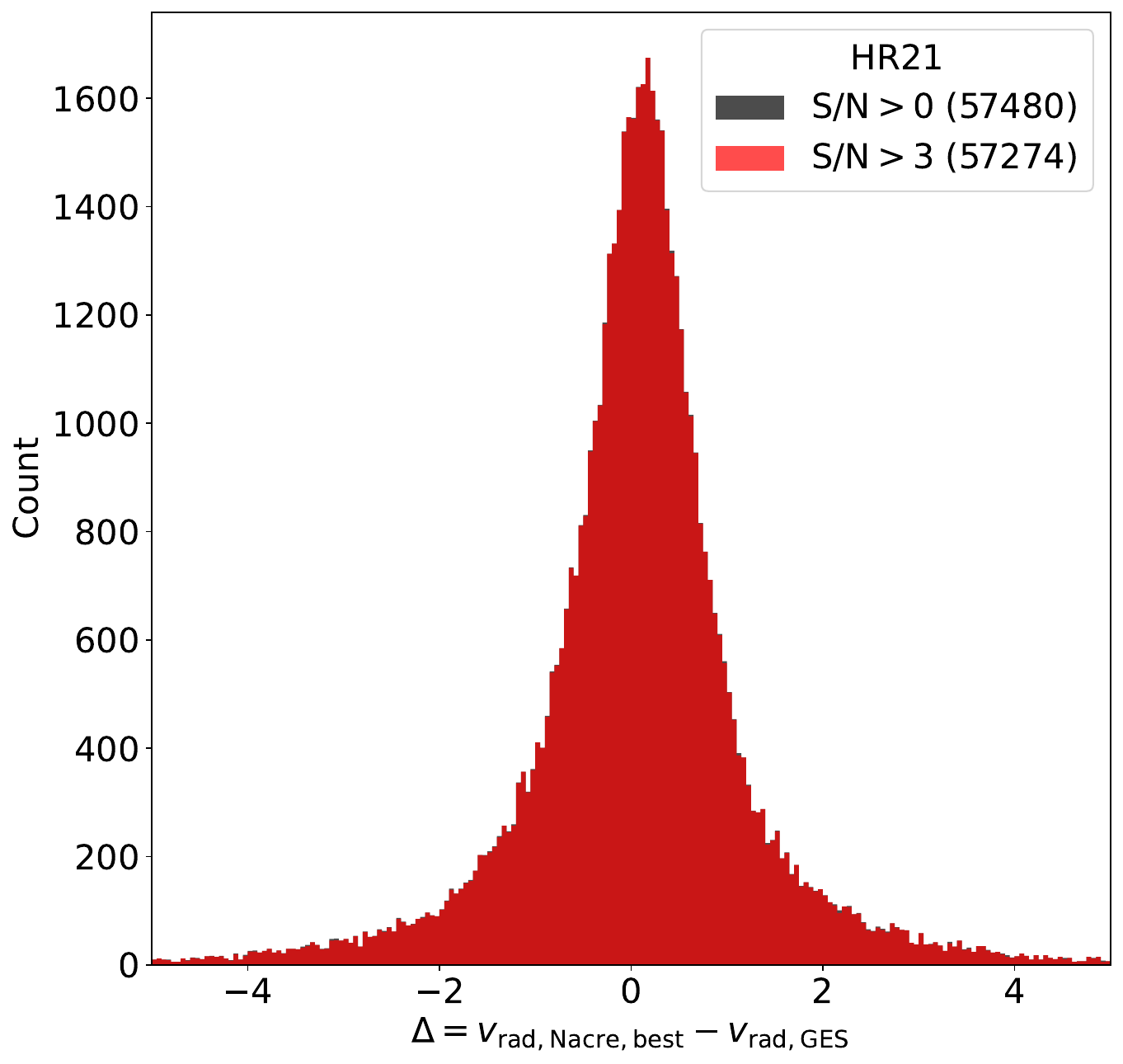}
  \caption{\label{Fig:iDR5_Nacre_vs_GES_velocity} Distribution of $\Delta = v_{\mathrm{rad,Nacre,best}} - v_{\mathrm{rad,GES}}$ {for HR10 (left) and HR21 (right) observations where $v_{\mathrm{rad,GES}}$ is the recommended \Gaia-ESO radial velocity. Black: $\snr \ge 0$; red: $\snr \ge 3$. The mean, standard deviation and interquartile range of each distribution, expressed in \si{\kilo\metre\per\second}, are: $(-0.24, 1.47, 0.67)$ for HR10 with $\snr > 0$; $(-0.25, 1.23, 0.56)$ for HR10 with $\snr > 3$; $(0.07, 3.77, 1.05)$ for HR21 with $\snr > 0$; $(0.07, 3.77, 1.03)$ for HR21 with $\snr > 3$. The number of objects in a given selection is indicated in the legend box. For HR21 (right), the two histograms are almost identical and the black one is not visible.}}
\end{figure*}

In order to check the reliability of {the \nacre radial velocities}, we also compared them to {the \Gaia-ESO recommended radial velocities $v_{\mathrm{rad,GES}}$. Figure~\ref{Fig:iDR5_Nacre_vs_GES_velocity} displays the distribution (observation-wise) of the differences $v_{\mathrm{rad,Nacre,best}} - v_{\mathrm{rad,GES}}$. The mean difference between $v_{\mathrm{rad,Nacre,best}}$ and $v_{\mathrm{rad,GES}}$ is \SI{-0.27}{\kilo\metre\per\second} for HR10, while it amounts to \SI{0.07}{\kilo\metre\per\second} for HR21. These numbers demonstrate that the agreement of our velocity scale to the two \Gaia-ESO velocity scales is rather good: only HR10 \nacre radial velocities are affected by a small offset that is not corrected for in the Table~\ref{Tab:iDR5_SB2_details}.}

\begin{figure*}
  \centering
  \includegraphics[width=0.85\columnwidth]{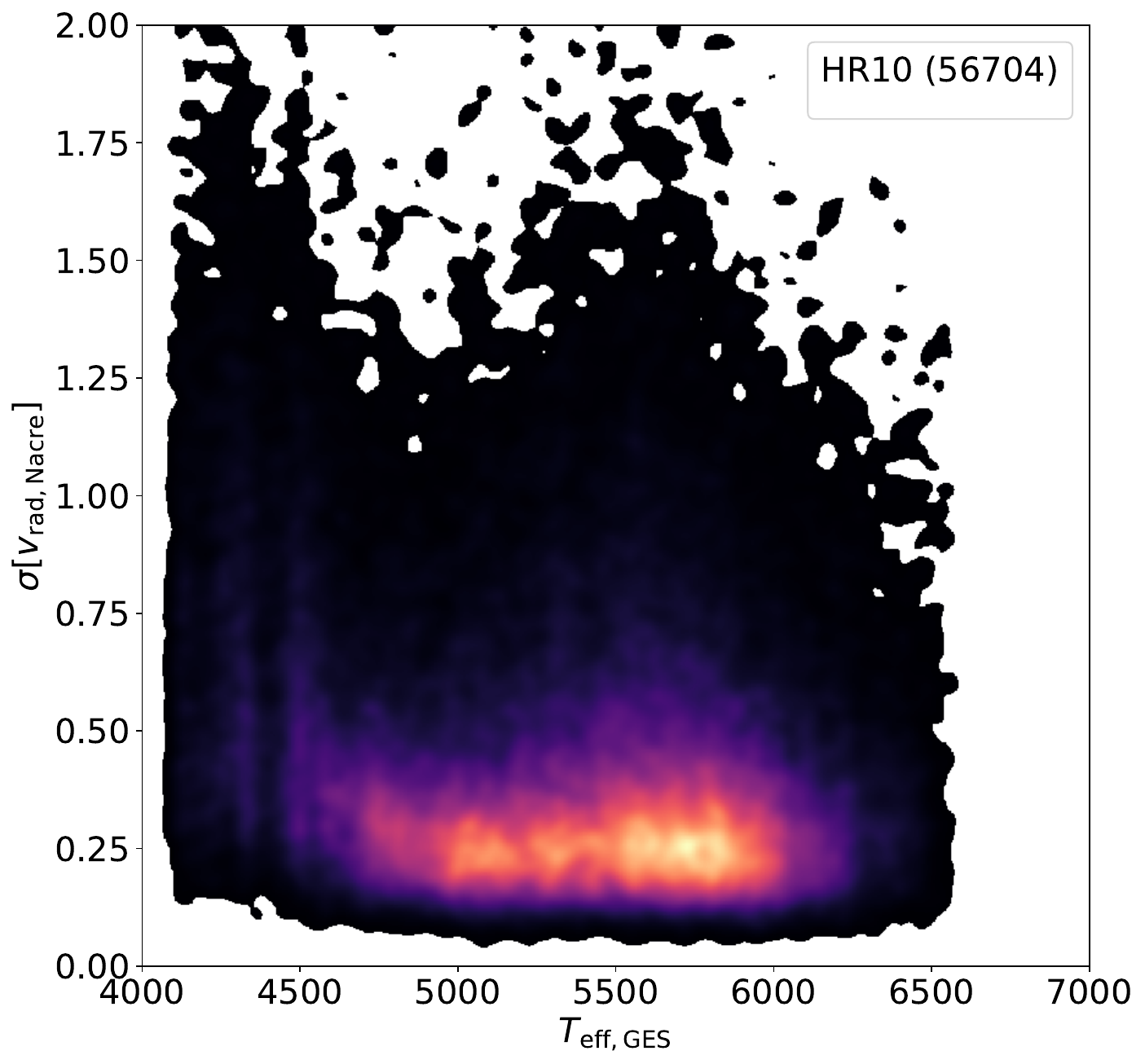}
  \includegraphics[width=0.85\columnwidth]{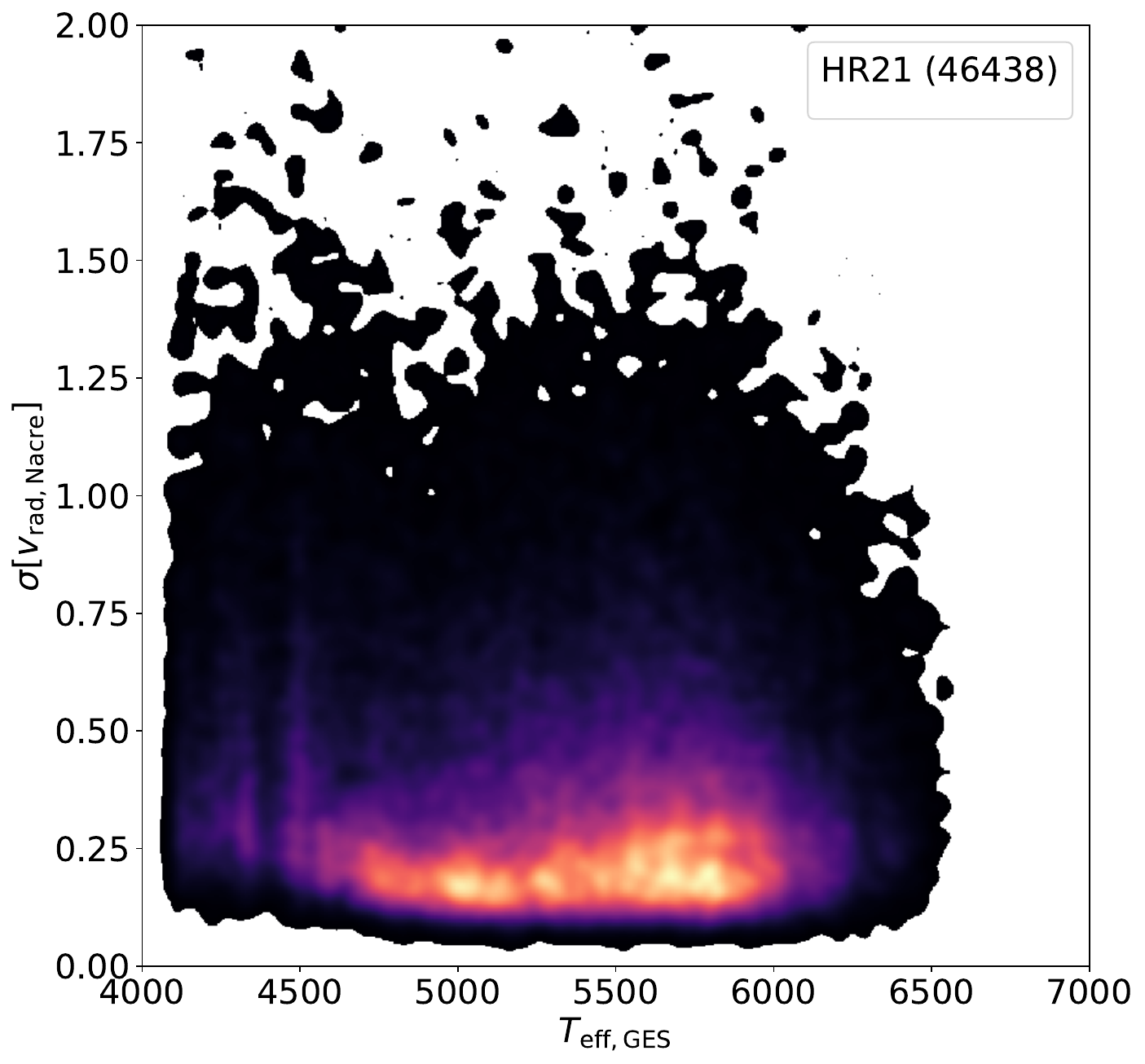}
  
  \includegraphics[width=0.85\columnwidth]{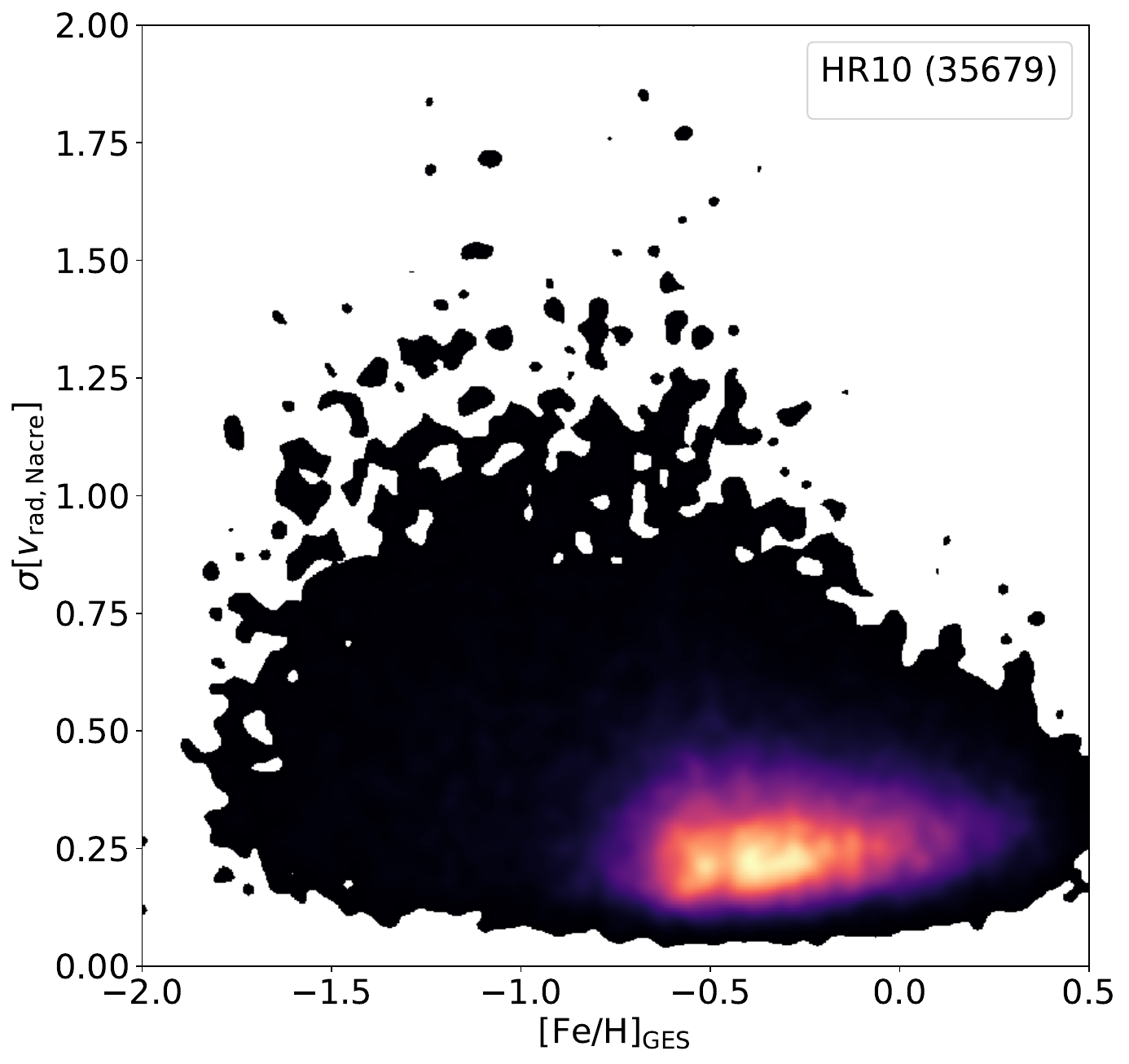}
  \includegraphics[width=0.85\columnwidth]{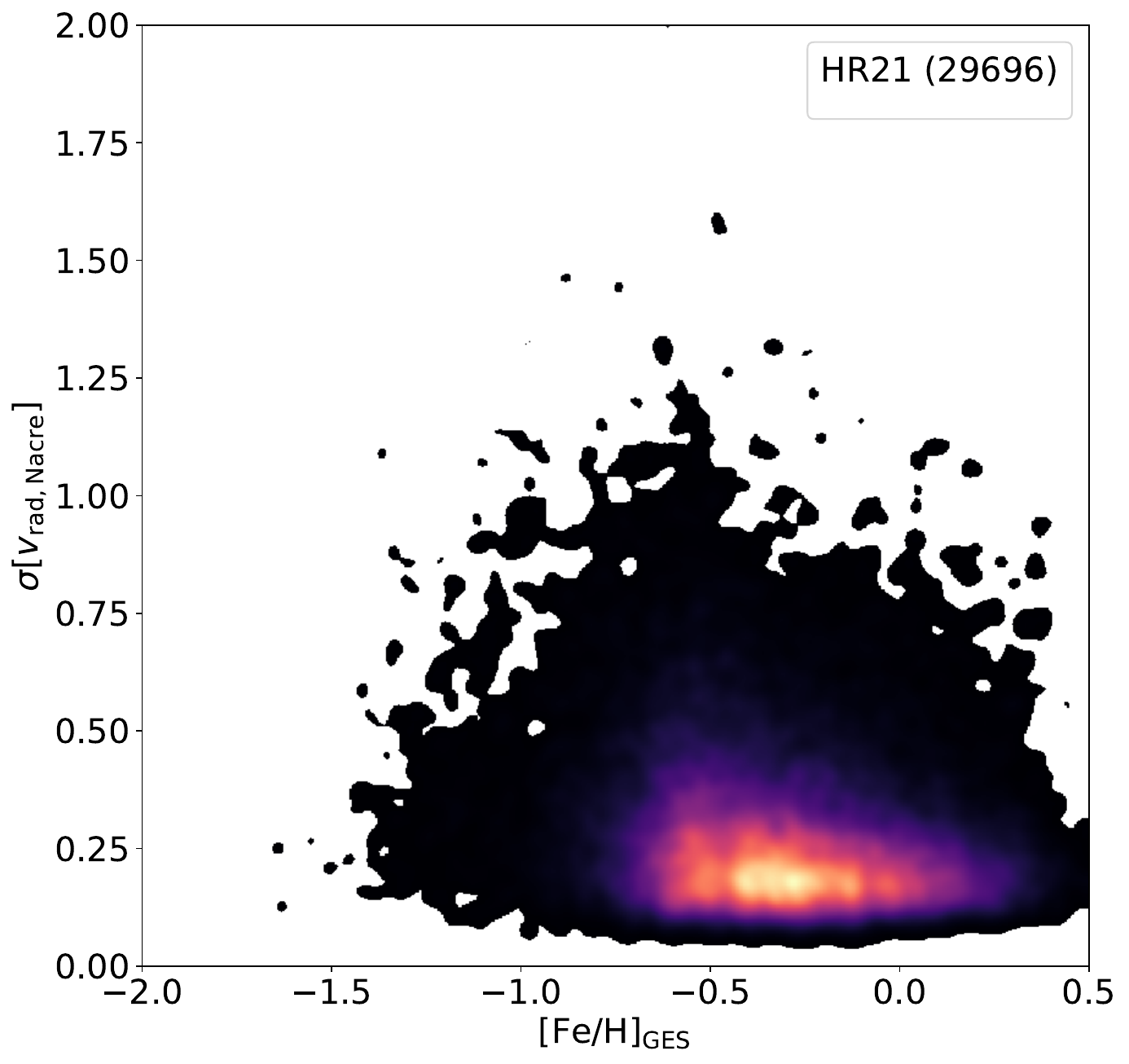}
  \caption{\label{Fig:Density_map_sigma_vs_GES_Teff_FeH} {Density map of $\sigma[v_{\mathrm{rad,Nacre}}]$ as a function of $T_{\mathrm{eff,GES}}$ (top) and $\abratio{Fe}{H}_{\mathrm{GES}}$ (bottom) for HR10 (left) and HR21 (right) observations. $T_{\mathrm{eff,GES}}$ and $\abratio{Fe}{H}_{\mathrm{GES}}$ are respectively the \Gaia-ESO recommended temperature and Fe abundance. The number of objects in a given selection is indicated in the legend box. White areas denote zones of the plane where there is no data.}}
\end{figure*}

{Figure~\ref{Fig:Density_map_sigma_vs_GES_Teff_FeH} shows the density map of $\sigma[v_{\mathrm{rad,Nacre}}]$ as a function of $T_{\mathrm{eff,GES}}$ (recommended \Gaia-ESO effective temperature) and $\abratio{Fe}{H}_{\mathrm{GES}}$ (recommended \Gaia-ESO Fe abundance). We note that $\sigma[v_{\mathrm{rad,Nacre}}]$ increases toward cold effective temperature for both HR10 and HR21; the trend $\sigma[v_{\mathrm{rad,Nacre}}]$ vs. $T_{\mathrm{eff,GES}}$ tend to be flat for temperatures hotter than \SI{5000}{\kelvin} but the scatter around this flat trend becomes larger above \SI{5500}{\kelvin}. $\sigma[v_{\mathrm{rad,Nacre}}]$ increases for $\abratio{Fe}{H}_{\mathrm{GES}} \le -1$ for both setups, and for HR10 only, $\sigma[v_{\mathrm{rad,Nacre}}]$ increases for $\abratio{Fe}{H}_{\mathrm{GES}} \ge 0$. The cross-correlation requires the presence of spectral features to work: the precision of this technique thus decreases for metal-poor and/or hot objects.}

\subsection{Error budget}

Once a local maximum of the CCF is associated to a stellar component, \doe performs a local fit of the CCF (with the help of the Python numpy's function \emph{numpy.optimize.curve\_fit}) by a Gaussian (or by a multi-Gaussian {function} in case of multiple stellar components) in order to retrieve the position of this maximum, \ie the radial velocity of the stellar component. The fitting function returns the parameters defining the Gaussian as well as the associated covariance matrix. The diagonal element of the covariance matrix corresponding to the velocity $r_{vv}$ directly provides the $1\sigma$ uncertainty $\mathrm{e}_{\mathrm{fit}} = \sqrt{r_{vv}}$ on the radial velocity due to the fitting procedure. A {typical value} of this source of uncertainty is \SI{0.01}{\kilo\metre\per\second}.

\begin{figure*}
  \centering
  \includegraphics[width=0.85\columnwidth]{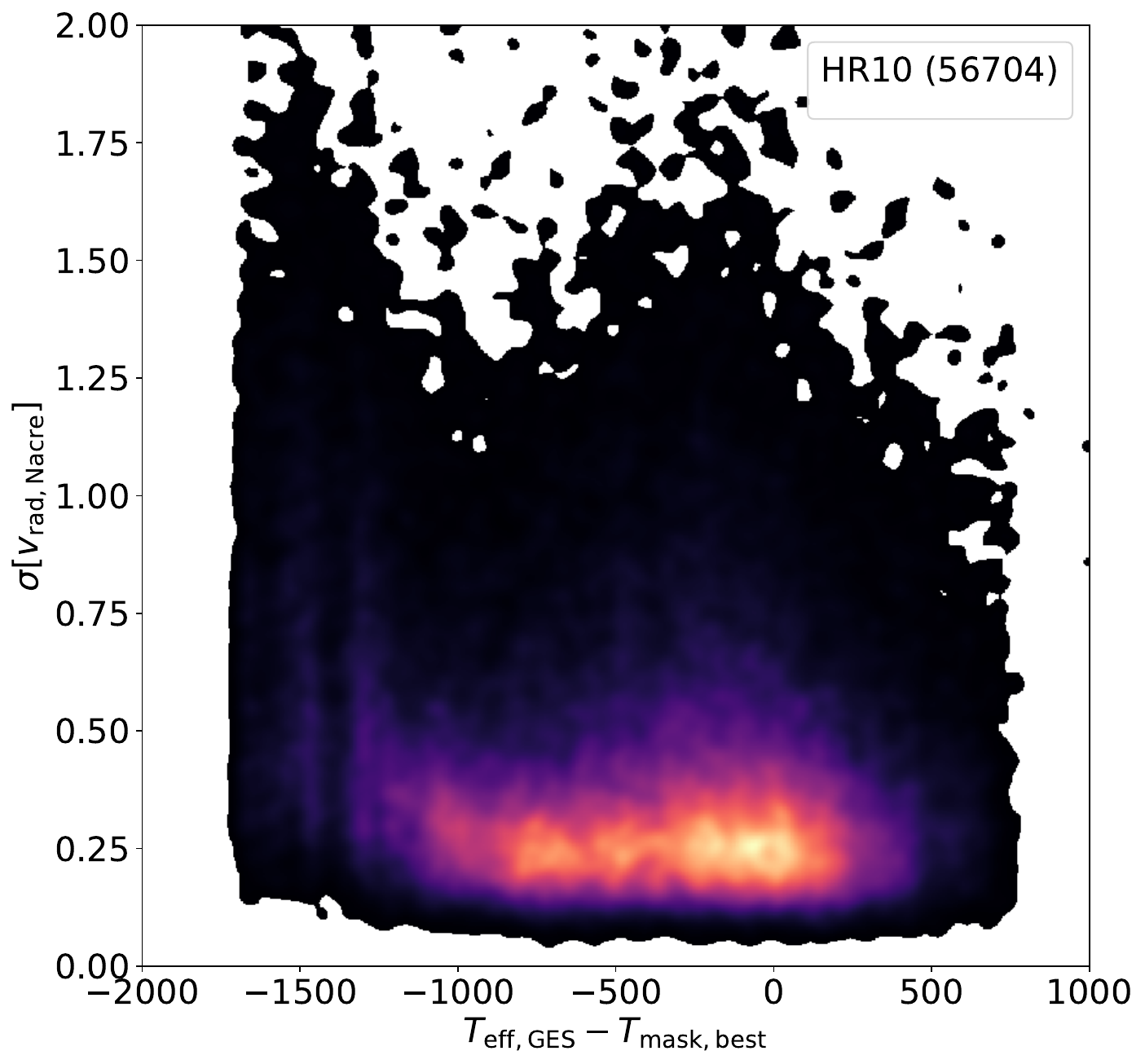}
  \includegraphics[width=0.85\columnwidth]{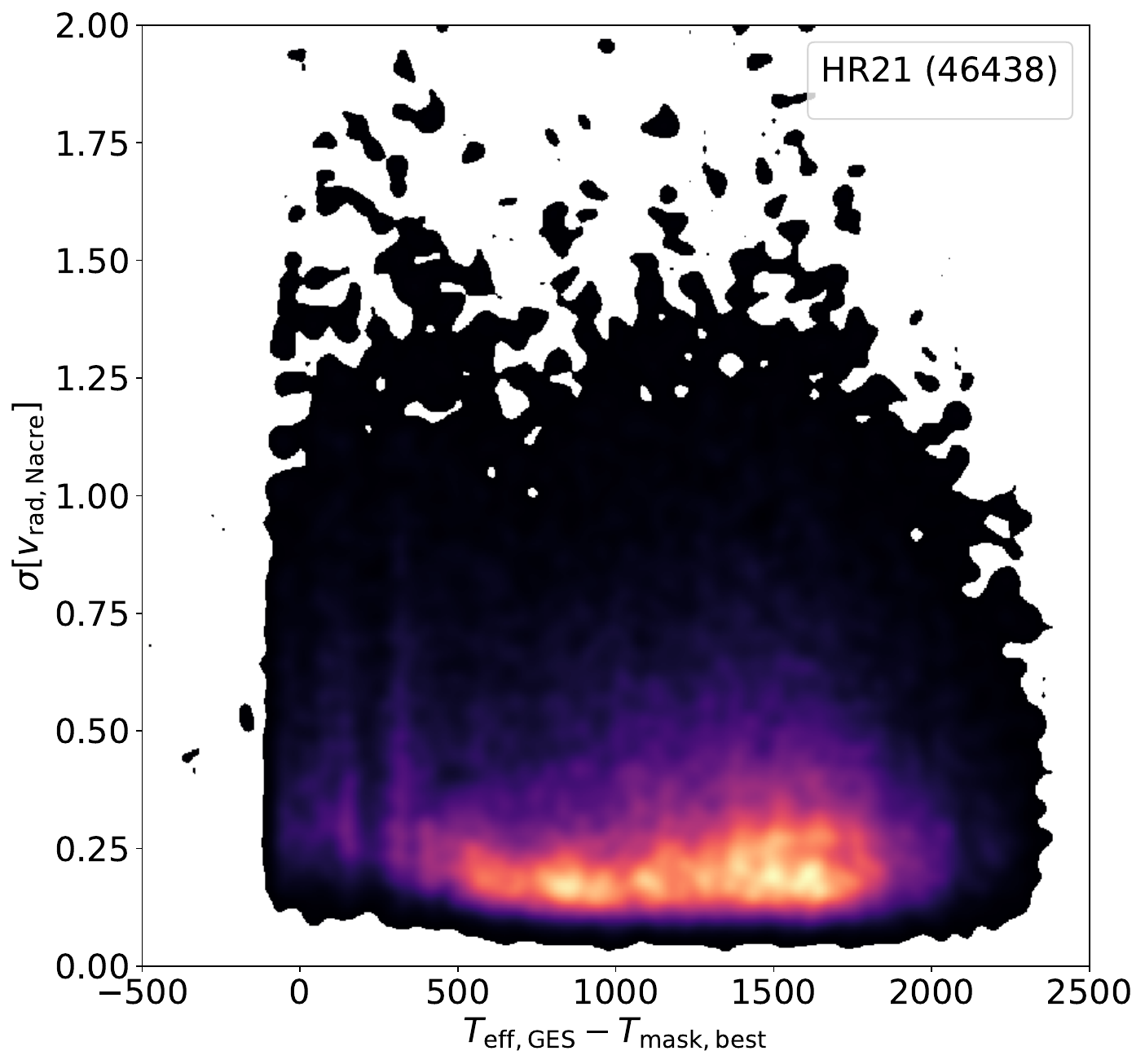}
  \caption{\label{Fig:Density_map_sigma_vs_mismatch} {Density map of $\sigma[v_{\mathrm{rad,Nacre}}]$ as a function of $T_{\mathrm{eff,GES}} - T_{\mathrm{mask,best}}$ for HR10 (left) and HR21 (right) observations where $T_{\mathrm{eff,GES}}$ is the recommended \Gaia-ESO effective temperature and $T_{\mathrm{mask,best}}$ is the temperature associated to the best-mask \nacre CCF. The number of objects in a given selection is indicated in the legend box. White areas denote zones of the plane where there is no data.}}
\end{figure*}

Figure~\ref{Fig:Density_map_sigma_vs_mismatch} illustrates the dispersion $\sigma[v_{\mathrm{rad,Nacre}}]$ as a function of $T_{\mathrm{eff,GES}} - T_{\mathrm{mask,best}}$ where $T_{\mathrm{eff,GES}}$ is the effective temperature of the star and $T_{\mathrm{mask,best}}$ is the effective temperature of {the mask of the best-mask \nacre CCF}. Since we plot single observations, a given star may contribute more than once to this map. The star's temperature is the recommended effective temperature provided by the \Gaia-ESO working group dedicated to the final homogenisation (WG15) at the end of the iDR5 analysis round. Therefore, in Fig.~\ref{Fig:Density_map_sigma_vs_mismatch}, only the targets with a recommended temperature are displayed. The quantity $T_{\mathrm{eff,GES}} - T_{\mathrm{mask,best}}$ can be interpreted as a measure of the spectral mismatch between the analysed star and the best template spectrum. {The mask called 'alphaCenA' ($T_{\mathrm{eff}} \approx \SI{5800}{\kelvin}$) yields the best-mask \nacre CCF for the vast majority of the HR10 observations, while the mask called 'betaAra' ($T_{\mathrm{eff}} \approx \SI{4200}{\kelvin}$) yields the best-mask \nacre CCF for the vast majority of the HR21 observations. This explains why the quantity $T_{\mathrm{eff,GES}} - T_{\mathrm{mask,best}}$ tends to be negative for HR10 and positive for HR21.} 

We also note that the diagram does not exhibit any clear relation between $\sigma[v_{\mathrm{rad,Nacre}}]$ and $T_{\mathrm{eff,GES}} - T_{\mathrm{mask,best}}$: the distribution is rather flat, with a high density of points around the line $\sigma[v_{\mathrm{rad,Nacre}}] \approx \SI{0.15}{\kilo\metre\per\second}$ while an extended poorly populated tail exists at any $T_{\mathrm{eff}} - T_{\mathrm{mask,best}}$. Thus, one can conclude that the uncertainty due to the spectral mismatch on a single radial velocity measurement typically amounts at least to \SI{0.15}{\kilo\metre\per\second}. Some vertical stripes are visible: they are {likely} due to the sampling of the temperature grids used by the \Gaia-ESO analysis nodes.

\begin{figure*}
  \centering
  \includegraphics[width=0.85\columnwidth]{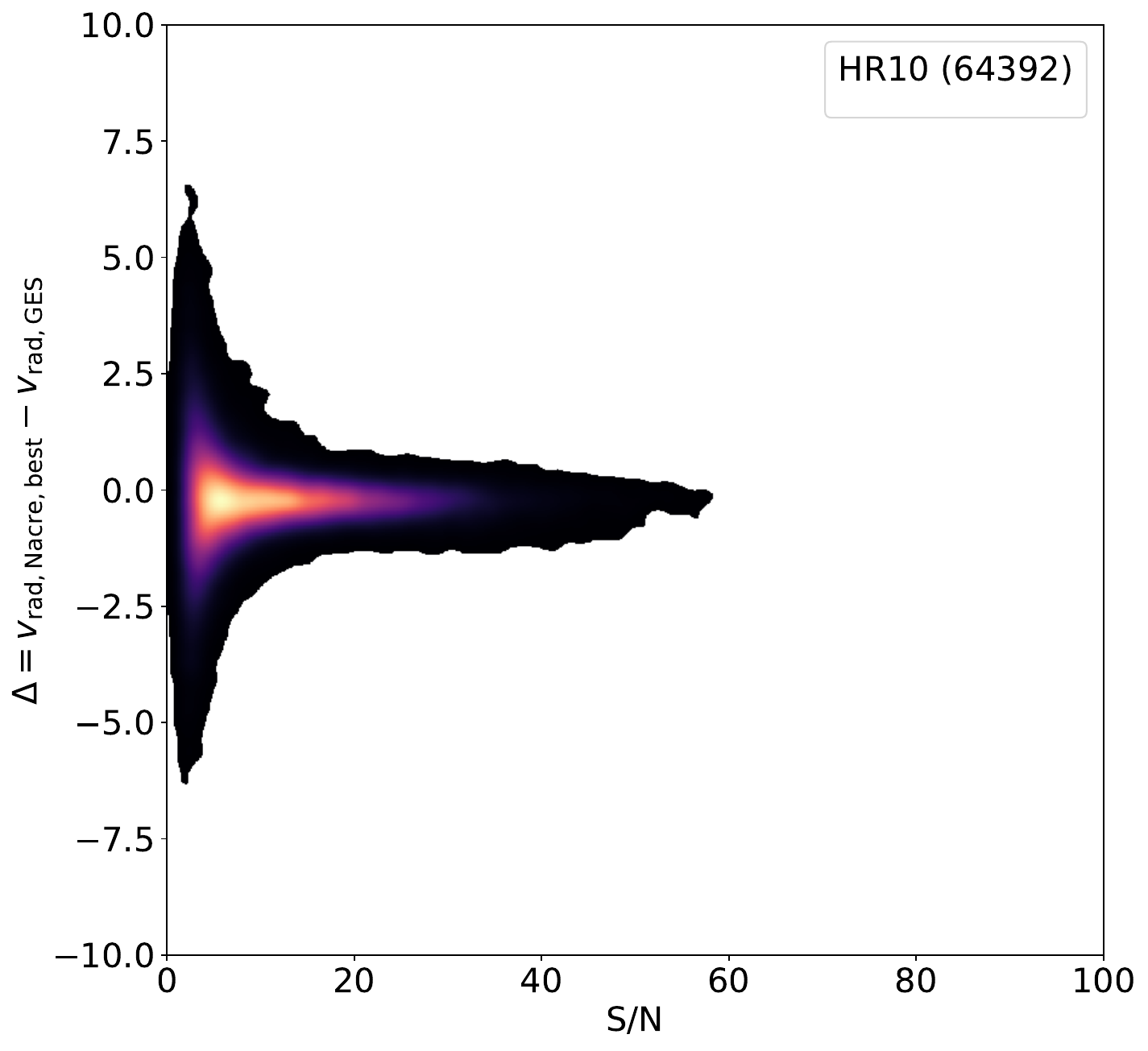}
  \includegraphics[width=0.85\columnwidth]{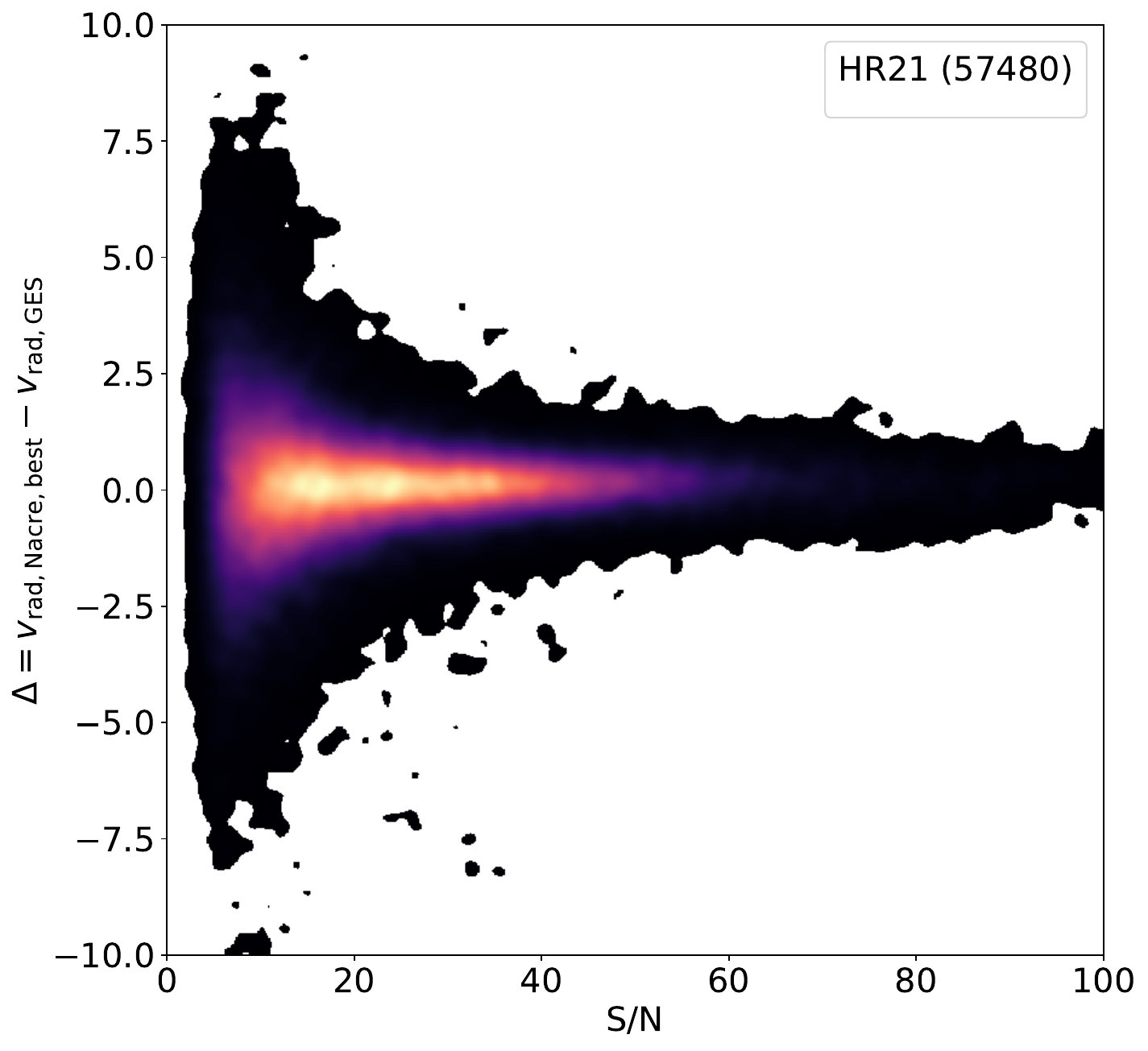}
  
  \includegraphics[width=0.85\columnwidth]{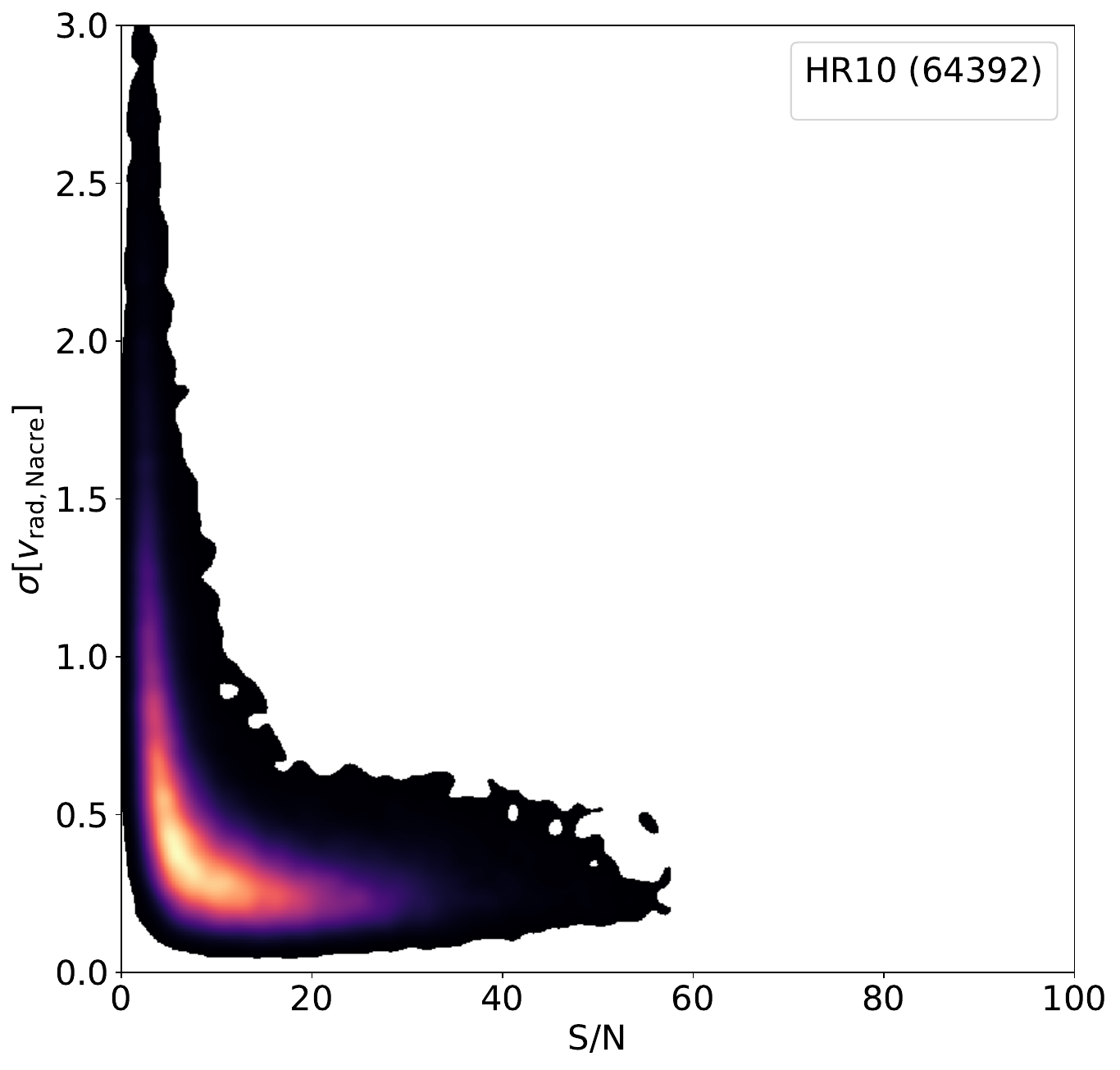}
  \includegraphics[width=0.85\columnwidth]{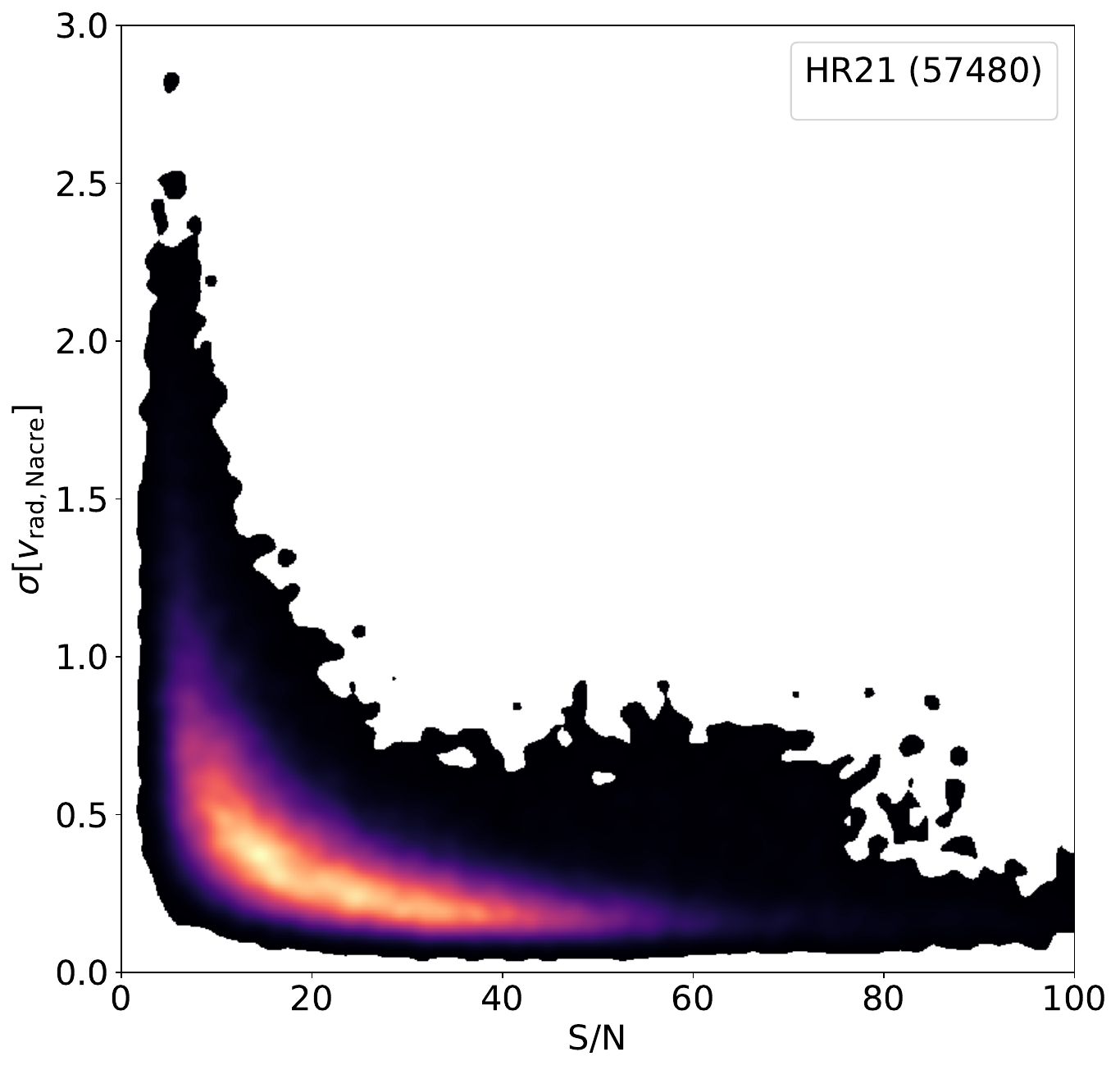}
  \caption{\label{Fig:Density_map_sigma_vs_SNR} Density map of $\Delta = v_{\mathrm{rad,Nacre,best}} - v_{\mathrm{rad,GES}}$ (top) and $\sigma[v_{\mathrm{rad,Nacre}}]$ (bottom) as a function of $\snr$ for HR10 (left) and HR21 (right) observations. {The number of objects in a given selection is indicated in the legend box. White areas denote zones of the plane where there is no data.}}
\end{figure*}

{Figure~\ref{Fig:Density_map_sigma_vs_SNR} shows the density map of $\Delta = v_{\mathrm{rad,Nacre,best}} - v_{\mathrm{rad,GES}}$ and $\sigma[v_{\mathrm{rad,Nacre}}]$ as a function of \snr where $v_{\mathrm{rad,GES}}$ is the recommended \Gaia-ESO radial velocity. There is no trend between $\Delta$ and \snr but the scatter increases with decreasing \snr: a low \snr does not introduce significant biases in $v_{\mathrm{rad,Nacre,best}}$ but degrades its precision. The latter is also shown by the decreasing trend of $\sigma[v_{\mathrm{rad,Nacre}}]$ with \snr, displayed in the bottom panel of Fig.~\ref{Fig:Density_map_sigma_vs_SNR}. We also notice that the radial velocity precision increases faster with the \snr for HR10 than for HR21 due to the fact that there are more weak lines in HR10 than in HR21.}

\section{Results}
\label{Sec:Results}

\subsection{Double-lined spectroscopic binaries}

Our analysis of the iDR5 sub-sample (as defined in Sect.~\ref{Sect:iDR5subsample}) with the help of the \nacre CCFs and \doe, has led to the discovery of \num{322} SB2 among the \num{37565} Milky Way field stars forming the parent sample. The cross-match of the SB2 sample with \Gaia DR3 is provided in Table~\ref{Tab:iDR5_SB2_simple_list} and the radial-velocity time-series, together with CCF-related quantities, are provided in Table~\ref{Tab:iDR5_SB2_details}.

\begin{figure}
  \centering
  \includegraphics[width=\columnwidth]{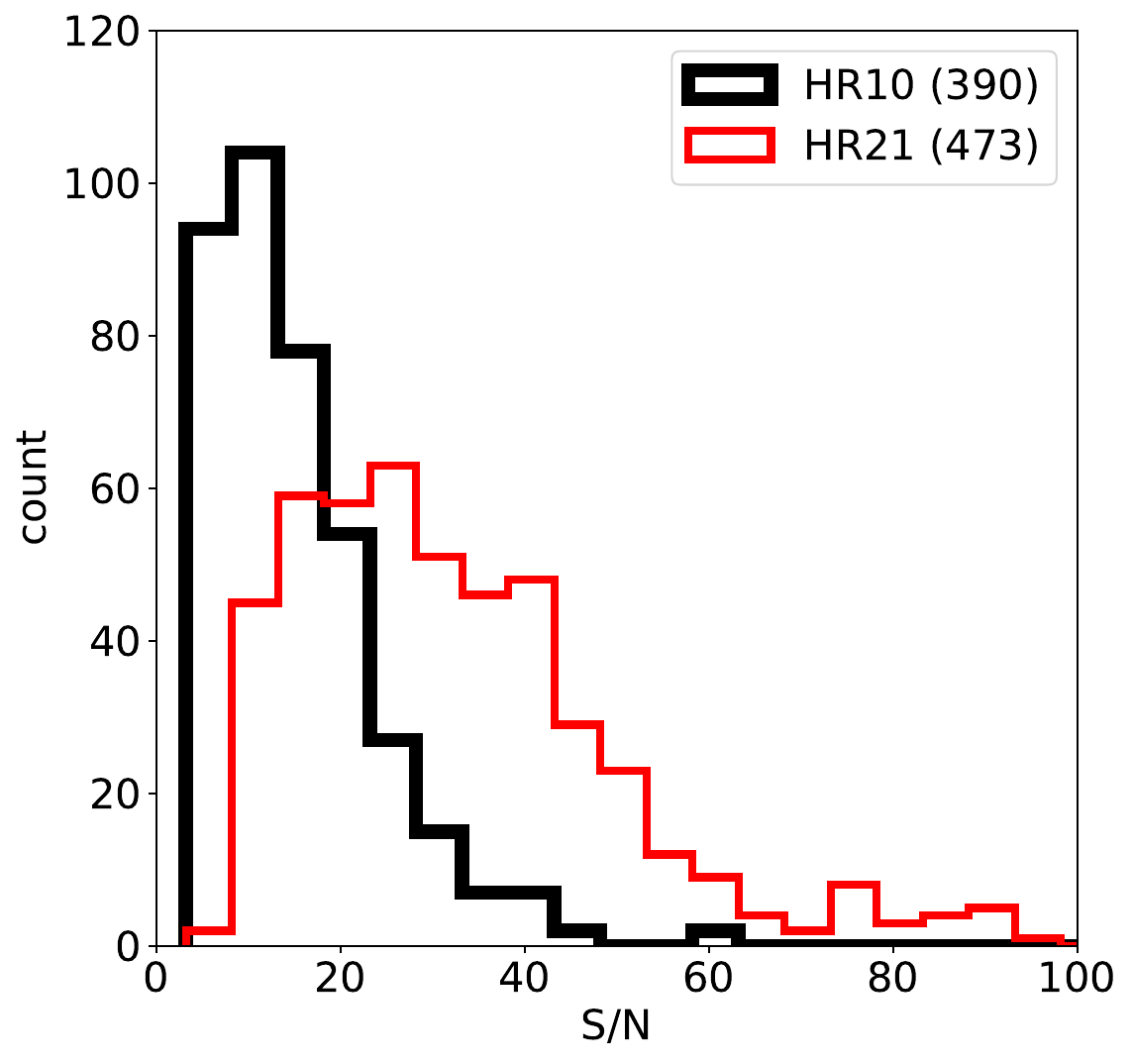}
  \caption{\label{Fig:iDR5_SB2_SNR_distribution} $\snr$ distributions for HR10 (black thick line) and HR21 (red thin line) observations triggering an SB2 detection (binwidth = 5). The histogram is observation-wise, so a given star may appear more than once (in the same or another $\snr$ bin) depending on the observations where the binary nature was detected in. The number of objects in a given selection is indicated in the legend box.}
\end{figure}

Figure~\ref{Fig:iDR5_SB2_SNR_distribution} illustrates the distribution (per observation; {\ie a given object may count more than once}) of the $\snr$ of the spectra used to compute the CCFs in the case of SB2 detections. The $\snr$ of the target spectrum can be as low as $3-4$ {for HR10 and $5-6$ for HR21} and it still allows the detection of multiple stellar components in the CCF. The HR10 and the HR21 distributions have a different shape: the HR10 (resp., HR21) distribution peaks at $\snr \sim 15$ (resp., $\snr \sim 25$); {the mean \snr is \num{15} for HR10 and \num{32} for HR21.} The {differences in the} $\snr$ distribution of the SB2 detections for both setups {are mainly due to the different} properties of their parent distribution, \ie the $\snr$ distribution of their parent HR10 and HR21 samples, respectively. We have indeed a mean $\snr \sim 10$ for all of the investigated HR10 spectra and a mean $\snr \sim 22$ for all of the investigated HR21 spectra.

{We note also that \SI{65}{\percent} (resp., \SI{86}{\percent}) of the investigated HR10 spectra have an \snr lower than \num{10} (resp., \num{20}), while it is the case for only \SI{29}{\percent} (resp., \SI{57}{\percent}) of the investigated HR21 spectra. Logically, we have more detections at \snr lower than 20 for HR10 (\num{297} out of \num{390}, \ie \SI{76}{\percent}) than for HR21 (\num{133} out of \num{473}, \ie \SI{28}{\percent}). Similarly, in the \snr range $[0, 20]$, we have \num{297} SB2 detections among \num{69475} HR10 observations (\SI{0.43}{\percent}), while we have \num{133} SB2 detections among \num{46047} HR21 observations (\SI{0.29}{\percent}). These statistics show that we probably miss more SB2 detections at low \snr with HR21 than we do with HR10 because of a low \snr. This could be due to the fact that the HR10 wavelength range hosts more (weak) metallic lines for FGK stars than the HR21 wavelength range does, such that even at low \snr, all those absorption lines build up a meaningful HR10 CCF.}

{The ratio of the number of HR10 SB2 detections (observation-wise) to the number of analysed HR10 observations ($\num{390} / \num{80490}$, \ie \SI{4.8}{\percent}) is lower than the ratio of the number of HR21 SB2 detections to the number of analysed HR21 observations ($\num{473} / \num{80237}$, \ie \SI{5.8}{\percent}). We already commented on the fact that low values of \snr probably decrease the SB2 detection efficiency in HR21 observations compared to HR10 observations. For \snr larger than 20 (referred to as `high \snr' in the next sentences), we do not see significant differences in the detection efficiency between HR10 and HR21: we have \num{93} SB2 detections among \num{10323} HR10 observations (\SI{0.91}{\percent}), while we have \num{340} SB2 detections among \num{34053} HR21 observations (\SI{1.0}{\percent}). This suggests that the excess of HR21 spectra at high \snr compared to HR10 explains why we have more SB2 detections among HR21 observations than among HR10 observations. In other words, the overall lower quality of HR10 spectra (compared to that of HR21 spectra) hampers the detection of SB2 and may cost up to about \num{100} HR10 SB2 detections (assuming we can apply the detection rate measured above for the HR21 observations to the HR10 case).}

\begin{figure}
  \centering
  \includegraphics[width=\columnwidth]{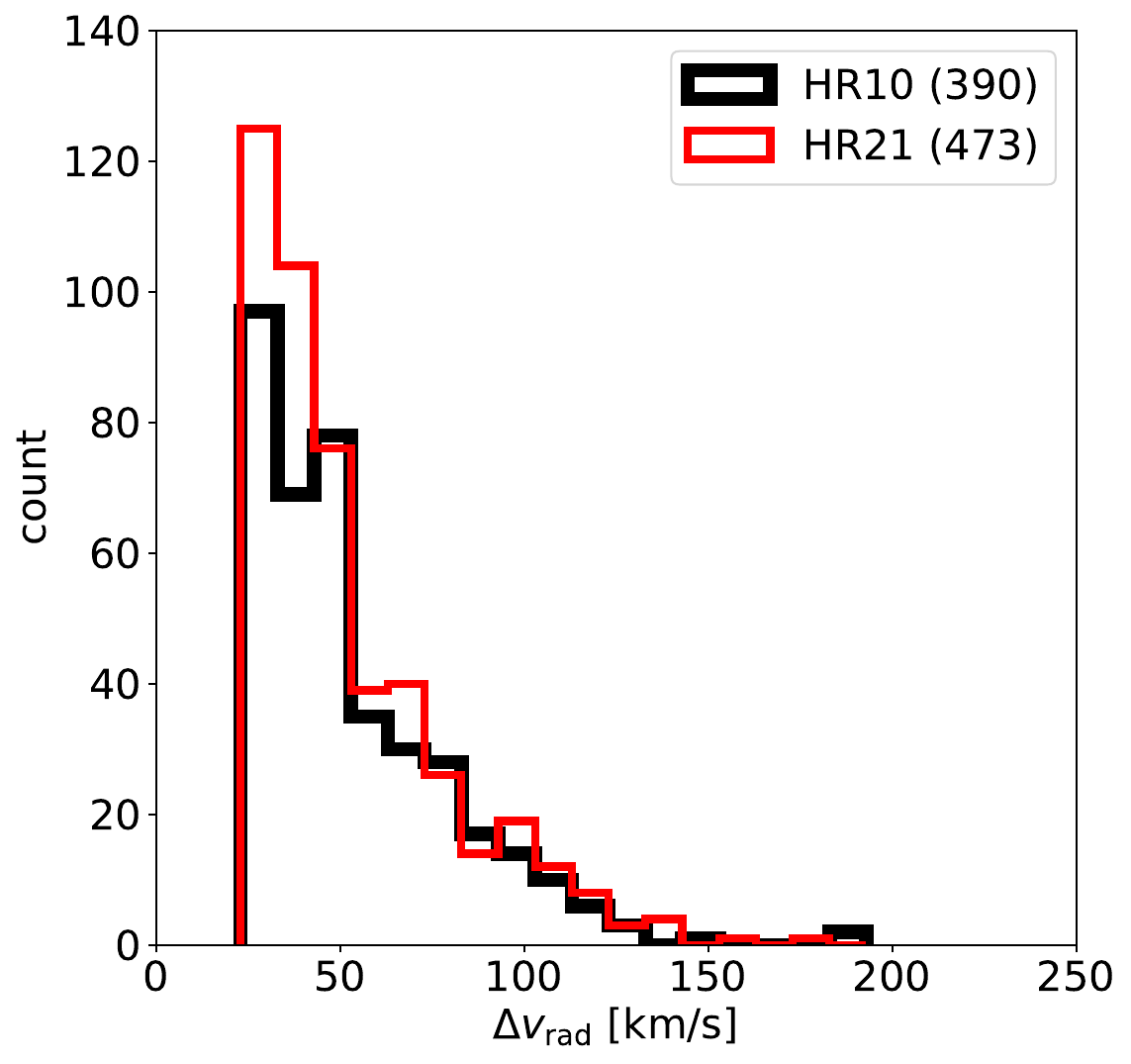}
  \caption{\label{Fig:iDR5_SB2_deltaV_distribution} Same as Fig.~\ref{Fig:iDR5_SB2_SNR_distribution} for $\Delta v_{\mathrm{rad}}$ in \si{\kilo\metre\per\second} (binwidth = \SI{10}{\kilo\metre\per\second}). The number of objects in a given selection is indicated in the legend box.}
\end{figure}

Figure~\ref{Fig:iDR5_SB2_deltaV_distribution} displays the distribution (per observation) of the radial velocity differences (\ie difference between the two components of the binary system) for both setups. The two distributions are very similar. In each case, the smallest detectable $\Delta v_{\mathrm{rad}}$ is about $\sim \SI{25}{\kilo\metre\per\second}$, they peak at $\Delta v_{\mathrm{rad}} \sim \num{25}-\SI{30}{\kilo\metre\per\second}$ and they decrease at the same rate towards higher $\Delta v_{\mathrm{rad}}$. {The difference between the HR10 and the HR21 resolutions does not translate into a difference between the smallest detectable $\Delta v_{\mathrm{rad}}$ for each setup (see also Sect.~\ref{Sec:Comparison_Merle2017}
).}

In conclusion, thanks to the \nacre CCFs, our analysis is almost not sensitive to setup effects (spectral range, spectral resolution) {in terms of smallest detectable $\Delta v_{\mathrm{rad}}$}, {which was the goal of our effort in developing a new detection technique; another demonstration of the improved SB$n$-detection efficiency is given in the subsection~\ref{Sec:Comparison_Merle2017} with a comparison to \cite{2017A&A...608A..95M}.} {Our technique remains sensitive to the \snr of the underlying spectra: though the minimum $\snr$ required for a detection can be as low as $3 - 4$, poorer-quality spectra still mean lower number of SB2 detections. It remains also sensitive to the spectral content: a low number of weak metallic lines hampers the detection of SB2.}

\subsection{Triple-lined and quadruple-lined spectroscopic binaries}

\begin{figure*}
  \centering
  \includegraphics[width=\columnwidth]{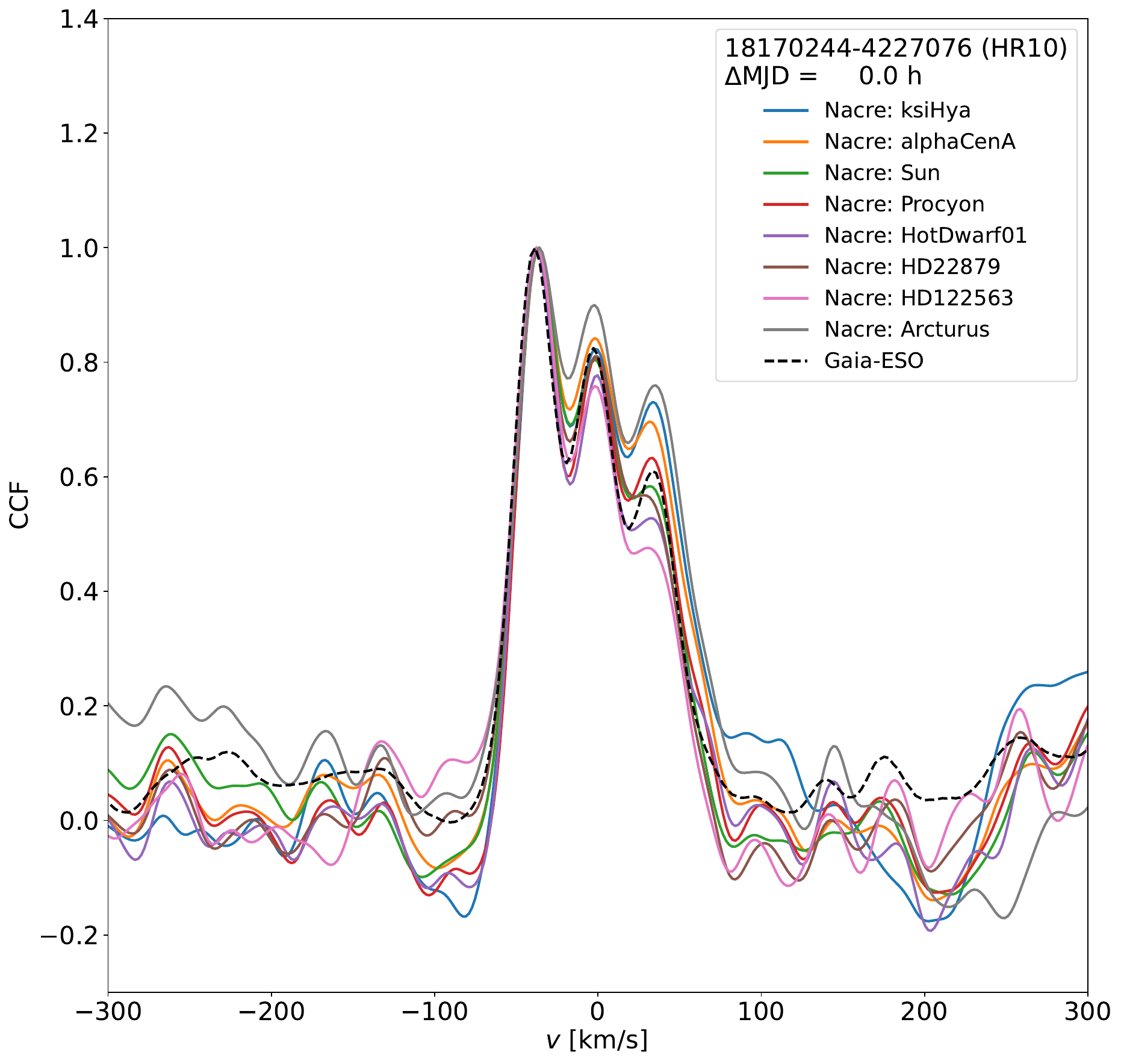}
  \includegraphics[width=\columnwidth]{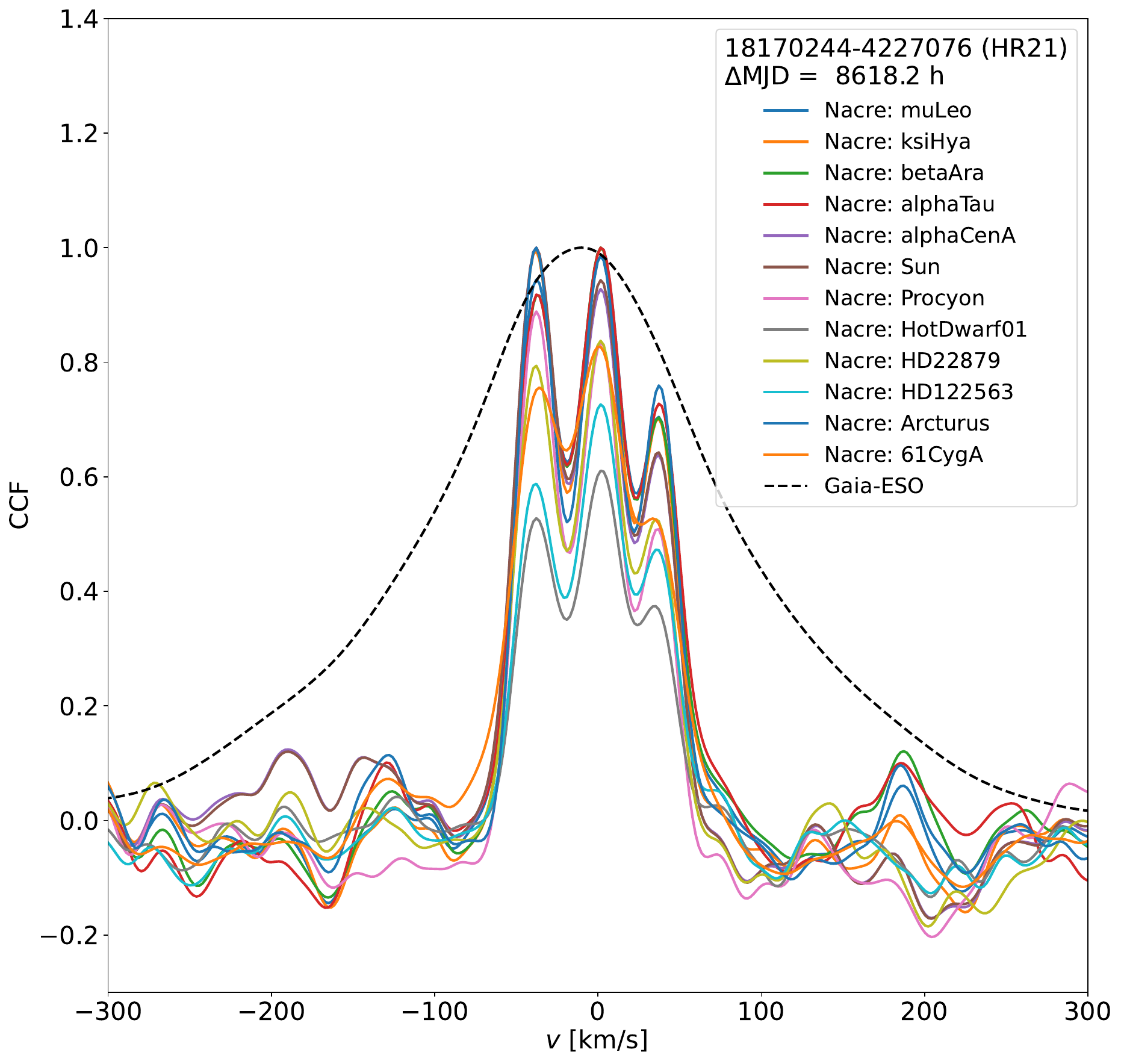}
  \caption{\label{Fig:iDR5_example_SB3} Examples of CCFs for an SB3 (CNAME 18170244-4227076). Left: HR10. Right: HR21. In the HR21 setup, we see how the new masks nicely allow the detection of the three components (coloured CCFs), totally hidden in the \Gaia-ESO CCF (dashed black line). The three components were already detected in HR10 with the \Gaia-ESO CCF.}
\end{figure*}

We find \num{12} SB3, of which two may be SB4, among the \num{37565} Milky Way field stars. Seven stars were identified both in HR10 and HR21 as probable SB3 candidates, as illustrated in Figure~\ref{Fig:iDR5_example_SB3}. Three tentative SB3 are also listed; they constitute tentative detections since the three peaks are only detected in only one of the two setups. In addition, we provide a tentative detection of two SB4 candidates. The data are summarised in Table~\ref{Tab:iDR5_SB3_SB4_simple_list}. They are all faint objects; this is probably the reason why none of them is identified as non-single star by \Gaia DR3 \citep{GaiaDR3Arenou}. {Besides, we note that} the \Gaia mean RVs are not provided, except for the brightest target (CNAME 19243943+0048136, $G \sim 14.8$). The present spectroscopic-binary study {still brings a very new piece of science in the \Gaia era} since the overlap with \Gaia DR3 {in terms of advanced \Gaia products (\eg radial velocities, multiplicity or atmospheric parameters)} is very small.

\begin{table*}
  \normalsize
  \centering
  \caption{\label{Tab:iDR5_SB3_SB4_simple_list} Cross-match of the \Gaia-ESO SB3 and SB4 samples, and \Gaia DR3: \Gaia parallax; magnitudes; Renormalised Unit Weight Error (\texttt{RUWE}); radial velocities.}
  \begin{tabular}{@{}lcS[table-format=2.3]@{\,\( \pm \)\,}S[table-format=2.3]S[table-format=2.3]S[table-format=2.3]S[table-format=2.3]S[table-format=3.2]@{}}
    \toprule
    {\Gaia-ESO CNAME} & {\Gaia source ID} & \multicolumn{2}{c}{$\varpi$} & {$G$} & {$G_{\mathrm{BP}} - G_{\mathrm{RP}}$} & {\texttt{RUWE}} & {$v_{\mathrm{rad}}$}\\
    & & \multicolumn{2}{c}{mas} & {\si{\mag}} & {\si{\mag}} & & {\si{\kilo\metre\per\second}}\\
    \midrule
    \midrule
    Probable SB3 \\
    \midrule
    08202324-1402560\tablefootmark{a}        &       5723402401367964032     &        0.687  &  0.036      &       15.667        &        1.111        &        0.954        &          nan\\
    08393088-0033389\tablefootmark{b}        &       3073616933620127488     &        0.433  &  0.032      &       15.137        &        0.757        &        1.078        &          nan\\
    08402079-0025437                         &       3075120756289226240     &        0.589  &  0.041      &       15.379        &        0.995        &        1.068        &          nan\\
    15120307-4049481                         &       6005054498237096832     &        2.882  &  0.633      &       16.210        &        1.152        &       11.683        &          nan\\
    15420717-4407146\tablefootmark{a}        &       5989143813386060416     &        0.936  &  0.085      &       17.082        &        1.875        &        1.006        &          nan\\
    18170244-4227076\tablefootmark{a}        &       6724489665257001472     &        0.514  &  0.037      &       15.781        &        1.098        &        0.939        &          nan\\
    21393685-4659598                         &       6563596479474777600     &        0.575  &  0.032      &       15.576        &        0.990        &        1.079        &          nan\\
    \midrule
    Tentative SB3 \\
    \midrule
    00195847-5423227\tablefootmark{a}        &       4923333999386498048     &        0.945  &  0.411      &       15.577        &        0.919        &       15.259        &          nan\\
    15161563-4125518\tablefootmark{a}        &       6004304635606156928     &        0.388  &  0.107      &       16.644        &        1.280        &        1.993        &          nan\\
    16003634-0745523\tablefootmark{c}        &       4349436487865043840     &        0.051  &  0.066      &       16.765        &        1.097        &        1.033        &          nan\\
    \midrule
    Tentative SB4 \\
    \midrule
    18263326-3146508                         &       4045322974757410560     &        0.623  &  0.669      &       16.937        &        1.741        &       10.934        &          nan\\
    19243943+0048136\tablefootmark{a,b}      &       4263752096494076160     &        0.789  &  0.027      &       14.792        &        1.373        &        0.834        &       -8.220\\
    \bottomrule
  \end{tabular}
  \tablefoot{
    \tablefoottext{a}{\cite{2017A&A...608A..95M}}
    \tablefoottext{b}{\cite{2020A&A...635A.155M}}
    \tablefoottext{c}{Potential binary from proper motion anomaly \citep{2019A&A...623A..72K}}
  }
\end{table*}

Some stars actually display multiple-peaked CCFs without being spectroscopic binaries; this is for example the case of Cepheids \citep{2016MNRAS.463.1707A}. However, among the SB3 and SB4 discovered in the present study, only one (CNAME 08402079-0025437) shows signs of photometric variability (with the \Gaia flag \texttt{phot\_variable\_flag} being raised). Most of the SB3 and SB4 of Table~\ref{Tab:iDR5_SB3_SB4_simple_list} can thus be considered as genuine multiple stars {whose number of stellar components (3, 4 or more) will have to be confirmed by future follow-up}.

Three SB3 candidates, namely CNAME 08202324-1402560, 18170244-4227076 and 15420717-4407146, were already detected as {spectroscopic binaries by \cite{2017A&A...608A..95M} but only from HR10 observations: the first two already as SB3, while the third one as SB2}. With the new masks, they are confirmed as SB3 also with HR21 {observations}. The three other SB3 are new candidates which were never detected previously and are unknown on Simbad. Among the three tentative SB3, {two of them, namely} 00195847-5423227 and 15161563-412551, were detected as SB2 in the previous release by \citet{2017A&A...608A..95M}. The third one shows proper motion anomaly \citep{2019A&A...623A..72K}, strengthening the multiple nature of this target.

Additionally, two targets could be SB4 candidates, namely CNAME 18263326-3146508 and 19243943+0048136. Their triple nature is indubitable, and a fourth weak component is observed in their CCFs; they would deserve additional monitoring at higher resolution to confirm their quadruple nature. 18263326-3146508 is not referenced on Simbad but has a very large \Gaia \texttt{RUWE} (Renormalised Unit Weight Error). 19243943+0048136 is in the CoRoT field and was already detected as spectroscopic binary \citep{2017A&A...608A..95M,2020A&A...635A.155M}. {We stress that the reliable detection of up to four stellar components in GIRAFFE observations is another direct demonstration of the improved detection efficiency endowed by the \nacre masks: this is a new result compared to our previous study \citep{2017A&A...608A..95M}.}

Only three targets among twelve show large \Gaia \texttt{RUWE}. This shows that even if a large value of the \texttt{RUWE} points toward the possible multiple nature of sources, it does not catch all of them, especially the distant ones for which the astrometric signal is not detectable, as already well shown in \citet{2019MmSAI..90..395J} and \citet{2021AJ....162..184K}. {These three large-\texttt{RUWE} targets can be deemed as a likely SB3, a tentative SB3 and a tentative SB4}; they {decidedly} deserve a spectroscopic follow-up since spectroscopy+astrometry would allow to derive precise masses.

Finally, we stress that the derived atmospheric parameters (either by \Gaia-ESO or by \Gaia) are highly unreliable for such composite spectra {unless} the components {have very similar atmospheric parameters} and the same {radial velocities}. \Gaia provided such atmospheric parameters for nine targets of Table~\ref{Tab:iDR5_SB3_SB4_simple_list} but {they are} unreliable for such spectroscopic binaries. {Appendices~\ref{SecApp:_atlas_SB3} and \ref{SecApp:_atlas_SB4} provides the reader with an atlas of CCFs and spectra (classified by increasing MJD for a given system) for all of the detected SB3 and SB4. All available epochs, independently of the number of detected components, are shown; the only requirement is that the computed CCFs are interpretable. This means that according to the stellar waltz occurring between the system's stars, the reader will see one, two or three (or four for the tentative SB4) stellar components in the CCFs.} We emphasise the fact that some of the shown CCFs for the detected SB3 and SB4 are more difficult to interpret than the CCFs of SB2 because the height of the peaks associated to some of the $N - 1$ stellar components may be comparable to the typical size of the noise. All these systems have been flagged by our pipeline as unusual and have been manually inspected before being classifyied as SB3 or SB4. As an example, we provide a detailed discussion of the CCFs available for the CNAME 08393088-0033389 (Sec.~\ref{Sec:Appendix_08393088_0033389}) and 08402079-0025437 (Sec.~\ref{Sec:Appendix_08402079_0025437}) highlighting the arguments leading to their identification as SB3.

\subsection{Locus in the colour-magnitude diagram and comparison to \Gaia}

{We cross-match the \Gaia-ESO parent sample (\ie \num{37565} objects) and the \Gaia DR3 catalogue \citep{2022arXiv220800211G,2022arXiv220605989B}: Table~\ref{Tab:Gaia_crossmatch_count_vs_quantities} gives the count of objects with a published \Gaia value for different physical quantities of interest for this study. We use the photometric bands of \Gaia DR3 \citep{2021A&A...649A...3R} to place our SB2, SB3 and SB4 in the colour-magnitude diagram (CMD) displayed in Fig.~\ref{Fig:iDR5_SB2_SB3_SB4_CMD} where $\mathcal{M}_{\mathrm{G}}$ is the absolute magnitude in the $G$ band computed using the apparent $G$ magnitude and the parallax, and $G_{\mathrm{BP}} - G_{\mathrm{RP}}$ is the colour index. We also use the estimated line-of-sight extinction and reddening given by the fields \texttt{ag\_gspphot} and \texttt{ebpminrp\_gspphot} in the \Gaia source catalogue \citep{2022arXiv220606138A} to construct the extinction-free photometric quantities, denoted by the index $0$. Moreover, to plot the CMDs, we apply the following quality filters to clean the parent sample: $\texttt{parallax\_over\_error} \ge 5$ and $\texttt{RUWE} \le 1.4$; after filtering, we are left with \num{21880} (resp., \num{20730}) objects for the reddened (resp., dereddened) CMD. On the other hand, as for the spectroscopic multiples we have detected, we do not apply any filters on their \Gaia quantities: if $\mathcal{M}_{\mathrm{G}}$ and $G_{\mathrm{BP}} - G_{\mathrm{RP}}$ (or their dereddened version) exist, then we show the object in the CMD (SB2 as red disk, SB3 as cyan diamond, and SB4 as magenta plus). In the extinction-free CMD, we miss \num{20} SB2, three SB3 and one SB4 because of missing $A_G$ and $\mathrm{E}(\mathrm{BP} - \mathrm{RP}) = A_{\mathrm{BP}} - A_{\mathrm{RP}}$. Figure~\ref{Fig:iDR5_SB2_SB3_SB4_CMD} shows both the reddened (left) and dereddened (right) CMD.}

{Before discussing the CMDs, a word of caution is needed. We remind here that the line-of-sight extinction is estimated by the General Stellar Parameterizer from Photometry (GSP-Phot), part of the astrophysical parameters inference system \citep[Apsis;][]{2013A&A...559A..74B}. As explained in \citet{2022arXiv220606138A}, the Bayesian estimation of $A_G$, $A_{\mathrm{BP}}$ and $A_{\mathrm{RP}}$ depends on the choice of synthetic spectral energy distributions (SEDs) integrated over the $G$ (resp., BP, RP) passband during the forward-modelling of the magnitude in the $G$ (resp., BP, RP) band. GSP-Phot explicitly targets non-variable single stars and as such, the used synthetic SEDs are computed for single stars only. Therefore, the inaccuracy and imprecision of $A_G$, $A_{\mathrm{BP}}$ and $A_{\mathrm{RP}}$ might be larger for the objects of our SB$n$ samples than for the other objects of the parent-sample. Another Apsis module, namely Multiple Star Classifier \citep[MSC;][]{2022arXiv220605864C}, tries to estimate $A_G$ (among other parameters) by assuming that any \Gaia source is an unresolved co-eval binary system with a given flux ratio. We note that nearly all of the SB$n$ candidates reported in this article have a \Gaia MSC $A_G$. Unfortunately, MSC does not provide an estimate of the reddening and, even if it was available, using both GSP-Phot and MSC estimates to correct the photometry of an object depending on its suspected degree of multiplicity may also introduce other uncontrolled biases in the dereddened CMD. Because of the above issue, in the next paragraphs, we discuss the photometric properties of the SB$n$ with respect to their counterpart in the single-star main-sequence using both the reddened and dereddened CMD.}

\begin{table*}
  \footnotesize
  \centering
  \caption{\label{Tab:Gaia_crossmatch_count_vs_quantities} Count of objects with a published given \Gaia physical quantity or fulfilling a given condition. Columns are: the sample name; the count before cross-matching; the count after cross-matching; the number of object with, respectively, a \Gaia parallax $\varpi$, or a $G$ magnitude, or a BP-RP colour, or an extinction for the $G$, BP and RP bands, or a radial velocity; the count of objects fulfilling respectively the condition $\texttt{RUWE} \le 1.4$ or $\texttt{parallax\_over\_error} \ge 5$.}
  \begin{tabular}{@{}lrrrrrrrrr@{}}
    \toprule
    Sample & before & after & $\varpi$ & $G$ & $G_{\mathrm{BP}} - G_{\mathrm{RP}}$ & $A_G$ and & $v_{\mathrm{rad}}$ & \multicolumn{2}{c}{Applied filters}\\
    \cmidrule{9-10}
     & & & & & & $\mathrm{E}(\mathrm{BP} - \mathrm{RP})$ & & \texttt{RUWE} & rel. unc. $\varpi$\\
    \midrule
    \midrule
    full parent-sample & 37565 & 37260 & 37101 & 37260 & 37245 & 35278 & 1053 & 35951 & 23457\\
    SB2 sample         &   322 &   322 &   322 &   322 &   320 &   300 &   16 &   291 &   242\\
    SB3 sample         &    10 &    10 &    10 &    10 &    10 &     7 &    0 &     7 &     6\\
    SB4 sample         &     2 &     2 &     2 &     2 &     2 &     1 &    1 &     1 &     1\\
    \bottomrule
  \end{tabular}
  \normalsize
\end{table*}

{We note that the SB2s detected in the present paper are mainly distributed along a sequence, shifted upward (so toward brighter magnitudes) in the uncorrected and in the extinction-corrected CMD, parallel to the main-sequence outlined by the parent sample in black. In order to estimate the mean shift between the SB2 main-sequence and the parent-sample main-sequence, we perform respectively a quadratic and quartic fit of these two sequences over the colour range $[0.7, 3.1]$. To this end, we rebin in both colour and magnitude the two sequences before fitting the polynomial. A Monte-Carlo simulation ($N_{\mathrm{draw}} = \num{801}$) is used to propagate the uncertainties on the \Gaia parallaxes, $G$ fluxes, and BP and RP fluxes (uncertainties on the extinction and reddening are not propagated, though) in order to estimate the uncertainty on the locus of the centroid of each sequence. It is illustrated by the green and blue thick lines in Fig.~\ref{Fig:iDR5_SB2_SB3_SB4_CMD}. On the dereddened CMD, the mean difference in extinction-free absolute magnitude between each sequence, over the colour range $[0.7, 3.1]$ is $\num{0.78} \pm \SI{0.04}{\mag}$. This should be compared to the difference of \SI{0.75}{\mag} expected between the absolute magnitude of an SB2 formed of two twin stars and the magnitude of each star taken separately. We note the existence of a small group of SB2 objects fainter than expected: they are below the blue trend and we even see a dozen of them around $G_{\mathrm{BP}} - G_{\mathrm{RP}} \approx 1$ (and similarly, around $\left(G_{\mathrm{BP}} - G_{\mathrm{RP}}\right)_0 \approx 1$). In the reddened CMD, the two SB2 at coordinates $(0.969, 7.94)$ and $(0.983, 7.21)$ are respectively 14001598-4223091 and 18492379-4232588.}

{In the extinction-free CMD of Fig.~\ref{Fig:iDR5_SB2_SB3_SB4_CMD}, we note that six out of seven SB3 are on the bright side of the SB2 sequence (\ie above the mean trend). These six objects tend to be \SI{1.03}{\mag} (std. dev. = \SI{0.35}{\mag}) brighter than their counterpart in the parent-sample; for a system made of three identical stars, we expect an absolute magnitude brighter by \SI{1.20}{\mag} compared to the magnitude of the star alone. In the reddened CMD, two SB3, namely CNAME 15120307-4049481 (coordinates $(1.15, 8.5)$ in the CMD) and 00195847-5423227 (coordinates $(0.92, 5.45)$), are surprisingly faint; \Gaia DR3 provides extinction estimates for only 00195847-5423227 and this object is still fainter than the other SB3 in the dereddened CMD.}

{In the reddened CMD, the two tentative SB4 are brighter than the SB2 and SB3 lying in the same colour range. After dereddening, only one SB4 remains -- 19243943+0048136 -- in the extinction-free CMD: it is moved to the hot-edge of the main-sequence at the coordinates $(0.57, 2.80)$. This area of the CMD is less populated by the parent sample and it is outside the range where we parameterised the parent-sample main-sequence. If we compare the magnitude of 19243943+0048136 to the magnitudes of all objects found in a narrow colour-interval of \SI{0.01}{\mag}, centred on the colour of this SB4, then we find a mean difference of \SI{0.52}{\mag} with a standard deviation of \SI{1.09}{\mag} (computed with \num{61} objects). We remind that the difference in absolute magnitude is \SI{1.51}{\mag} for a system made of four identical stars, compared to the absolute magnitude of each star separately. We note that the computed mean difference is much smaller than the theoretical one but given the weakly constrained mean difference, this test might not be very conclusive.}

{In the previous paragraphs, we noted that some of the detected SB2, SB3 and SB4 have a suspicious position in the reddened and dereddened CMD. These objects tend to have also a RUWE larger than \num{1.4}. The SB2 forming the small group around $G_{\mathrm{BP}} - G_{\mathrm{RP}} \approx 1$ tend to have a large RUWE (\eg 01194949-6003364 with a RUWE of \num{3.979}, 00195847-5423227 with a RUWE of \num{15.259}, 15230178-4204383 with a RUWE of \num{10.961}). It is also true for some other SB2 below the blue trend like 02003583-0053539 at reddened coordinates (1.404, 6.49) and with a RUWE of \num{3.982}. And in particular, the two SB2 14001598-4223091 and 18492379-4232588 (outliers in the reddened CMD; not shown in the dereddened CMD) have a RUWE of respectively \num{3.908} and \num{11.559}. The two SB3 00195847-5423227 (outlier in both CMD) and 15120307-4049481 (only shown in the reddened CMD) have a RUWE of respectively \num{15.259} and \num{11.683}. One of the two SB4, 18263326-3146508, has a RUWE of \num{10.934}, while the other, 19243943+0048136 (the only one shown in both version of the CMD), has a small RUWE (\num{0.834}). The large values of RUWE indicate a likely unreliable astrometry: thus the parallax could be inaccurate and the object could be further away and intrinsically brighter than computed. Therefore, inaccurate \Gaia astrometry due to the non-single nature of the sources could be one of the reason why a fraction of our spectroscopic binaries are significantly fainter than other spectroscopic binaries at similar $G_{\mathrm{BP}} - G_{\mathrm{RP}}$.}

{We note that \SI{5}{\percent} of the SB2 have a \Gaia radial velocity \citep{2022arXiv220605902K}, while \SI{3}{\percent} of the stars in the parent-sample have one. Given the larger uncertainty on the former value, we conclude that the two samples are equally missing a \Gaia estimate for the radial velocity and that the faintness of the source -- rather than its degree of multiplicity -- is the main explanation for the absence of a \Gaia radial velocity. Among the iDR5 sample under investigation, \Gaia DR3 flags \num{27} stars out of the \num{37260} as non-single star (field \texttt{non\_single\_star} of the \Gaia main catalogue; \citealt{GaiaDR3Arenou}): four are reported as eclipsing binaries (EB) and \num{23} as astrometric binaries (AB). None of them are identified as SB in the present study: the \Gaia EB might be too faint to have \Gaia radial velocities, while the \Gaia AB could be wide long-period binaries that cannot be caught by a spectroscopic survey not designed for radial velocity monitoring as it is the case for the \Gaia-ESO. In addition, we also cross-matched the \num{803} \Gaia-ESO SB1 candidates from \citet{2020A&A...635A.155M} with \Gaia DR3 and only nine are flagged as non-single star by \Gaia DR3: three as AB, five as SB, and one as EB. The reason of the small number of objects in our SB1 \citep[from][]{2020A&A...635A.155M} and SB$n \ge 2$ (this work) samples flagged by \Gaia as SB is that the \Gaia SB detection is limited to targets brighter than $G \sim 13$, while the median $V$ magnitude of \Gaia-ESO targets is around \num{15}. Despite the wealth of data brought to the community by the \Gaia mission, ground-based spectroscopic surveys still bring newness in the field of SB detection and characterisation.}

{Other indicators of stellar multiplicity can be found in the \Gaia `Astrophysical parameters' tables: \texttt{classprob\_dsc\_combmod\_binarystar} and \texttt{classprob\_dsc\_specmod\_binarystar} that are the probabilities from the Discrete Source Classifier \citep[DSC;][]{2022arXiv220606710D} of being a binary star, probability obtained using either the combined spectro-photo-astrometric information (classifier `combmod') or the (BP/RP) spectrometric information alone (classifier `specmod'); \texttt{flags\_msc} that is a flag indicating (when set to 1) an unreliable inference result from MSC. \citet{2022arXiv220606710D} shows that the completeness and purity for the class `binary' is poor with the two classifiers: only \SI{0.2}{\percent} of the unresolved binaries of their validation data-set are recovered (see their Table~3). Given this performance index, we do not expect more than a couple SB2 to be correctly flagged by DSC. Only one SB$n$ (an SB2) has \texttt{classprob\_dsc\_combmod\_binarystar} larger than \num{0.01}. About \num{80} SB2 (resp., eleven SB2) have \texttt{classprob\_dsc\_specmod\_binarystar} larger than \num{0.01} (resp., \num{0.10}) and no SB2 has this probability larger than \num{0.3}. Three SB3 have a probability of at least \SI{1}{\percent} to be binary according to \texttt{classprob\_dsc\_specmod\_binarystar}, while this field gives a null probability of binarity for the two detected SB4. Using the census from \citet{2020A&A...635A.155M}, we find three SB1 out of \num{803} with a probability \texttt{classprob\_dsc\_combmod\_binarystar} between \num{1} and \SI{10}{\percent}. We conclude that it is not possible to use the fields \texttt{classprob\_dsc\_combmod\_binarystar} and \texttt{classprob\_dsc\_specmod\_binarystar} for a robust selection of binary stars, at least in the \Gaia-ESO magnitude regime. Eleven SB$n$ have no MSC parameters; four are flagged as unreliable by MSC; \num{319} have \texttt{flags\_msc} set to the default value of 0.}

{If we look at the \Gaia column \texttt{ipd\_frac\_multi\_peak} (IFMP) giving the percentage of windows with a successful Image Parameters Determination (IPD) and for which the IPD has identified a double peak, then we find: \num{59} SB2 with a non-null IFMP, only \num{12} SB2 have an IFMP larger than \SI{10}{\percent} and the largest IFMP value is \SI{71}{\percent} for 14001598-4223091; \num{3} SB3, 00195847-5423227 ($\mathrm{IFMP} = \SI{30}{\percent}$), 15120307-4049481  ($\mathrm{IFMP} = \SI{16}{\percent}$) and 21393685-4659598 ($\mathrm{IFMP} = \SI{2}{\percent}$), have a non-null IFMP; one SB4, 18263326-3146508, has a non-null IFMP ($\mathrm{IFMP} = \SI{60}{\percent}$). If we look at the \Gaia column \texttt{ipd\_gof\_harmonic\_amplitude} (IGHA) whose a large value indicates that the source has some non-isotropic spatial structure, then we find \num{162} SB2 (resp., \num{63}), two SB3 (resp., two) and two SB4 (resp., two) with an IGHA larger than \num{0.022} (resp., \num{0.051}), where \num{0.022} (resp., \num{0.051}) is the median IGHA  of the parent sample (resp., the median IGHA increased by its interquartile range). We conclude that a future release of \Gaia should be able to correctly flag at least some of the SB$n$ uncovered in this work as binaries.}

\begin{figure}
    \centering
    \includegraphics[width=\linewidth]{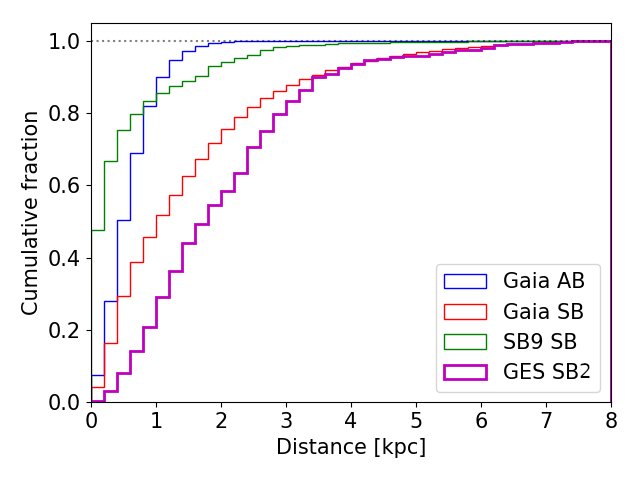}
    \caption{Comparison of the cumulative distributions of \Gaia-ESO SB2 (magenta; this work), \Gaia astrometric binaries (blue) and spectroscopic binaries \citep[red;][]{GaiaDR3Arenou} and spectroscopic binaries from the \sbnine catalogue \citep[green;][]{2004A&A...424..727P} as a function of the distance from us. The \sbnine catalogue was cross-matched using the CDS X-match service \citep{2020ASPC..522..125P} with \Gaia DR3 \citep{GaiaDR3Arenou}.}
    \label{fig:cumul_histo_SB_VB}
\end{figure}

\begin{figure*}
  \centering
  \includegraphics[width=0.9\columnwidth]{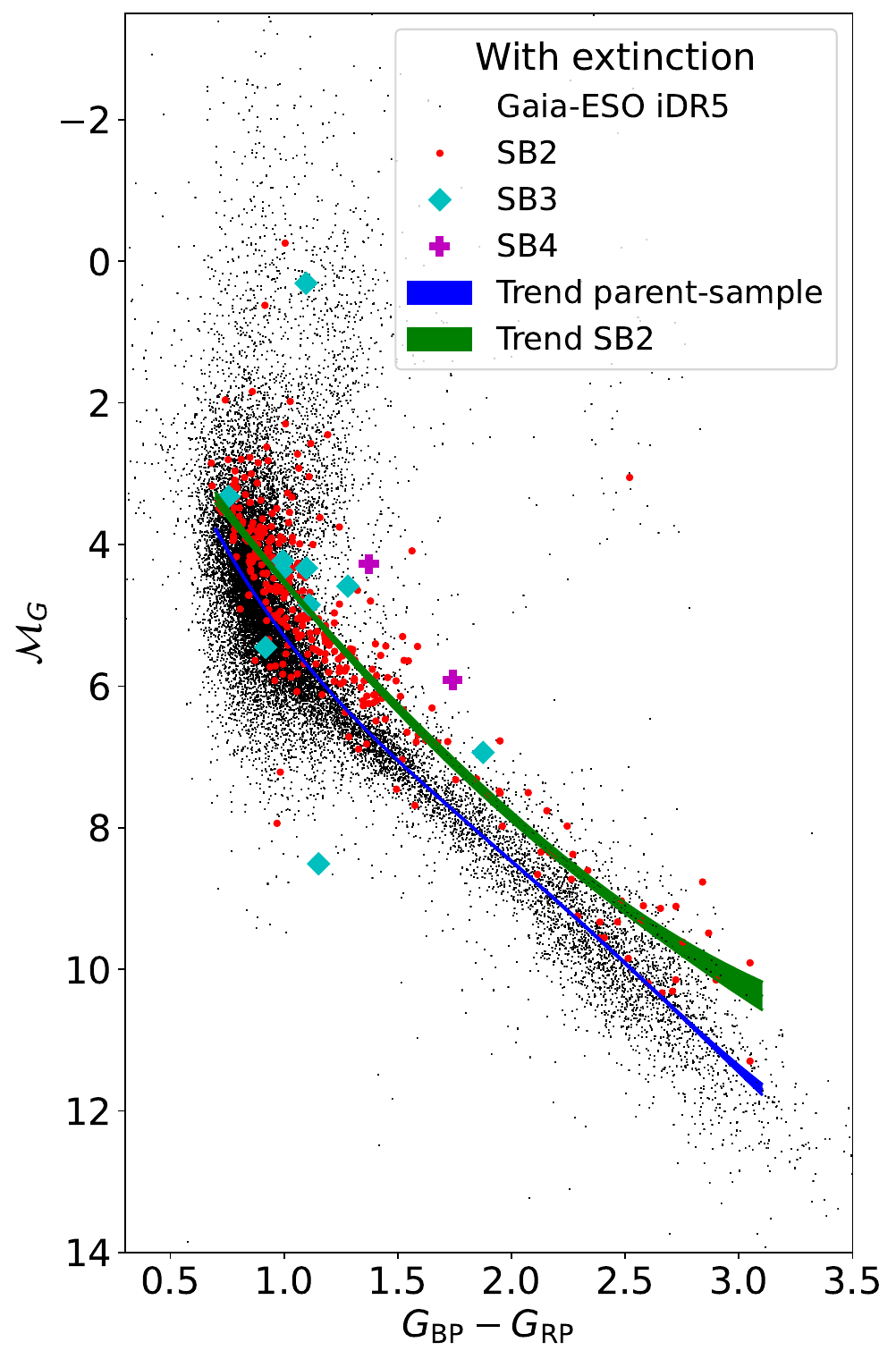}
  \includegraphics[width=0.9\columnwidth]{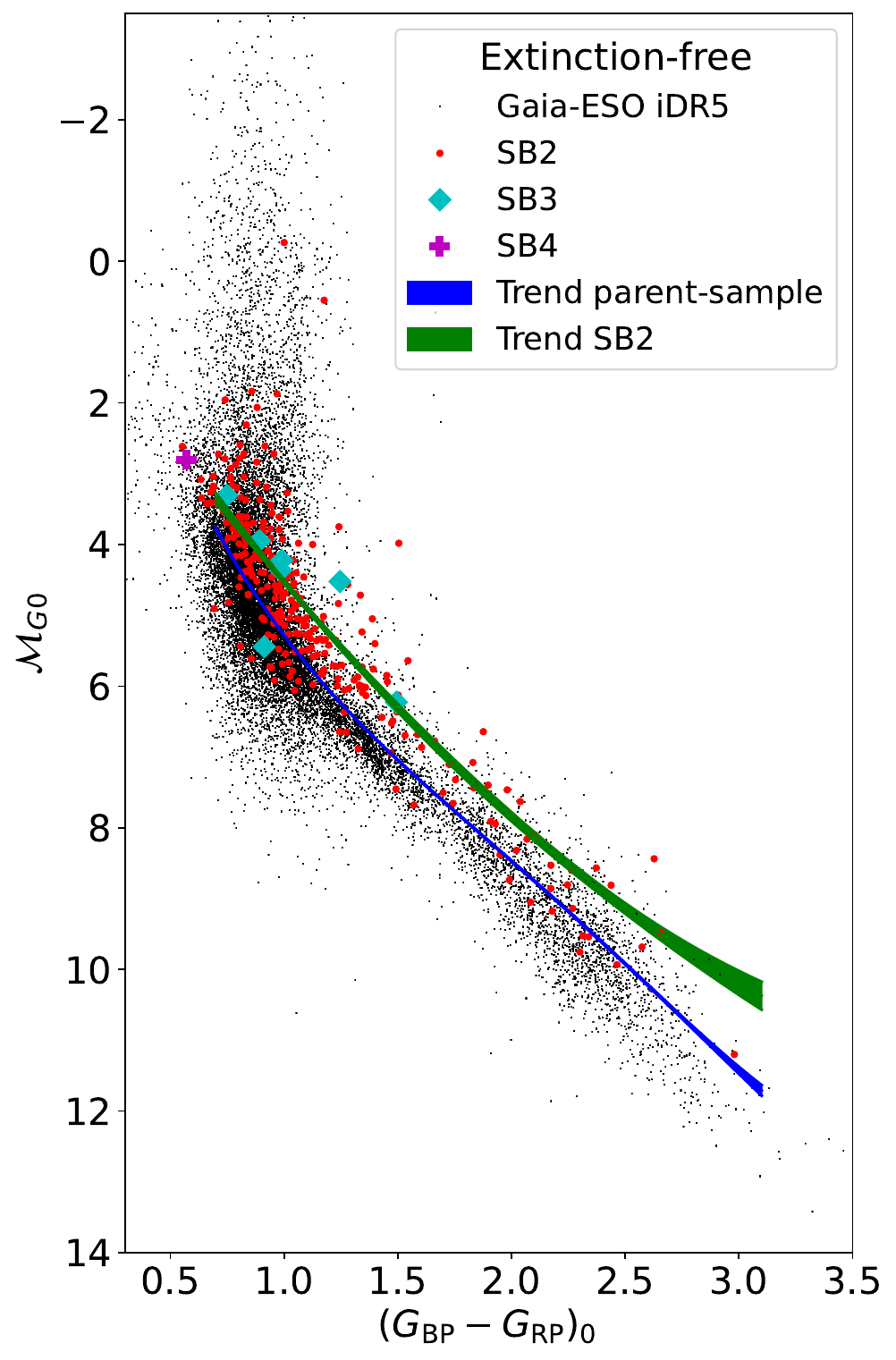}
  \caption{\label{Fig:iDR5_SB2_SB3_SB4_CMD} Colour-magnitude diagram of the \Gaia-ESO parent-sample (black dot). Spectroscopic binaries are overplotted: SB2 (red disc), SB3 (cyan diamond), and SB4 (magenta plus). $\mathcal{M}_{\mathrm{G}}$ is the absolute magnitude in the $G$ band. The blue and green thick lines are respectively a quartic and quadratic fit of the rebinned parent-sample main-sequence and of the rebinned SB2 sequence. The width of the lines materialises the $1-\sigma$ uncertainty on the locus of these two trends only due to the uncertainties on the parallaxes and photometry; the thickness of these lines is therefore not related to the vertical scatter of the datapoints. Left: photometric quantities are not corrected for the extinction. Right: photometric quantities are corrected for the extinction by using $A_G$ and $\mathrm{E}(G_{\mathrm{BP}} - G_{\mathrm{RP}})$ from \Gaia DR3.}
\end{figure*}

{Figure~\ref{fig:cumul_histo_SB_VB} shows the cumulative distributions of \Gaia-ESO SB2 discussed in this study, of all SB from the \sbnine catalogue and of \Gaia ABs and SBs\footnote{\label{note:gaia_binaries}We consider \Gaia binaries constrained only by astrometry or only by spectroscopy, and not binaries characterised by two or more techniques.} \citep{GaiaDR3Arenou} as a function of their \Gaia distance. We see that nearly \SI{100}{\percent} of the \Gaia DR3 AB are detected within \SI{2}{\kilo\parsec} from the Sun while \SI{30}{\percent} of the \Gaia SB (\ie $\sim \num{65000}$ systems) are at a distance larger than \SI{2}{\kilo\parsec} and up to \SI{8}{\kilo\parsec}. We note that the \Gaia-ESO SB2 also sample large distances: \SI{40}{\percent} of them are at a distance larger than \SI{2}{\kilo\parsec}; for distances larger than \SI{3.5}{\kilo\parsec}, the cumulative distribution of \Gaia-ESO SB2 and \Gaia SB overlap.}

{To summarise this section, the extinction-free \Gaia CMD shows that the SB2, SB3 and SB4 we detected in the iDR5 parent-sample tend to be brighter than the average magnitude of stars from the parent-sample in the same colour range: this photometric argument strengthens the spectroscopic detection. In particular, the sequence outlined by the SB2 lies \SI{0.78}{\mag} above the sequence of the bulk sample. The cross-match with \Gaia DR3 establishes that none of the SB2, SB3 and SB4 listed in this work are flagged as non-single star by the \Gaia main source catalogue. Thus, future ground-based spectroscopic surveys, like WEAVE or 4MOST, whose resolution is similar to the \Gaia-ESO GIRAFFE observations, will play an essential role in the profiling of the Milky Way spectroscopic binary population.}

\subsection{Comparison to \cite{2017A&A...608A..95M}}
\label{Sec:Comparison_Merle2017}

In this subsection, we compare the results regarding our SB2 detection to the results in \cite{2017A&A...608A..95M}. \cite{2017A&A...608A..95M} ran \doe on the \Gaia-ESO CCFs, while in the present study, we run \doe on the \nacre CCFs. The comparison of the SB2 reported here to those reported in \cite{2017A&A...608A..95M} will thus allow us to judge the efficiency's gain offered by the \nacre CCFs. However, the comparison is not straightforward because a) the present study focuses on the Milky Way field stars observed in HR10 and HR21 while \cite{2017A&A...608A..95M} analysed the whole \Gaia-ESO dataset and b) the present study relies on iDR5 data while \cite{2017A&A...608A..95M} used iDR4 data. {Hence an observation \emph{exhibiting a binary signature in the present work} will be considered for the comparison only if it fulfils the following criteria: a) the target was already part of the iDR4 release; b) the target is observed at least once with either HR10 or HR21; c) the target is a Milky Way field star (its field is thus labelled as \texttt{GES\_MW\_XX\_YY} in the \Gaia-ESO notation). We note that \num{116734} HR10 and HR21 spectra (\num{27344} objects) with a field \texttt{GES\_MW\_XX\_YY} are found in iDR4; when this study's iDR5 parent-sample is restricted to the MJDs covered by iDR4, it is downsized from \num{160727} spectra (\num{37565} objects) to \num{114850} (\num{27006} objects). The set of spectra is smaller than that of the fourth data-release since some targets were removed from the \Gaia-ESO target list between the two internal releases. \cite{2017A&A...608A..95M} identified \num{342} SB2 within the full iDR4, \ie including also observations obtained with other setups than GIRAFFE HR10 and HR21 and observations of cluster stars. In \cite{2017A&A...608A..95M}, \num{158} SB2 out of the \num{342} fall within our selection criteria (HR10 or H21, \texttt{GES\_MW\_XX\_YY}); but three objects, namely CNAME 09594300-4054056, 09594650-4059014 and 10004160-4053496, shall be removed since they are not in the iDR5 target list. As a result, from \cite{2017A&A...608A..95M}, we keep \num{187} HR10 and \num{152} HR21 observations exhibiting a binary signature, which corresponds to \num{155} SB2. We will refer to this sample as `iDR4 analysis'. And from this study, we keep \num{278} HR10 and \num{333} HR21 observations exhibiting a binary signature, which corresponds to \num{227} SB2. This sub-sample is seen as a re-analysis\footnote{We neglect the fact that the spectra available in iDR4 may have slightly changed in iDR5.} of iDR4 HR10 and HR21 spectra with the \nacre CCFs in lieu of the \Gaia-ESO CCFs. That being said, for the sake of clarity, we will now refer to this sample as `iDR4 re-analysis'.}

Taking the number of SB2 detections in both setups at face value clearly demonstrates that the use of the \nacre CCFs allows us to significantly increase the number of detections when we re-analyse the iDR4 sample. {The number of detections (in terms of observations) is increased by $611 / 339 \approx 1.8$, while the number of detected SB2 is multiplied by $227 / 155 \approx \num{1.5}$.} In particular, it is striking to see that the number of SB2 detections based on HR21 spectra is {more than} doubled thanks to our new way of computing the cross-correlation functions. {We note however that \num{29} of the \num{155} iDR4 SB2 are absent from the iDR4 re-analysis. These objects are characterised by observations with low \snr: \num{43} out of the \num{117} exposures available for these \num{29} targets have $\snr \leq 5$, \num{74} have $\snr \leq 10$. The good exposures, most of them obtained with HR21, can have \snr as high as \num{80} but a visual inspection of the spectra points at low metallicity and/or hot ($\ge \SI{6000}{\kelvin}$) objects. As a consequence, the strongest lines visible in HR21 spectra are the \ion{Ca}{II} triplet, \ie the lines that we discard to avoid excessively broad CCFs. The HR10 and HR21 CCFs of these metal-poor/hot objects tend to have large correlation noise, making the interpretation of secondary maxima as stellar component(s) difficult.}

\begin{figure*}
  \centering
  \includegraphics[height=0.85\columnwidth]{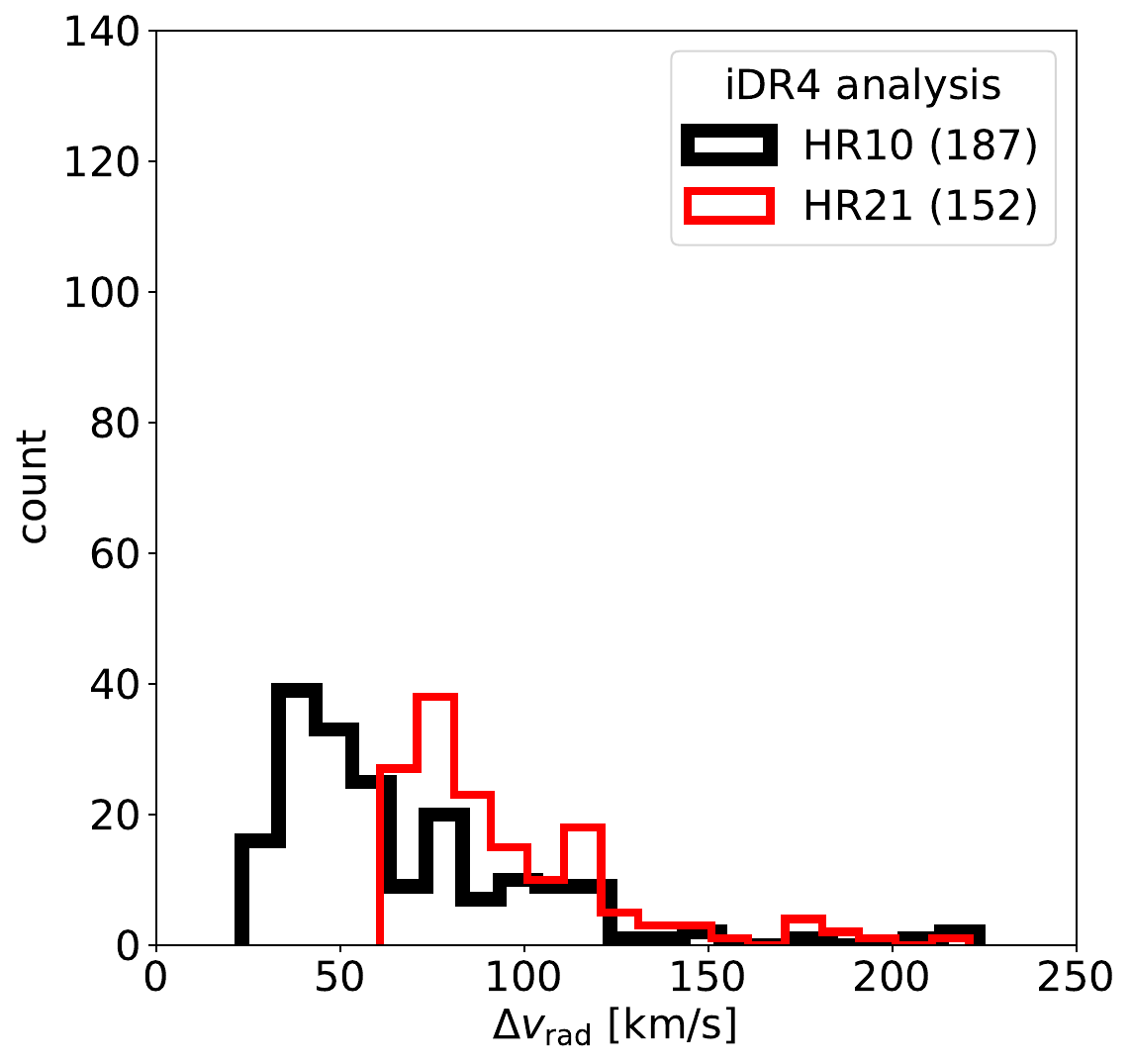}
  \includegraphics[height=0.85\columnwidth]{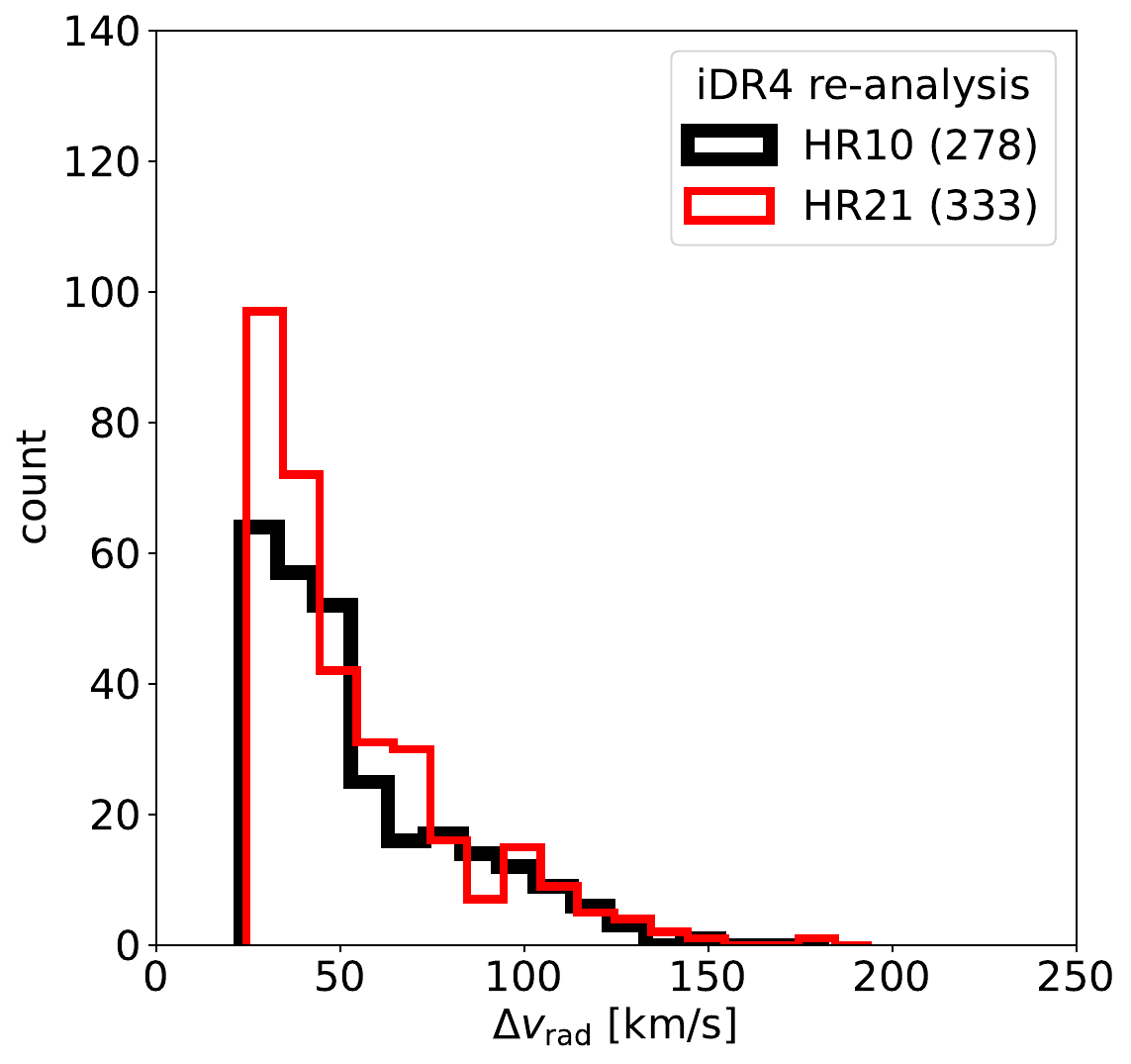}
  \caption{\label{Fig:deltaV_distribution_iDR4_iDR4re} $\Delta v_{\mathrm{rad}} $ distributions for HR10 (black thick line) and HR21 (red thin line) observations triggering an SB2 detection. Left: earliest iDR4 analysis described in \cite{2017A&A...608A..95M}; right: iDR4 re-analysis (this study). We are interested in the individual exposures, therefore an SB2 with more than one epoch will be counted more than once and may appear in different bins depending on the change of its components velocities from one epoch to another.}
\end{figure*}

Figure~\ref{Fig:deltaV_distribution_iDR4_iDR4re} shows the distribution (per observation) of the radial-velocity differences between the two components of SB2 CCFs for the iDR4 sample (left) and the re-analysed iDR4 sample (right). For the iDR4 sample, we note that the distribution for HR21 (red) starts at $\Delta v_{\mathrm{rad}} \sim \SI{60}{\kilo\metre\per\second}$ while the distribution for HR10 (black) starts at $\Delta v_{\mathrm{rad}} \sim \SI{25}{\kilo\metre\per\second}$. This was already noted in \cite{2017A&A...608A..95M}: the \Gaia-ESO CCFs for HR21 tend to be broader than their HR10 counterpart which hampers the detection of SB2 with small $\Delta v_{\mathrm{rad}}$. On the other hand, for the iDR4 re-analysis, we note that the $\Delta v_{\mathrm{rad}}$ distributions for HR10 and HR21 roughly start at the same value $\Delta v_{\mathrm{rad}} \sim \SI{25}{\kilo\metre\per\second}$. This fact also demonstrates the gain offered by the \nacre CCFs: we are now able to sample the same $\Delta v_{\mathrm{rad}}$ with both GIRAFFE setups. The HR10 $\Delta v_{\mathrm{rad}}$ distribution of the iDR4 re-analysis resembles the one of the primitive iDR4 analysis, except that the bins are more populated. {This is another check that our \nacre masks return reliable velocities and do not introduce \emph{new} biases in the velocity measurements (\eg new bias that could have been caused by the limited number of lines in the \nacre masks).} The mean and standard deviation of the difference $\Delta_{\mathrm{4-5}} = v_{\mathrm{iDR4}} - v_{\mathrm{iDR5}}$ between the velocities of SB2 common to \cite{2017A&A...608A..95M} and the iDR4 re-analysis (this study) are: \SI{0.1}{\kilo\metre\per\second} and \SI{1.3}{\kilo\metre\per\second} for the first component and \SI{-0.04}{\kilo\metre\per\second} and \SI{1.8}{\kilo\metre\per\second} for the second component.

\begin{figure}
  \centering
  \includegraphics[width=0.85\columnwidth]{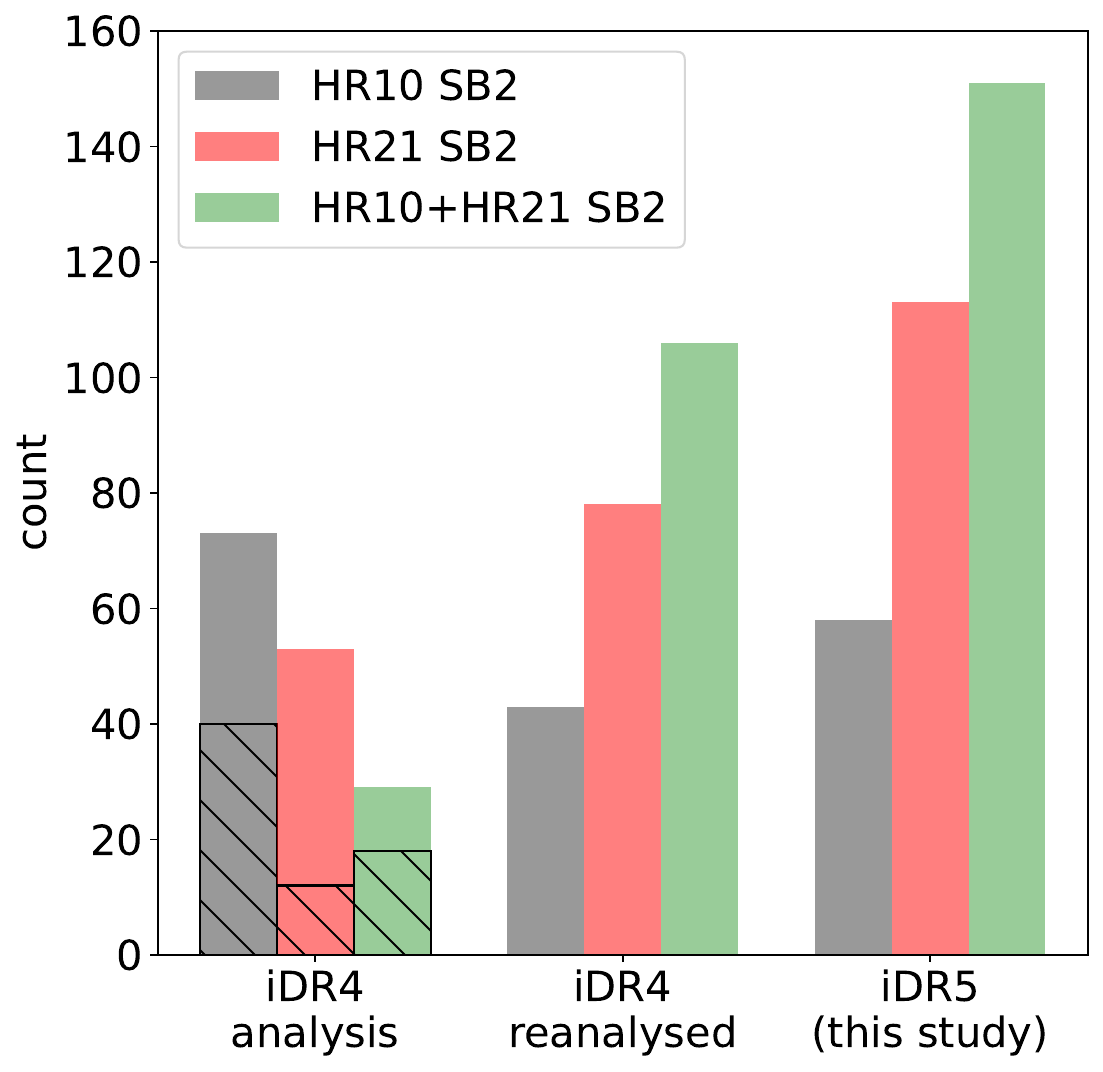}
  \caption{\label{Fig:SB2_counts_iDR4_iDR4re_iDR5} {SB2 count depending on the setup(s) triggering the SB2 detection. Black: `HR10 SB2', \ie SB2 detected from HR10 observations only. Red: `HR21 SB2', \ie SB2 detected from HR21 observations. Green: `HR10+21 SB2', \ie SB2 detected from both HR10 and HR21 observations. Left: earliest iDR4 analysis by \cite{2017A&A...608A..95M}. Middle: iDR4 re-analysis with \nacre masks. Right: iDR5 analysis (this study). In the left group, the hatched bars overplotted indicate the SB2 objects detected from HR10 only (resp., from H21 only; resp., from HR10 and HR21) in \cite{2017A&A...608A..95M} and that are now detected thanks to both setups in the iDR4 re-analysis. The difference of height between the hatched bar and that of the green bar is explained by about ten iDR4 SB2, formerly detected in both HR10 and HR21, and now detected only with one of the two setups (the best-mask CCF did not exist for \snr reasons in some cases).}}
\end{figure}

Figure~\ref{Fig:SB2_counts_iDR4_iDR4re_iDR5} compares the number of detected SB2 in {the earliest iDR4 analysis and its re-analysis}. {Here we look at three subgroups: `HR10 SB2' (resp., `HR21 SB2') contains SB2 only detected with HR10 (resp., HR21) observations, and `HR10+21 SB2' contains SB2 detected with both setups.} Once again, the comparison of the histogram for the iDR4 analysis and re-analysis shows that the use of the \nacre CCFs radically improves the outcome of the analysis. While in the iDR4 analysis, the largest (resp., smallest) number of detection occurred in the `HR10 SB2' group (resp., `HR10+21 SB2' group), it is the opposite for the re-analysis. It means that most of the discovered SB2 in the re-analysis have at least one detection in each setup, which implies that the new list of SB2 is more robust. The striking increase in the number of objects in the `HR10+21 SB2' group between the initial iDR4 analysis and its re-analysis is explained by the hatched bars: \num{41} `HR10 SB2' and \num{15} `HR21 SB2' in iDR4 have become `HR10+21 SB2' in the re-analysis. It demonstrates that the \nacre CCFs indeed increase the detection efficiency in the HR21 observations. We note that the \nacre CCFs also slightly increase the detection efficiency in HR10 observations.

\subsection{SB2 detection rate}

{Using the Ninth Catalogue of Spectroscopic Binaries (\sbnine) by \cite{2004A&A...424..727P,2009yCat....102020P}, we attempt to estimate the SB detection rate given the quality (mainly, resolution and \snr) of the \Gaia-ESO data and the observational strategy followed by \Gaia-ESO (\ie epochs at which \Gaia-ESO collected spectra). The \sbnine catalogue\footnote{\url{https://sb9.astro.ulb.ac.be/}} is built from heterogeneous sources: it is a compilation of the orbits of spectroscopic multiple systems published by various authors up to 2021. The latest version (2021-03-02) contains \num{1642} SB2 orbits corresponding to \num{1294} SB2 systems (some systems have several sets of orbital parameters). The \sbnine catalogue attempts to provide the spectral types and luminosity classes for each SB2 spectroscopic component. After parsing the columns giving the tentative spectral types and luminosity classes, we find that \num{788} SB2 have at least one of the two components being a FGK (set~\num{0}) star and the remaining SB2 systems have uncharacterised components or both components have a spectral class different than FGK (\eg O star, white dwarf, Wolf-Rayet star). Among the \num{788} SB2 of set~\num{0}, \num{326} (\num{324} with full set of orbital parameters) SB2 have a FGK main-sequence star as the primary component (set~\num{1}). \num{108} of \num{326} systems have a secondary component that is also FGK. Finally, \num{54} (\num{52} with full set of orbital parameters) SB2 are (quasi-)twin stars (set~\num{2}). For some systems, more than one orbit is reported by \sbnine; in that case, we simply keep the first orbit in the list since our aim is to have a realistic set of SB2 orbits and not the \emph{most likely} orbit of the system under consideration. In this section, we use both the set \num{1} and \num{2} to get two different estimates of the SB2 detection rate within \Gaia-ESO iDR5.}

{Our input \Gaia-ESO sample mainly comprises FGK dwarf and giant stars; however, the locus of the \Gaia-ESO SB2 in the \Gaia CMD shows that the identified SB2 are on the SB2 main sequence. Therefore, for the test described in this section, we decide to use the SB2 listed in the \sbnine catalogue for which at least one component is an FGK main-sequence star, \ie the sets~\num{1} and \num{2}.}

{To estimate the detection rate of SB2 given the \Gaia-ESO observational constraints, we proceed as follows. We list all observed CNAMEs with all their corresponding epochs for the iDR5 parent sample. Thus, one or more epochs (generally, two and four) are attached to a given CNAME. We also associate to each epoch the \snr of the corresponding HR10 or HR21 spectrum. For a given set of $N$ \sbnine systems, we randomly associate $N$ CNAMEs (and their corresponding $M$ epochs and $M$ \snr). Then, for each of the $N$ \sbnine systems, we use their orbital parameters to compute the velocities of the two components at each of the $M$ \Gaia-ESO epochs. For each epoch, we compute the radial velocity $\Delta v = v_2 - v_1$ difference between the two components. If $|\Delta v| > \SI{30}{\kilo\metre\per\second}$ and $\snr > 5$ and $|v_{1,2}| < \SI{250}{\kilo\metre\per\second}$, then we assume a detection. If one of the condition is not valid, we assume that we miss the detection of the two individual components. These criteria correspond to the limitations due to the spectral resolution of the \Gaia-ESO spectra or due to our filters to assess the quality of our CCFs for false-positive or very uncertain detections. We emphasise that these simulations only test the SB2 detection from a kinematical aspect (the check is made on the velocity difference) and not from a spectroscopic aspect: we do not generate composite spectra to verify whether the two components are detectable at the same wavelength ranges (\ie at the HR10 and HR21 wavelengths). This choice is partly constrained by the fact that the stellar components of the \sbnine systems are not always fully characterised, preventing us from computing the corresponding synthetic spectrum of a given system. If we restrict ourselves to fully-characterised systems formed of two main-sequence stars (not necessarily (quasi-)twin stars), we are left with only \num{88} systems, which will result in a more imprecise detection rate than what can be achieved with set~\num{1}. We repeat the simulations \num{501} times which is enough to ensure a numerical convergence of our estimated percentages on the unit digit. Since the set of \Gaia-ESO CNAMEs is much larger than the set 1 and 2 of SB2 systems, it is equivalent to run the simulations spectral type by spectral type or with all spectral types together. For the sake of simplicity, we perform separated simulations to estimate per-spectral-type rates and we perform simulations on the full sets (1 or 2) for global rates.}

\begin{figure*}
  \centering
  \includegraphics[width=0.82\columnwidth]{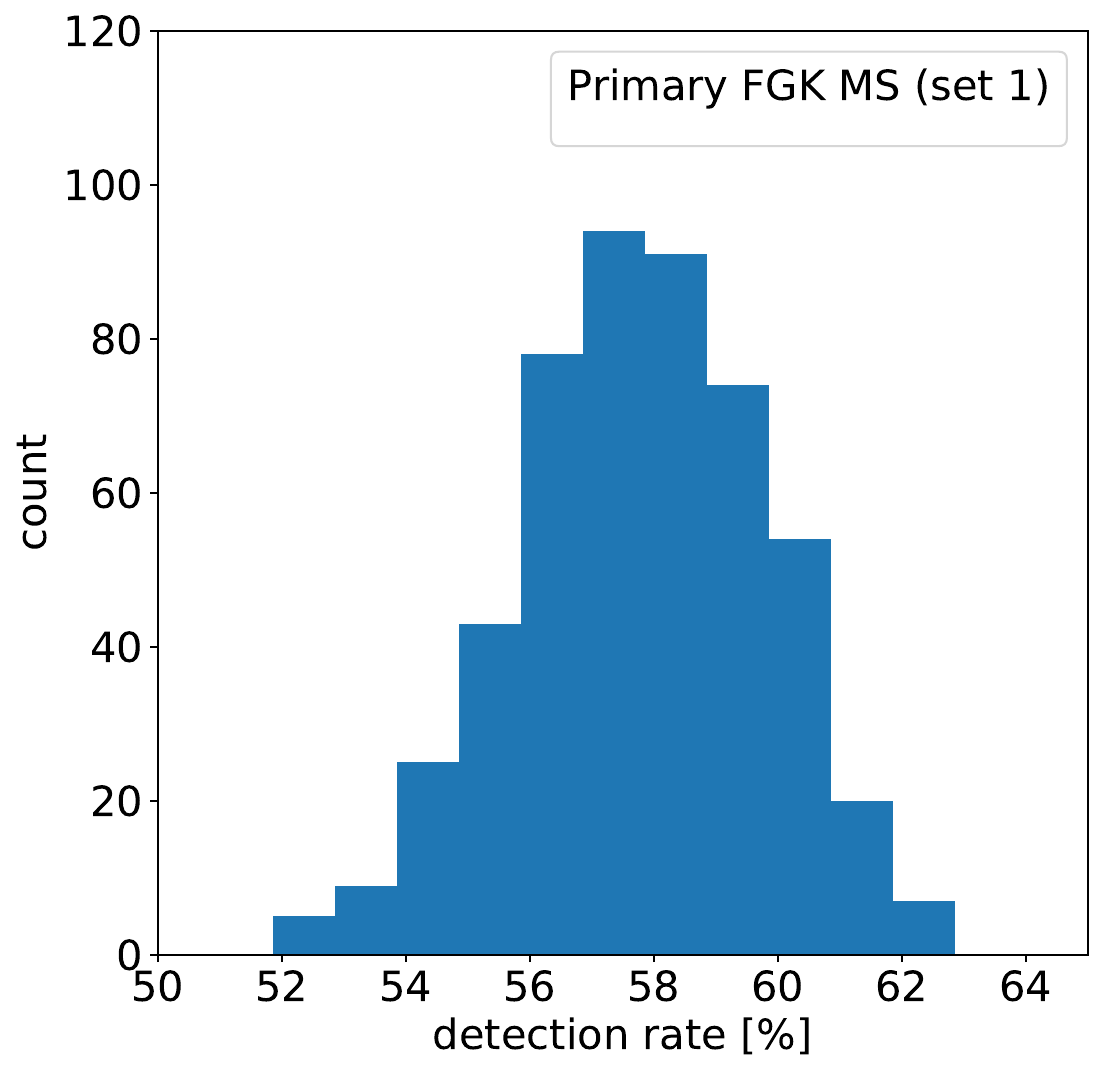}
  \includegraphics[width=0.85\columnwidth]{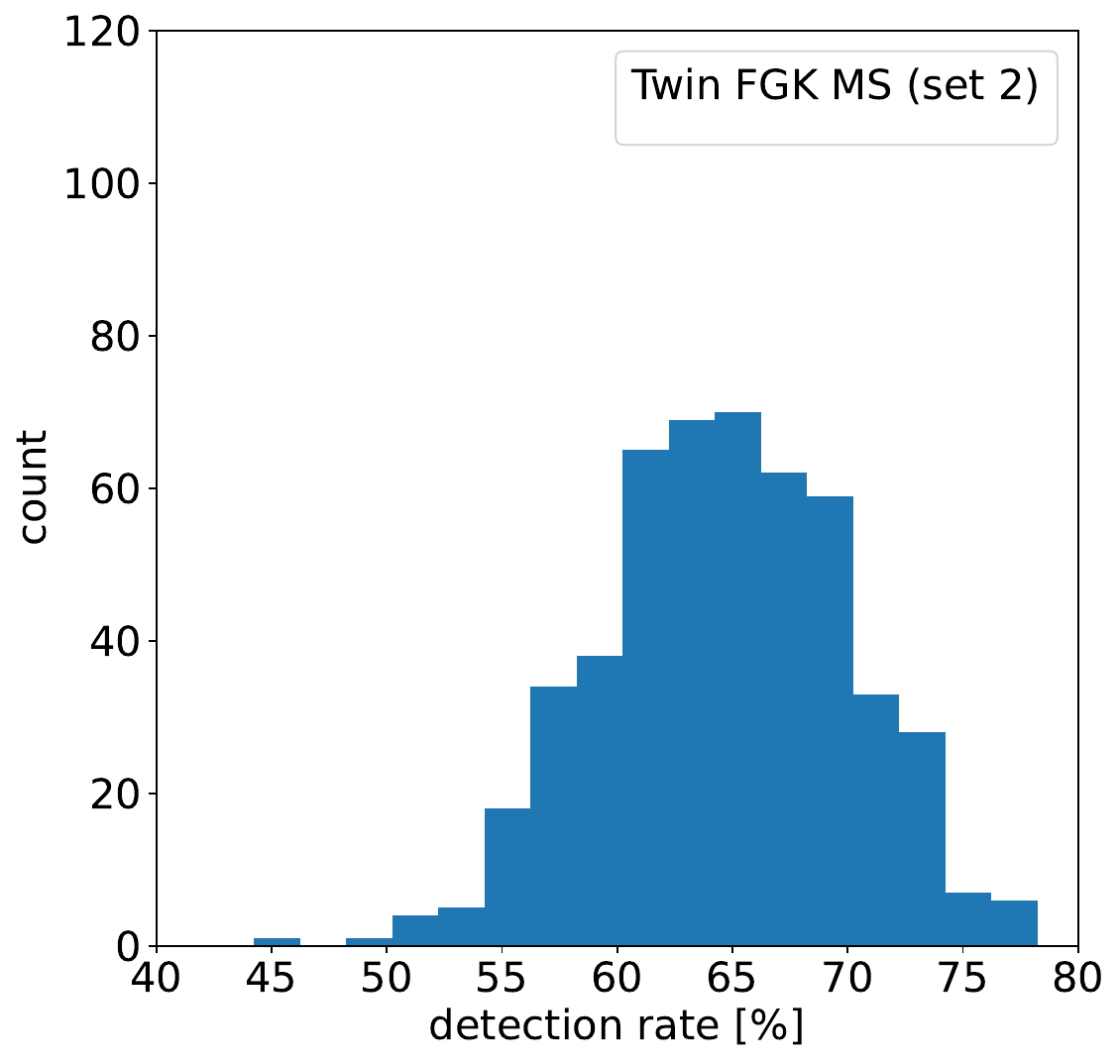}
  \caption{\label{Fig:Detection_rate_spectral_type} Distribution of the detection rates for the set~\num{1} (left; $\mathrm{binwidth} = \SI{1}{\percent}$) and \num{2} (right; $\mathrm{binwidth} = \SI{2}{\percent}$). The $y$-axis gives the count of simulations with a detection rate within the range defined by the bin coordinates. For instance, for the set~\num{1}, $\approx \num{94}$ out of \num{501} simulations have a detection rate between $\approx \SI{57}{\percent}$ and $\approx \SI{58}{\percent}$.}
\end{figure*}

{Figure~\ref{Fig:Detection_rate_spectral_type} shows the distribution of the detection rate for each set (1 and 2) after the \num{501} simulations. We estimate a mean detection rate of $\num{57.8} \pm \SI{2.0}{\percent}$ from the set~\num{1} and of $\num{64.8} \pm \SI{5.2}{\percent}$ from the set~\num{2}. If we apply these rates to the number of detected SB2 in our \Gaia-ESO sample, we find a number of SB2 between \num{497} and \num{557}. Given the input \Gaia-ESO sample contains about \num{37565} targets, we derive an SB2 frequency (among the field stars) in the range $[0.0132, 0.0148]$.}

{Combining the results from \citet{2020A&A...635A.155M} and from this study, it is possible to compute the spectroscopic-binary-star fraction $F_{\mathrm{binary}}$ and the spectroscopic-binary frequency $f_{\mathrm{binary}}$. \citet{2020A&A...635A.155M} found \num{803} SB1 (their selection with a $3-\sigma$ confidence) among about \num{43500} \Gaia-ESO targets. Their SB1 is downsized to \num{581} when restricted to the parent sample used in this study. Applying their detection rate of \SI{25}{\percent}, we obtain a theoretical number of \num{2324} SB1. On the other hand, the theoretical number of SB2 is \num{528} when we apply an average detection rate of \SI{61}{\percent} to the number of detected SB2. This yields a spectroscopic-binary-star fraction, expressed in percentage, of $F_{\mathrm{binary}} \approx \SI{7.6}{\percent}$. This value is smaller than the binary-star fraction of \SI{30}{\percent} quoted by \citet{2017ApJS..230...15M} for solar-type stars (their Table~13); this can be understood by the fact that their statistics is not restricted to spectroscopically-detected binaries, while we are.}

\subsection{Mass-ratio distribution}
\label{Sec:Mass_ratio}

{The time series of radial-velocity pairs for the detected SB2s can be used in order to derive the mass ratio $q$ of the system. Indeed, the mass ratio $q$ is given by the absolute value of the slope of the $v_{\mathrm{s}}$ vs. $v_{\mathrm{p}}$ relation, where $v_{\mathrm{p}}$ and $v_{\mathrm{s}}$ are respectively the velocity of the primary and the secondary components (\eg \citealt{2017PASP..129h4201F}). Table~\ref{Tab:iDR5_SB2_details} shows that \num{148} SB2 have at least three epochs, most of them (\num{120}) having four epochs. To measure the slope of the $v_{\mathrm{s}}$ vs. $v_{\mathrm{p}}$ relation, we first need to assign correctly each epoch-velocity at its respective component and then perform a linear regression in the $v_{\mathrm{p}} - v_{\mathrm{s}}$ plane.}

{The height of the CCF peak of each component can be a help to follow each component in the CCFs along time. We remind that the selection of the \nacre best mask does not rely on a spectral matching criterion but on a contrast criterion. As a consequence, there is no a priori assurance that the same mask is chosen as best mask for all of the epochs of a given SB2. And indeed, it is empirically confirmed in Table~\ref{Tab:iDR5_SB2_details} where we note that alphaCenA is picked as the best mask for the vast majority of HR10 observations, while betaAra is picked as the best mask for all of the HR21 observations. To solve this issue, for a given object and a given epoch, we compute the average $K_i$ of the heights $H_{i,j}$ of the peak $i$, measured in the $j = 1 .. N_{\mathrm{good}}$ CCFs used for the analysis: the peak with the largest average height gets numbered 1, the other is numbered 2. We repeat this operation for all of the epochs of a given SB2 in order to have a series of velocity $v_{\mathrm{rad,1}}$ (resp., $v_{\mathrm{rad,2}}$) tracking the same stellar component. At this stage, we use the numbering (component 1, component 2) and not the labelling (primary, secondary) since this procedure is not sufficient to pinpoint the primary component (\ie the most massive in order to have $q < 1.$). Table~\ref{Tab:iDR5_SB2_details} gives the average height of the first-component' peak $K_1 = \langle k_1 \rangle_{\mathrm{CCF}}$, the normalised average heights $k_i = K_i / K_1$. By construction, $k_1$ is equal to 1 all the time, except when we manually correct the component assignation (see below). Though we do not make use of those quantities in the present study, Table~\ref{Tab:iDR5_SB2_details} gives also the height of the first-component peak $H_1$, the normalised heights $h_i = H_i / H_1$, and the widths $w_i$ of the components' peak, all measured in the best-mask CCF. We then inspect the automatic assignations to identify possible errors; this is specified in the column 'mode' of Table~\ref{Tab:iDR5_SB2_details}. For instance, with the automatic ordering, the component 1 of 00410028-3707024 would see its velocity jumping from \SI{21.9}{\kilo\metre\per\second} to \SI{45.5}{\kilo\metre\per\second} in about \SI{26}{\minute}, hence an acceleration of \SI{54.6}{\kilo\metre\per\second\tothe{2}}, at odds with what is observed for the other systems. For 02035132-4215039, we think that the automatic ordering is correct: we have $v_{\mathrm{rad},1} = \SI{43.61}{\kilo\metre\per\second}$ and $v_{\mathrm{rad},2} = \SI{-21.21}{\kilo\metre\per\second}$ at $\mathrm{MJD} = 56912.325754$, and $v_{\mathrm{rad},1} = \SI{-4.29}{\kilo\metre\per\second}$ and $v_{\mathrm{rad},2} = \SI{36.71}{\kilo\metre\per\second}$ at $\mathrm{MJD} = 56968.141005$. We could be tempted to swap the velocities for one of the two epochs but this is hardly justified because the two epochs are \num{56} days away and $k_2$ is always significantly lower than 1. There are also doubtful cases like 03103980-5007403. For this system, the automatic ordering assigns the velocities $v_{\mathrm{rad},1} = \SI{39.97}{\kilo\metre\per\second}$ and $v_{\mathrm{rad},2} = \SI{-12.89}{\kilo\metre\per\second}$ at $\mathrm{MJD} = 56310.061037$, and $v_{\mathrm{rad},1} = \SI{-14.02}{\kilo\metre\per\second}$ and $v_{\mathrm{rad},2} = \SI{42.02}{\kilo\metre\per\second}$ at $\mathrm{MJD} = 56312.038016$. The two epochs are about two days away, hence an acceleration of \SI{-1.14}{\kilo\metre\per\second\tothe{2}} with the automatic assignation. If we were to manually swap the velocities for one of the two epochs, the acceleration would become \SI{0.04}{\kilo\metre\per\second\tothe{2}}. Both are plausible accelerations: the former would still be smaller (in absolute value) than the acceleration seen for the component 1 of 07205928-0033063 between $\mathrm{MJD} = 55998.069876$ and $\mathrm{MJD} = 55998.087887$ ($k_2$ is significantly different from 1 to trust the classification); the latter would be similar to what is seen for the component 1 of 05190040-5402093 between $\mathrm{MJD} = 56312.097403$ and $\mathrm{MJD} = 56314.091117$ (same remark on its $k_2$). Besides, for 03103980-5007403, $k_2 \approx 0.9$, so too close to 1 to help in the decision making.}

\begin{figure}
  \centering
  \includegraphics[width=0.85\columnwidth]{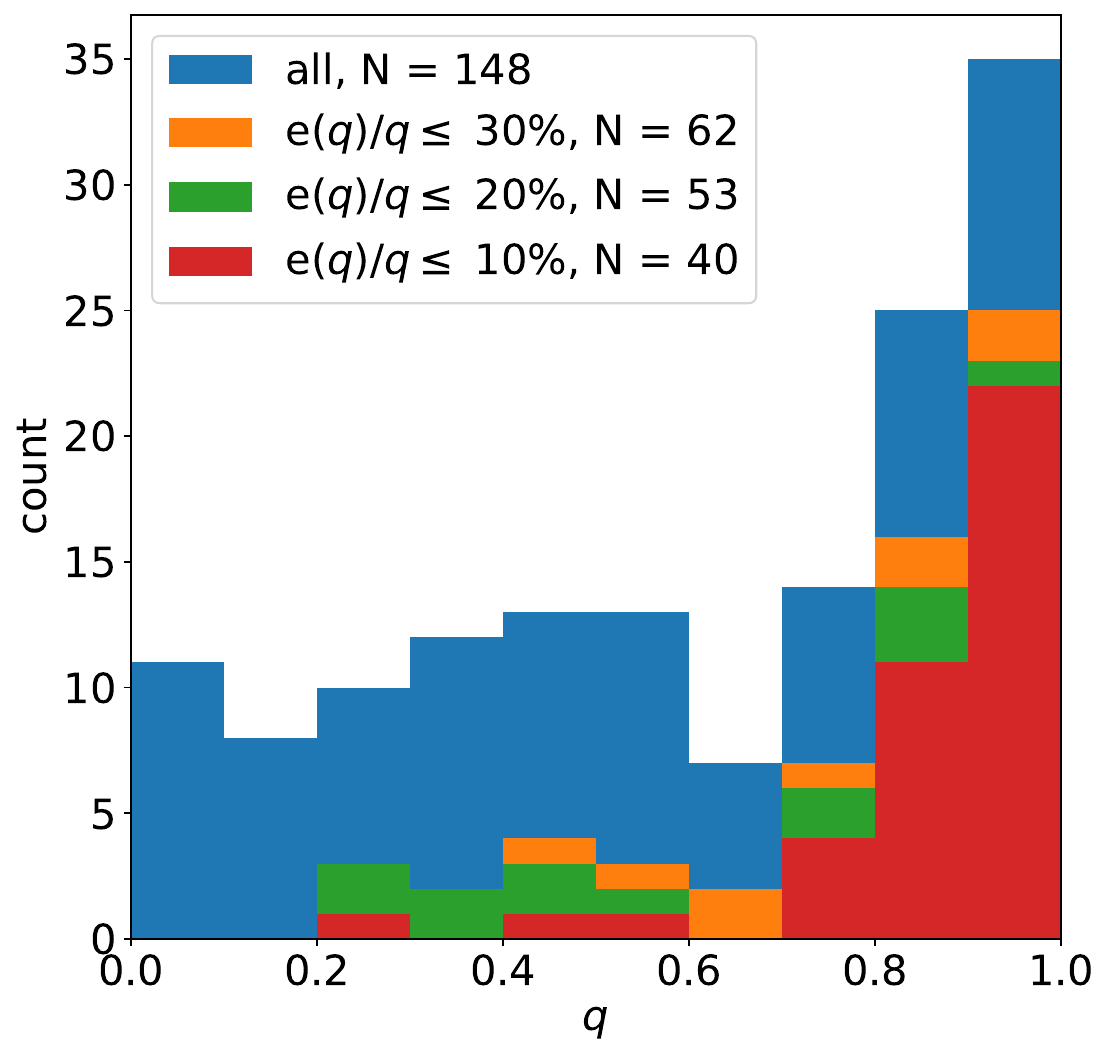}
  \caption{\label{Fig:Mass_ratio_distribution} Mass ratio ($q$) distribution for the iDR5 SB2 sample. Blue: whole sample of \num{148} $q$ estimates; orange: only systems with a relative uncertainty on $q$ lower than \SI{30}{\percent}; green: only systems with a relative uncertainty on $q$ lower than \SI{20}{\percent}; red: only systems with a relative uncertainty on $q$ lower than \SI{20}{\percent}. The number of systems for each cut is indicated in the legend box. {The four histograms are superimposed; thus, for instance, the visible green part of the bars only shows the system with a relative uncertainty on $q$ between \SI{10}{\percent} and \SI{20}{\percent}.}}
\end{figure}

{We now assume that $v_{\mathrm{p}} = v_{\mathrm{rad},1}$ and $v_{\mathrm{s}} = v_{\mathrm{rad},2}$, and perform the linear regression with the help of the scipy routine \emph{scipy.odr.ODR}, which allows to propagate the uncertainties on both coordinates. We remind here that the radial velocities are adopted from the best-mask CCF. The \emph{final} statistical uncertainty on the radial velocity is taken as the standard deviation of the velocities measured on all of the masks and corrected for small statistics using the Student's distribution. We force a minimal uncertainty of \SI{1}{\kilo\metre\per\second}. The absolute value of the slope gives the mass ratio $q$ of the system. If $q$ is larger than unity, then we assume that we made a mistake by considering that the component 1 should play the role of the primary component: we then swap the roles and redo the linear regression. Figure~\ref{Fig:Mass_ratio_distribution} displays the distribution of the mass ratios for \num{148} systems of the iDR5 SB2 sample; the different colors correspond to different levels of relative uncertainty: the blue histogram displays all \num{148} systems, while the red histogram displays only the \num{40} best constrained $q$ (with a relative uncertainty below \SI{10}{\percent}). The mean {mass-ratio} uncertainty $\mathrm{e}(q)$ on $q$ is \num{0.24} and the median uncertainty is \num{0.19}. However, $\mathrm{e}(q)$ can be as low as $\approx 0$: this indicates an ill-constrained regression and/or unrealistic errors on the radial velocities. We recommend to use a minimal uncertainty of \num{0.2} on $q$ for all systems. The blue distribution (all estimates) shows a thick tail for $q \lessapprox 0.7$ and is skewed toward $q \approx 1$. The tail for small mass ratios quickly disappear when we impose a more and more stringent quality criterion -- a smaller and smaller relative uncertainty -- on the estimated $q$. The red distribution, that should contain the systems with the most robust estimate of $q$, has most of the $q$ in the range $[0.7, 1.]$: this is expected for SB2 made of FGK main-sequence stars and it is in good agreement with \eg \citet{2019MmSAI..90..359B} and \citet{2020Obs...140....1B}. Indeed, in order to detect the two stellar components in the same spectral range, the two stars should be of similar spectral types and luminosities, and therefore, they should have a similar mass, hence $q \approx 1$.} Moreover, the red histogram is consistent with the shift in luminosity measured in the \Gaia CMD: the uncovered SB2 population is likely dominated by systems with $q$ larger than \num{0.9}, producing a vertical magnitude shift close to \num{0.75}. We therefore advise the end-user to use the published mass-ratios $q$ after applying a filter on the relative uncertainty on $q$.

\section{Summary and conclusion}

{The main findings of this study are:}

\begin{itemize}
\item {we craft a dozen cross-correlating masks, called \nacre, to obtain CCFs as narrow as possible, and as such, to detect radial-velocity components with a radial-velocity difference as low as possible;}
\item {the analysis of \num{160727} \Gaia-ESO iDR5 GIRAFFE FGK spectra with the \nacre masks and the automated \doe extremum-finding tool leads to the detection of \num{322} SB2, ten SB3 and two tentative SB4;}
\item {\nacre masks and \doe performances are excellent: the smallest detectable $\Delta v_{\mathrm{rad}}$ is around \SI{25}{\kilo\metre\per\second} for both HR10 (optical range, $R = \num{21500}$) and HR21 (near-infrared range, $R = \num{18000}$) and so, $\Delta v_{\mathrm{rad}}$ is virtually insensitive to the spectral resolution and spectral range of the observations;}
\item {the performances are sensitive to the signal-to-noise ratio of the spectra: the \snr may however be as low as $3 - 4$ before preventing any detection;}
\item {as any cross-correlation-based technique, our technique relies on the existence of absorption lines in the spectra and is therefore sensitive to the spectral content: \nacre CCFs of metal-poor and hot objects tend to be difficult to interpret. Moreover, at a given \snr and for FGK targets, it is probably easier to detect an SB2 with HR10 than with HR21 since there are more (weak) metallic lines in the HR10 range than in the HR21 range for those stars;}
\item {this work shows that excluding pressure-broadened lines like the Paschen lines or resonant lines like the \ion{Ca}{II} triplet is key to improve the SB$n$ detection rate;}
\item {in the extinction-free \Gaia CMD, the uncovered SB2, SB3 and SB4 tend to be brighter than their counterpart of the parent-sample in the same colour range, advocating the multiple nature of these objects;}
\item {thanks to Monte-Carlo simulations, we estimate the true SB2 frequency to be in the range $[0.0132, 0.0148]$;}
\item {we find a mass-ratio distribution for the SB2 skewed toward $q = 1$, indicating that our SB2 sample is likely mainly made of main-sequence twin-star systems;}
\item {none of our SB2, SB3 and SB4 are reported as astrometric/eclipsing/spectroscopic binaries by \Gaia, establishing that ground-based spectroscopic surveys still hold an essential role in the \Gaia era to complete the census of binary stars in the Milky Way. In particular, the WEAVE and 4MOST massive spectroscopic surveys will work at a resolution similar to the GIRAFFE HR10 and HR21 resolutions. The present work can therefore find a future application in the context of these soon-to-start surveys.}
\end{itemize}

{A future publication will apply the \nacre masks and the approach described here to the final and public release of the \Gaia-ESO (namely, DR5, the fifth public data-release, corresponding to iDR6, the sixth internal data-release). In particular, this work will be extended to \Gaia-ESO HR15N observations, which targeted Milky Way stellar clusters.}

\section*{Acknowledgement}
\textbf{Institutional:}
MVdS thanks F.~Arenou for his reading of this article and his comments.
MVdS and SVE are supported by a grant from the Foundation ULB.
MVdS, LM and AB acknowledge partial support from Premiale 2016 MITiC and of the INAF funding for the WEAVE project.
MVdS, LM and CVV acknowledge support from the INAF Minigrant Checs.
TM is granted by the BELSPO Belgian federal research program FED-tWIN under the research profile Prf-2020-033\_BISTRO.
AJ is supported by FNRS-F.R.S. research project PDR T.0115.23.
GT acknowledges financial support from the Slovenian Research Agency (research core funding No. P1-0188).
FJE acknowledges financial support by ESA (SoW SCI-OO-SOW-00371).

\textbf{Consortium:}
Based on dataproducts from observations made with ESO Telescopes at the La Silla Paranal Observatory under programmes 188.B-3002, 193.B-0936, and 197.B-1074. These data products have been processed by the Cambridge Astronomy Survey Unit (CASU) at the Institute of Astronomy, University of Cambridge, and by the FLAMES/UVES reduction team at INAF/Osservatorio Astrofisico di Arcetri. These data have been obtained from the \Gaia-ESO Survey Data Archive, prepared and hosted by the Wide Field Astronomy Unit, Institute for Astronomy, University of Edinburgh, which is funded by the UK Science and Technology Facilities Council. This work was partly supported by the European Union FP7 programme through ERC grant number 320360 and by the Leverhulme Trust through grant RPG-2012-541. We acknowledge the support from INAF and Ministero dell’Istruzione, dell’Università e della Ricerca (MIUR) in the form of the grant "Premiale VLT 2012". The results presented here benefit from discussions held during the \Gaia-ESO workshops and conferences supported by the ESF (European Science Foundation) through the GREAT Research Network Programme.
This work has made use of data from the European Space Agency (ESA) mission Gaia (\url{https://www.cosmos.esa.int/gaia}), processed by the Gaia Data Processing and Analysis Consortium (DPAC, \url{https://www.cosmos.esa.int/web/gaia/dpac/consortium}). Funding for the DPAC has been provided by national institutions, in particular the institutions participating in the Gaia Multilateral Agreement.

\textbf{Databases:}
This research has made use of NASA’s Astrophysics Data System Bibliographic Services.
This research has made a heavy use of the SIMBAD database, the VizieR catalogue access tool and the cross-match service provided and operated at CDS, Strasbourg, France.
This work has made use of the VALD database, operated at Uppsala University, the Institute of Astronomy RAS in Moscow, and the University of Vienna.
This research has made use of the Washington Double Star Catalog maintained at the U.S. Naval Observatory.

\textbf{Softwares:}
This work has made use of python 3.x\footnote{\url{https://www.python.org}} \citep{vanrossum} and of the following python's modules: Astropy\footnote{\url{https://www.astropy.org}}, a community-developed core Python package and an ecosystem of tools and resources for astronomy \citep{2013A&A...558A..33A,2018AJ....156..123A,2022ApJ...935..167A}; matplotlib\footnote{\url{https://matplotlib.org/}} \citep{Hunter:2007}; numpy\footnote{\url{https://numpy.org/}} \citep{harris2020array}; scipy\footnote{\url{https://scipy.org/}} \citep{2020SciPy...NMeth}.

\bibliographystyle{aa}
\bibliography{binaries_mw_ges_dr5}

\clearpage

\appendix

\section{Cross-match of the SB2 sample and the \Gaia DR3}

\onecolumn
\begin{longtable}{@{}ccS[table-format=2.3]@{\,\( \pm \)\,}S[table-format=2.3]S[table-format=2.3]S[table-format=2.3]S[table-format=2.3]S[table-format=3.2]@{}}
  \caption{\label{Tab:iDR5_SB2_simple_list} Cross-match of the \Gaia-ESO SB2 sample and \Gaia DR3. Columns are: \Gaia parallax; magnitudes; \texttt{RUWE}; radial velocities.}\\
  \toprule
  {\Gaia-ESO CNAME} & {\Gaia source ID} & \multicolumn{2}{c}{$\varpi$} & {$G$} & {$G_{\mathrm{BP}} - G_{\mathrm{RP}}$} & {\texttt{RUWE}} & {$v_{\mathrm{rad}}$}\\
  & & \multicolumn{2}{c}{mas} & {\si{\mag}} & {\si{\mag}} & & {\si{\kilo\metre\per\second}}\\
  \midrule
  \midrule
  \endhead
  00040663-0101512	&	2449602745796628736	&	 0.657	&	0.133	&	17.802	&	 1.328	&	 1.001	&	   nan\\
  00041619-3659250	&	2307737811346103040	&	 2.845	&	0.122	&	18.038	&	 2.708	&	 1.060	&	   nan\\
  00195847-5423227	&	4923333999386498048	&	 0.945	&	0.411	&	15.577	&	 0.919	&	15.259	&	   nan\\
  00202300-5436167	&	4923128807324871808	&	 0.279	&	0.040	&	16.310	&	 0.728	&	 1.036	&	   nan\\
  00304621-0325305	&	2528600633323952384	&	 0.949	&	0.061	&	15.745	&	 1.311	&	 1.297	&	   nan\\
  00311681-5006169	&	4974352339645026944	&	 0.394	&	0.034	&	15.809	&	 0.782	&	 0.969	&	   nan\\
  00322605-4408092	&	4980384226073549952	&	 0.316	&	0.065	&	16.972	&	 1.092	&	 0.996	&	   nan\\
  00410028-3707024	&	5001051471262573312	&	 6.186	&	0.038	&	15.600	&	 2.389	&	 1.042	&	18.346\\
\bottomrule
\end{longtable}
\cleardoublepage
\twocolumn

\section{Velocity time-series for the SB2 sample}

\onecolumn
\footnotesize
\begin{landscape}
  \begin{longtable}{@{}ccS[table-format=2.3]@{\,\( \pm \)\,}S[table-format=2.3]S[table-format=2.3]@{\,\( \pm \)\,}S[table-format=2.3]@{\,\,}cc@{\,\,}c@{\,\,}cS[table-format=2.3]S[table-format=2.3]S[table-format=2.3]S[table-format=2.3]S[table-format=2.3]S[table-format=2.3]S[table-format=2.3]S[table-format=2.3]cc@{}}
    \caption{\label{Tab:iDR5_SB2_details} Velocity time-series for the iDR5 SB2 sample. Columns are: \Gaia-ESO CNAME; MJD (day); $v_{\mathrm{rad},1,2}$ the radial velocity of the component \num{1} and \num{2} in \si{\kilo\metre\per\second} and its $1-\sigma$ uncertainty; the first/second component assignation mode (A: automatic; M: manual; see Sec.~\ref{Sec:Mass_ratio}); the GIRAFFE setup; the best mask ID (1: alphaCenA; 2: betaAra; 3: muLeo: 4: Sun); $N_{\mathrm{good}}$ the number of good \nacre CCFs used for the analysis; the average CCF heights with $K_{1}$ being the absolute average height (average of all measured heights over $N_{\mathrm{good}}$ CCFs) of the first-component's CCF peak and $k_{1,2} = K_{1,2}$ being the normalised average heights of each component's peak (with the associated standard deviation around the mean); the best-mask CCF heights with $H_{1}$ being the absolute height of the first-component's CCF peak and $h_{1,2}$ being the normalised height of each component's peak; $w_{1,2}$ the width of each component's peak in \si{\kilo\metre\per\second}. For a given system, the observations are ordered by increasing MJD and each velocity column is meant to track a given component of the system. The first component is tentatively the most luminous component of the system. We force a minimal uncertainty of \SI{1}{\kilo\metre\per\second} on the velocities. $K_i$, $k_i$, $H_i$, $h_i$, $w_i$ are quantities obtained from the parameters of the Gaussian fit of the CCF peaks (see Sec.~\ref{Sec:CCF_analysis} and \ref{Sec:Mass_ratio}).}\\
    \toprule
    {\Gaia-ESO CNAME} & {MJD} & \multicolumn{5}{c}{velocities} & \multicolumn{3}{c}{Remarks on CCFs} & \multicolumn{5}{c}{Average CCF heights} & \multicolumn{5}{c}{Best-mask CCF heights and widths}\\
    \cmidrule(l){3-7} \cmidrule(l){8-10} \cmidrule(l){11-15} \cmidrule(l){16-20}
     & & \multicolumn{2}{c}{$v_{\mathrm{rad},1}$} & \multicolumn{2}{c}{$v_{\mathrm{rad},2}$} & {mode} & {setup} & {best} & {$N_{\mathrm{good}}$} & {$K_1$} & {$k_1$} & {std} & {$k_2$} & {std} & {$H_1$} & {$h_1$} & {$h_2$} & {$w_1$} & {$w_2$}\\
     & & \multicolumn{2}{c}{\si{\kilo\metre\per\second}} & \multicolumn{2}{c}{\si{\kilo\metre\per\second}} & & & & & & \multicolumn{2}{c}{} & \multicolumn{2}{c}{} & & & & {\si{\kilo\metre\per\second}} & {\si{\kilo\metre\per\second}} \\
    \midrule
    \midrule
    \endhead
    00040663-0101512	&	56197.087034	&	 28.83	&	 1.00	&	-20.96	&	 5.93	&	A	&	HR10	&	1	&	 5	&	 0.864	&	 1.000	&	 0.083	&	 0.409	&	 0.041	&	 0.883& 1.000	&	 0.339	&	15.430	&	15.200\\
    00040663-0101512	&	56197.107669	&	 30.28	&	 1.03	&	-22.99	&	 1.70	&	A	&	HR10	&	1	&	 7	&	 0.791	&	 1.000	&	 0.207	&	 0.553	&	 0.099	&	 0.861& 1.000	&	 0.401	&	16.110	&	18.790\\
    00040663-0101512	&	56205.144417	&	 68.48	&	 1.00	&	-63.47	&	 1.00	&	A	&	HR21	&	2	&	11	&	 1.082	&	 1.000	&	 0.023	&	 0.613	&	 0.056	&	 1.057& 1.000	&	 0.604	&	12.390	&	12.090\\
    00040663-0101512	&	56205.162404	&	 67.57	&	 1.00	&	-64.66	&	 1.00	&	A	&	HR21	&	2	&	 8	&	 1.077	&	 1.000	&	 0.011	&	 0.606	&	 0.064	&	 1.061& 1.000	&	 0.590	&	13.600	&	11.910\\
    00040663-0101512	&	56576.177839	&	-37.19	&	 1.00	&	 50.82	&	 1.31	&	A	&	HR10	&	1	&	 8	&	 0.927	&	 1.000	&	 0.022	&	 0.470	&	 0.047	&	 0.912& 1.000	&	 0.479	&	14.860	&	16.030\\
    \midrule
    00041619-3659250	&	56909.271801	&	  0.36	&	 1.00	&	-25.50	&	 1.00	&	A	&	HR21	&	2	&	 8	&	 0.895	&	 1.000	&	 0.016	&	 0.955	&	 0.085	&	 0.894& 1.000	&	 0.833	&	13.060	&	11.200\\
    \midrule
    00195847-5423227	&	56532.286507	&	 93.78	&	 1.00	&	152.50	&	 1.00	&	A	&	HR10	&	1	&	 8	&	 0.918	&	 1.000	&	 0.044	&	 0.964	&	 0.037	&	 0.904& 0.973	&	 1.000	&	13.790	&	15.680\\
    00195847-5423227	&	56532.307888	&	 94.11	&	 1.00	&	151.23	&	 1.00	&	A	&	HR10	&	1	&	 8	&	 0.922	&	 1.000	&	 0.057	&	 0.986	&	 0.050	&	 0.929& 0.967	&	 1.000	&	14.300	&	15.990\\
\bottomrule
  \end{longtable}
\end{landscape}
\normalsize
\cleardoublepage
\twocolumn

\section{Mass-ratios of SB2}

\onecolumn
\begin{longtable}{@{}cS[table-format=2.3]@{\,\( \pm \)\,}S[table-format=2.3]rc@{}}
  \caption{\label{Tab:iDR5_SB2_mass_ratios} List of mass-ratios for \num{148} SB2. Columns are: \Gaia-ESO CNAME; $q$ and its uncertainty; number of epochs $N_{\mathrm{epoch}}$; flag \texttt{component\_one\_is\_primary}. The uncertainty on $q$ results from the propagation of the uncertainties on the primary and secondary velocities through the linear regression. The last column indicates if the component 1 from Table~\ref{Tab:iDR5_SB2_details} is used as the primary component when performing the regression.}\\
  \toprule
  {\Gaia-ESO CNAME} & \multicolumn{2}{c}{$q$} & {$N_{\mathrm{epoch}}$} & {\texttt{component\_one\_is\_primary}}\\
  \midrule
  \midrule
  \endhead
  00040663-0101512	&	0.92	&	0.01	&	 5	&	False\\
  00595999-4700246	&	0.31	&	0.14	&	 6	&	False\\
  01000070-0100143	&	0.91	&	0.01	&	 4	&	False\\
  01194949-6003364	&	0.61	&	0.48	&	 4	&	True\\
  01301631-0320543	&	0.74	&	0.45	&	 4	&	True\\
  01390790-5403014	&	0.70	&	0.25	&	 4	&	True\\
  01393831-4648457	&	0.03	&	0.20	&	 4	&	True\\
  02005449-0055403	&	0.53	&	0.23	&	 4	&	False\\
\bottomrule
\end{longtable}
\cleardoublepage
\twocolumn

\section{Atlas of \Gaia-ESO SB3}
\label{SecApp:_atlas_SB3}

{In this section, we show the CCFs and the associated spectrum at all epochs for each of the detected SB3. In each figure, the left panel shows the \nacre (coloured lines) and \Gaia-ESO (dashed black line) CCFs, while the right panel shows the corresponding spectrum (zoomed on the flux range $[0.45, 1.1]$ and with the setup's full wavelength range displayed over three sub-panels). The displayed $x$- and $y$-ranges are the same for each kind of plot (CCF or spectra), such that the reader can compare directly two epochs. All the CCFs (\nacre or \Gaia-ESO) are normalised such that the highest peak has its maximum at 1. The legend box reminds the CNAME of the system, the setup and the elapsed time $\Delta \mathrm{MJD}$ since the first available epoch. The caption indicates the CNAME, epoch (MJD) and GIRAFFE setup, plus a numbering to ease the reading of the time-series. The CCFs are ordered by ascending MJD. According to the stellar waltz occurring in the examined system, one can follow the motion of the different stellar components along time: depending on the orbital phase, one can see one, two or three components.}
\cleardoublepage

\subsection{00195847-5423227}

\setlength\parindent{0cm}
\begin{minipage}{\textwidth}
  \centering
  \includegraphics[width=0.49\textwidth]{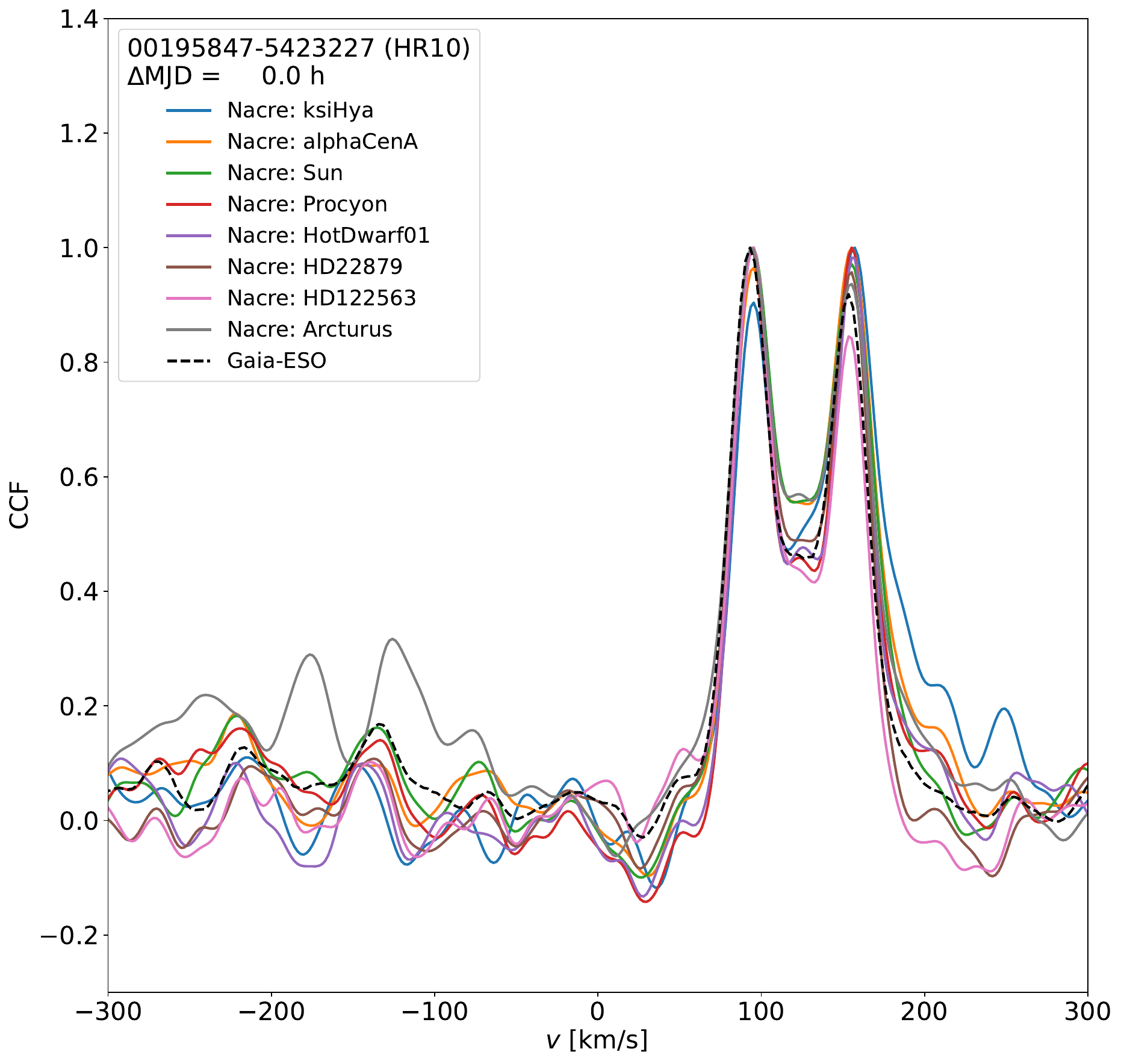}
  \includegraphics[width=0.49\textwidth]{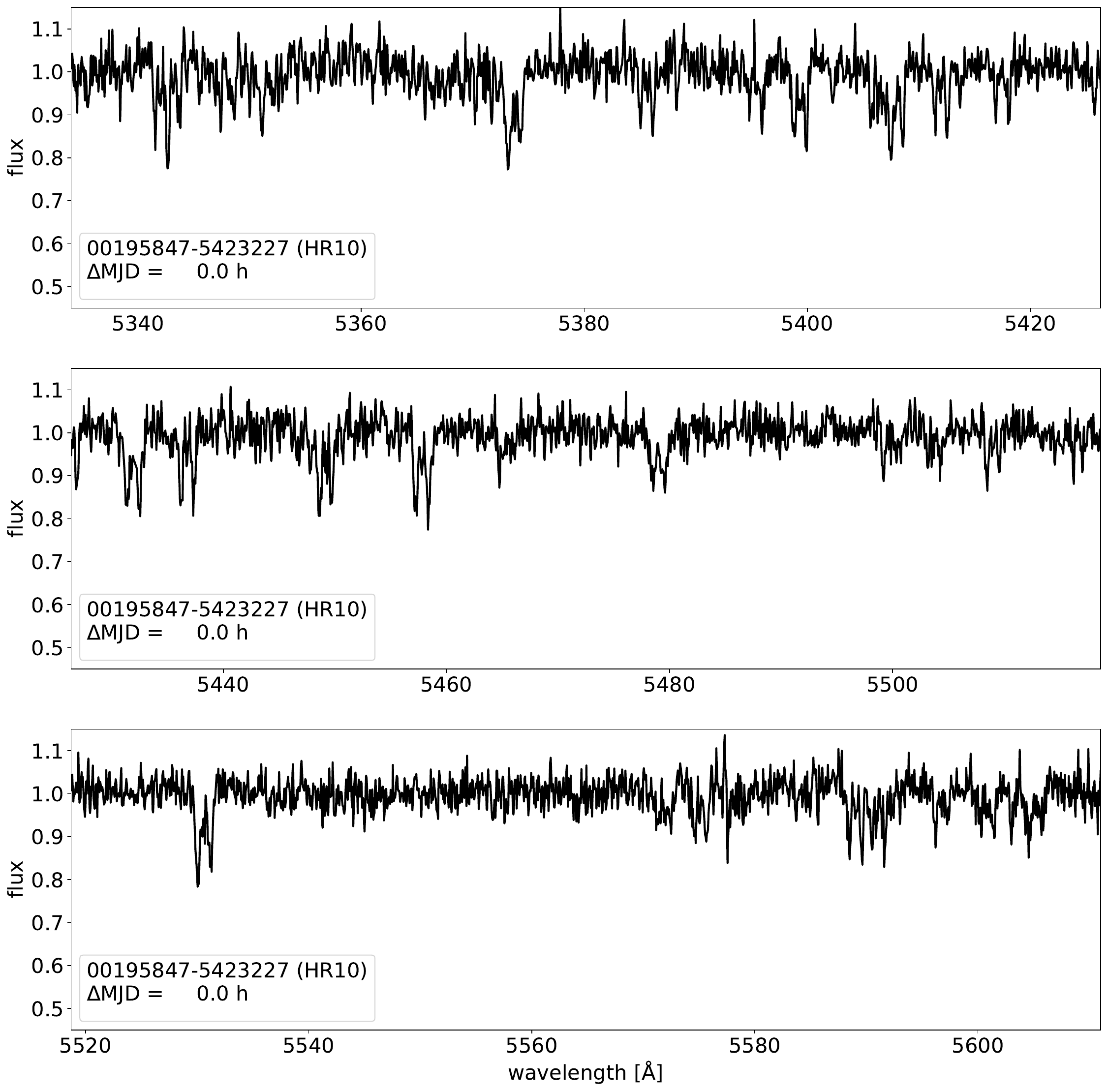}
  \captionof{figure}{\label{Fig:iDR5_SB3_atlas_00195847-5423227_1}(1/4) CNAME 00195847-5423227, at $\mathrm{MJD} = 56532.286507$, setup HR10.}
\end{minipage}
\begin{minipage}{\textwidth}
  \centering
  \includegraphics[width=0.49\textwidth]{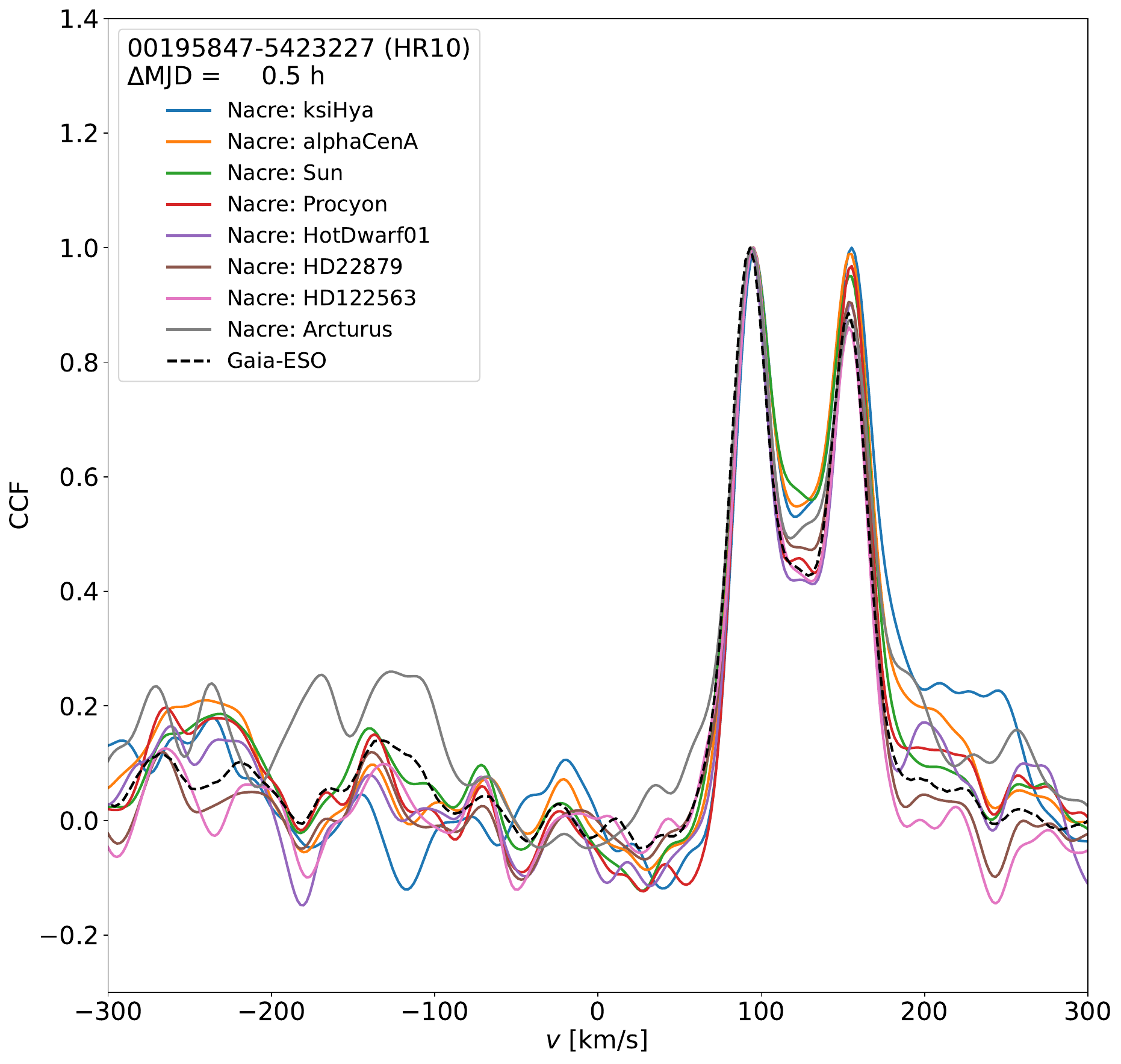}
  \includegraphics[width=0.49\textwidth]{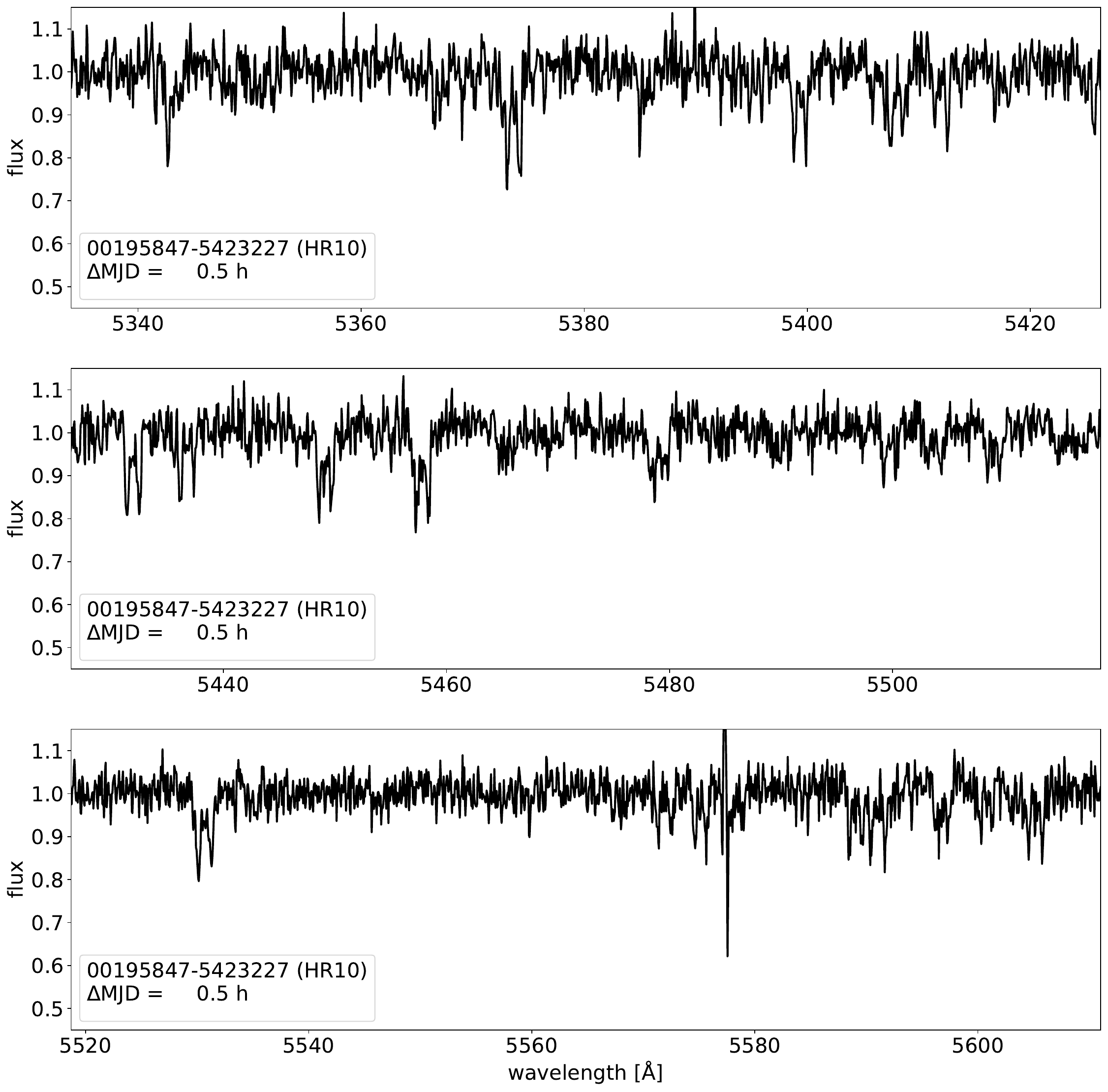}
  \captionof{figure}{\label{Fig:iDR5_SB3_atlas_00195847-5423227_2}(2/4) CNAME 00195847-5423227, at $\mathrm{MJD} = 56532.307888$, setup HR10.}
\end{minipage}
\clearpage
\begin{minipage}{\textwidth}
  \centering
  \includegraphics[width=0.49\textwidth]{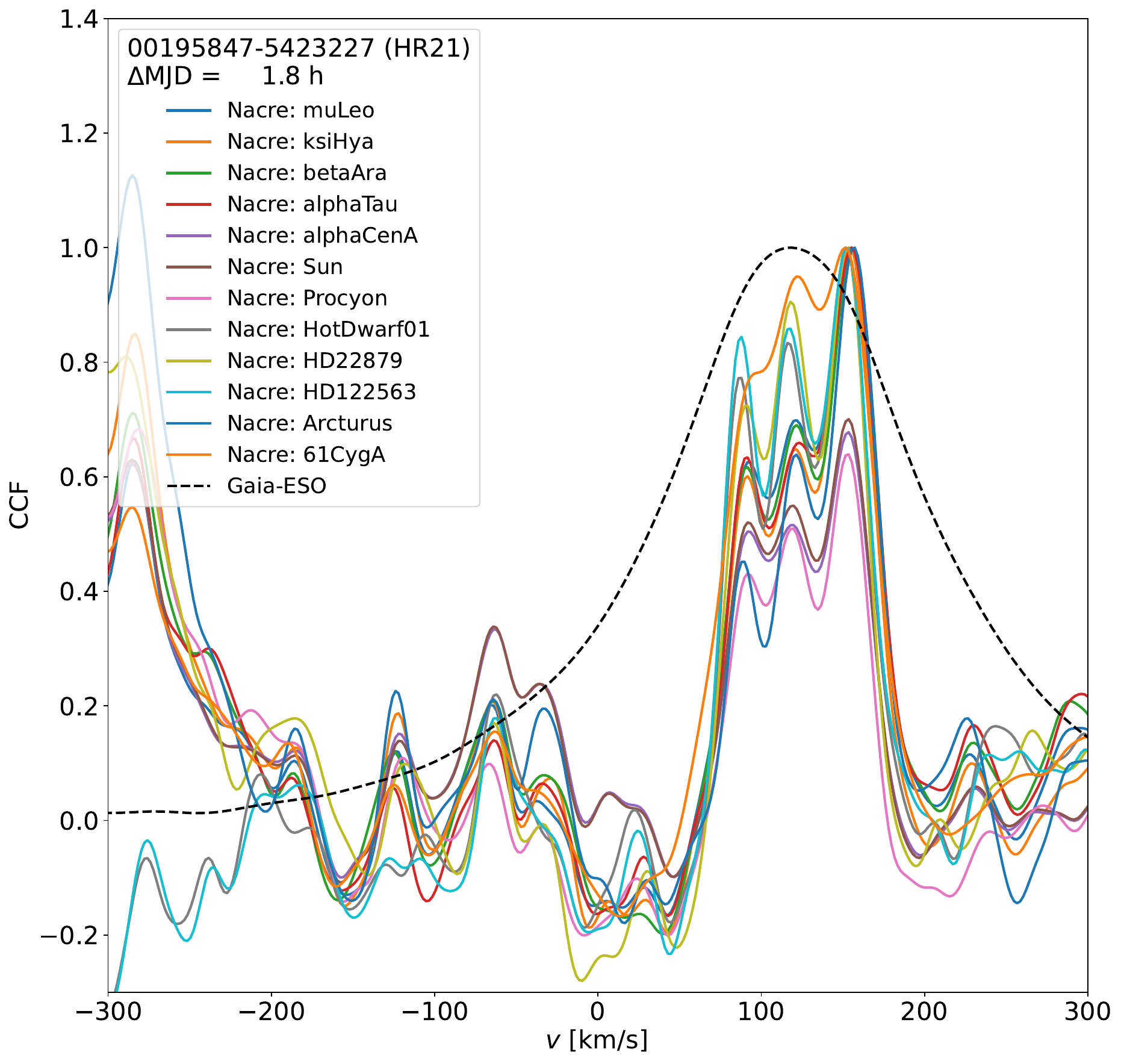}
  \includegraphics[width=0.49\textwidth]{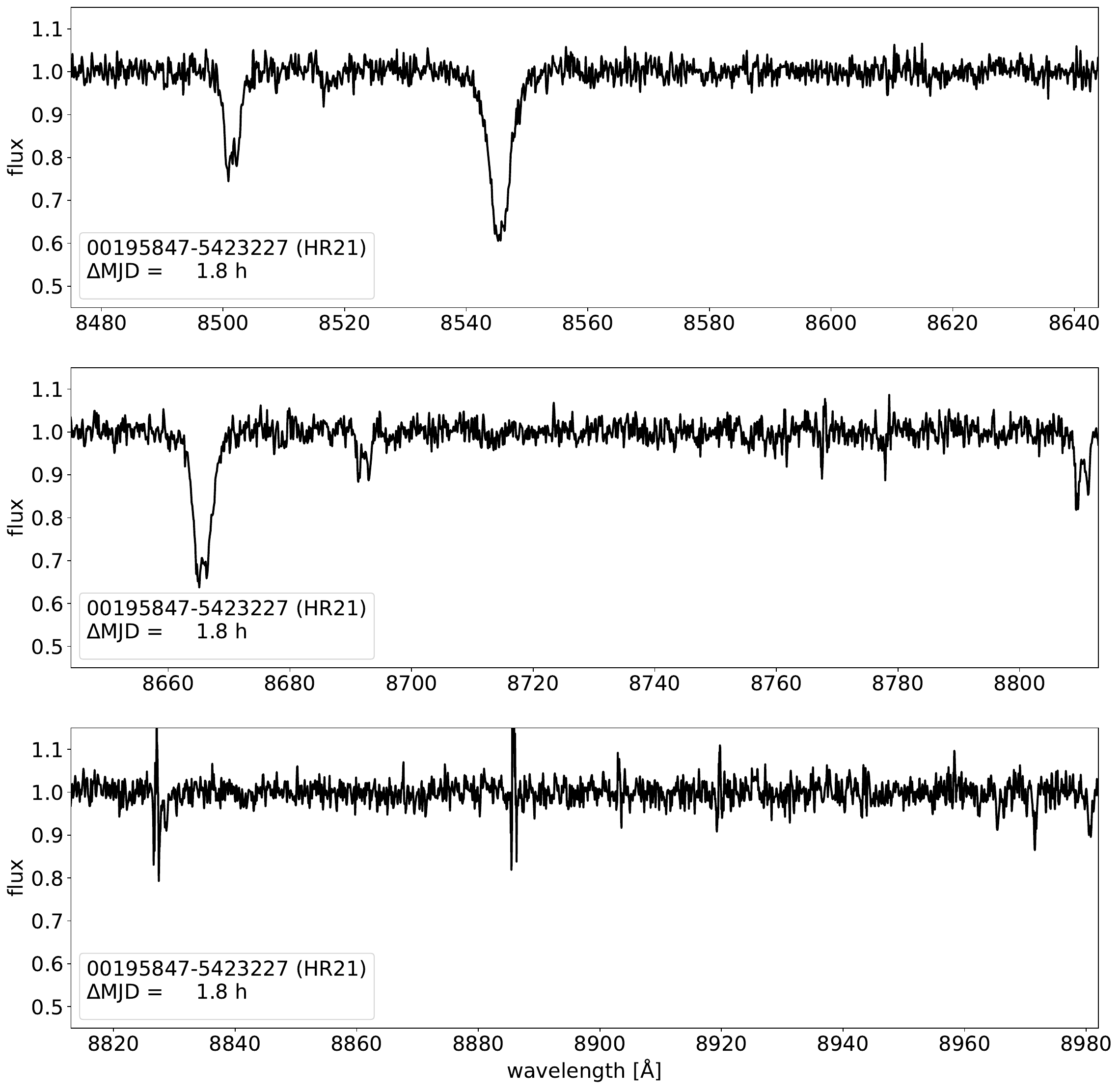}
  \captionof{figure}{\label{Fig:iDR5_SB3_atlas_00195847-5423227_3}(3/4) CNAME 00195847-5423227, at $\mathrm{MJD} = 56532.362038$, setup HR21.}
\end{minipage}
\begin{minipage}{\textwidth}
  \centering
  \includegraphics[width=0.49\textwidth]{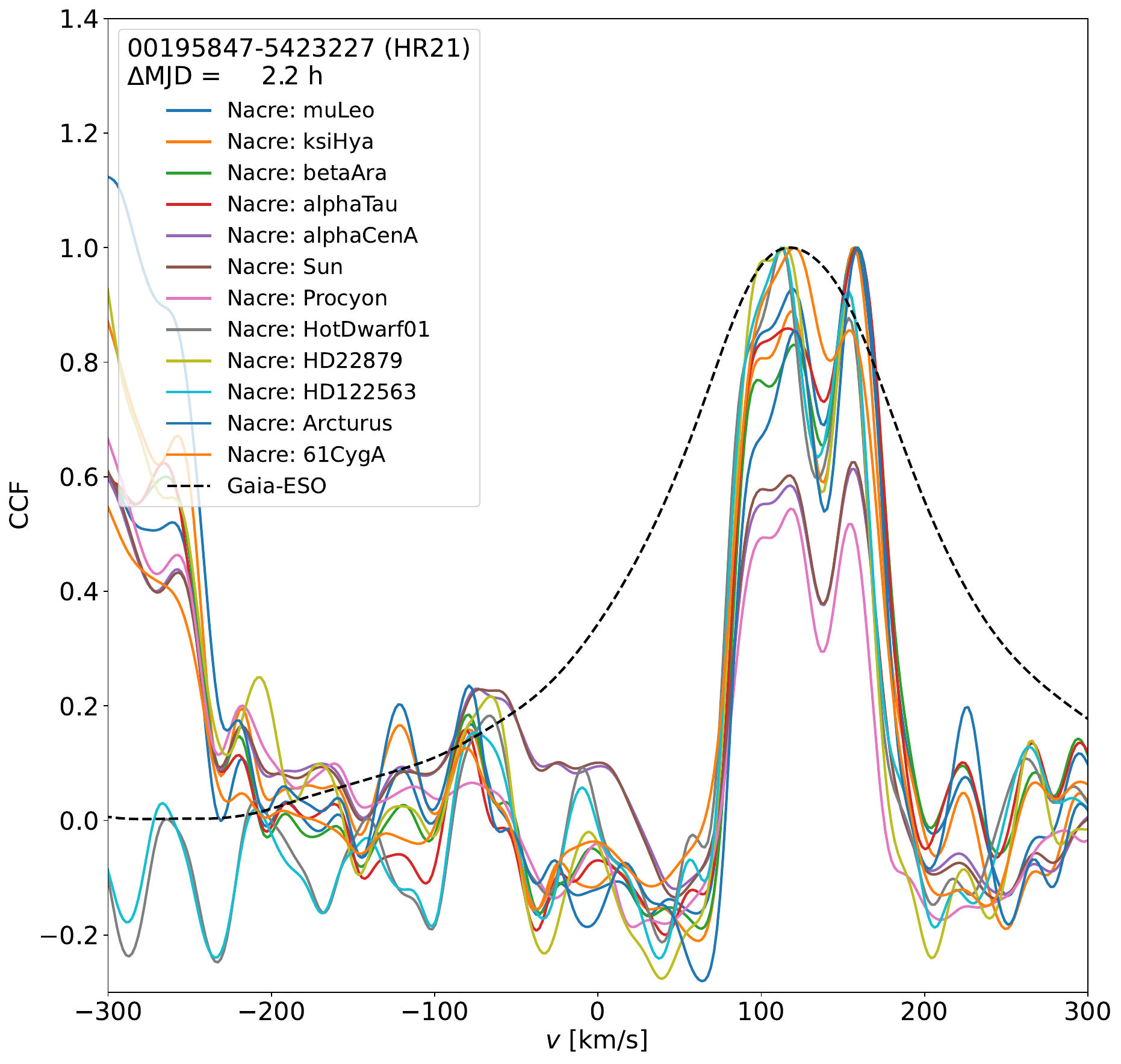}
  \includegraphics[width=0.49\textwidth]{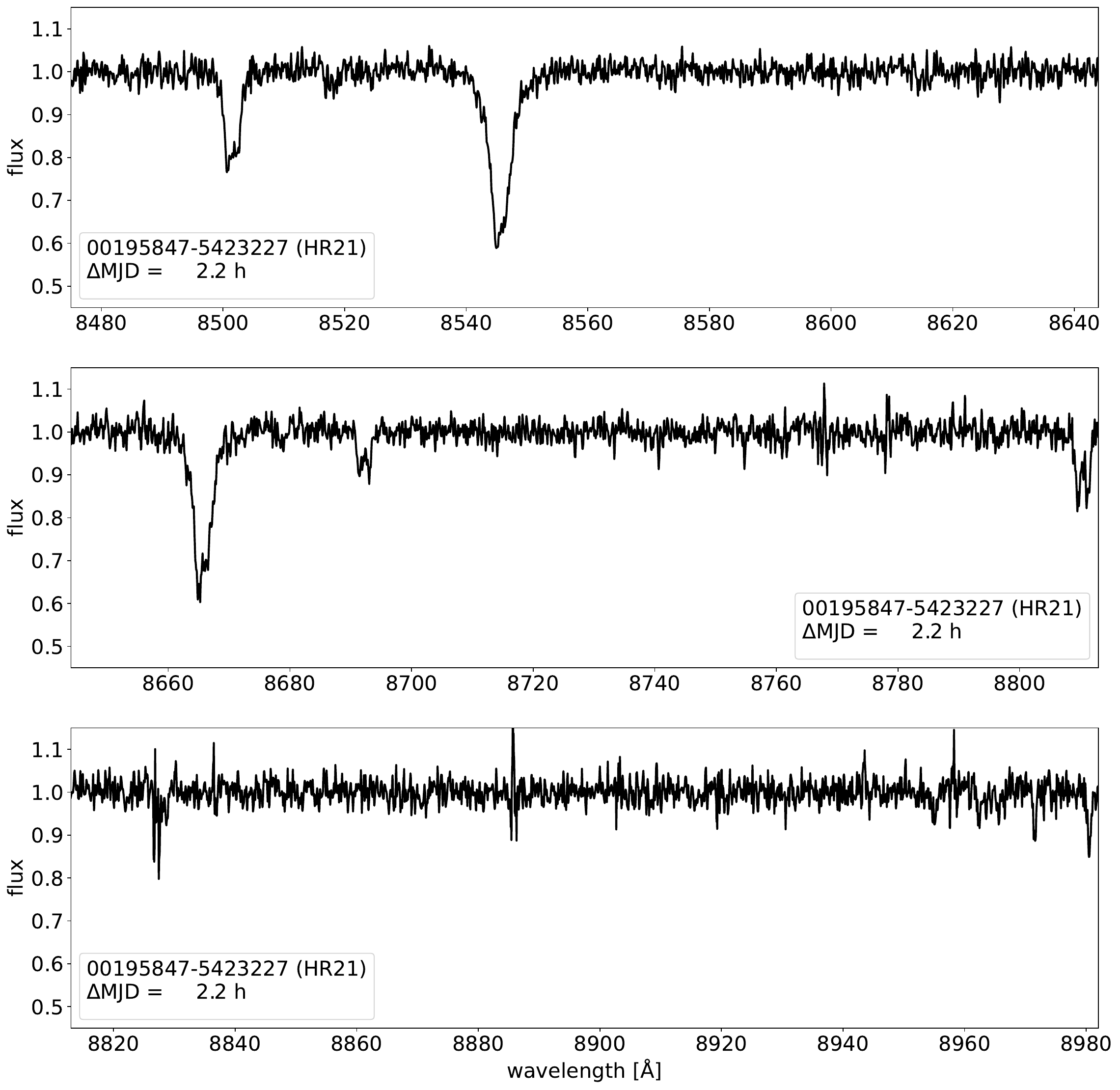}
  \captionof{figure}{\label{Fig:iDR5_SB3_atlas_00195847-5423227_4}(4/4) CNAME 00195847-5423227, at $\mathrm{MJD} = 56532.380019$, setup HR21.}
\end{minipage}
\cleardoublepage
\setlength\parindent{\defaultparindent}

\subsection{08202324-1402560}

\setlength\parindent{0cm}
\begin{minipage}{\textwidth}
  \centering
  \includegraphics[width=0.49\textwidth]{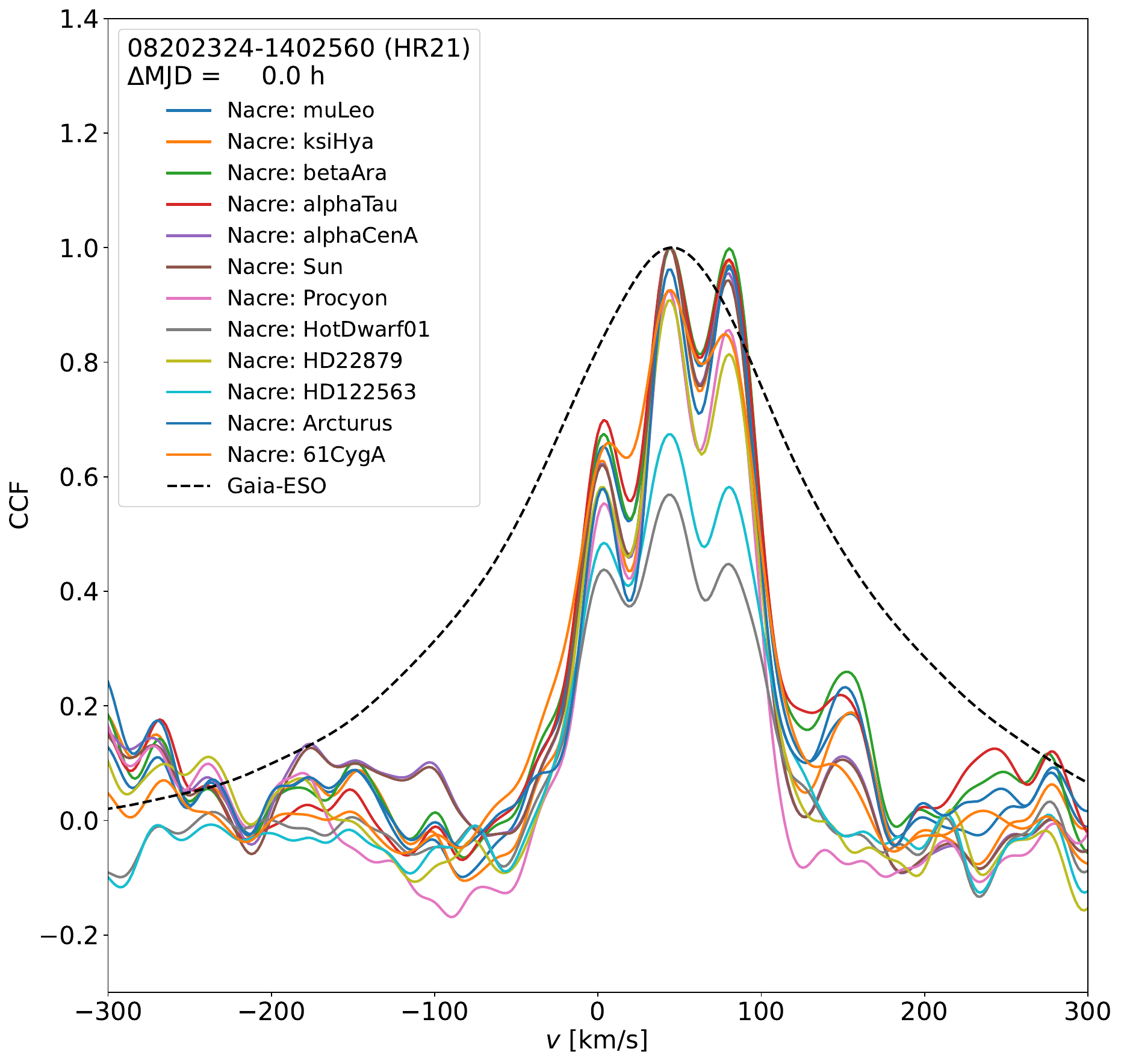}
  \includegraphics[width=0.49\textwidth]{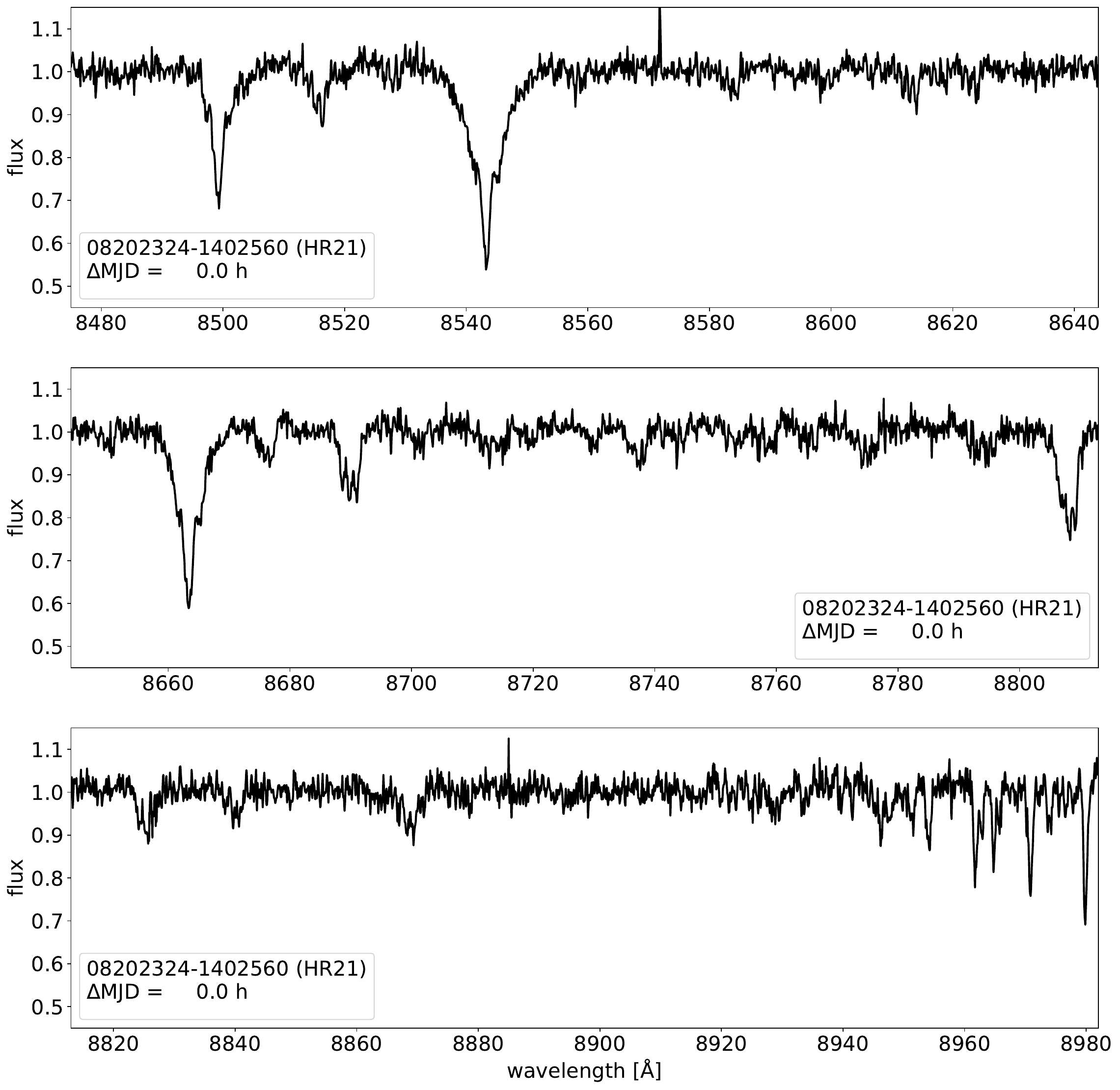}
  \captionof{figure}{\label{Fig:iDR5_SB3_atlas_08202324-1402560_1}(1/4) CNAME 08202324-1402560, at $\mathrm{MJD} = 56757.013357$, setup HR21.}
\end{minipage}
\begin{minipage}{\textwidth}
  \centering
  \includegraphics[width=0.49\textwidth]{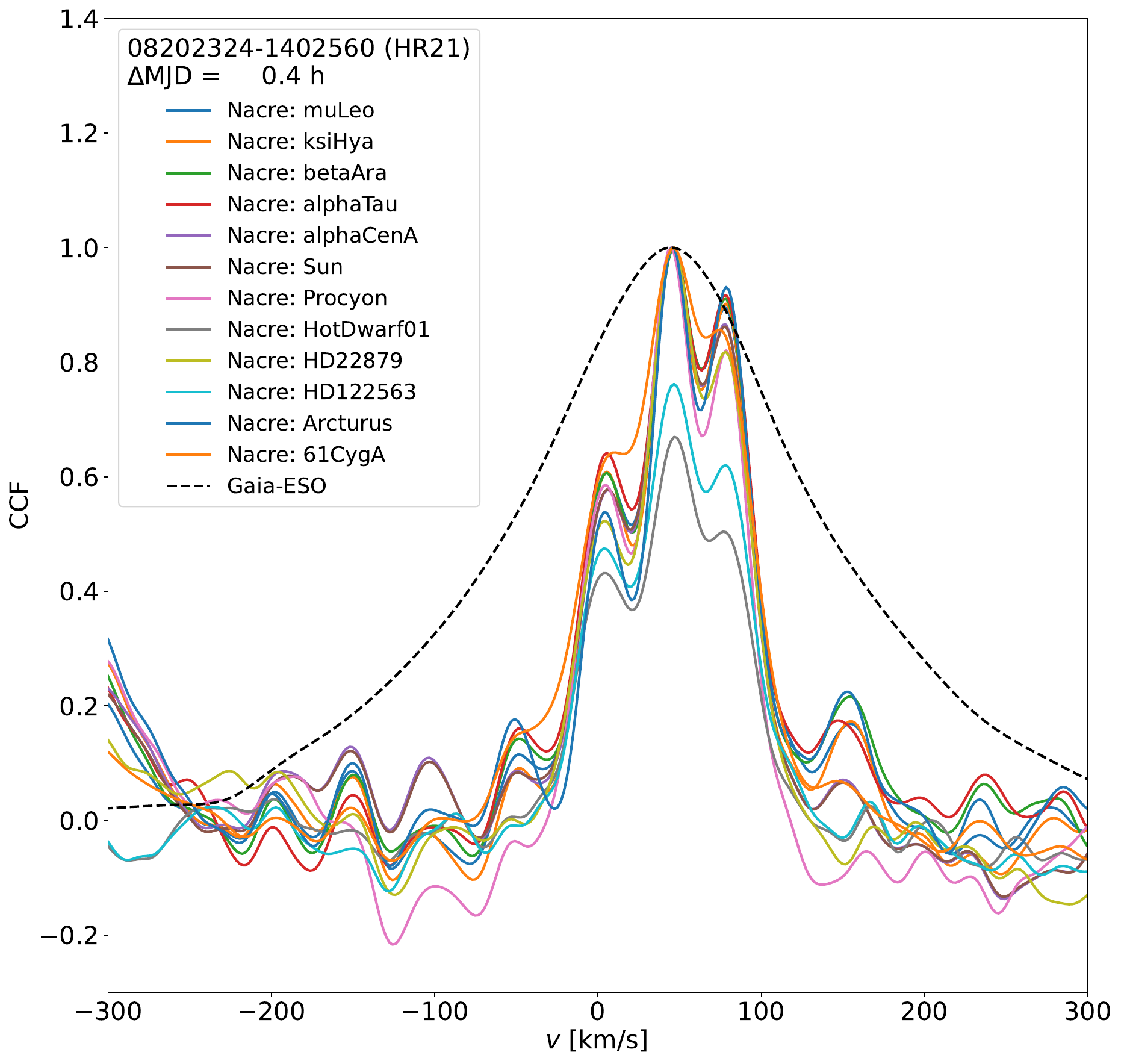}
  \includegraphics[width=0.49\textwidth]{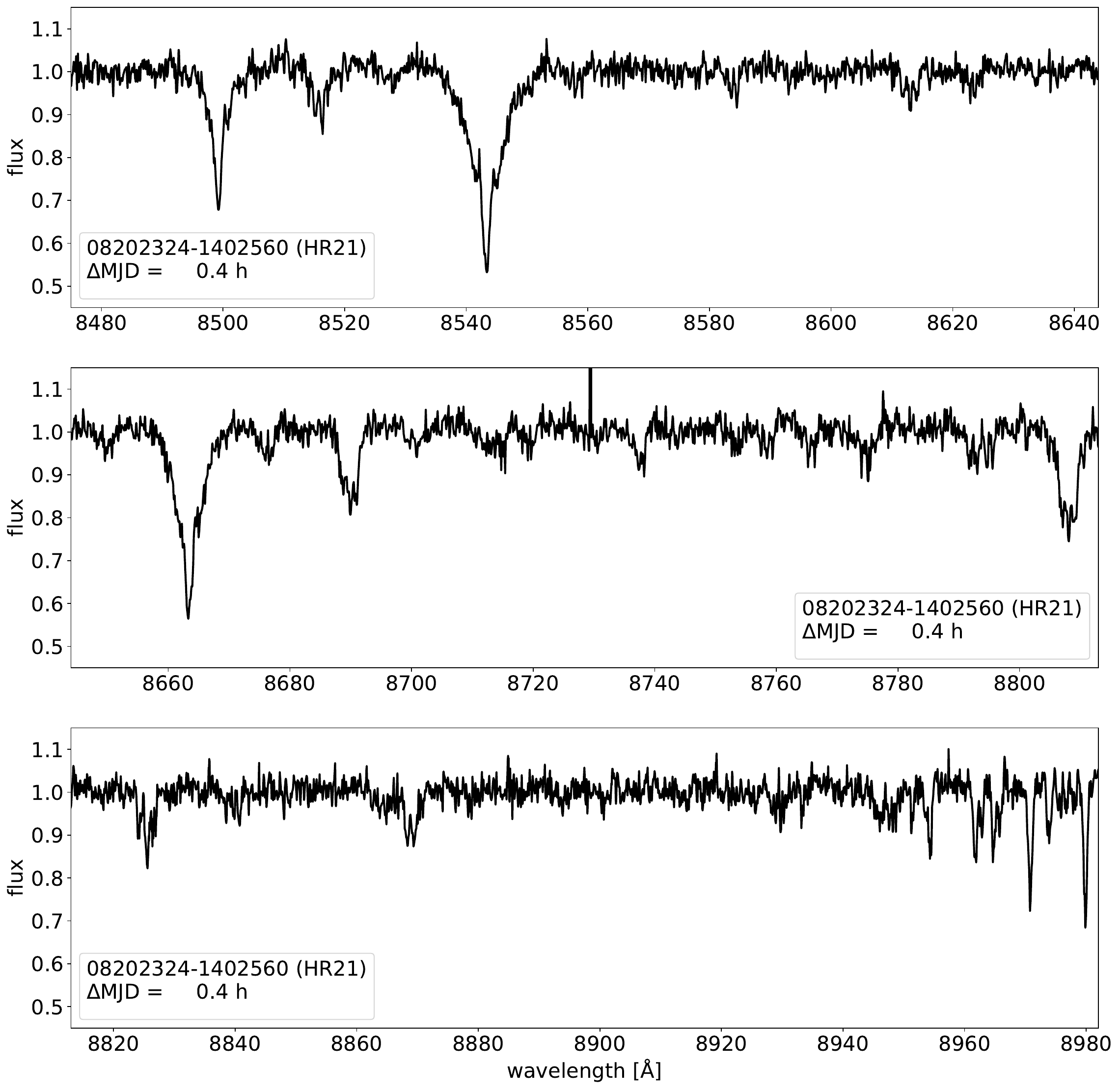}
  \captionof{figure}{\label{Fig:iDR5_SB3_atlas_08202324-1402560_2}(2/4) CNAME 08202324-1402560, at $\mathrm{MJD} = 56757.031351$, setup HR21.}
\end{minipage}
\clearpage
\begin{minipage}{\textwidth}
  \centering
  \includegraphics[width=0.49\textwidth]{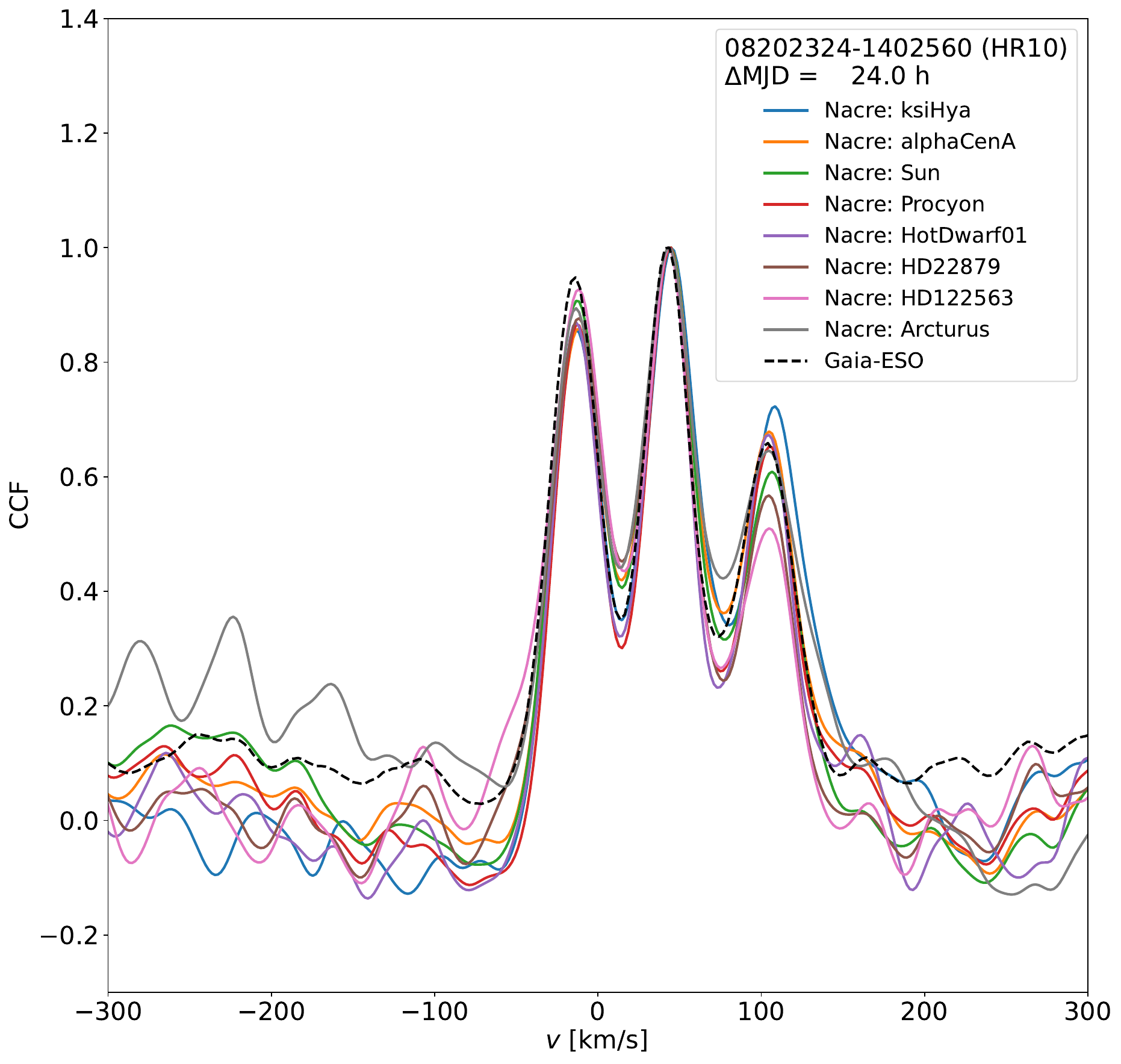}
  \includegraphics[width=0.49\textwidth]{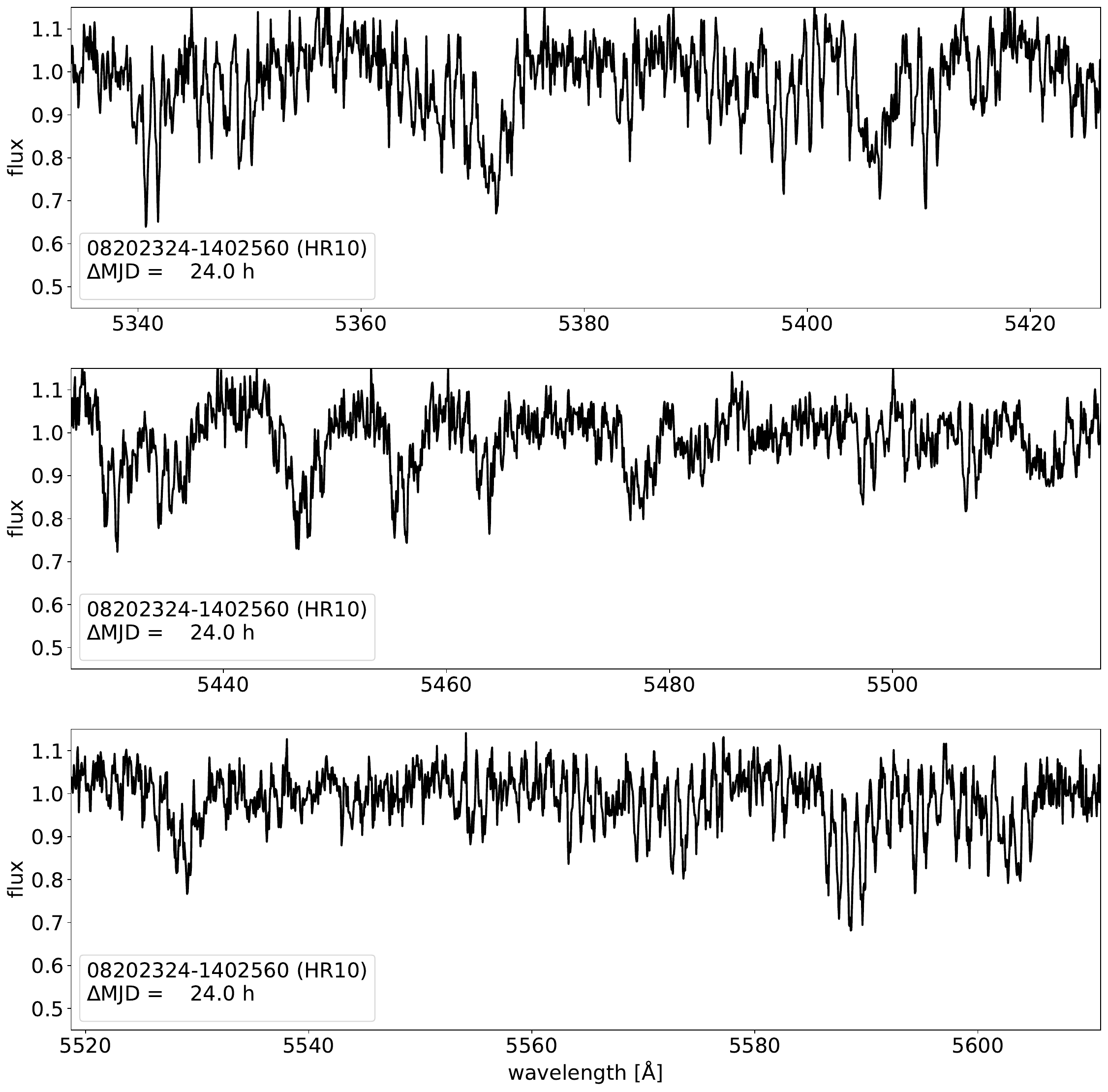}
  \captionof{figure}{\label{Fig:iDR5_SB3_atlas_08202324-1402560_3}(3/4) CNAME 08202324-1402560, at $\mathrm{MJD} = 56758.012499$, setup HR10.}
\end{minipage}
\begin{minipage}{\textwidth}
  \centering
  \includegraphics[width=0.49\textwidth]{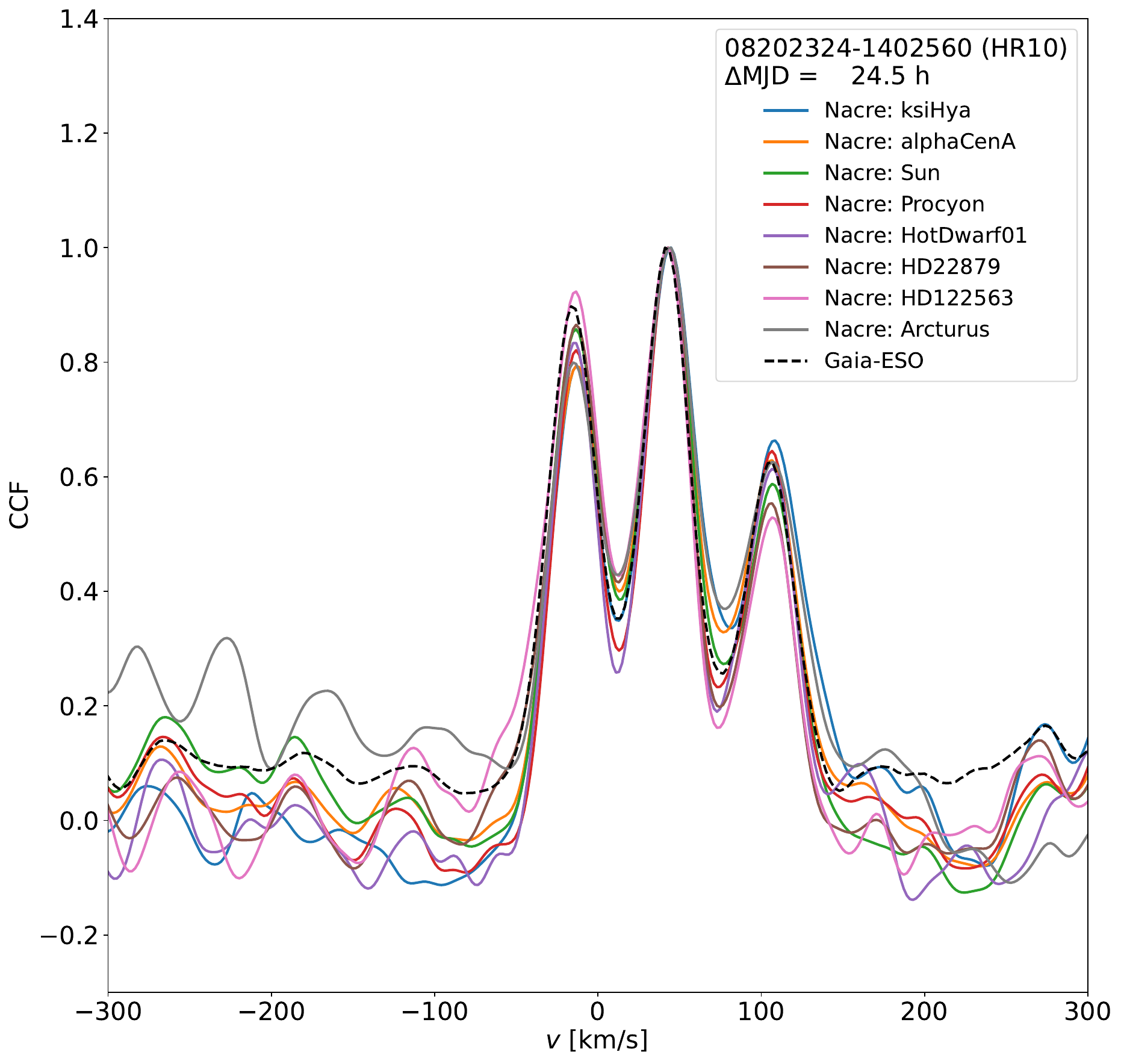}
  \includegraphics[width=0.49\textwidth]{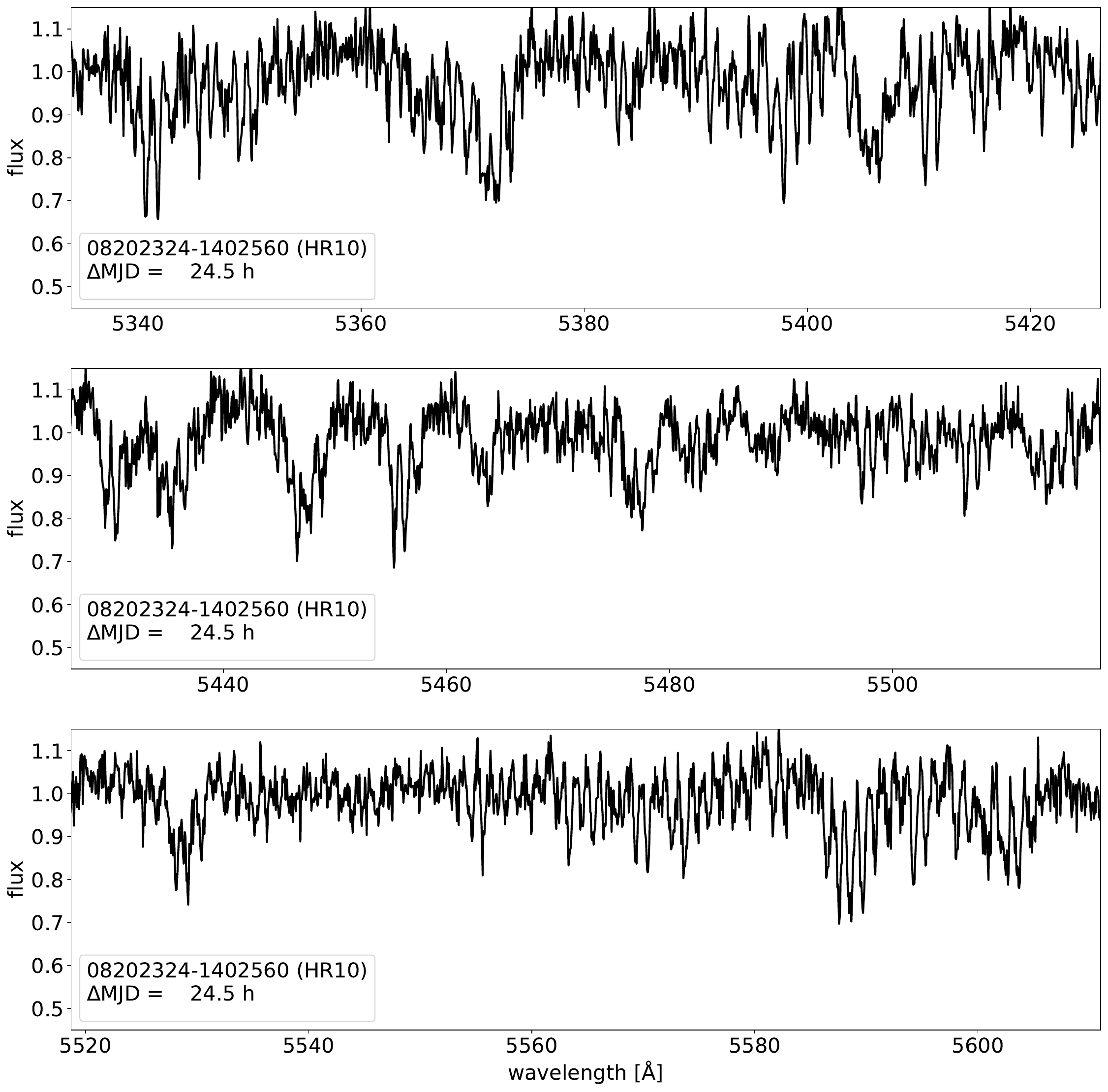}
  \captionof{figure}{\label{Fig:iDR5_SB3_atlas_08202324-1402560_4}(4/4) CNAME 08202324-1402560, at $\mathrm{MJD} = 56758.033832$, setup HR10.}
\end{minipage}
\cleardoublepage
\setlength\parindent{\defaultparindent}

\subsection{08393088-0033389}
\label{Sec:Appendix_08393088_0033389}

For CNAME 08393088-0033389 shown in this Section, our SB3 classification is based on Fig.~\ref{Fig:iDR5_SB3_atlas_08393088-0033389_1} and \ref{Fig:iDR5_SB3_atlas_08393088-0033389_2}. The three peaks visible on these figures stand clearly out of the correlation noise and are therefore interpreted as three potential stellar components. These secondary peaks are no longer visible in Fig.~\ref{Fig:iDR5_SB3_atlas_08393088-0033389_3} and \ref{Fig:iDR5_SB3_atlas_08393088-0033389_4}, but these spectra were acquired about 1508 hours after the first two (Fig.~\ref{Fig:iDR5_SB3_atlas_08393088-0033389_1} and \ref{Fig:iDR5_SB3_atlas_08393088-0033389_2}). The last two exposures (Fig.~\ref{Fig:iDR5_SB3_atlas_08393088-0033389_5} and \ref{Fig:iDR5_SB3_atlas_08393088-0033389_6}) are recorded about 1530 hours after the first two (Fig.~\ref{Fig:iDR5_SB3_atlas_08393088-0033389_1} and \ref{Fig:iDR5_SB3_atlas_08393088-0033389_2}) and now reveal four secondary peaks! It is unlikely that they are due to correlation noise, since peaks due to correlation noise should appear the same at all epochs whereas the peaks of Fig.~\ref{Fig:iDR5_SB3_atlas_08393088-0033389_3} and \ref{Fig:iDR5_SB3_atlas_08393088-0033389_4} are clean. Clearly this system deserves a follow-up to elucidate its exact nature (SB3 or even SB5?); it illustrates how complex it can be to characterize high-multiplicity systems.
\cleardoublepage

\setlength\parindent{0cm}
\begin{minipage}{\textwidth}
  \centering
  \includegraphics[width=0.49\textwidth]{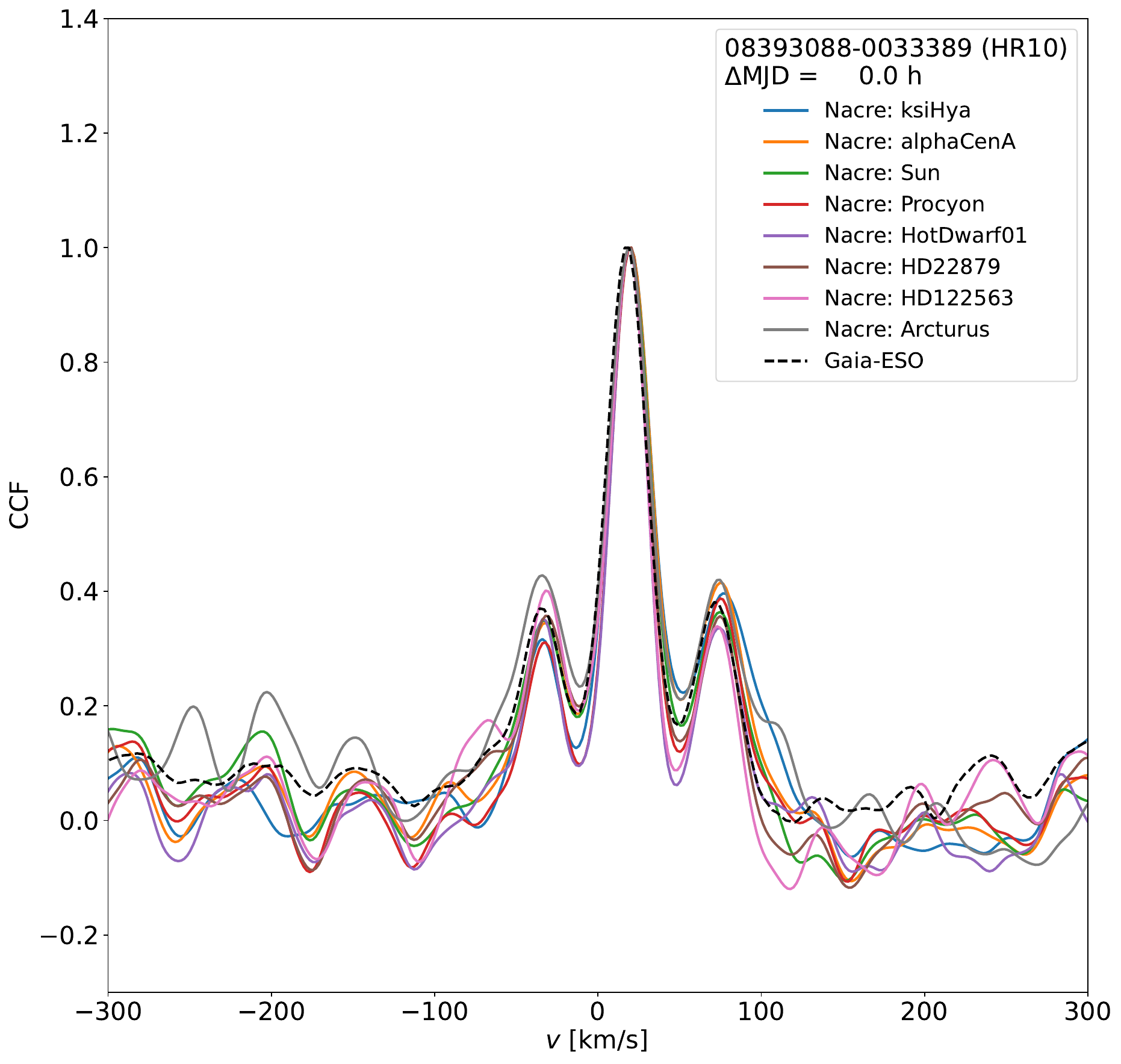}
  \includegraphics[width=0.49\textwidth]{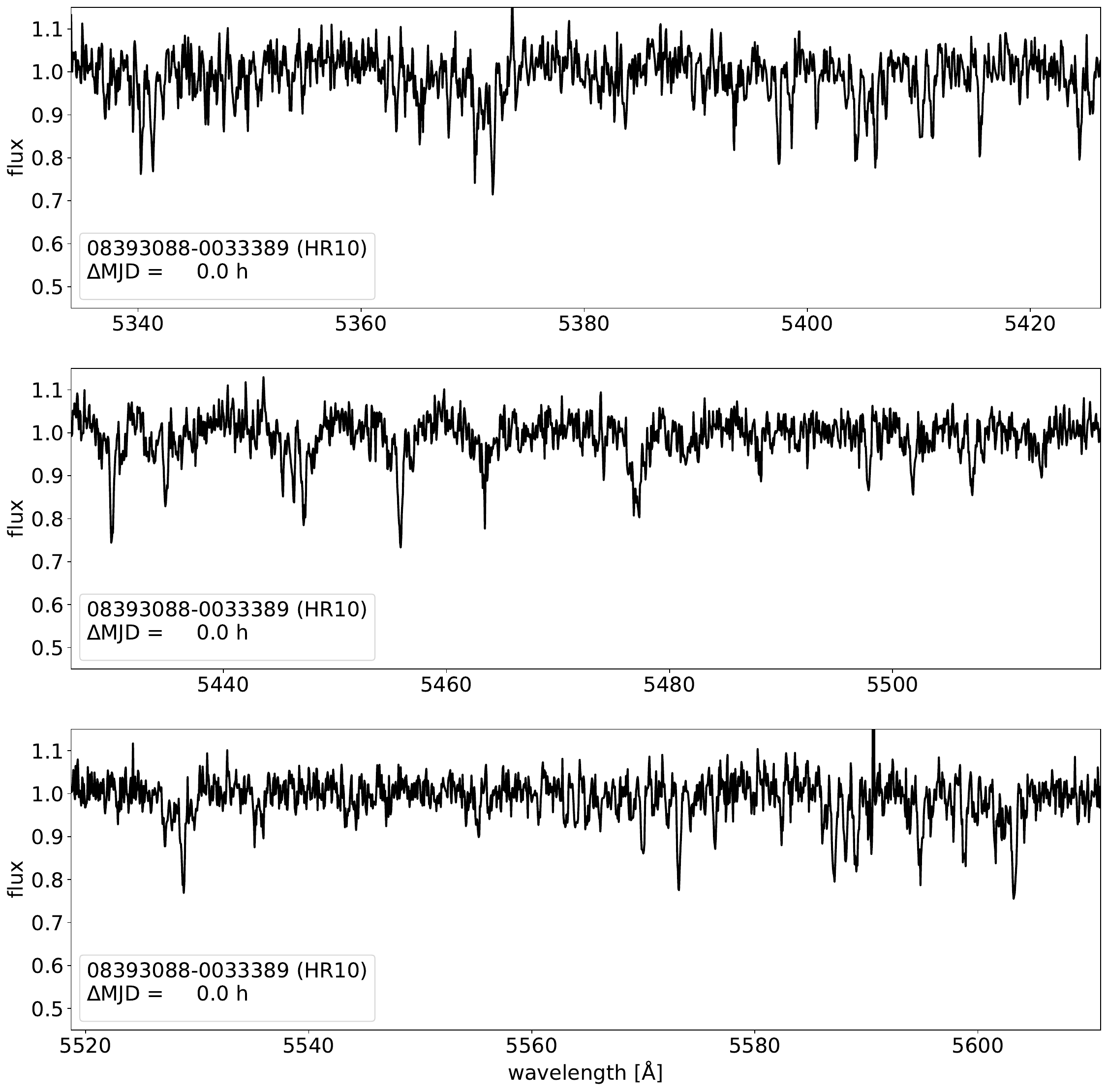}
  \captionof{figure}{\label{Fig:iDR5_SB3_atlas_08393088-0033389_1}(1/6) CNAME 08393088-0033389, at $\mathrm{MJD} = 56314.185014$, setup HR10.}
\end{minipage}
\begin{minipage}{\textwidth}
  \centering
  \includegraphics[width=0.49\textwidth]{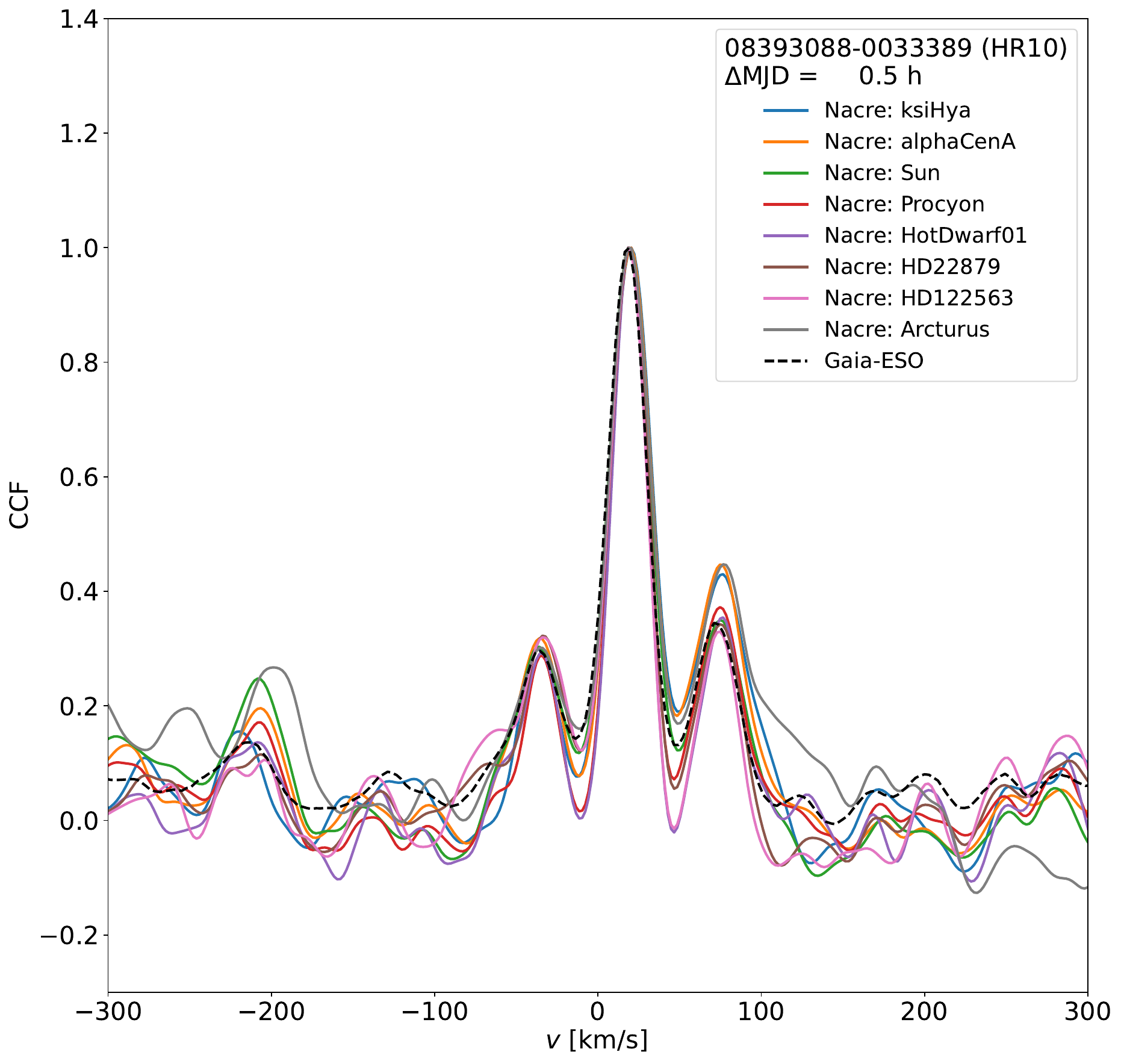}
  \includegraphics[width=0.49\textwidth]{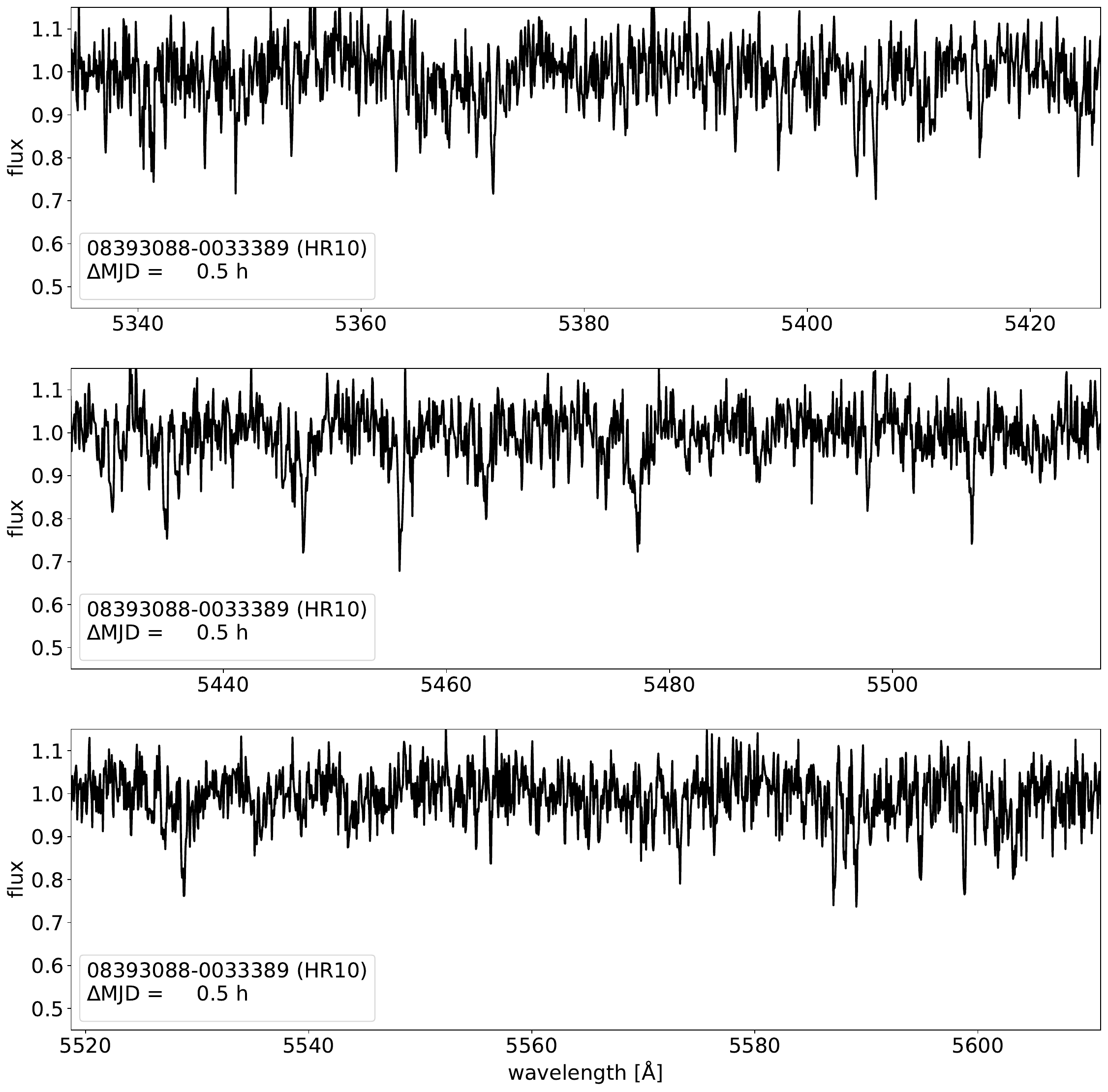}
  \captionof{figure}{\label{Fig:iDR5_SB3_atlas_08393088-0033389_2}(2/6) CNAME 08393088-0033389, at $\mathrm{MJD} = 56314.205499$, setup HR10.}
\end{minipage}
\clearpage
\begin{minipage}{\textwidth}
  \centering
  \includegraphics[width=0.49\textwidth]{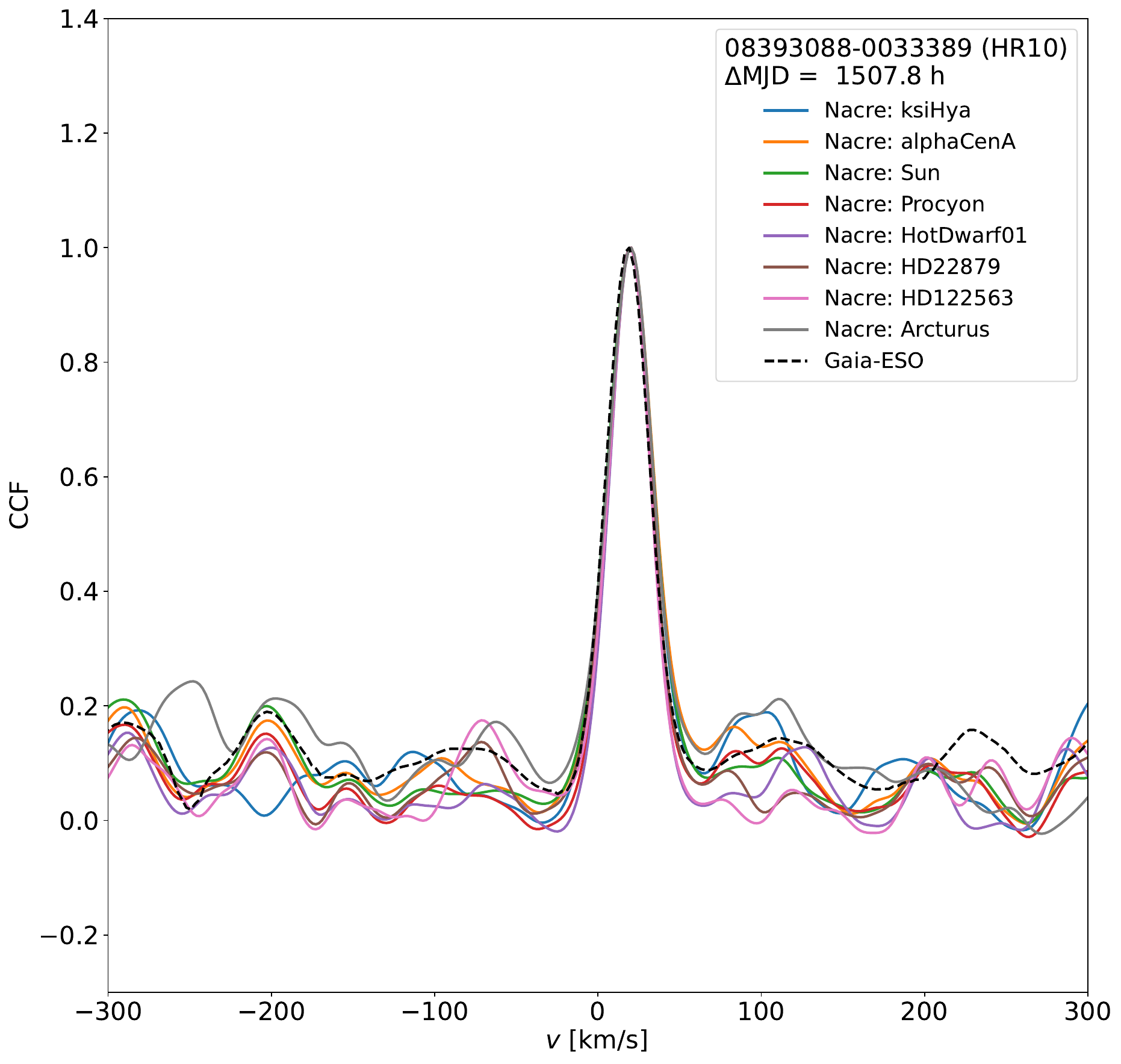}
  \includegraphics[width=0.49\textwidth]{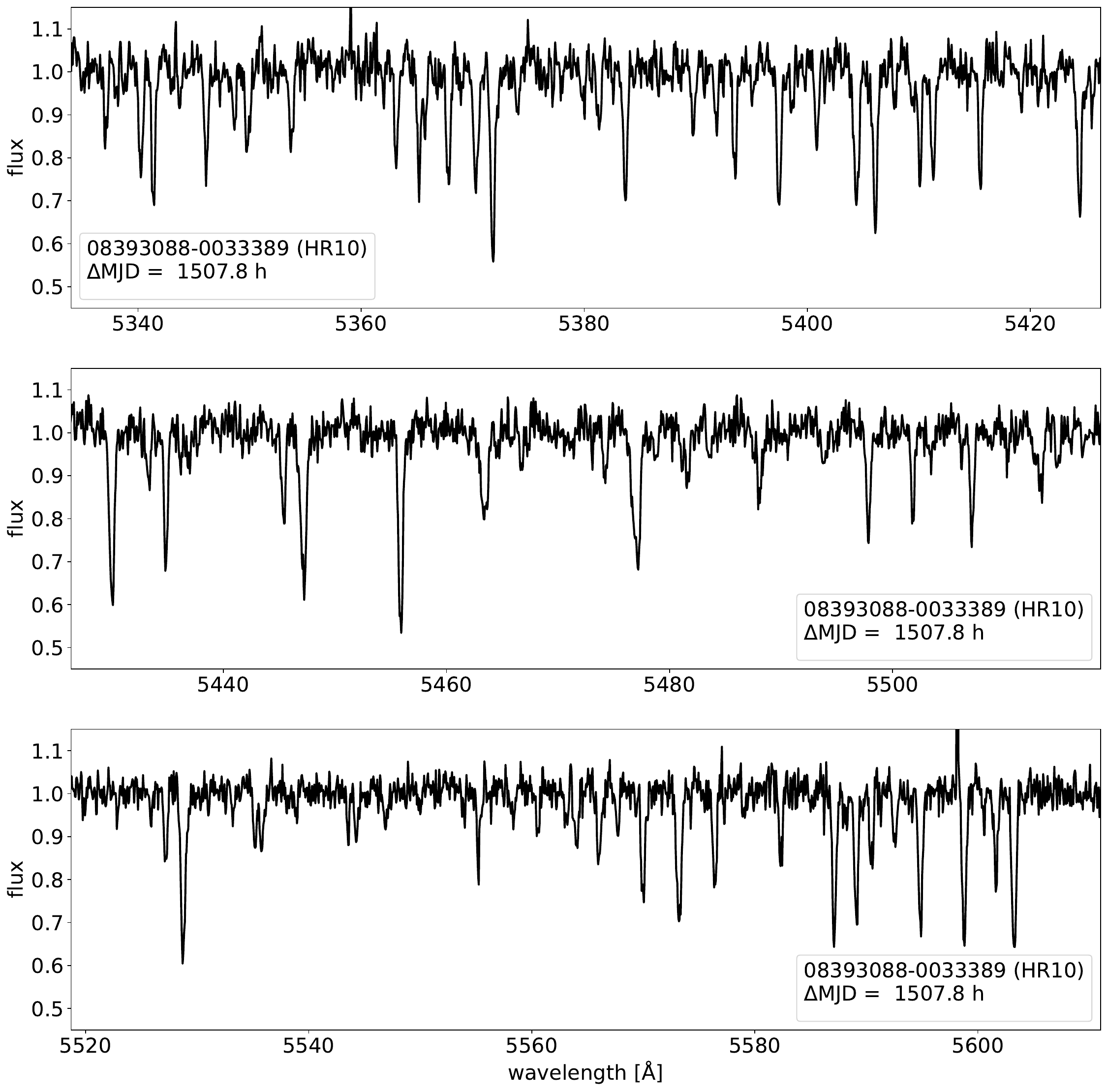}
  \captionof{figure}{\label{Fig:iDR5_SB3_atlas_08393088-0033389_3}(3/6) CNAME 08393088-0033389, at $\mathrm{MJD} = 56377.008829$, setup HR10.}
\end{minipage}
\begin{minipage}{\textwidth}
  \centering
  \includegraphics[width=0.49\textwidth]{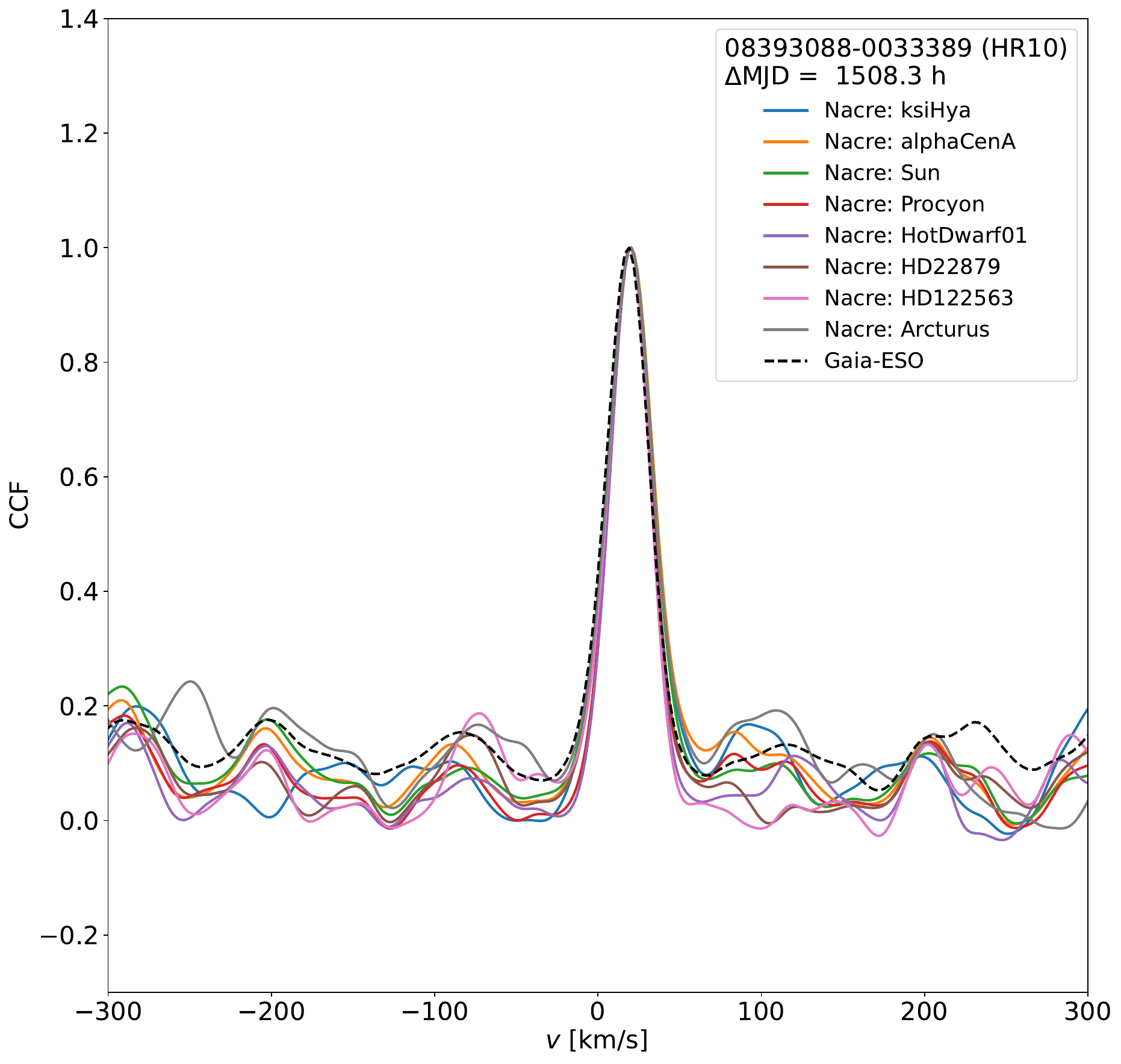}
  \includegraphics[width=0.49\textwidth]{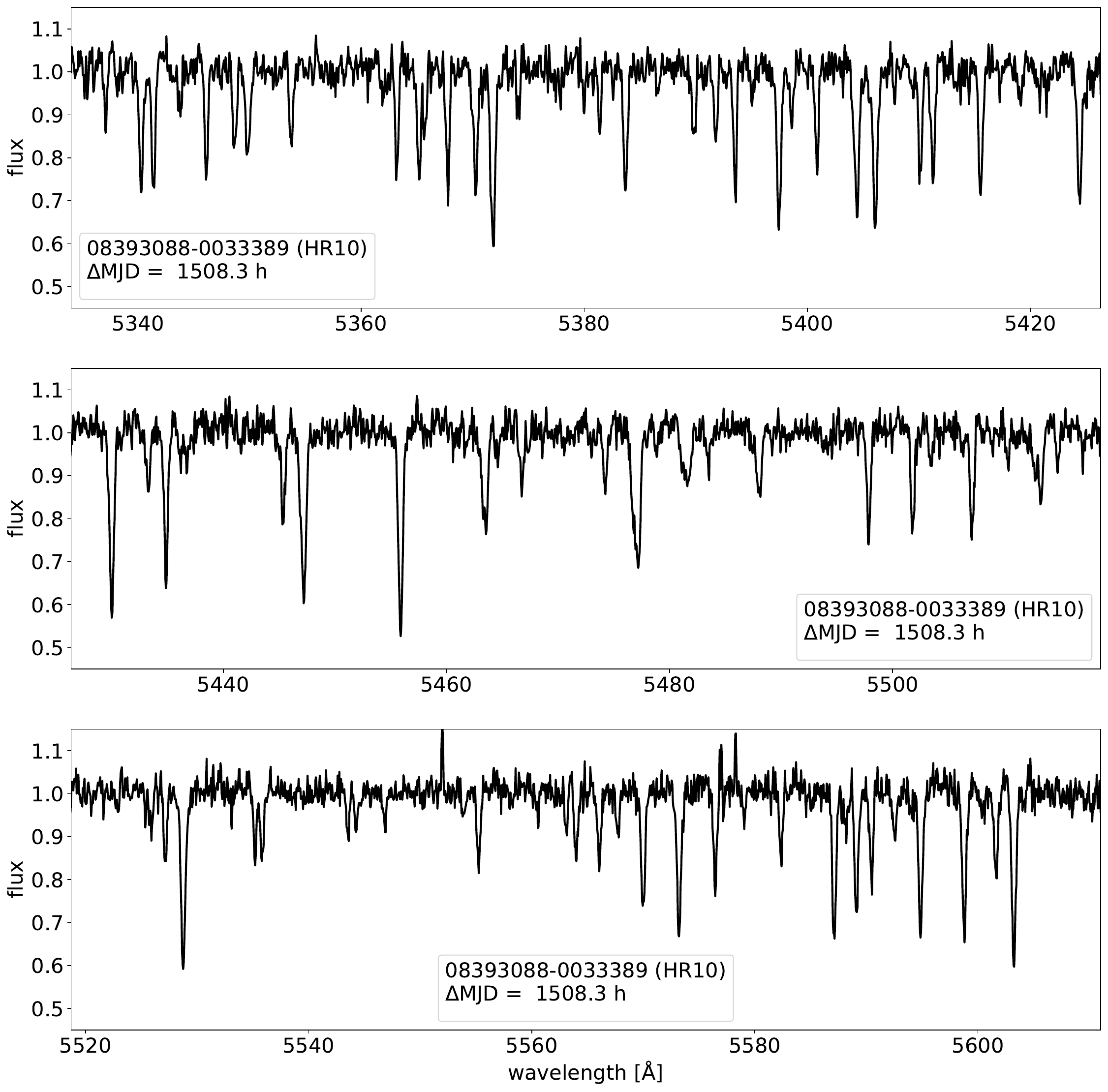}
  \captionof{figure}{\label{Fig:iDR5_SB3_atlas_08393088-0033389_4}(4/6) CNAME 08393088-0033389, at $\mathrm{MJD} = 56377.029952$, setup HR10.}
\end{minipage}
\clearpage
\begin{minipage}{\textwidth}
  \centering
  \includegraphics[width=0.49\textwidth]{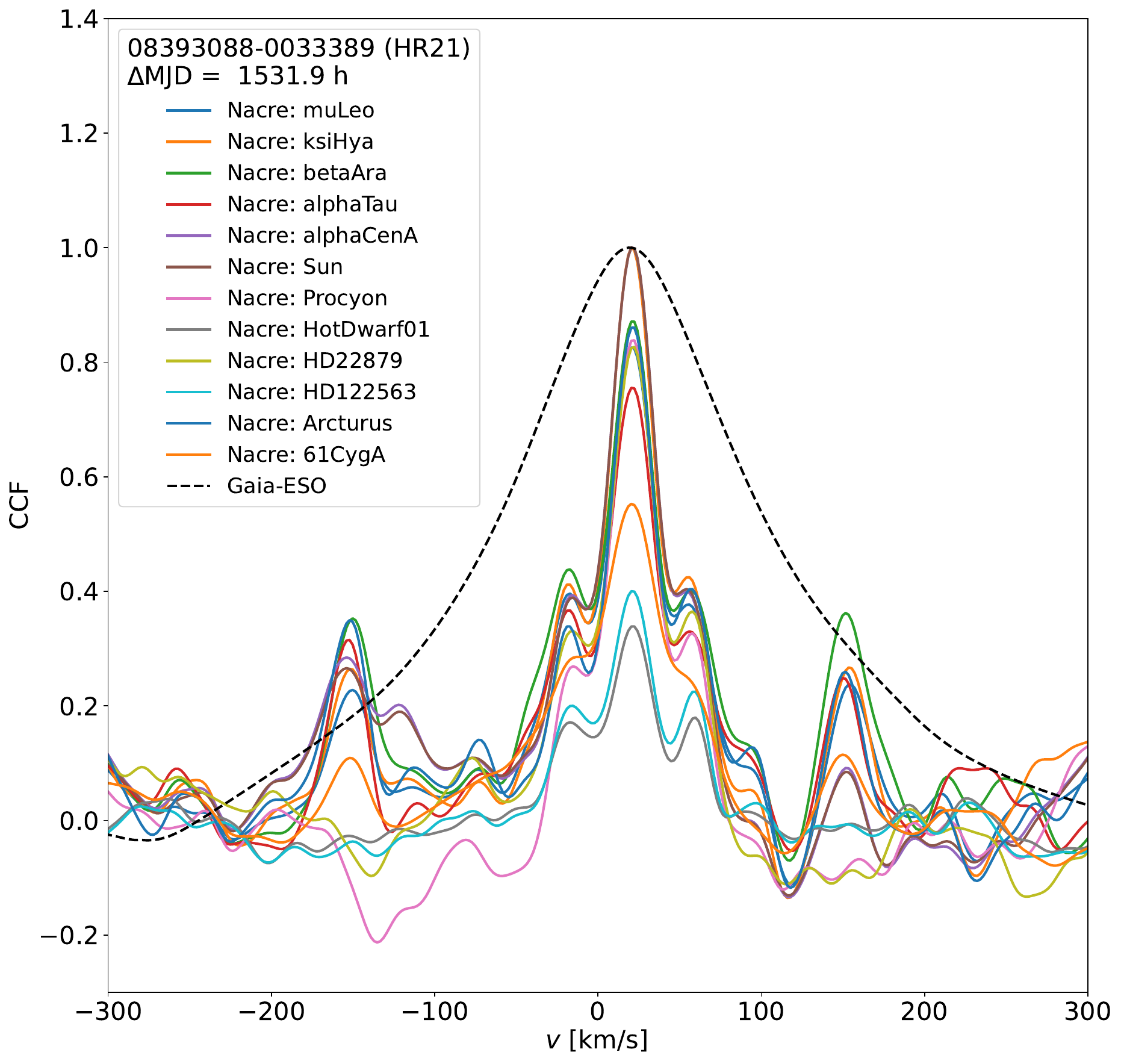}
  \includegraphics[width=0.49\textwidth]{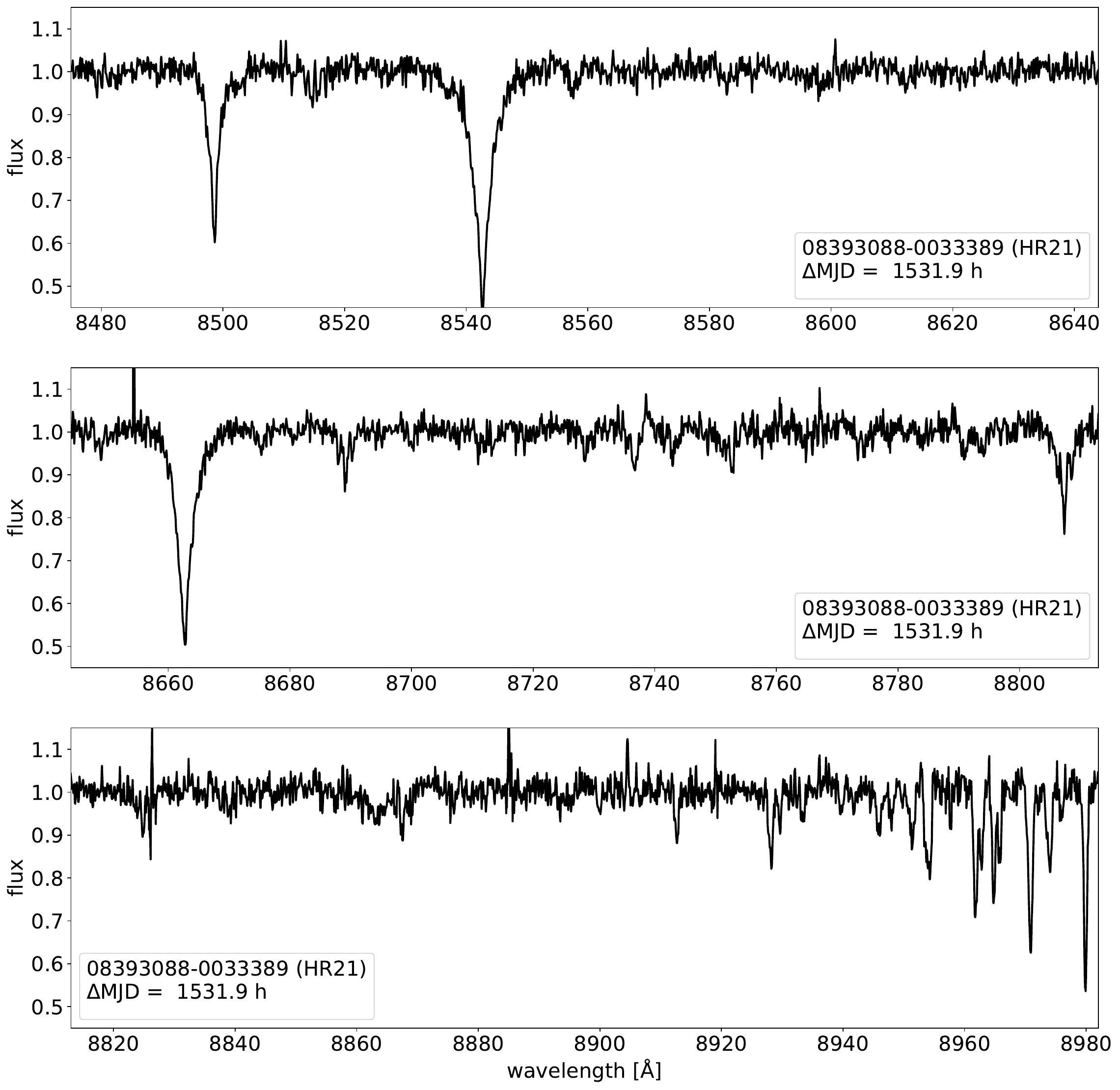}
  \captionof{figure}{\label{Fig:iDR5_SB3_atlas_08393088-0033389_5}(5/6) CNAME 08393088-0033389, at $\mathrm{MJD} = 56378.014598$, setup HR21.}
\end{minipage}
\begin{minipage}{\textwidth}
  \centering
  \includegraphics[width=0.49\textwidth]{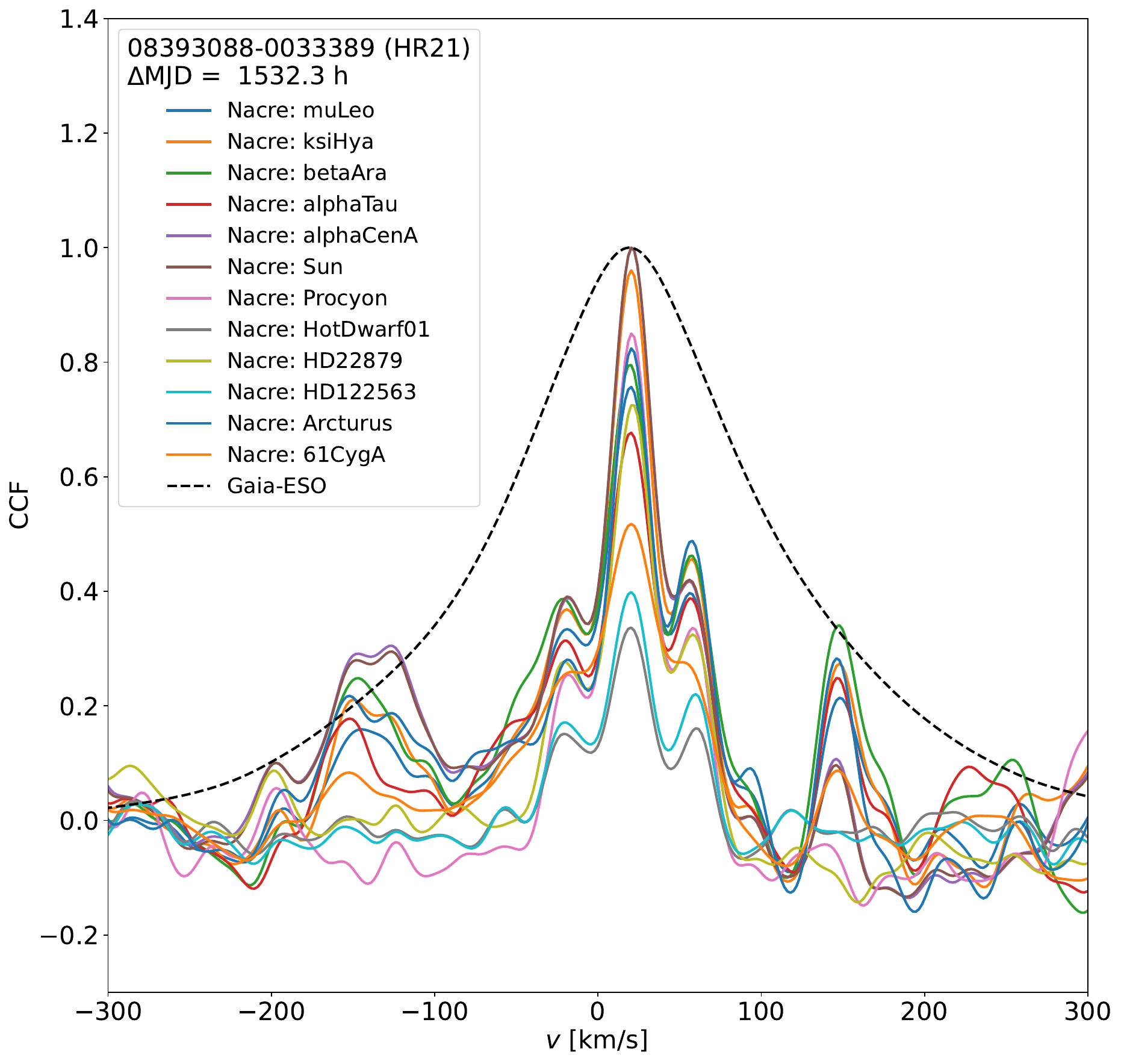}
  \includegraphics[width=0.49\textwidth]{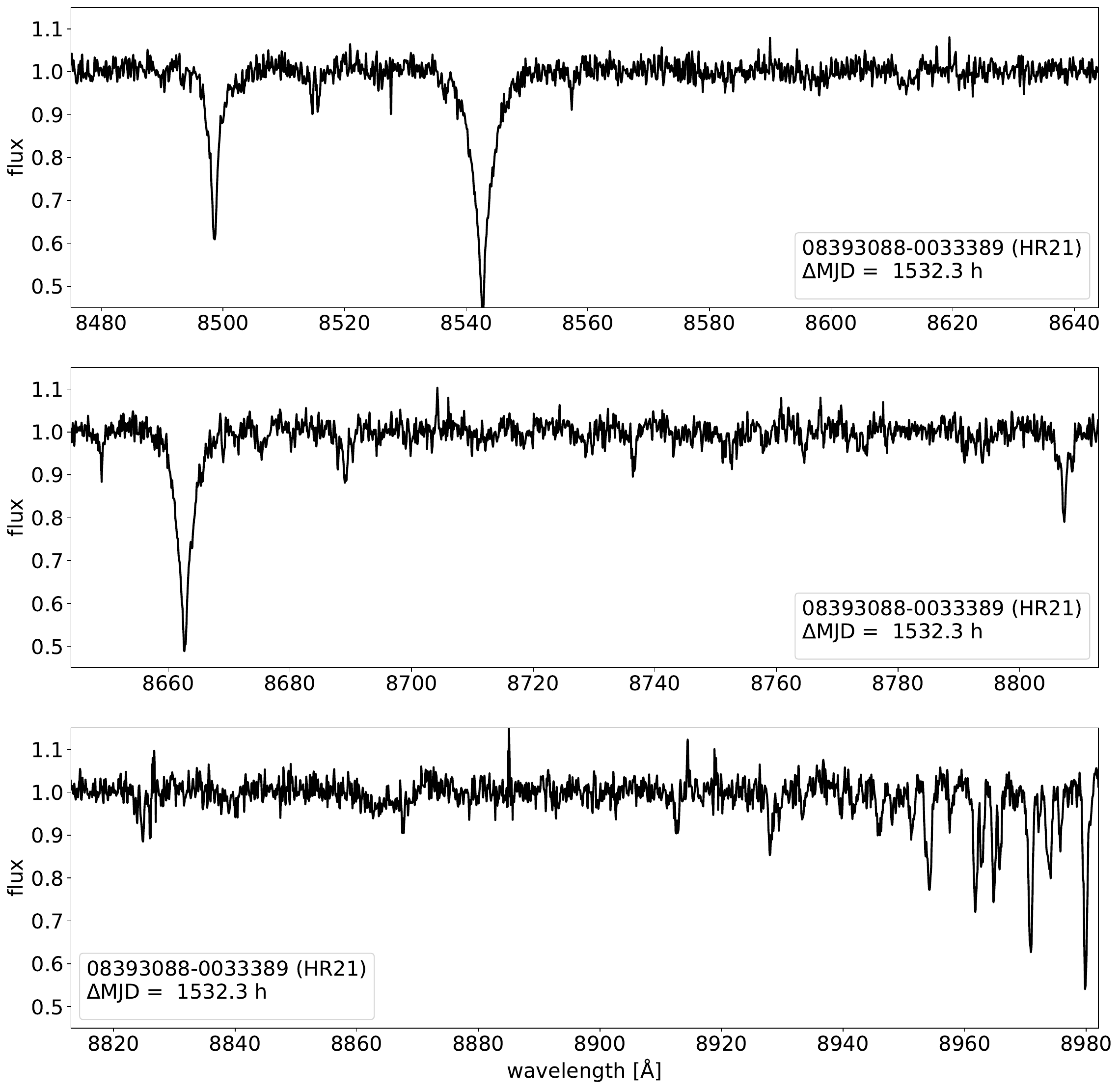}
  \captionof{figure}{\label{Fig:iDR5_SB3_atlas_08393088-0033389_6}(6/6) CNAME 08393088-0033389, at $\mathrm{MJD} = 56378.032579$, setup HR21.}
\end{minipage}
\cleardoublepage
\setlength\parindent{\defaultparindent}

\subsection{08402079-0025437}
\label{Sec:Appendix_08402079_0025437}

For CNAME 08402079-0025437 shown in this Section, our SB3 classification is based on Fig.~\ref{Fig:iDR5_SB3_atlas_08402079-0025437_1}, \ref{Fig:iDR5_SB3_atlas_08402079-0025437_2}, \ref{Fig:iDR5_SB3_atlas_08402079-0025437_5} and \ref{Fig:iDR5_SB3_atlas_08402079-0025437_6}. From Figures~\ref{Fig:iDR5_SB3_atlas_08402079-0025437_1} and \ref{Fig:iDR5_SB3_atlas_08402079-0025437_2}, we are confident in the existence of at least two stellar components. We may exclude the doubling as due to a shock from the variations occurring at later times, the CCF becoming either cleaner (single-peaked on Fig.~\ref{Fig:iDR5_SB3_atlas_08402079-0025437_3} and \ref{Fig:iDR5_SB3_atlas_08402079-0025437_4}) or more complex (multiply-peaked on Fig.~\ref{Fig:iDR5_SB3_atlas_08402079-0025437_5} and \ref{Fig:iDR5_SB3_atlas_08402079-0025437_6}, which is not at all expected for pulsating variables). The third peak on Fig.~\ref{Fig:iDR5_SB3_atlas_08402079-0025437_1} and \ref{Fig:iDR5_SB3_atlas_08402079-0025437_2} is admittedly just above the noise level. But from Fig.~\ref{Fig:iDR5_SB3_atlas_08402079-0025437_5} and \ref{Fig:iDR5_SB3_atlas_08402079-0025437_6}, we interpret the three peaks with a height larger than or equal to ~0.5 to be tracers of the stellar components. One might wonder, as for the previous system, whether the system might not even be of higher multiplicity, as judged by the peak with a height of 0.3.  As for the previous system, a dedicated spectroscopic follow-up is needed to assess the SB$n$ nature of 08402079-0025437.
\cleardoublepage

\setlength\parindent{0cm}
\begin{minipage}{\textwidth}
  \centering
  \includegraphics[width=0.49\textwidth]{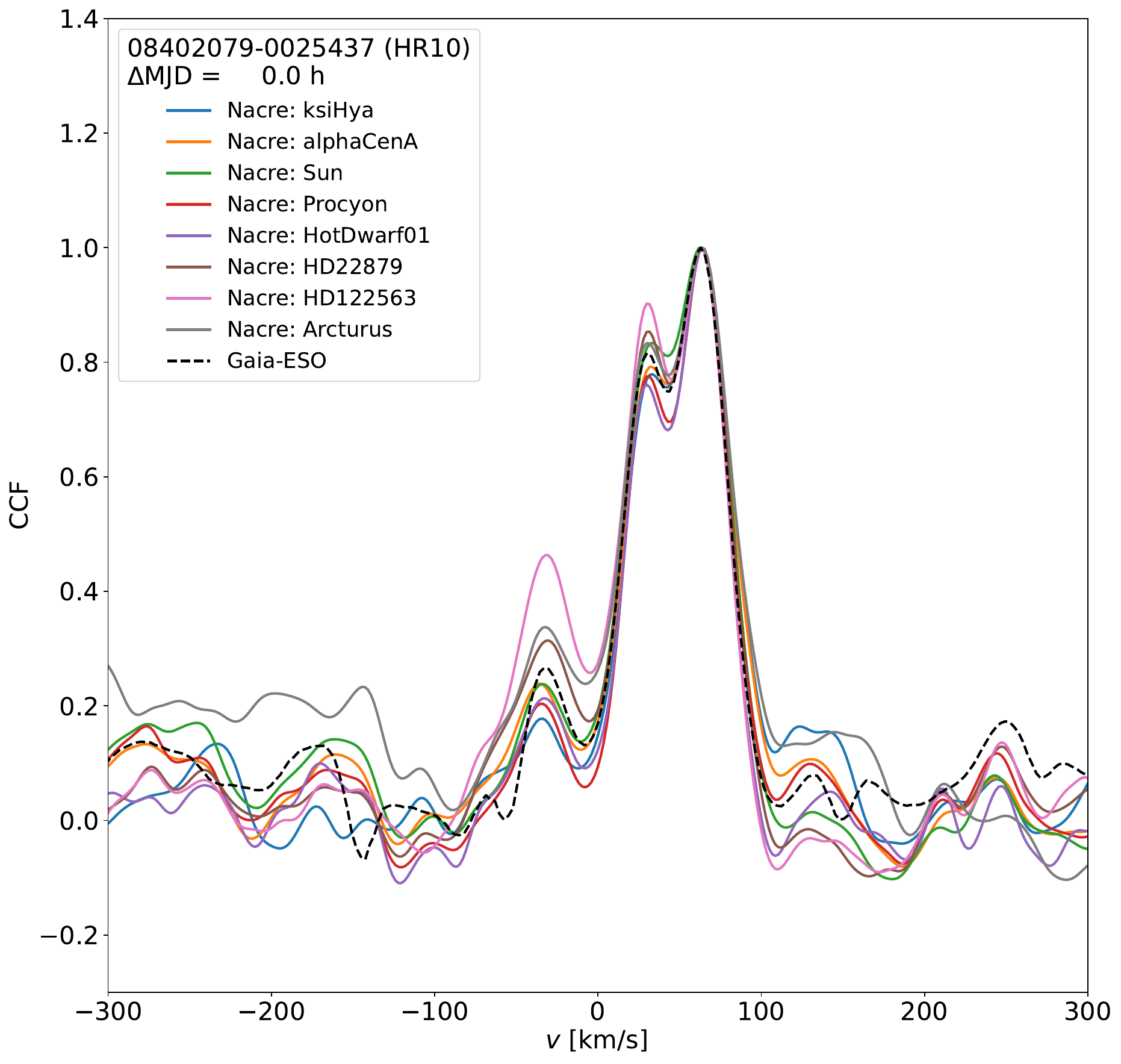}
  \includegraphics[width=0.49\textwidth]{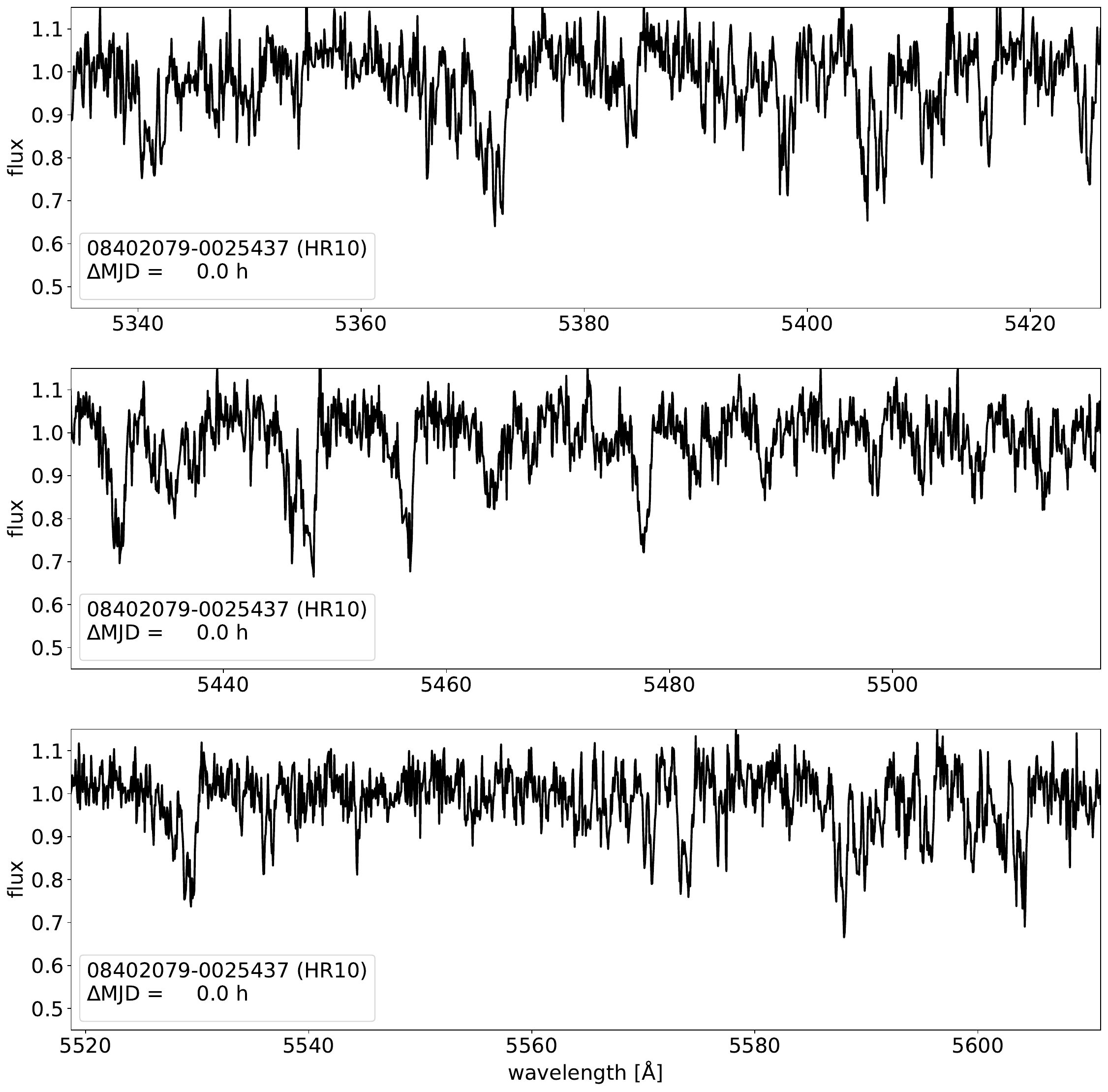}
  \captionof{figure}{\label{Fig:iDR5_SB3_atlas_08402079-0025437_1}(1/6) CNAME 08402079-0025437, at $\mathrm{MJD} = 56314.185014$, setup HR10.}
\end{minipage}
\begin{minipage}{\textwidth}
  \centering
  \includegraphics[width=0.49\textwidth]{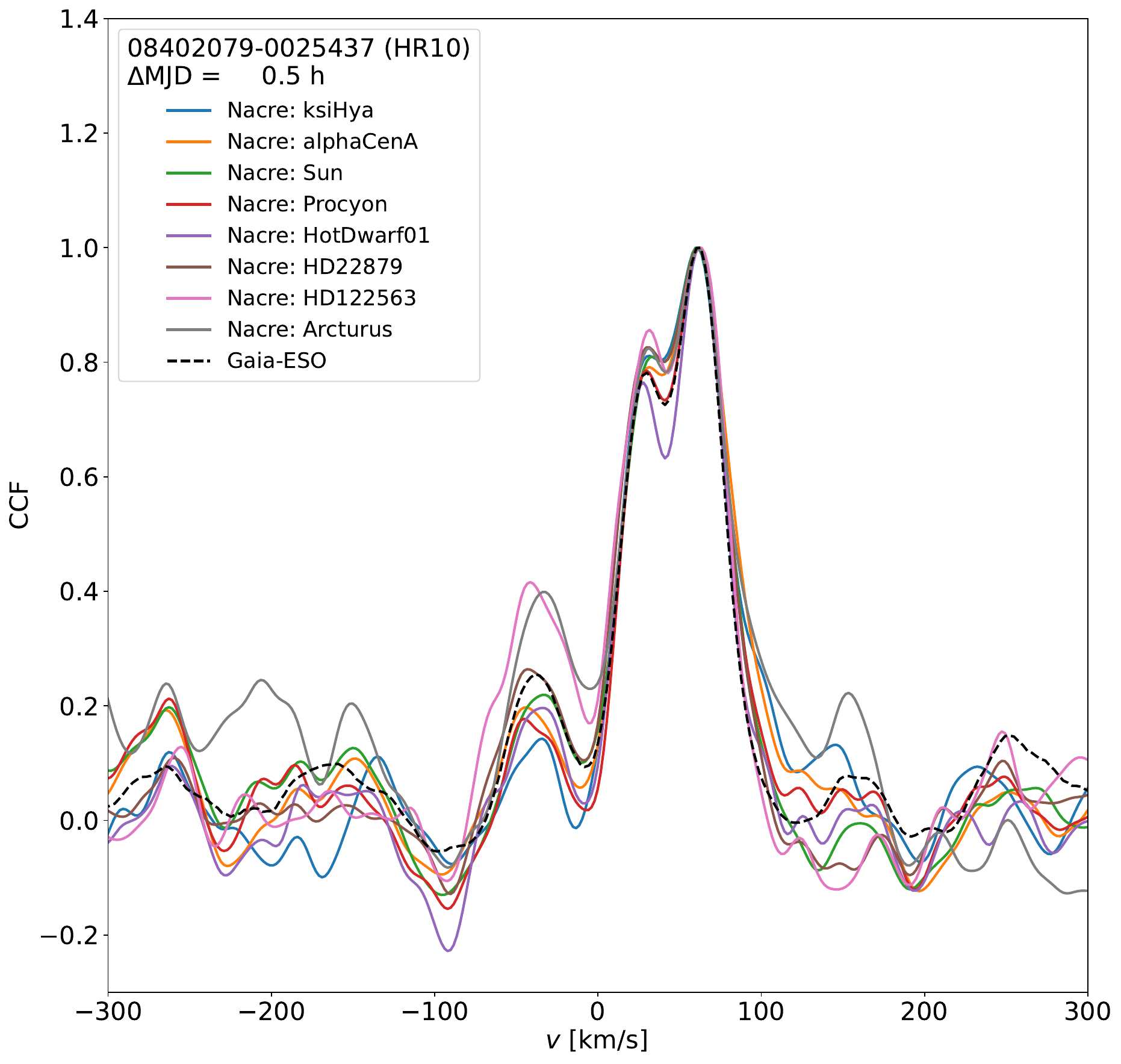}
  \includegraphics[width=0.49\textwidth]{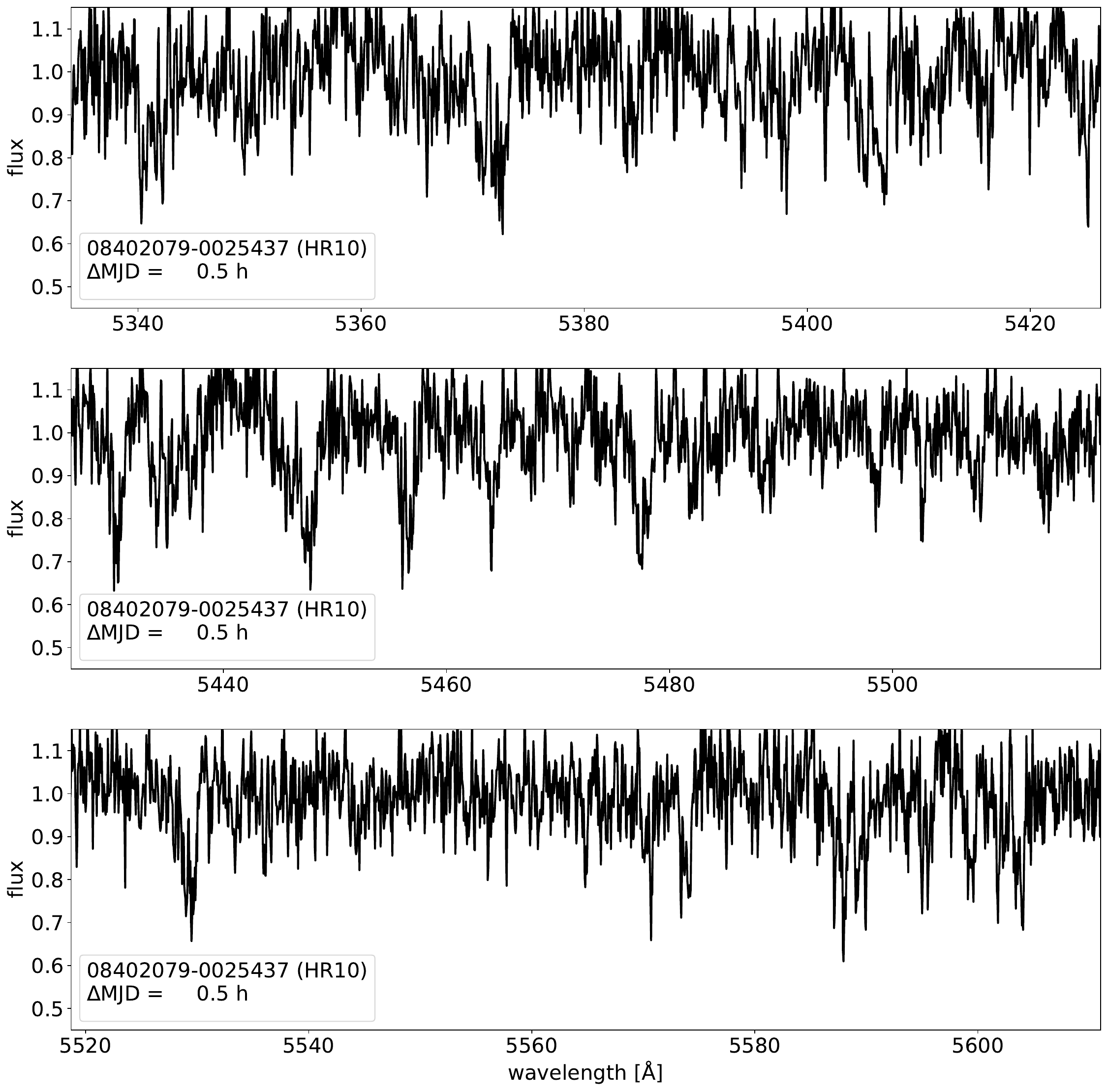}
  \captionof{figure}{\label{Fig:iDR5_SB3_atlas_08402079-0025437_2}(2/6) CNAME 08402079-0025437, at $\mathrm{MJD} = 56314.205499$, setup HR10.}
\end{minipage}
\clearpage
\begin{minipage}{\textwidth}
  \centering
  \includegraphics[width=0.49\textwidth]{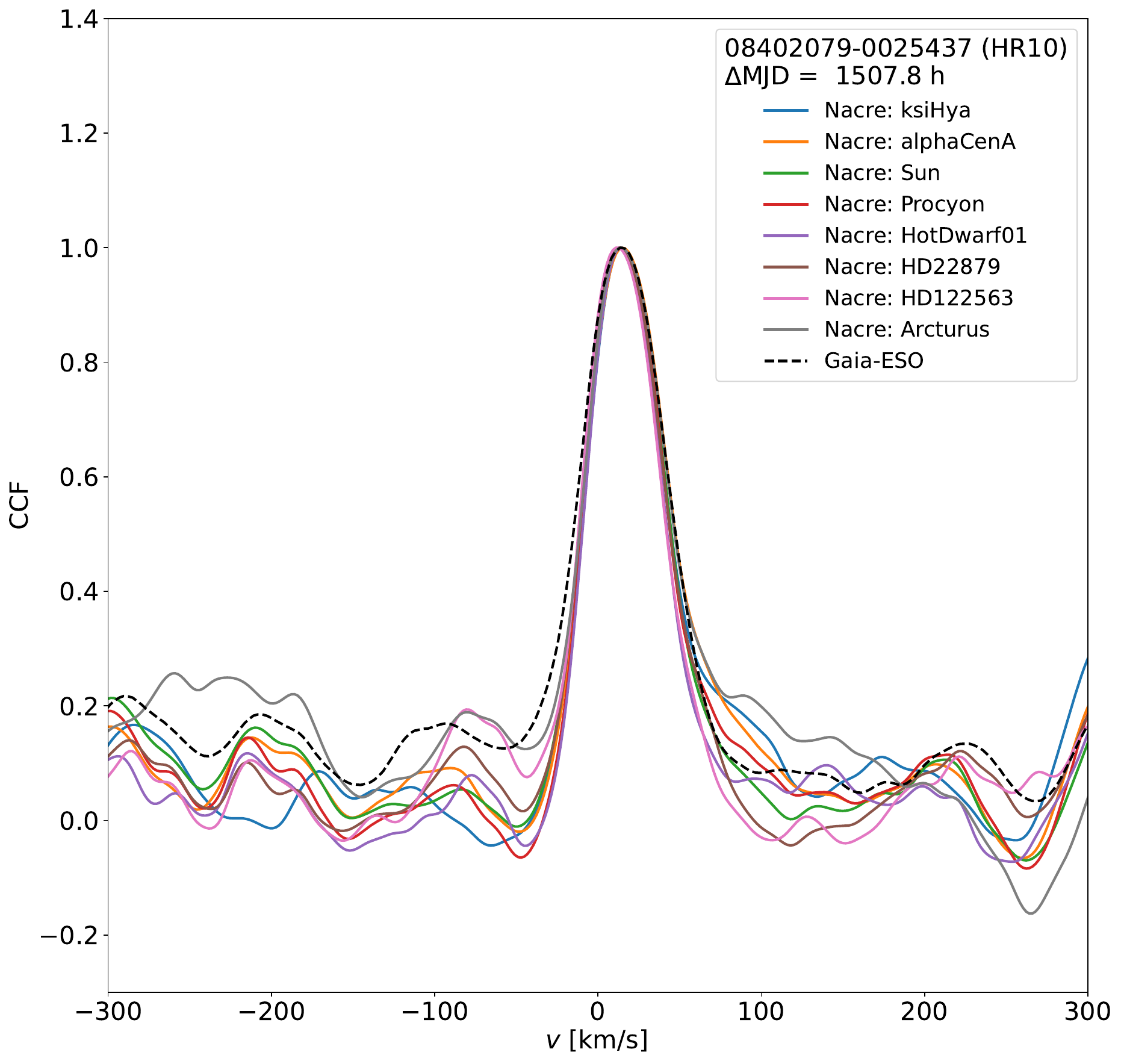}
  \includegraphics[width=0.49\textwidth]{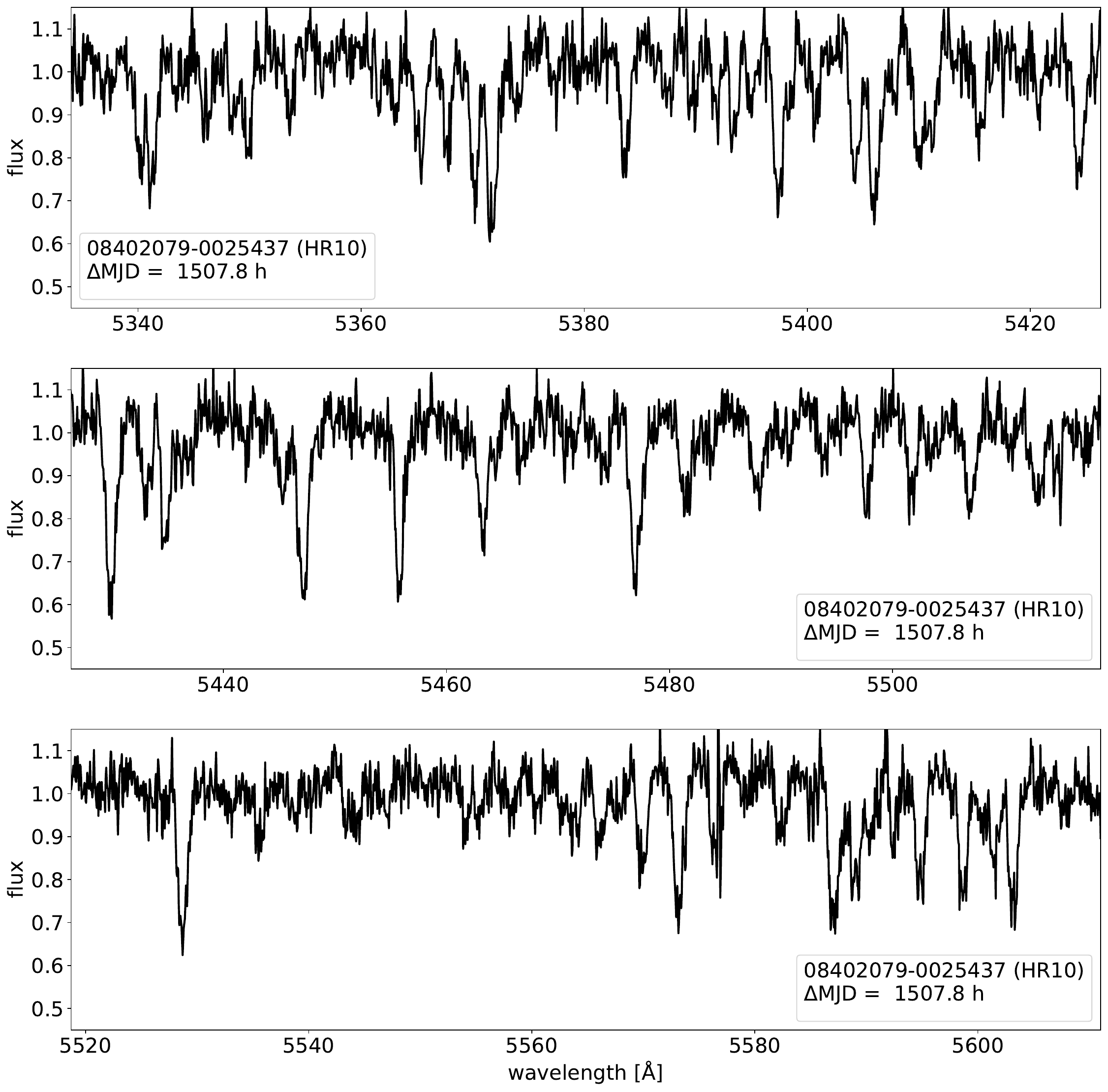}
  \captionof{figure}{\label{Fig:iDR5_SB3_atlas_08402079-0025437_3}(3/6) CNAME 08402079-0025437, at $\mathrm{MJD} = 56377.008829$, setup HR10.}
\end{minipage}
\begin{minipage}{\textwidth}
  \centering
  \includegraphics[width=0.49\textwidth]{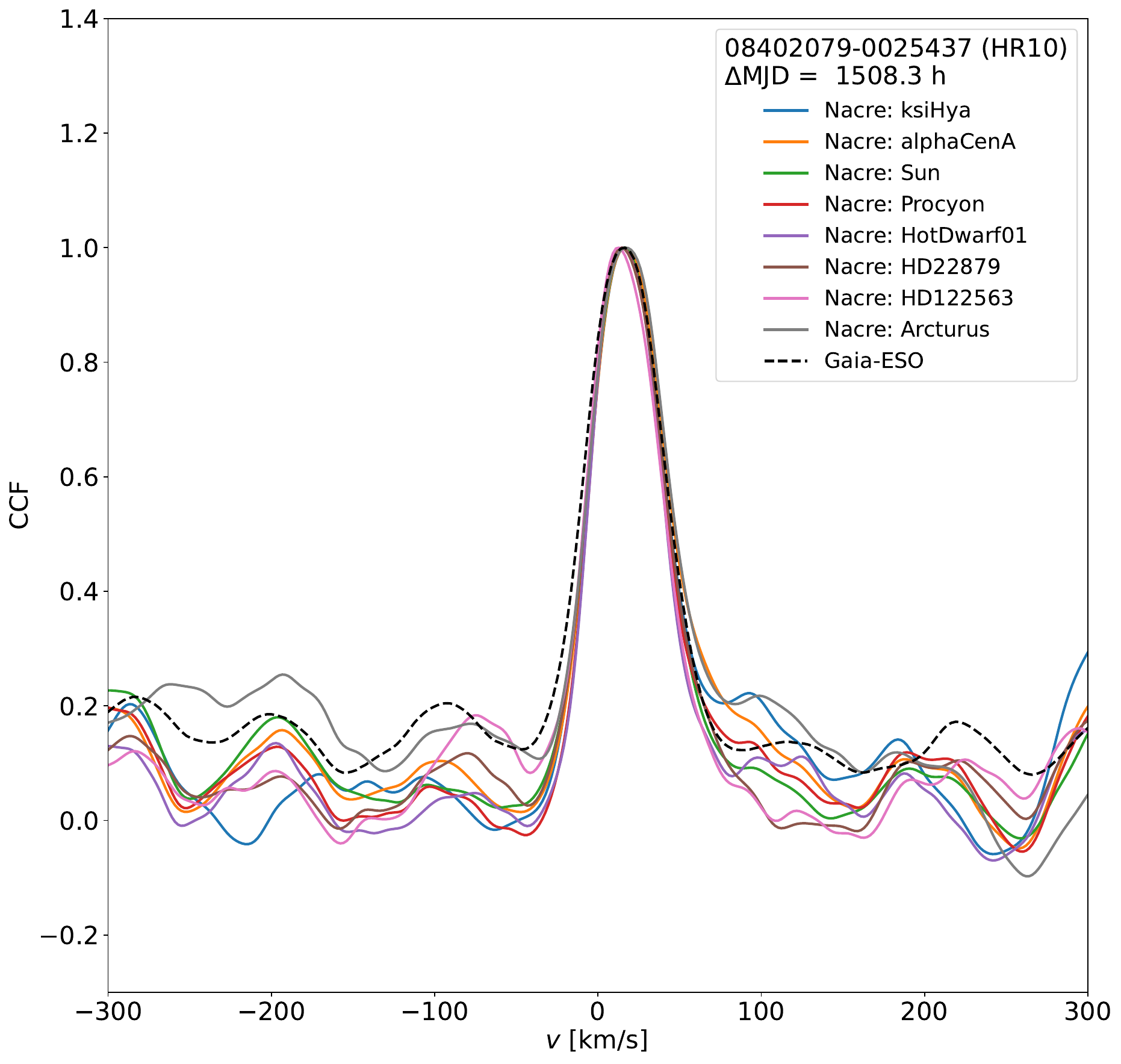}
  \includegraphics[width=0.49\textwidth]{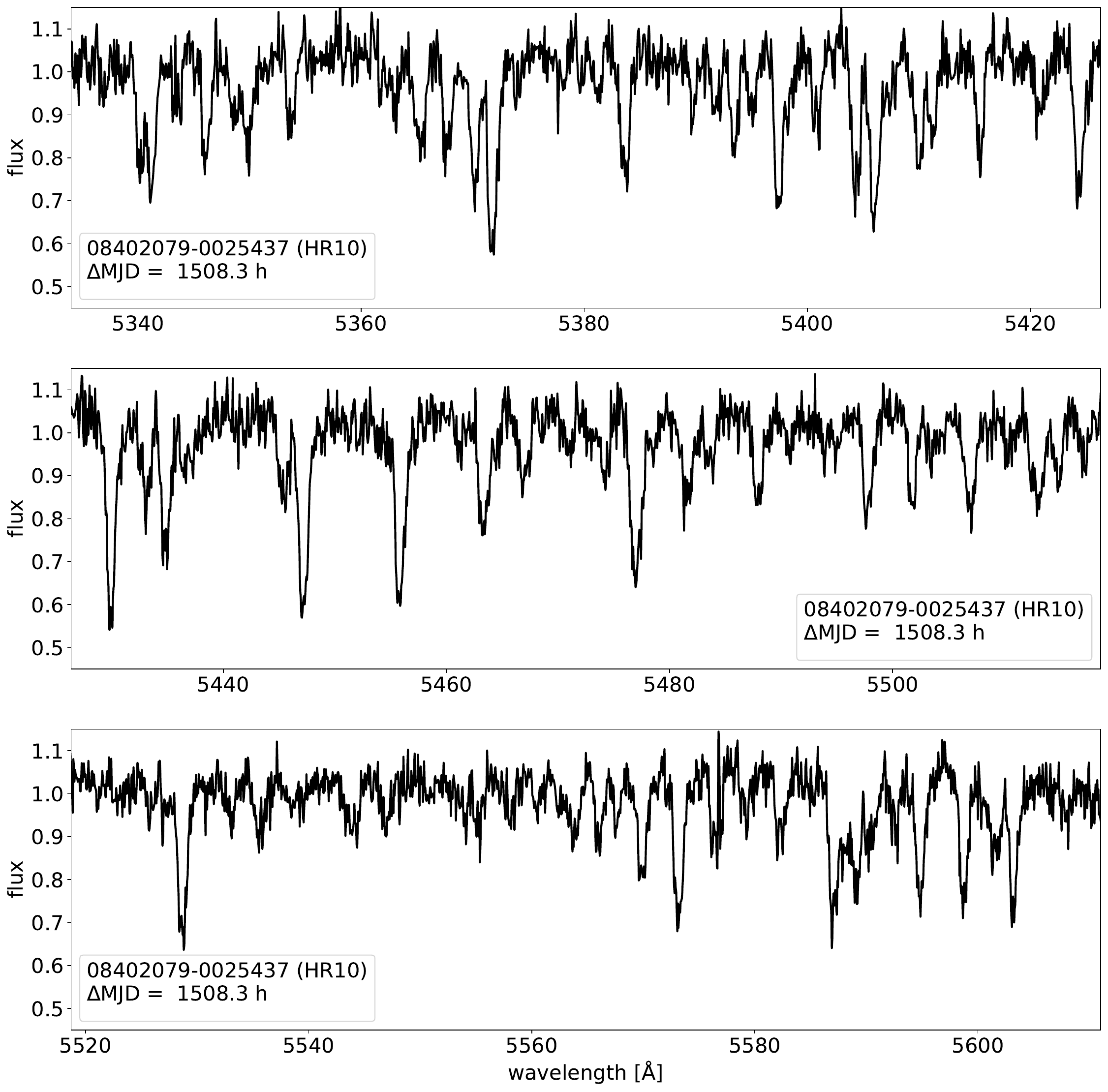}
  \captionof{figure}{\label{Fig:iDR5_SB3_atlas_08402079-0025437_4}(4/6) CNAME 08402079-0025437, at $\mathrm{MJD} = 56377.029952$, setup HR10.}
\end{minipage}
\clearpage
\begin{minipage}{\textwidth}
  \centering
  \includegraphics[width=0.49\textwidth]{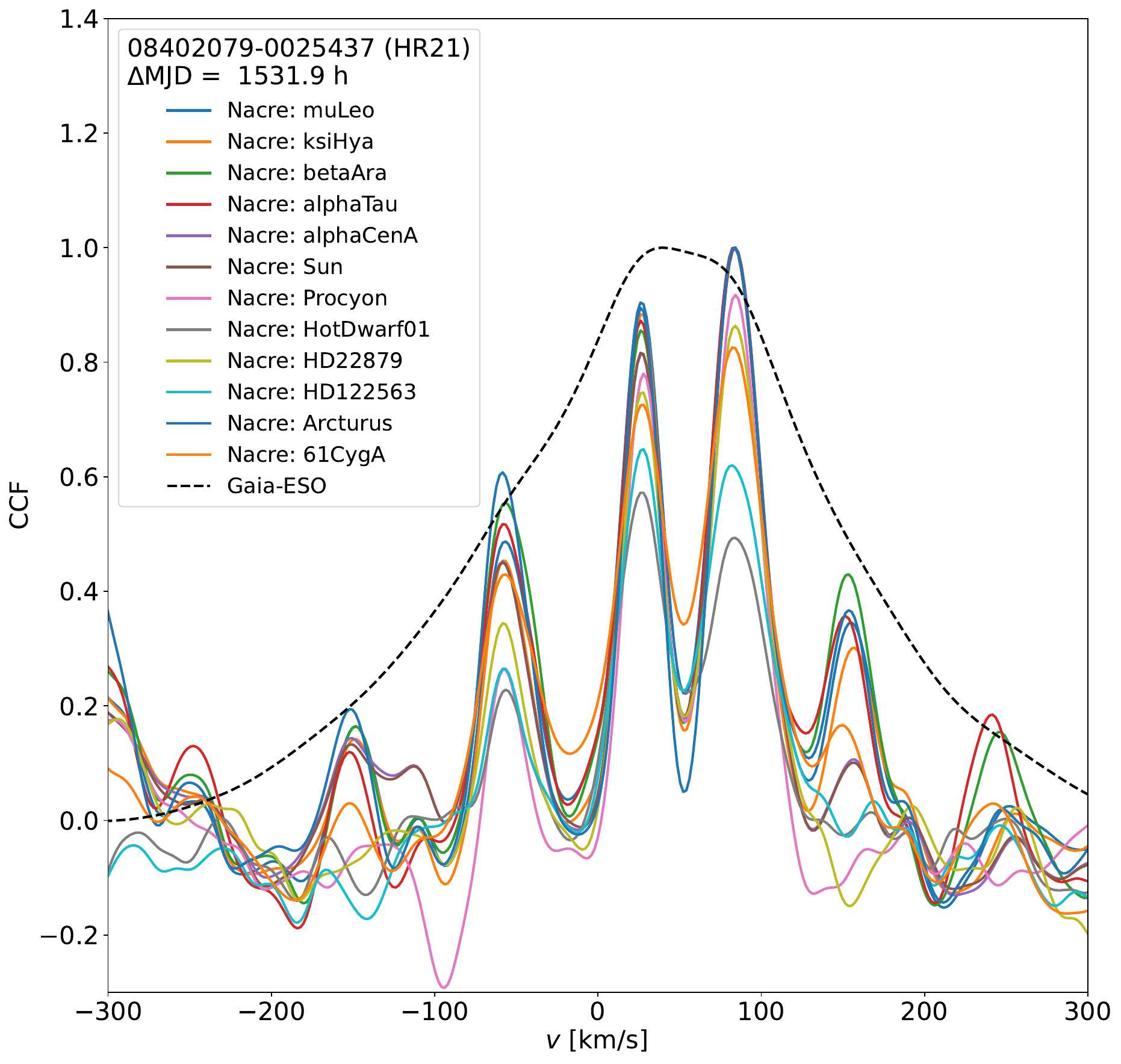}
  \includegraphics[width=0.49\textwidth]{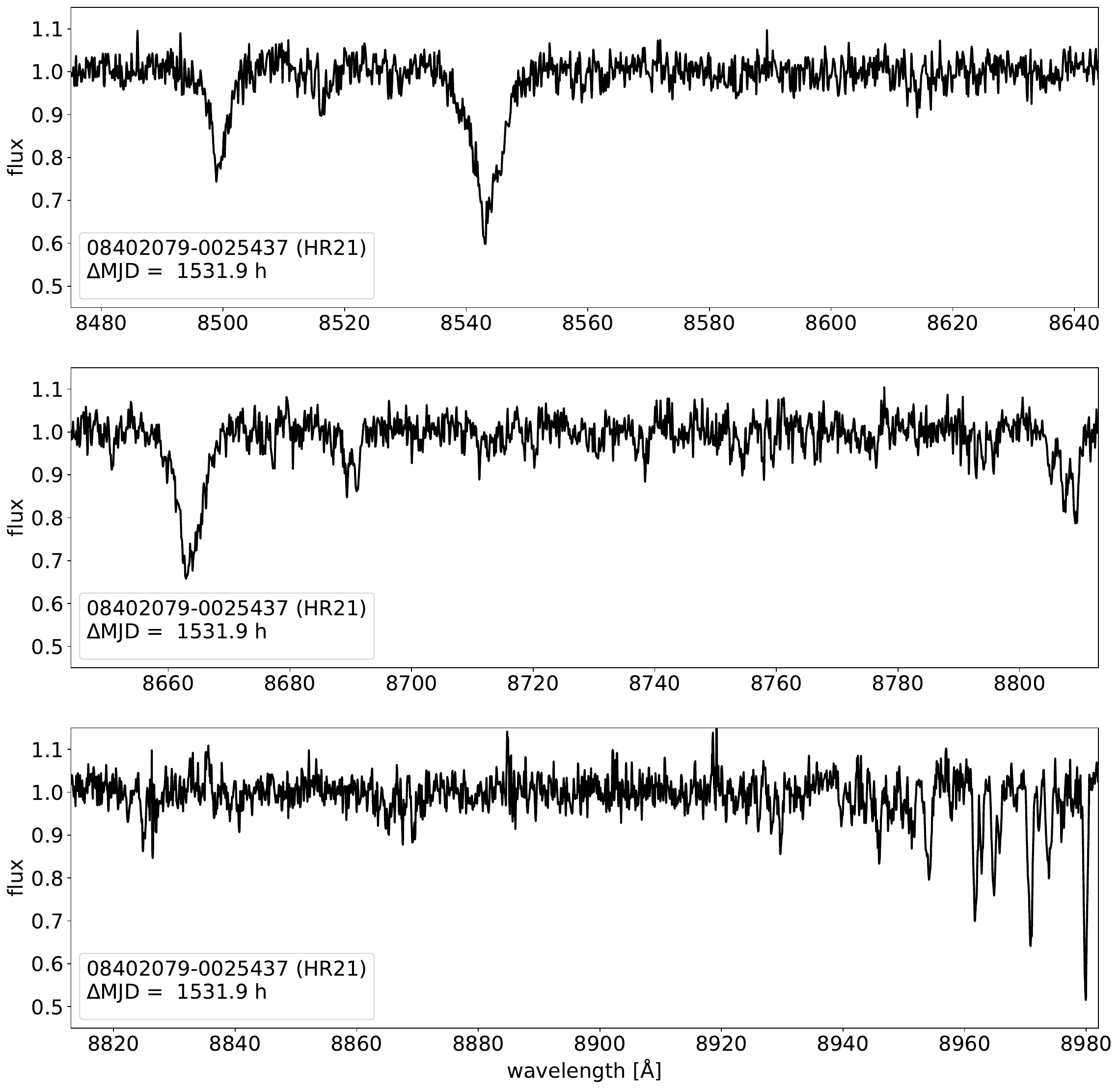}
  \captionof{figure}{\label{Fig:iDR5_SB3_atlas_08402079-0025437_5}(5/6) CNAME 08402079-0025437, at $\mathrm{MJD} = 56378.014598$, setup HR21.}
\end{minipage}
\begin{minipage}{\textwidth}
  \centering
  \includegraphics[width=0.49\textwidth]{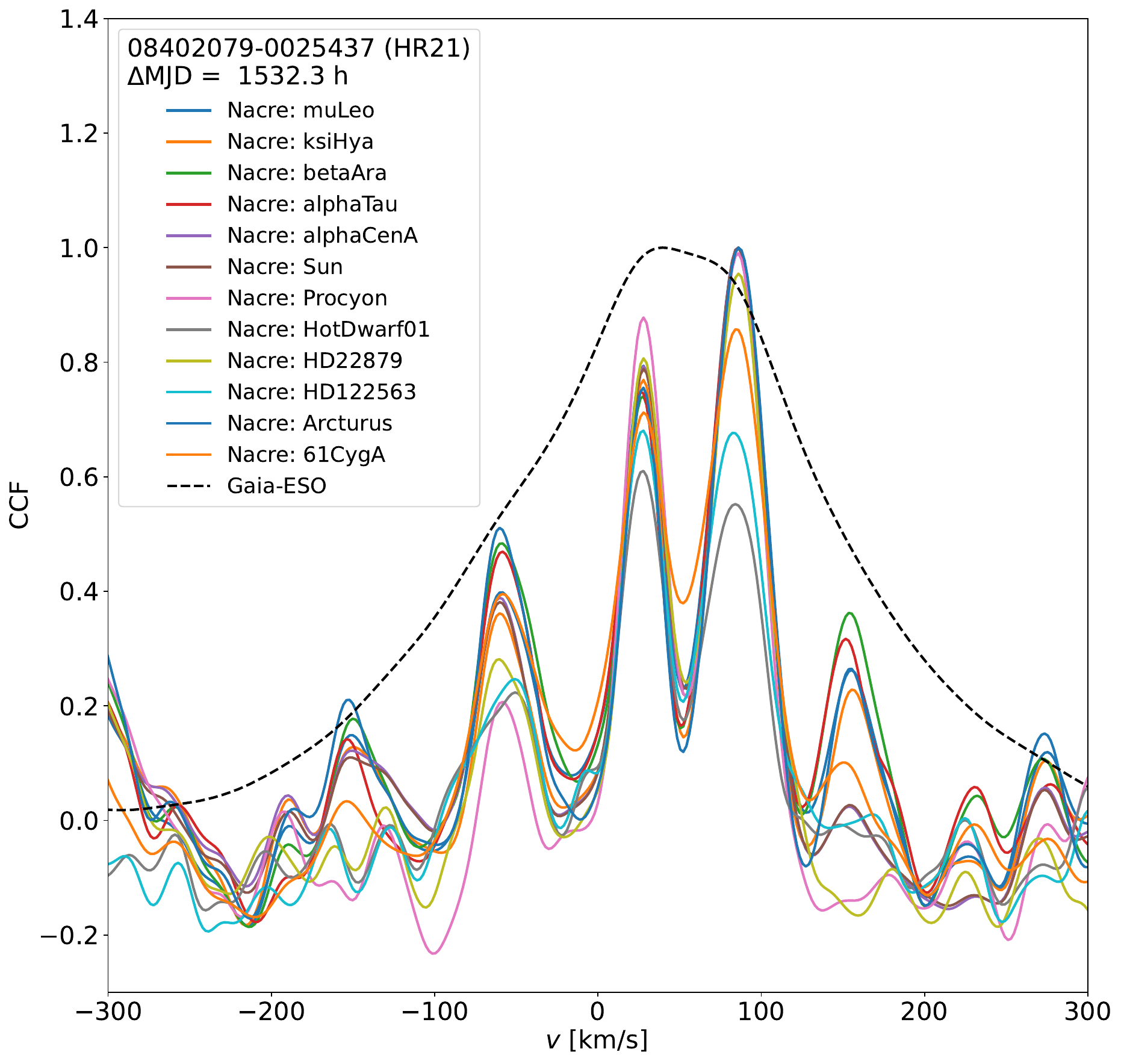}
  \includegraphics[width=0.49\textwidth]{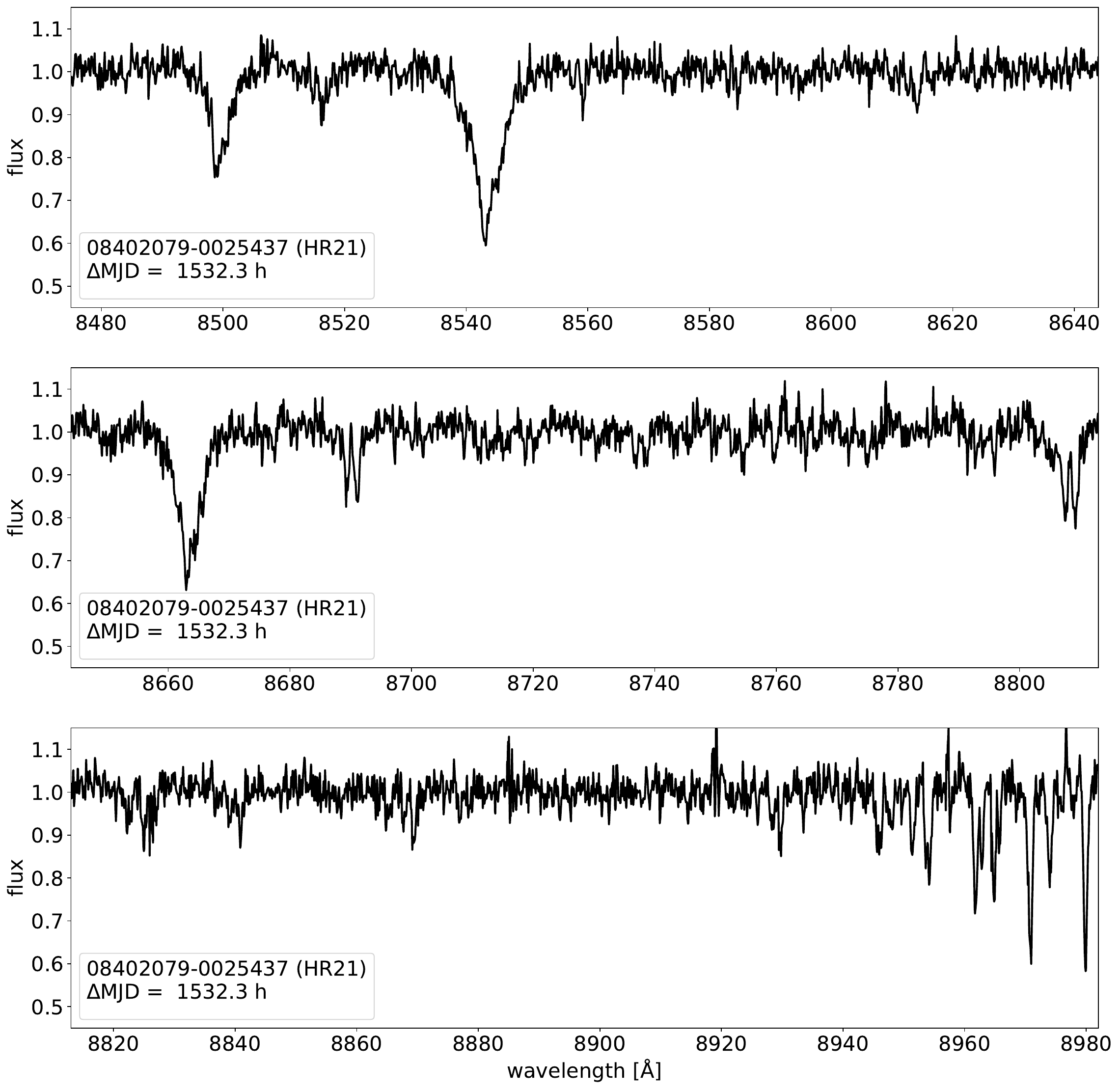}
  \captionof{figure}{\label{Fig:iDR5_SB3_atlas_08402079-0025437_6}(6/6) CNAME 08402079-0025437, at $\mathrm{MJD} = 56378.032579$, setup HR21.}
\end{minipage}
\cleardoublepage
\setlength\parindent{\defaultparindent}

\subsection{15120307-4049481}

\setlength\parindent{0cm}
\begin{minipage}{\textwidth}
  \centering
  \includegraphics[width=0.49\textwidth]{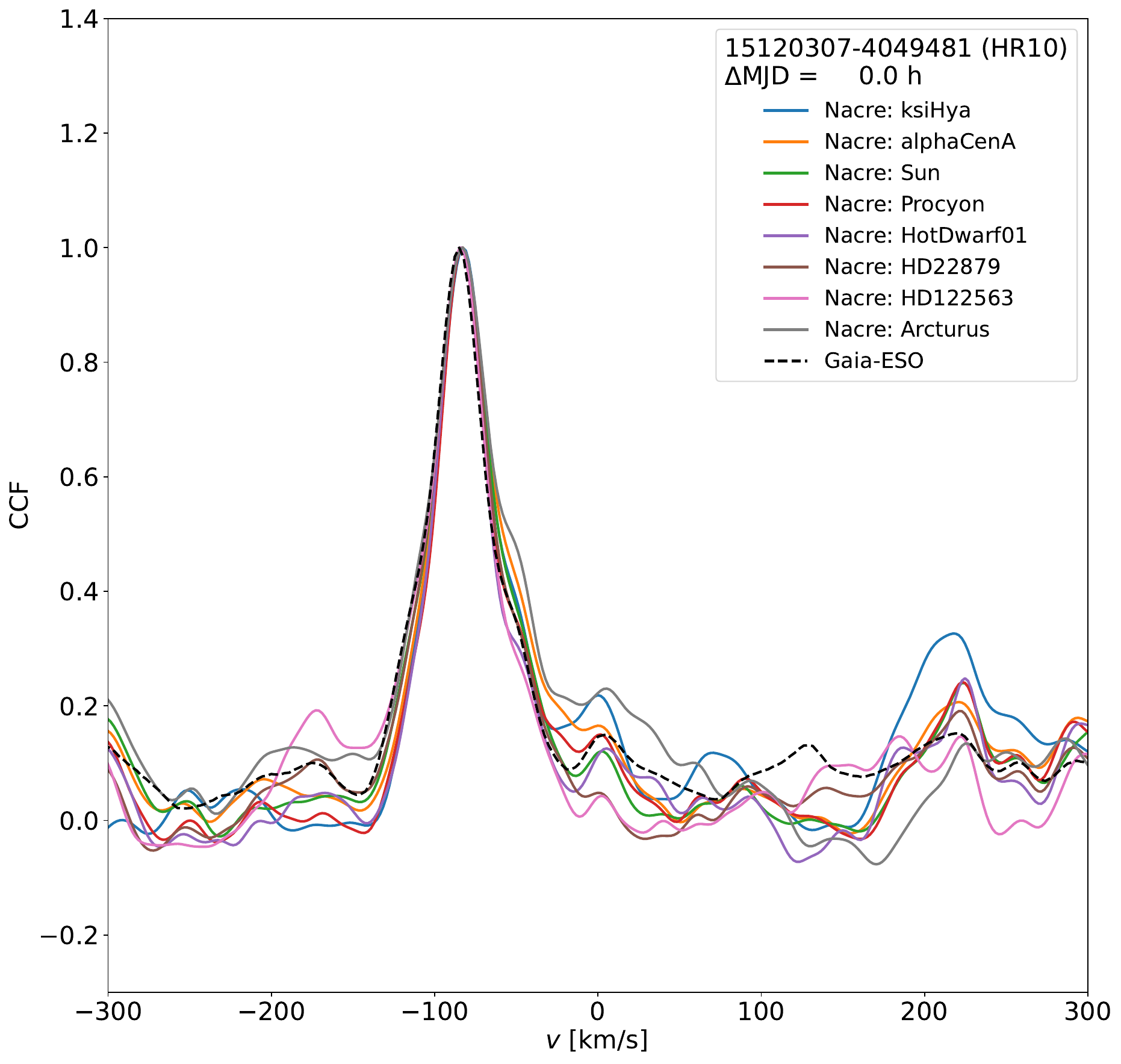}
  \includegraphics[width=0.49\textwidth]{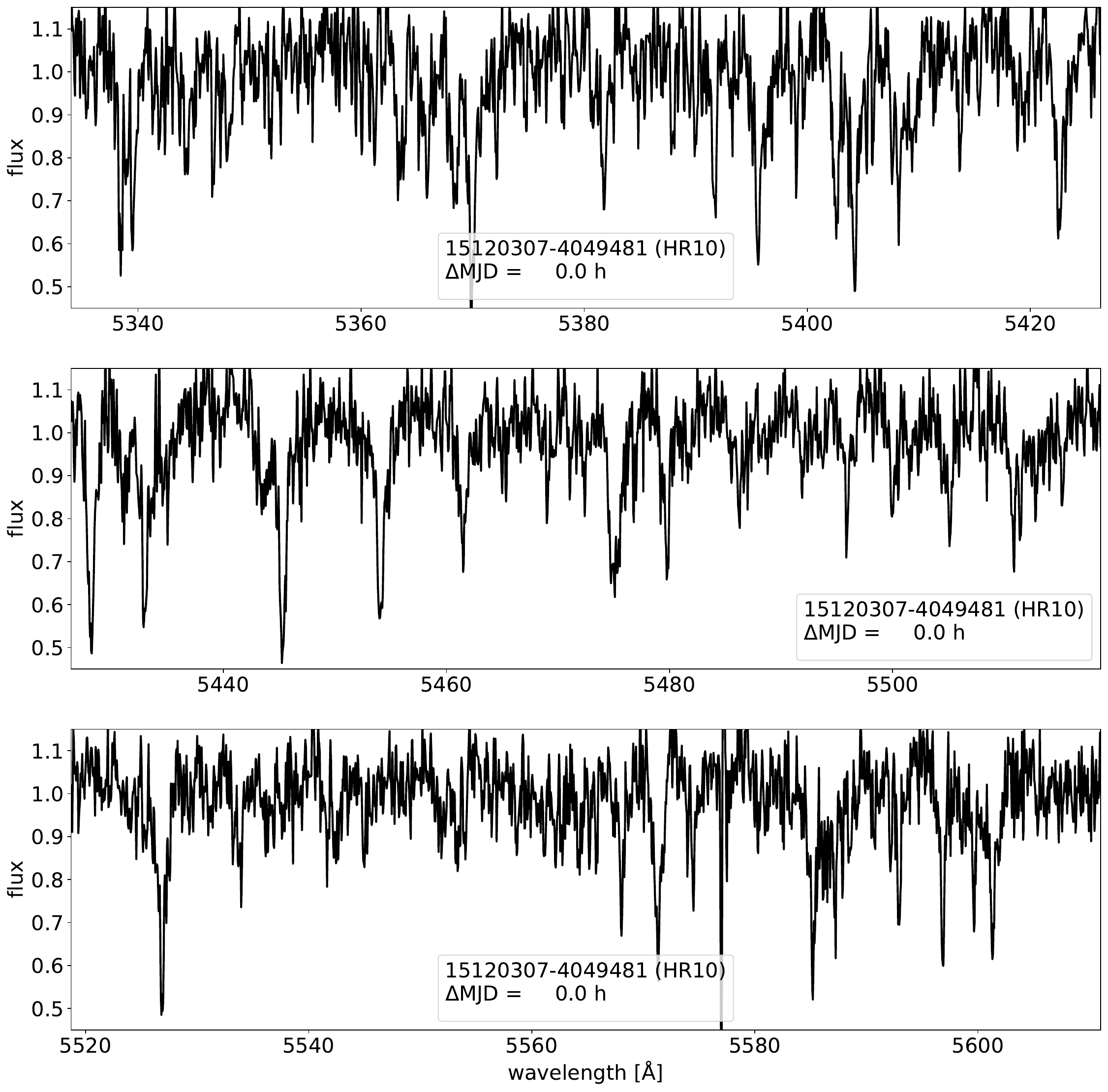}
  \captionof{figure}{\label{Fig:iDR5_SB3_atlas_15120307-4049481_1}(1/4) CNAME 15120307-4049481, at $\mathrm{MJD} = 56445.093053$, setup HR10.}
\end{minipage}
\begin{minipage}{\textwidth}
  \centering
  \includegraphics[width=0.49\textwidth]{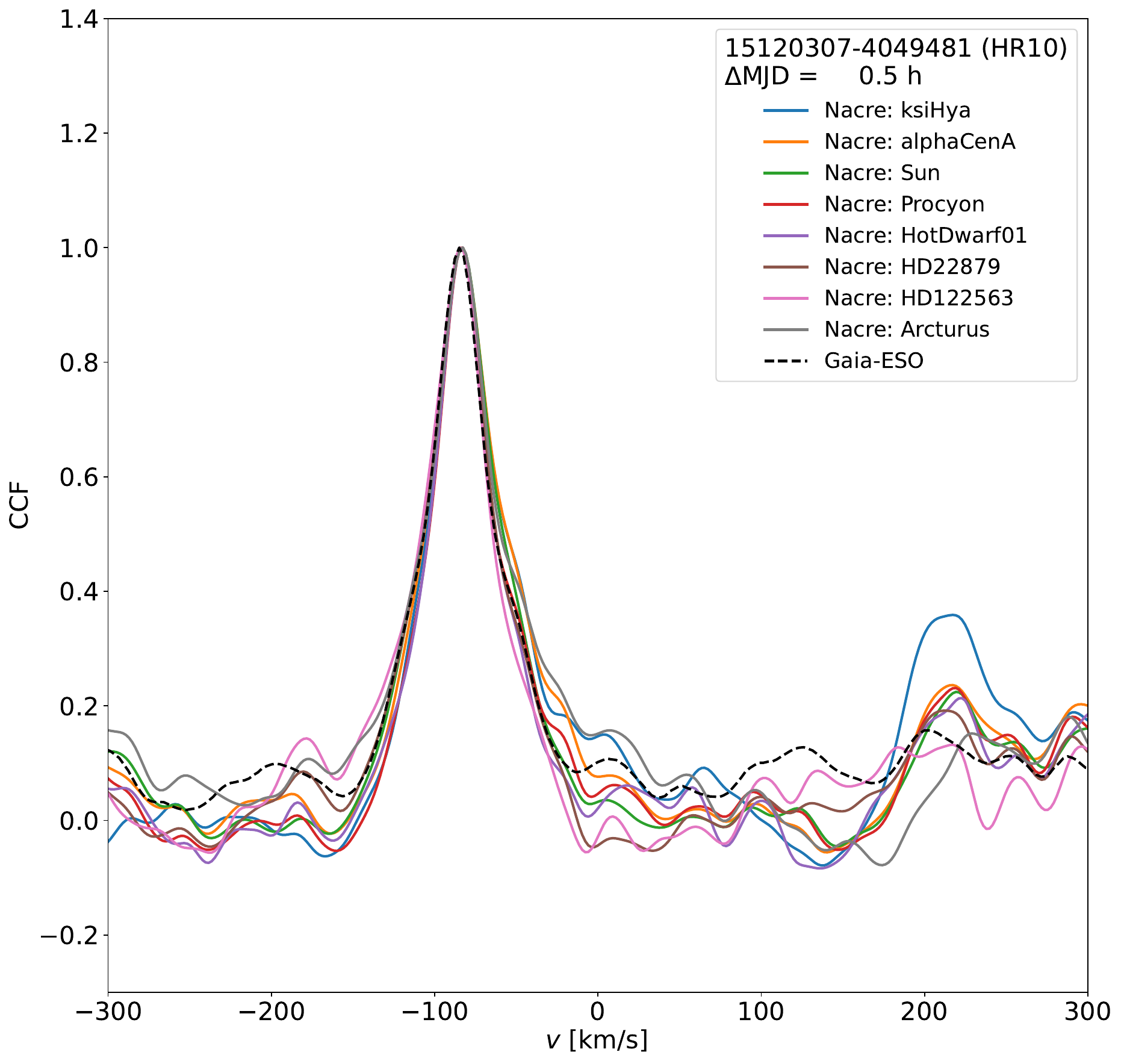}
  \includegraphics[width=0.49\textwidth]{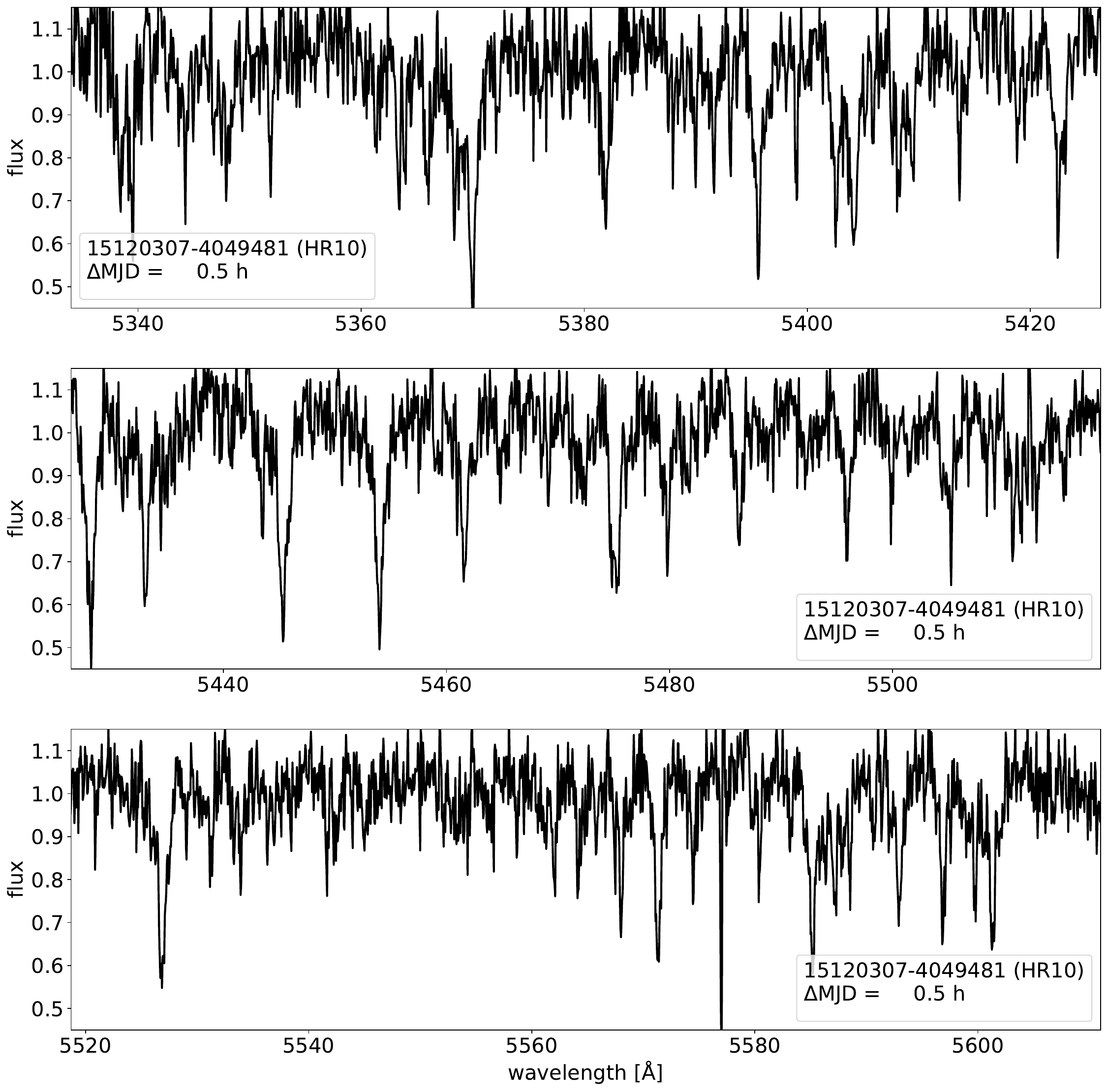}
  \captionof{figure}{\label{Fig:iDR5_SB3_atlas_15120307-4049481_2}(2/4) CNAME 15120307-4049481, at $\mathrm{MJD} = 56445.114255$, setup HR10.}
\end{minipage}
\clearpage
\begin{minipage}{\textwidth}
  \centering
  \includegraphics[width=0.49\textwidth]{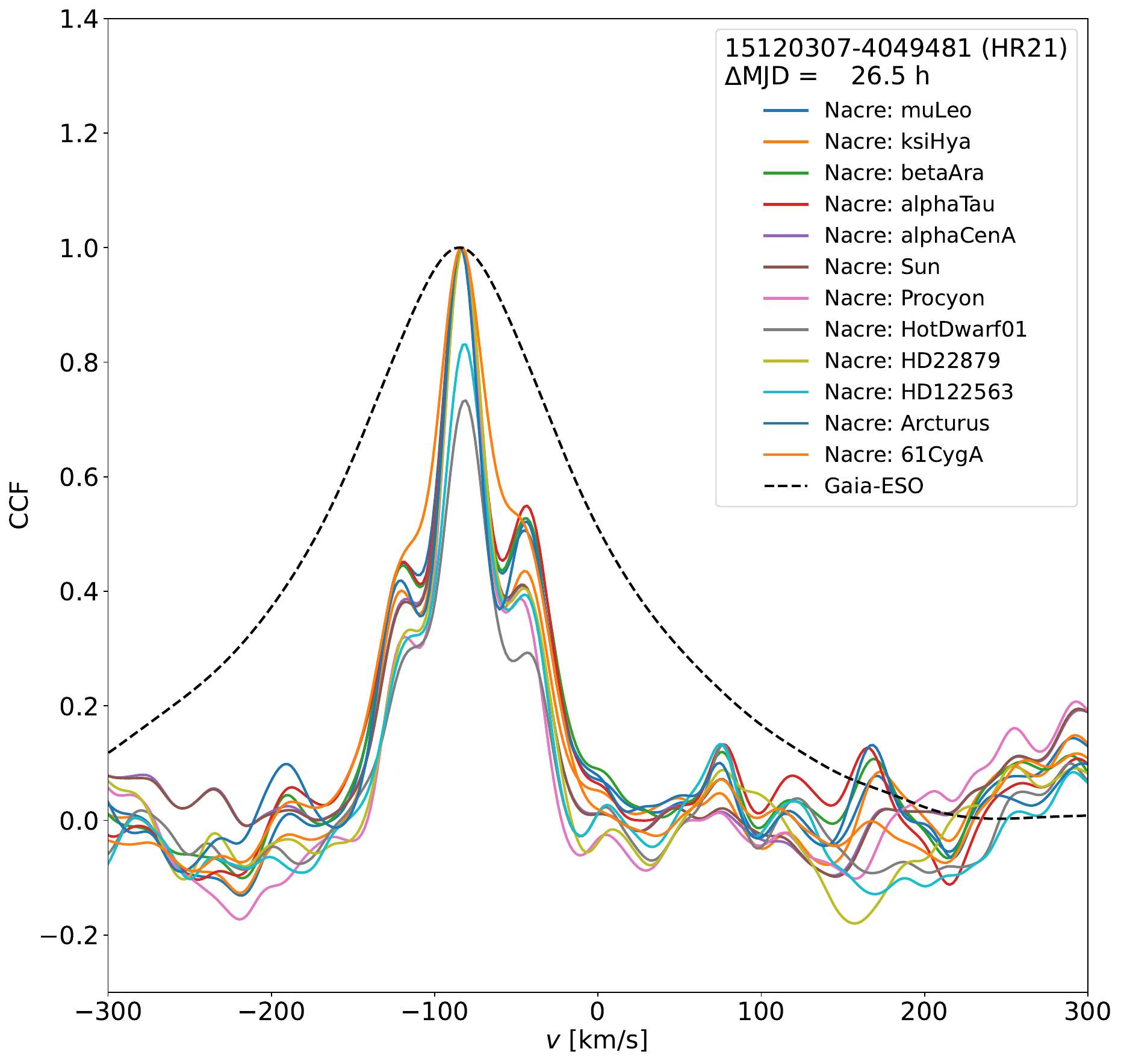}
  \includegraphics[width=0.49\textwidth]{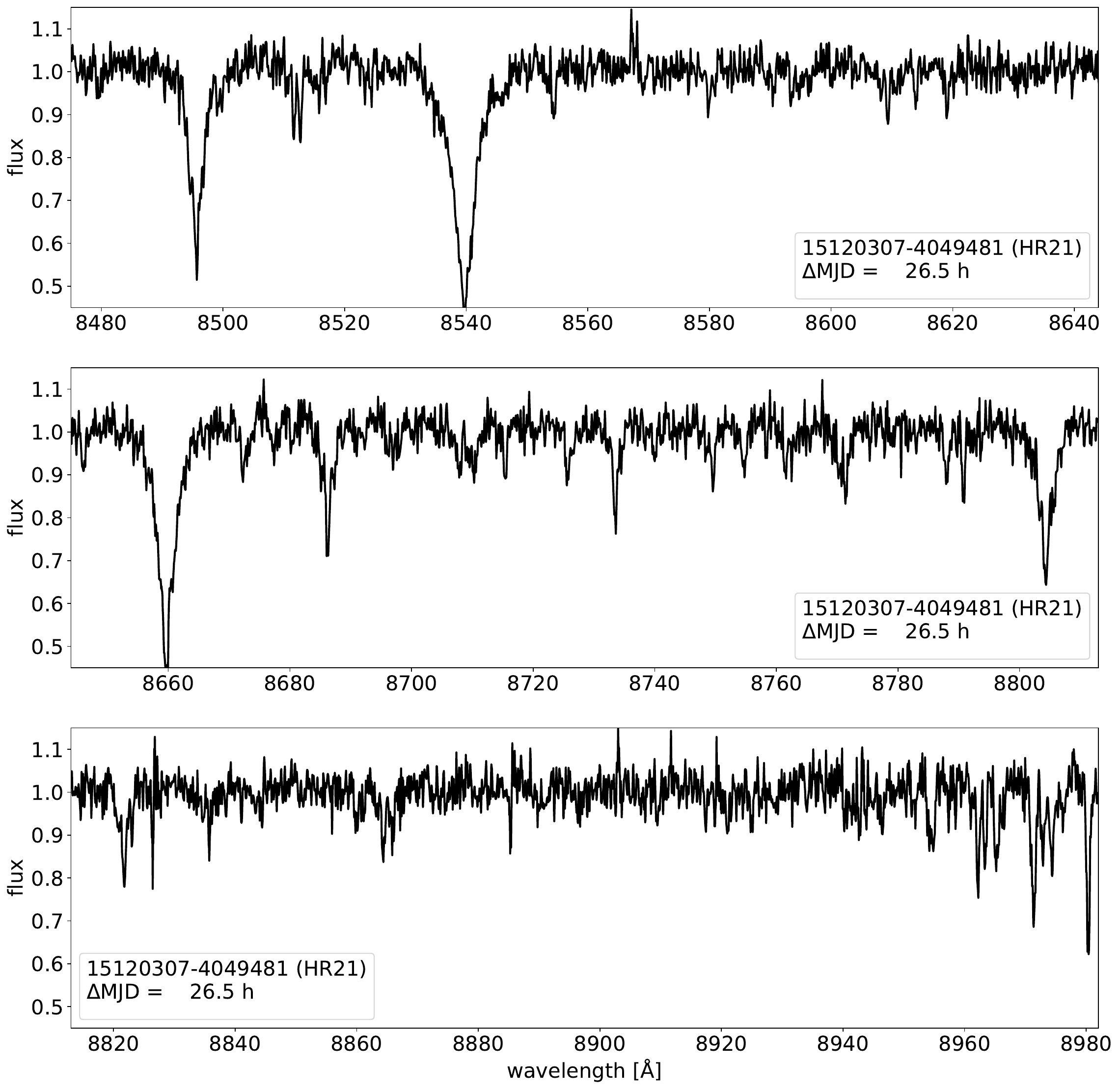}
  \captionof{figure}{\label{Fig:iDR5_SB3_atlas_15120307-4049481_3}(3/4) CNAME 15120307-4049481, at $\mathrm{MJD} = 56446.197221$, setup HR21.}
\end{minipage}
\begin{minipage}{\textwidth}
  \centering
  \includegraphics[width=0.49\textwidth]{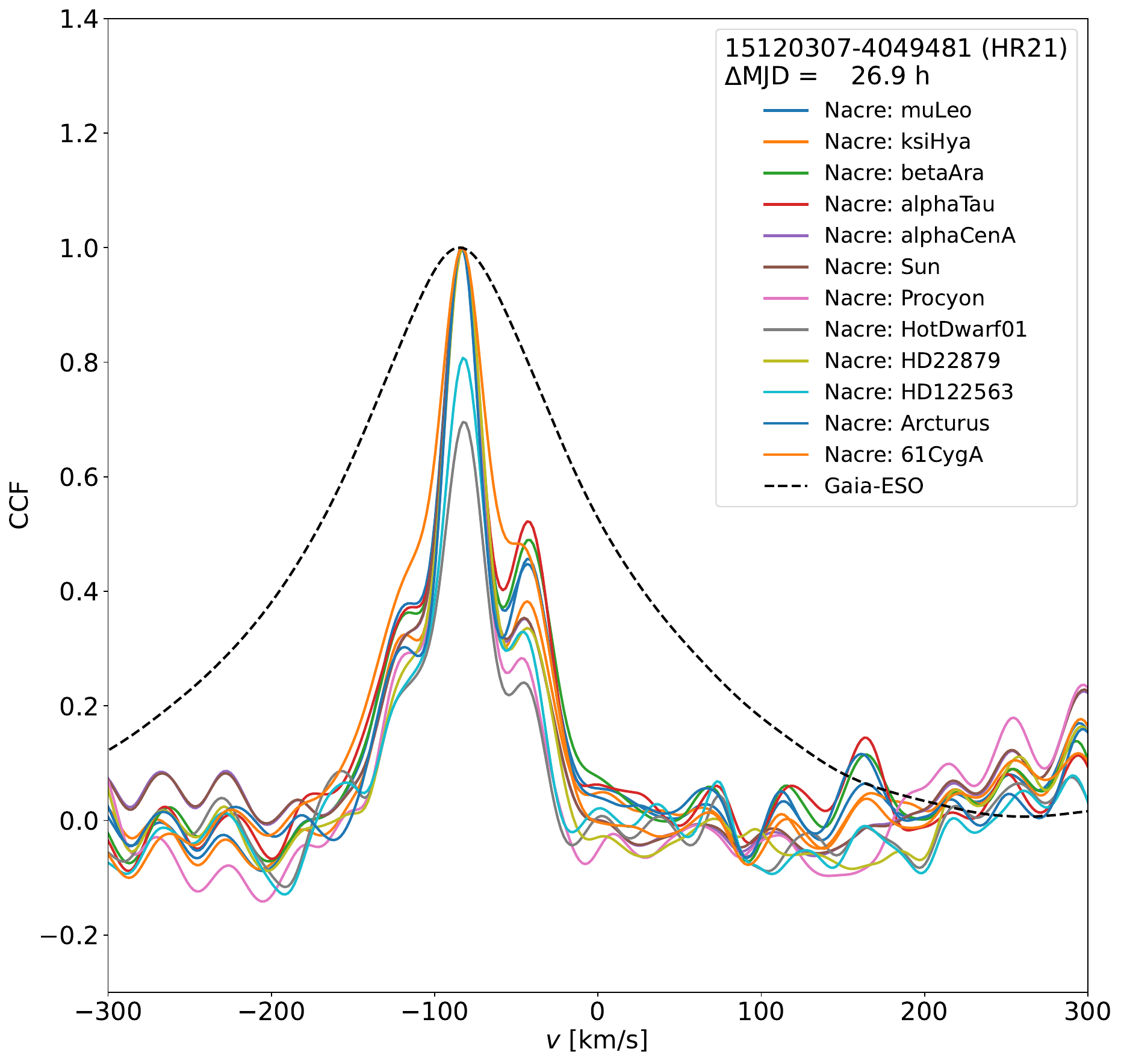}
  \includegraphics[width=0.49\textwidth]{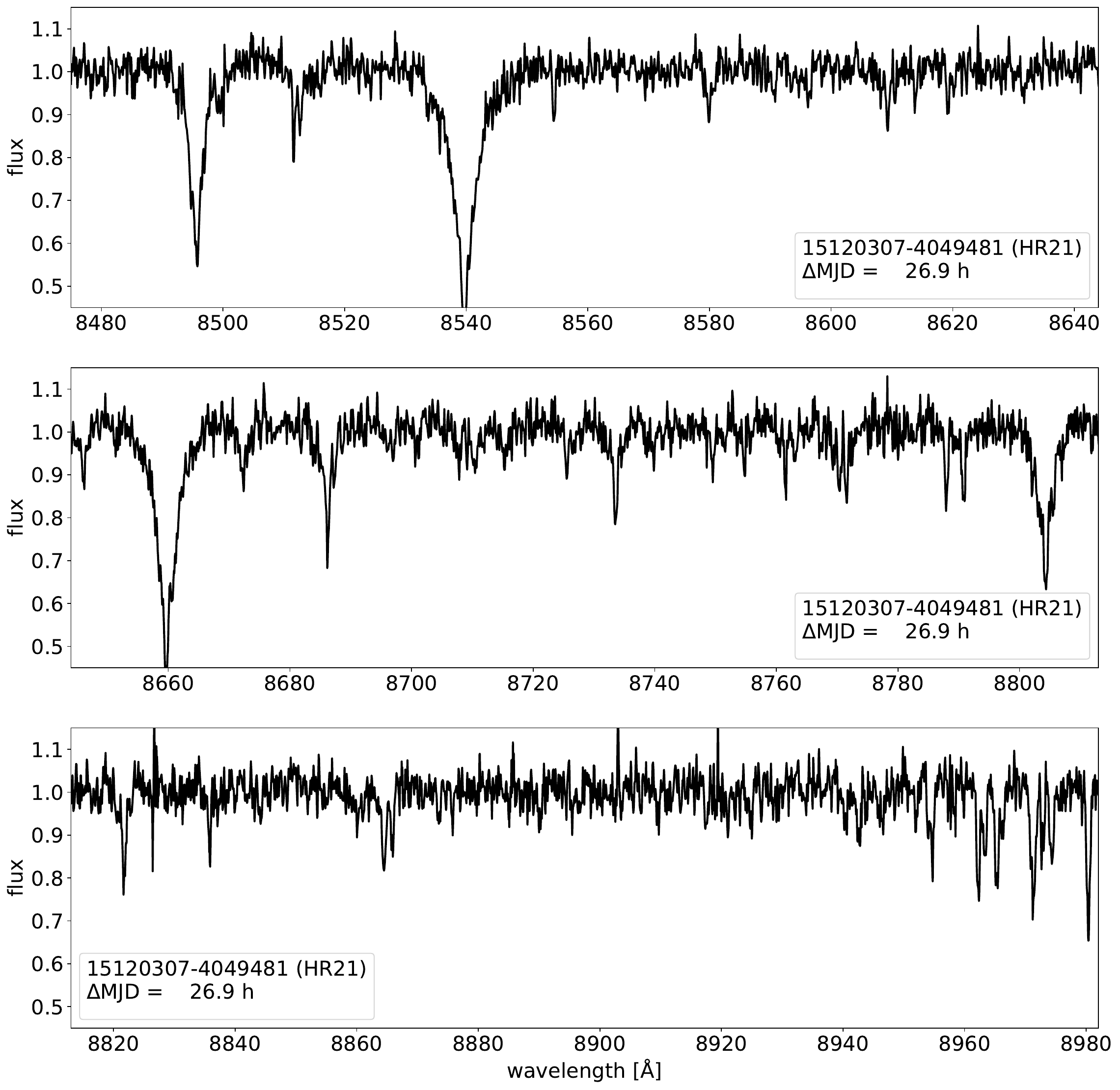}
  \captionof{figure}{\label{Fig:iDR5_SB3_atlas_15120307-4049481_4}(4/4) CNAME 15120307-4049481, at $\mathrm{MJD} = 56446.215217$, setup HR21.}
\end{minipage}
\cleardoublepage
\setlength\parindent{\defaultparindent}

\subsection{15161563-4125518}

\setlength\parindent{0cm}
\begin{minipage}{\textwidth}
  \centering
  \includegraphics[width=0.49\textwidth]{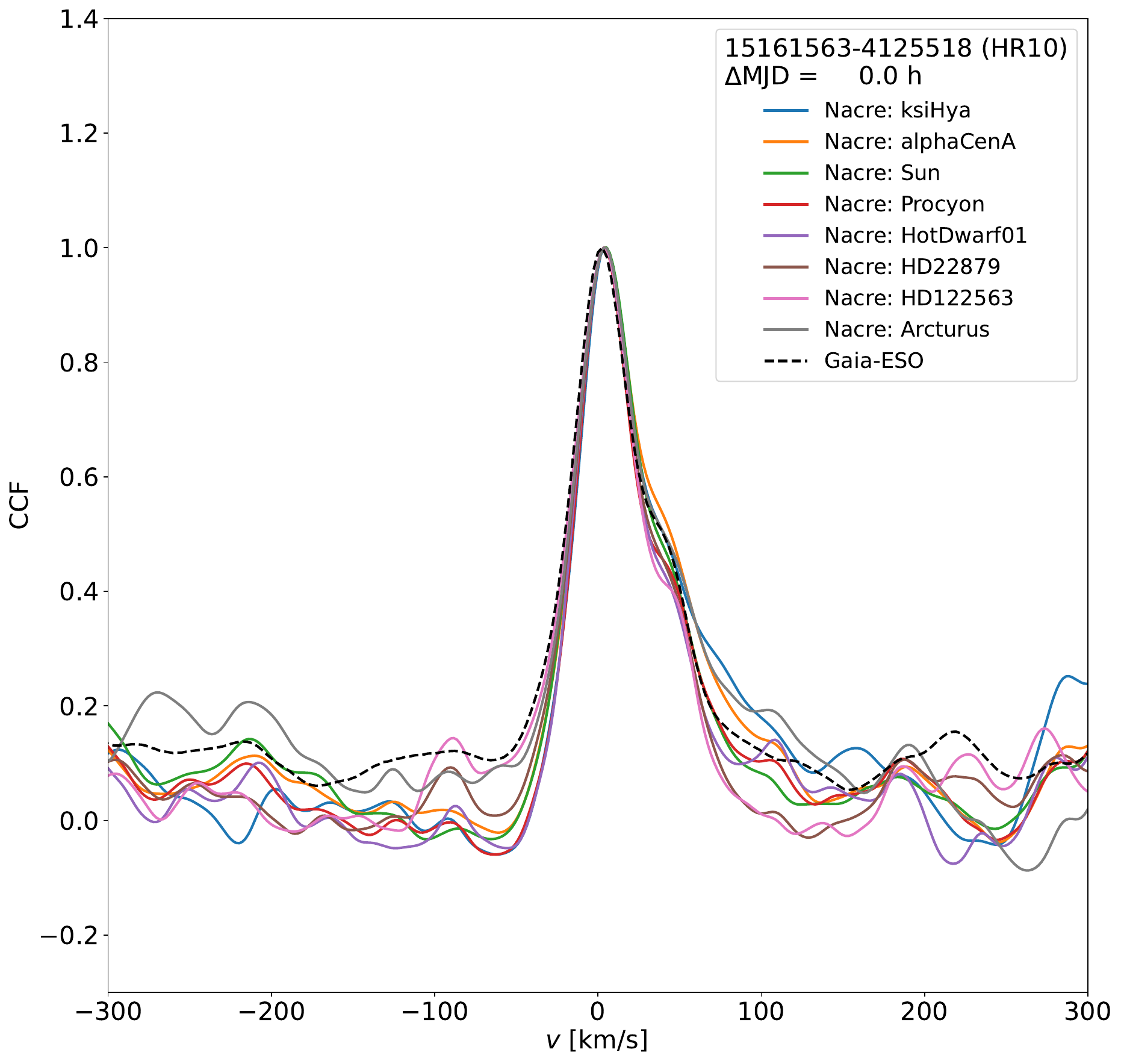}
  \includegraphics[width=0.49\textwidth]{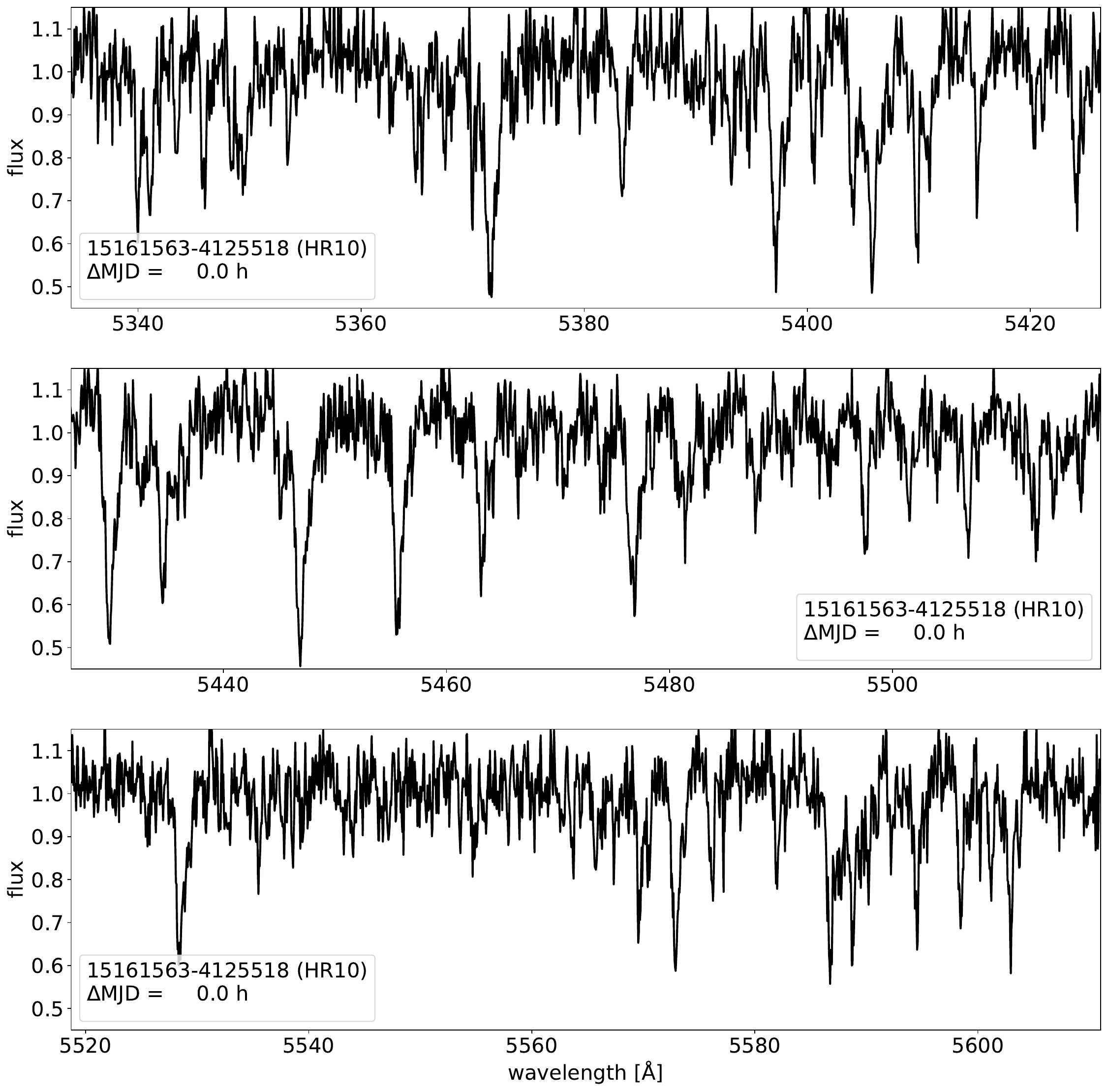}
  \captionof{figure}{\label{Fig:iDR5_SB3_atlas_15161563-4125518_1}(1/4) CNAME 15161563-4125518, at $\mathrm{MJD} = 56443.110397$, setup HR10.}
\end{minipage}
\begin{minipage}{\textwidth}
  \centering
  \includegraphics[width=0.49\textwidth]{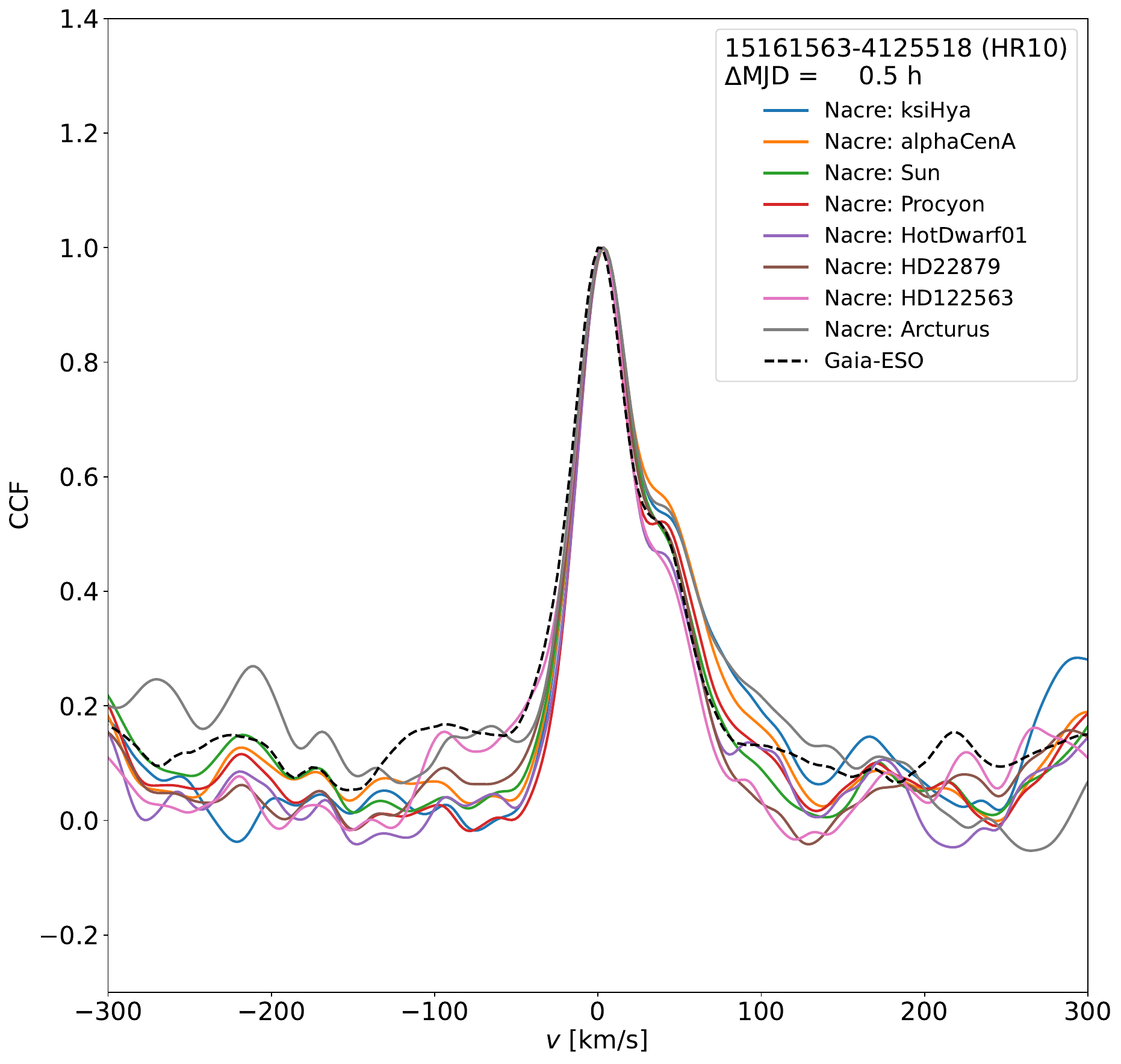}
  \includegraphics[width=0.49\textwidth]{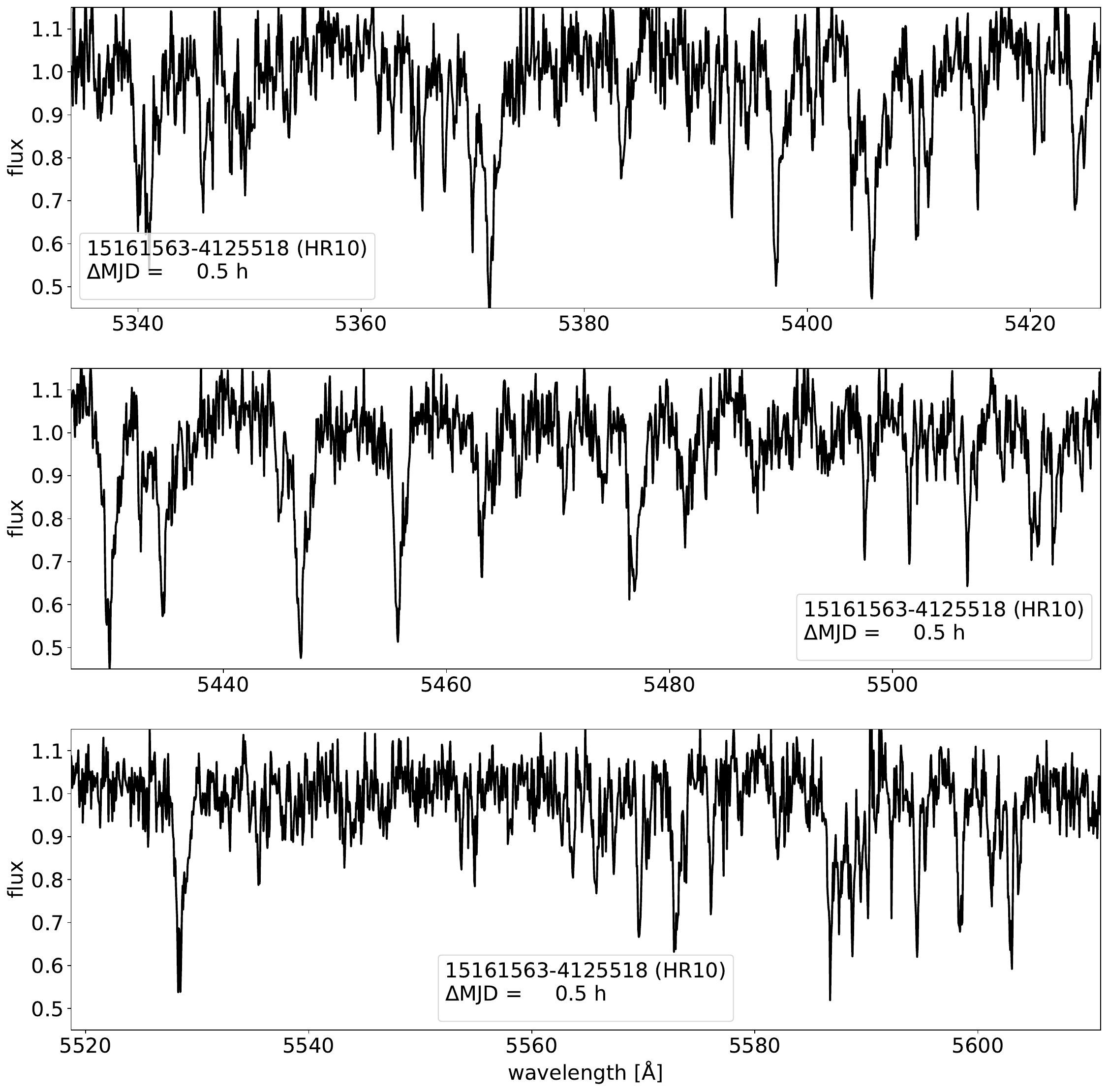}
  \captionof{figure}{\label{Fig:iDR5_SB3_atlas_15161563-4125518_2}(2/4) CNAME 15161563-4125518, at $\mathrm{MJD} = 56443.131571$, setup HR10.}
\end{minipage}
\clearpage
\begin{minipage}{\textwidth}
  \centering
  \includegraphics[width=0.49\textwidth]{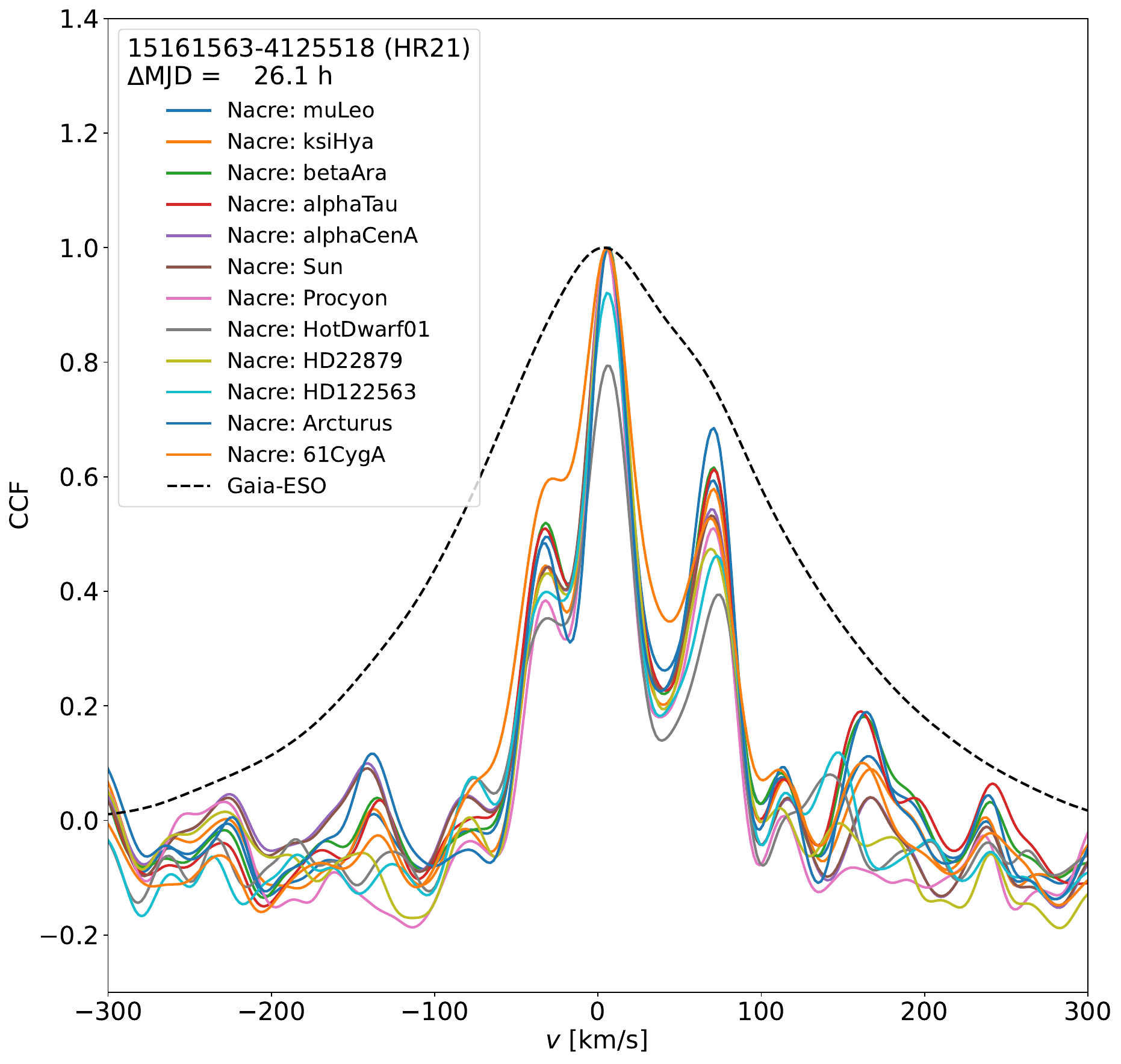}
  \includegraphics[width=0.49\textwidth]{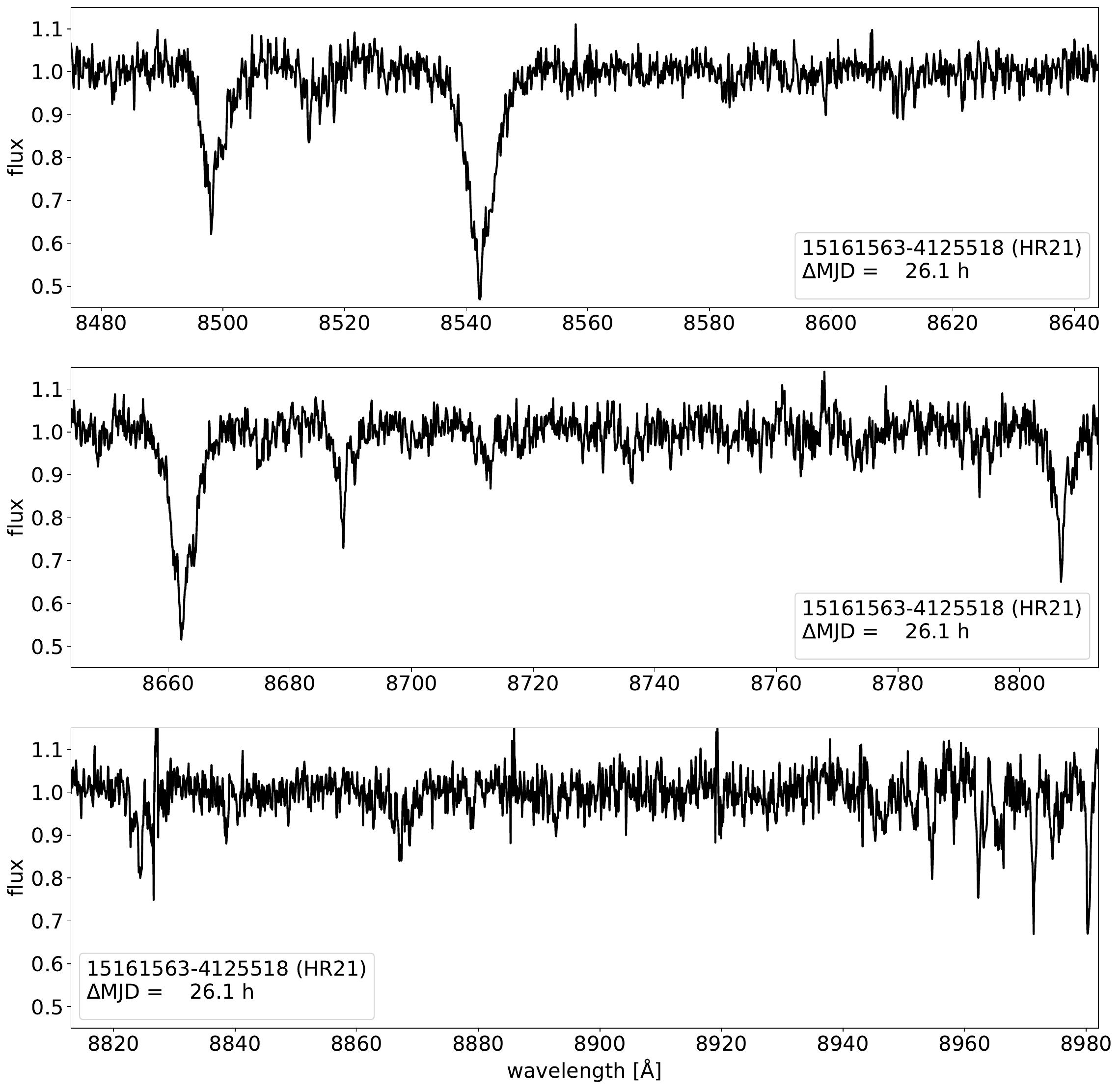}
  \captionof{figure}{\label{Fig:iDR5_SB3_atlas_15161563-4125518_3}(3/4) CNAME 15161563-4125518, at $\mathrm{MJD} = 56444.196398$, setup HR21.}
\end{minipage}
\begin{minipage}{\textwidth}
  \centering
  \includegraphics[width=0.49\textwidth]{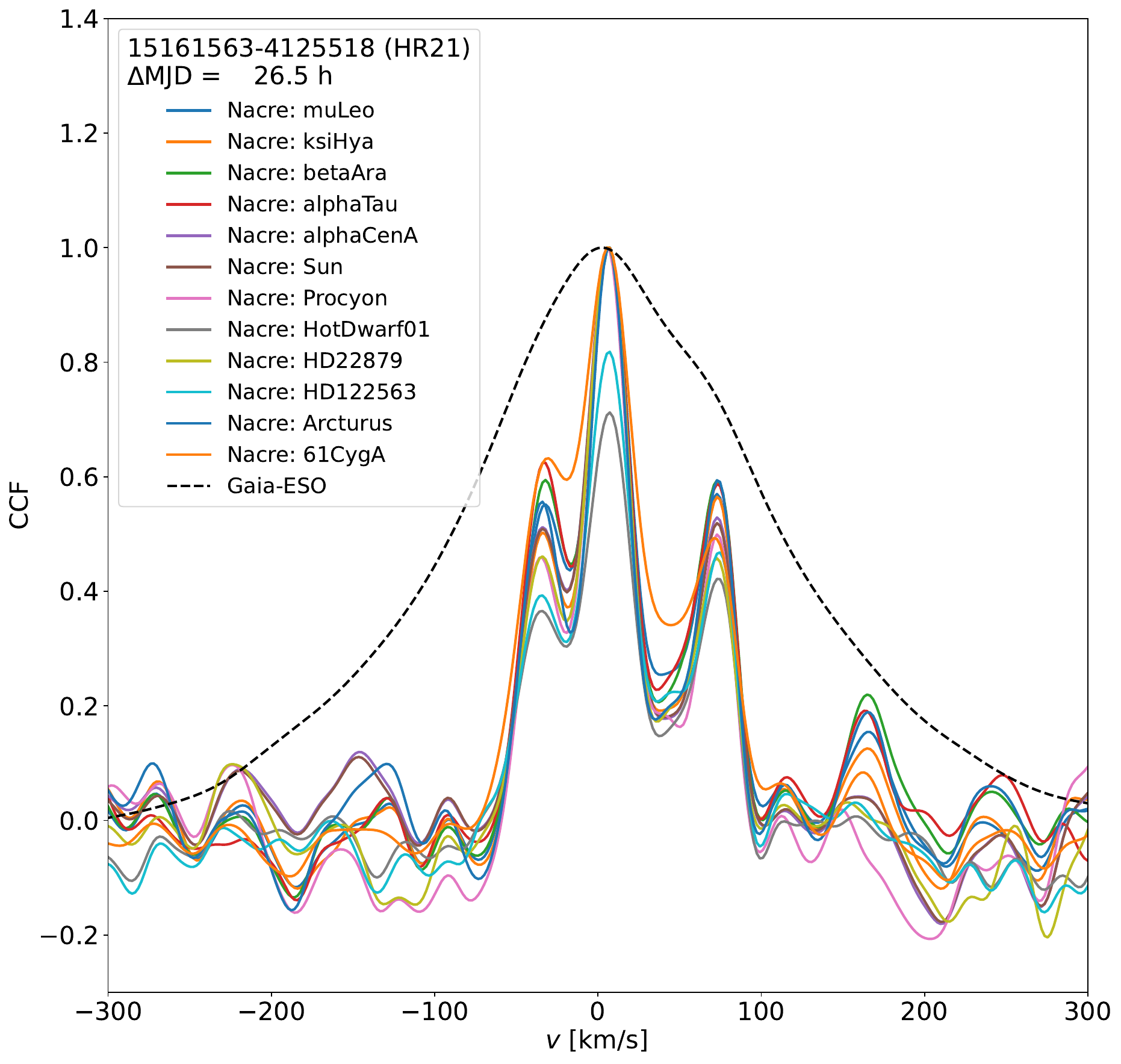}
  \includegraphics[width=0.49\textwidth]{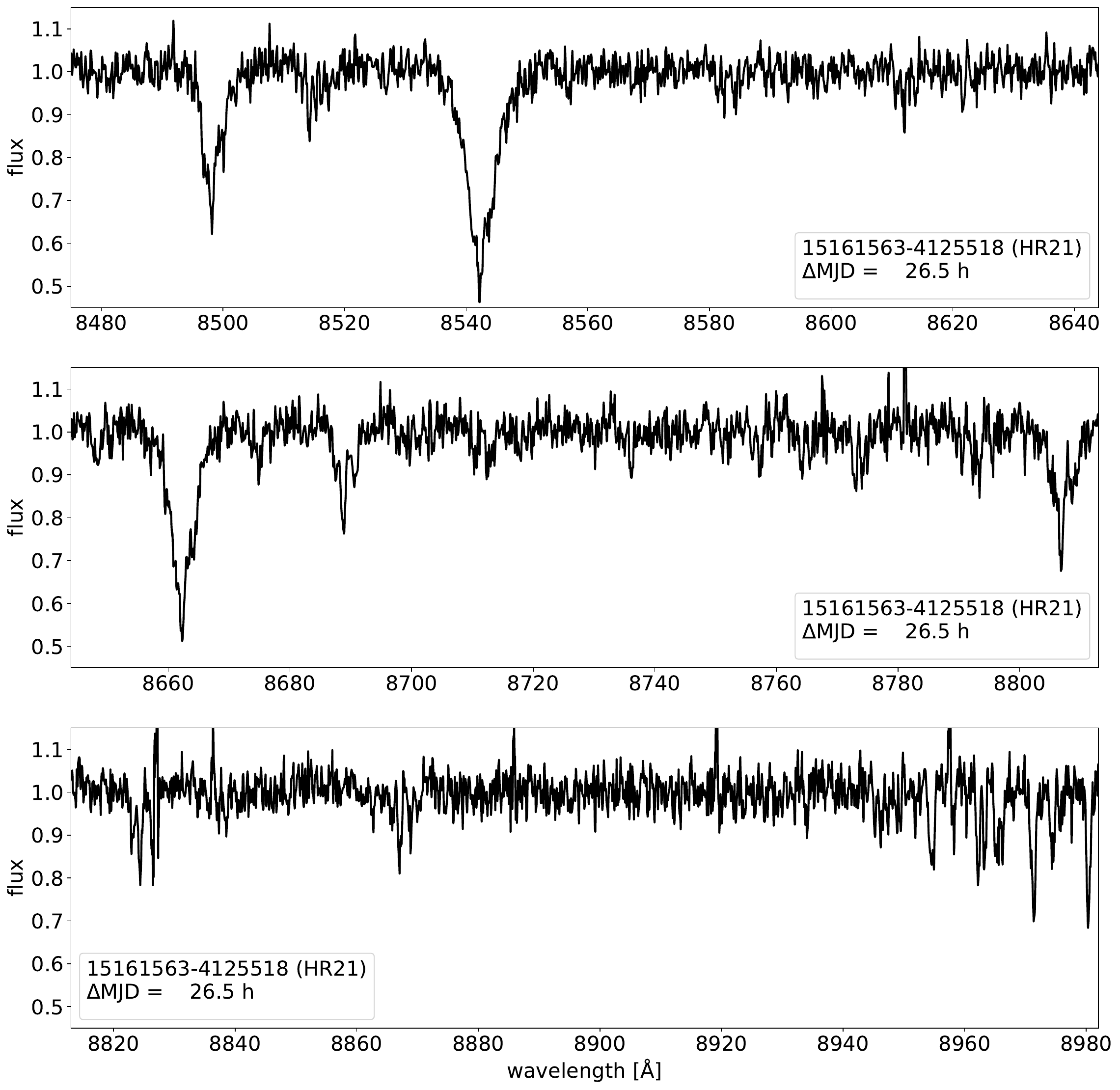}
  \captionof{figure}{\label{Fig:iDR5_SB3_atlas_15161563-4125518_4}(4/4) CNAME 15161563-4125518, at $\mathrm{MJD} = 56444.214383$, setup HR21.}
\end{minipage}
\cleardoublepage
\setlength\parindent{\defaultparindent}

\subsection{15420717-4407146}

\setlength\parindent{0cm}
\begin{minipage}{\textwidth}
  \centering
  \includegraphics[width=0.49\textwidth]{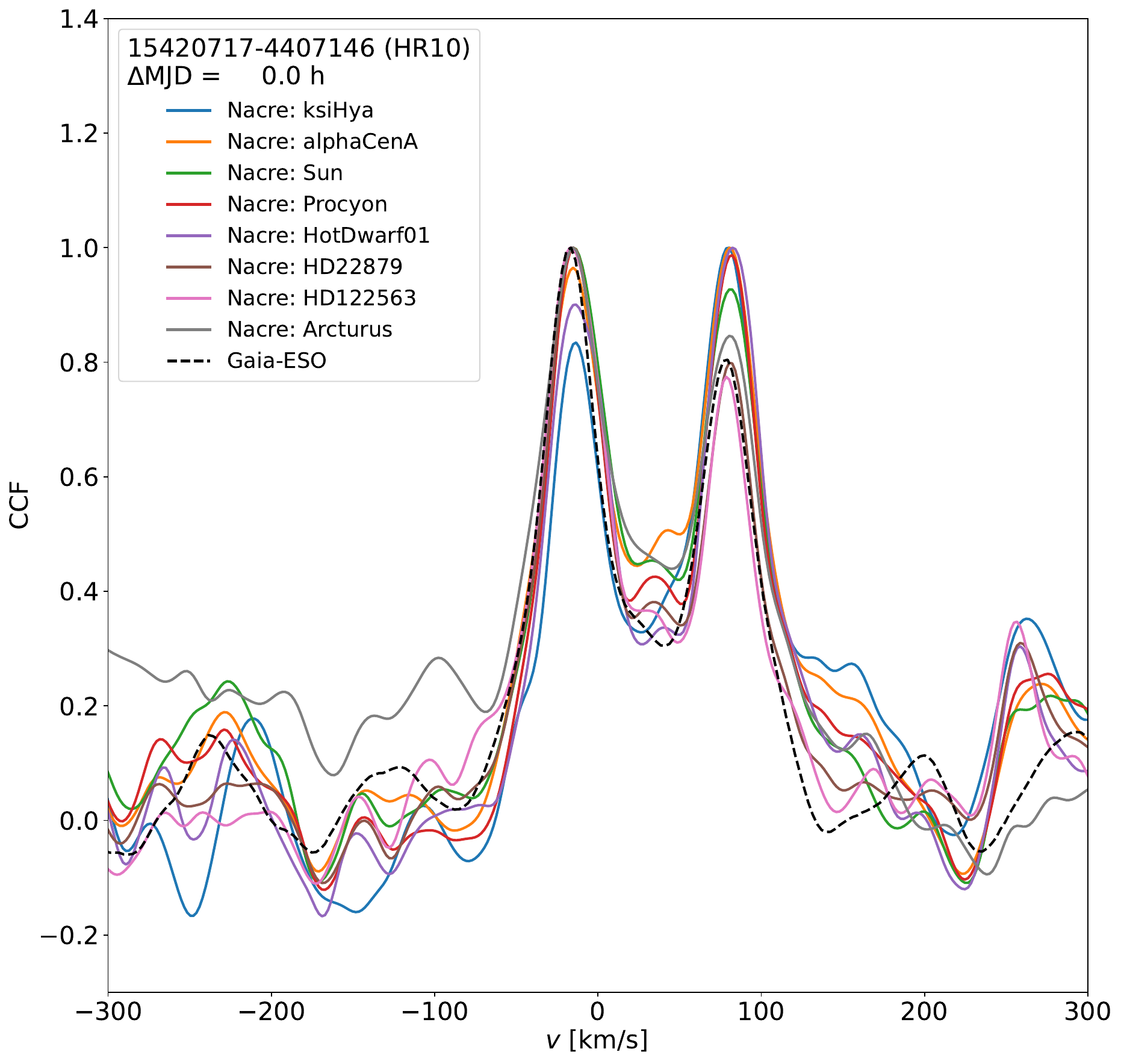}
  \includegraphics[width=0.49\textwidth]{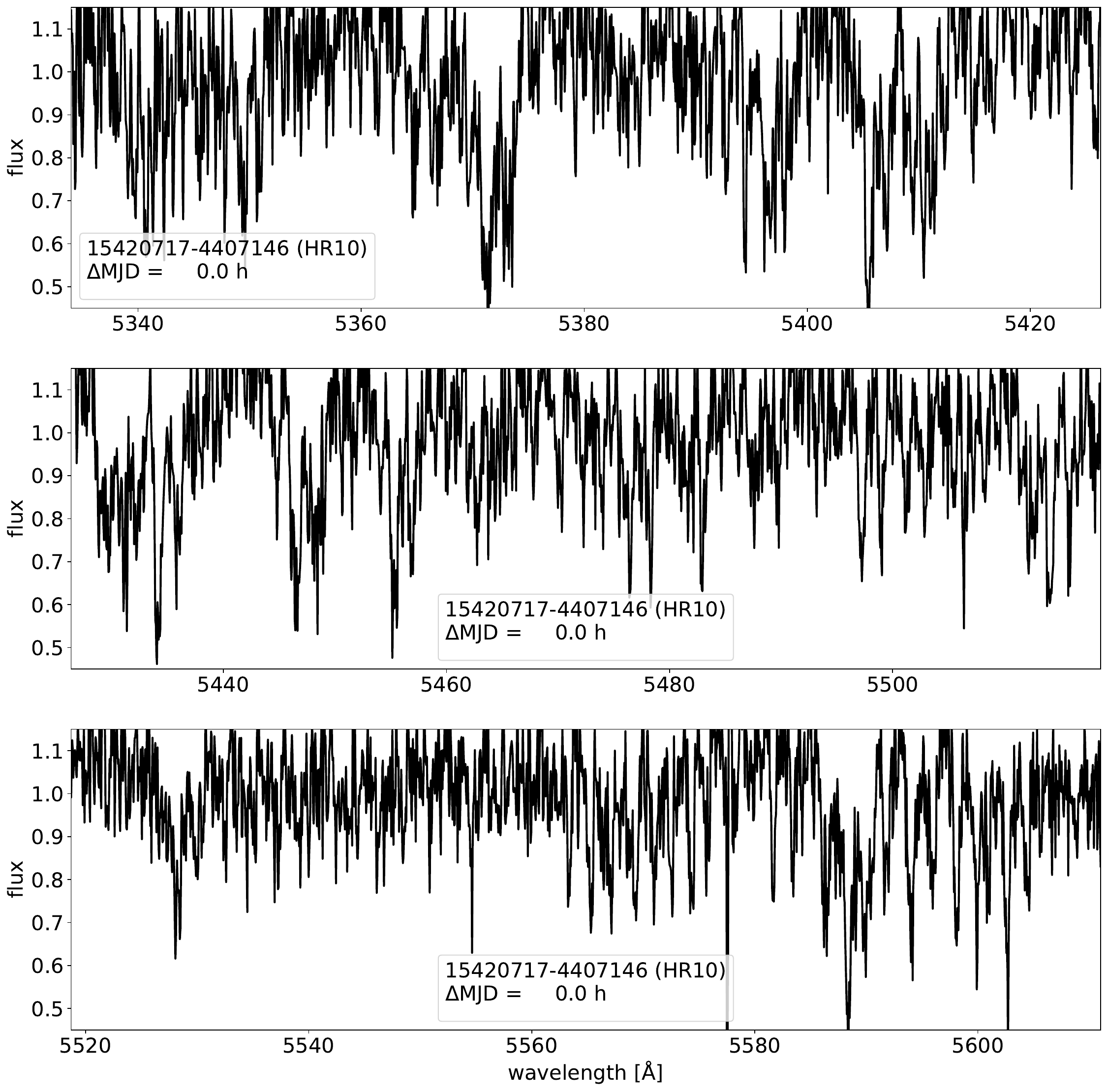}
  \captionof{figure}{\label{Fig:iDR5_SB3_atlas_15420717-4407146_1}(1/4) CNAME 15420717-4407146, at $\mathrm{MJD} = 56377.359329$, setup HR10.}
\end{minipage}
\begin{minipage}{\textwidth}
  \centering
  \includegraphics[width=0.49\textwidth]{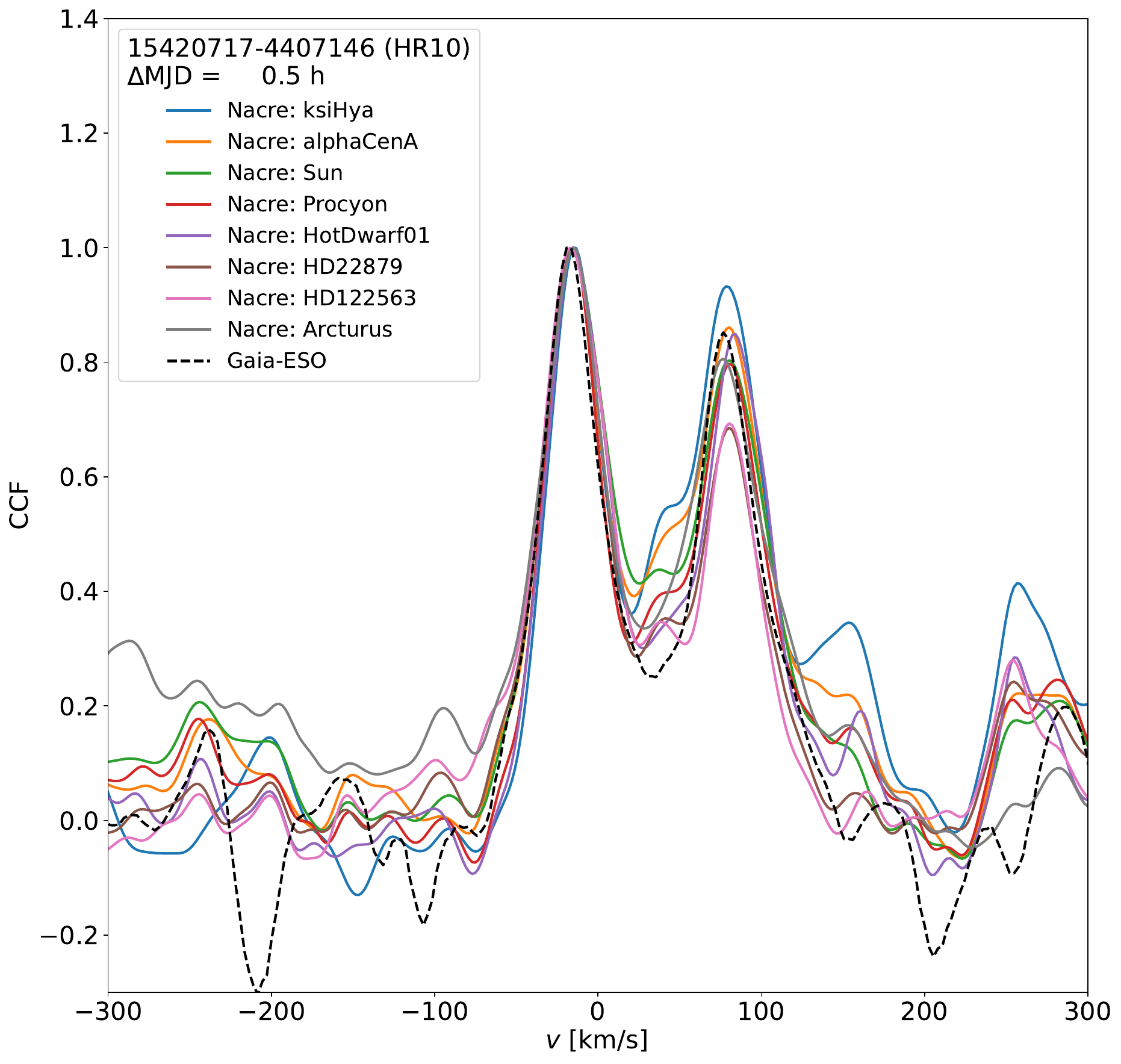}
  \includegraphics[width=0.49\textwidth]{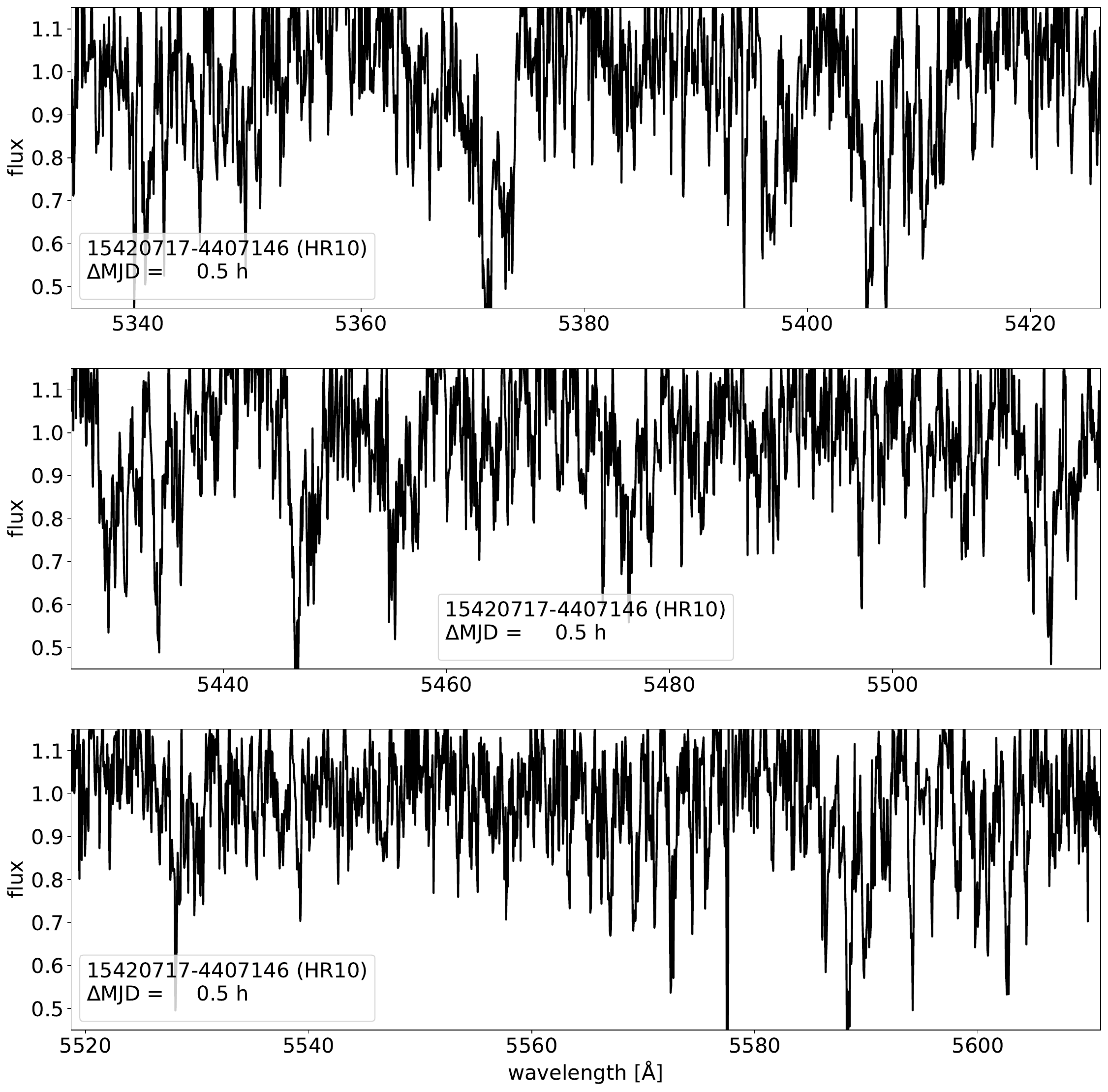}
  \captionof{figure}{\label{Fig:iDR5_SB3_atlas_15420717-4407146_2}(2/4) CNAME 15420717-4407146, at $\mathrm{MJD} = 56377.380557$, setup HR10.}
\end{minipage}
\clearpage
\begin{minipage}{\textwidth}
  \centering
  \includegraphics[width=0.49\textwidth]{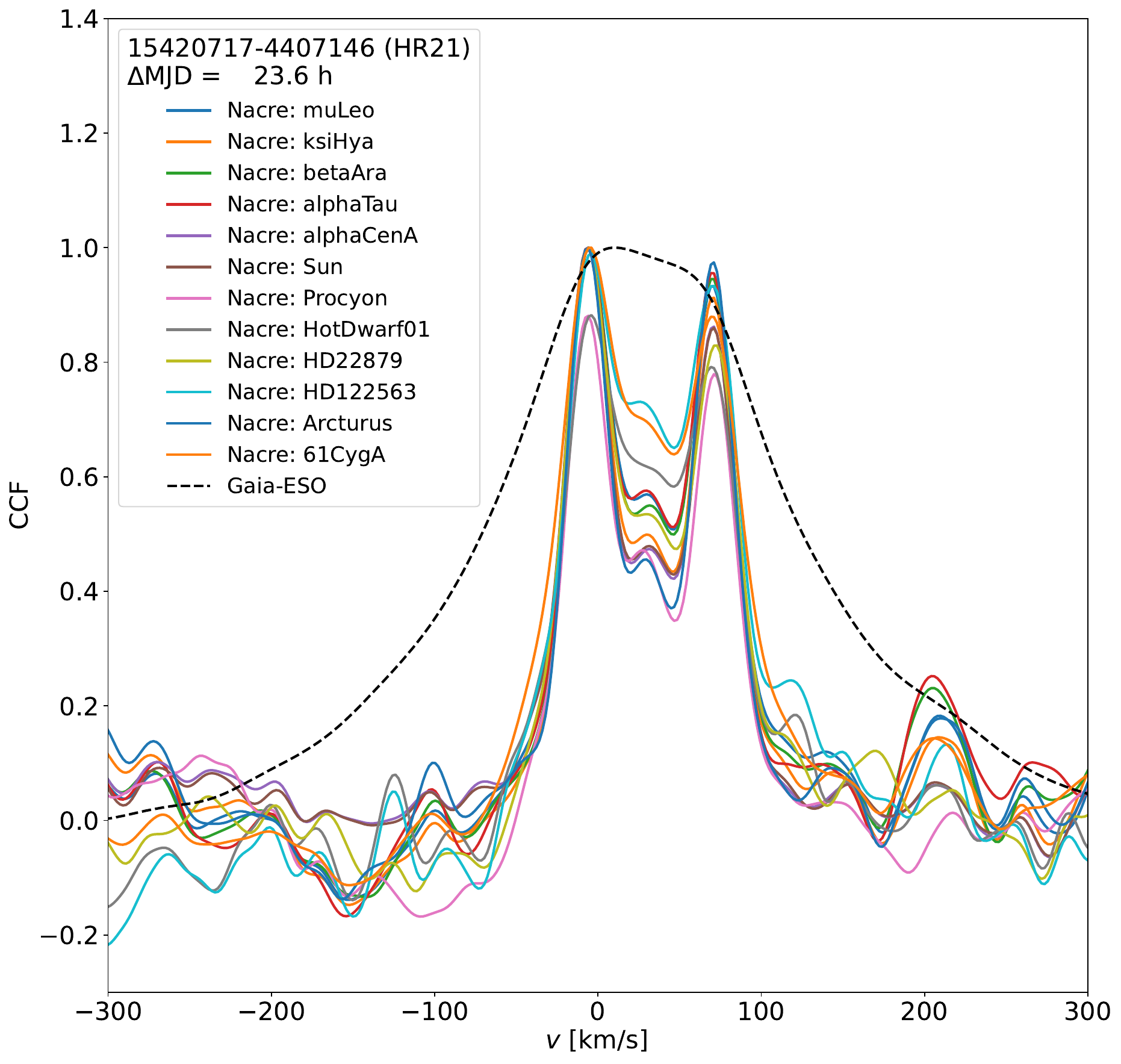}
  \includegraphics[width=0.49\textwidth]{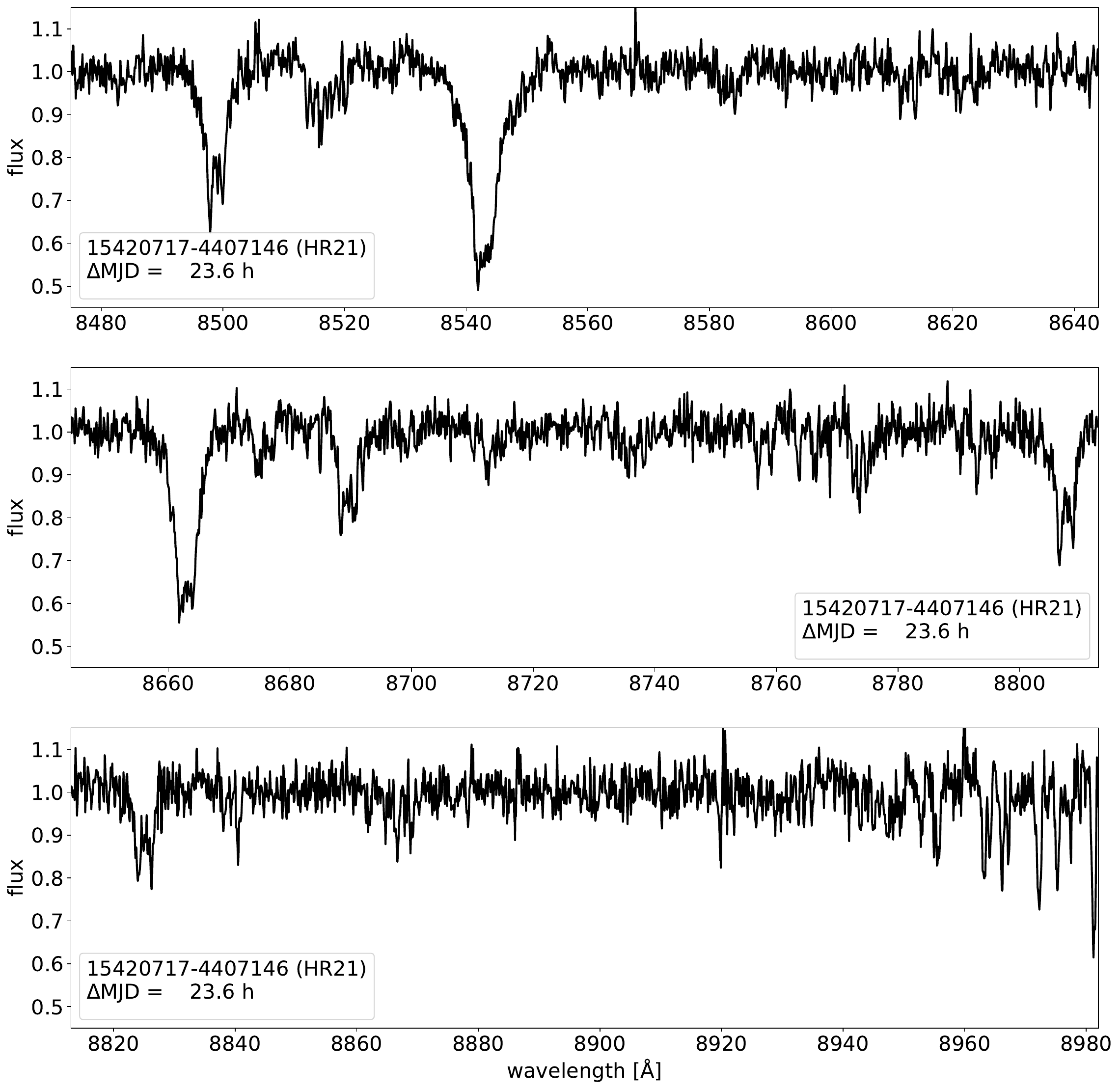}
  \captionof{figure}{\label{Fig:iDR5_SB3_atlas_15420717-4407146_3}(3/4) CNAME 15420717-4407146, at $\mathrm{MJD} = 56378.340677$, setup HR21.}
\end{minipage}
\begin{minipage}{\textwidth}
  \centering
  \includegraphics[width=0.49\textwidth]{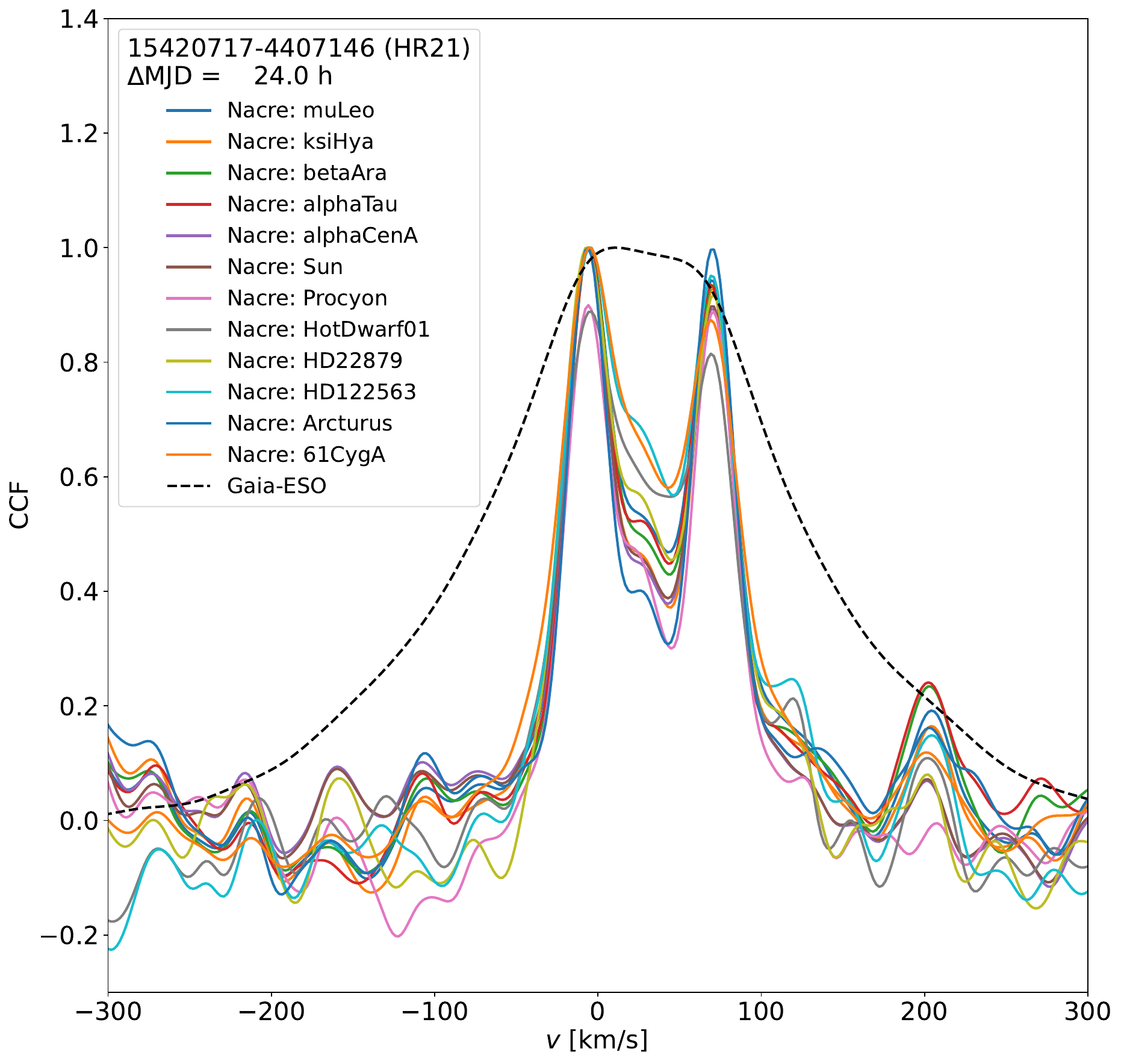}
  \includegraphics[width=0.49\textwidth]{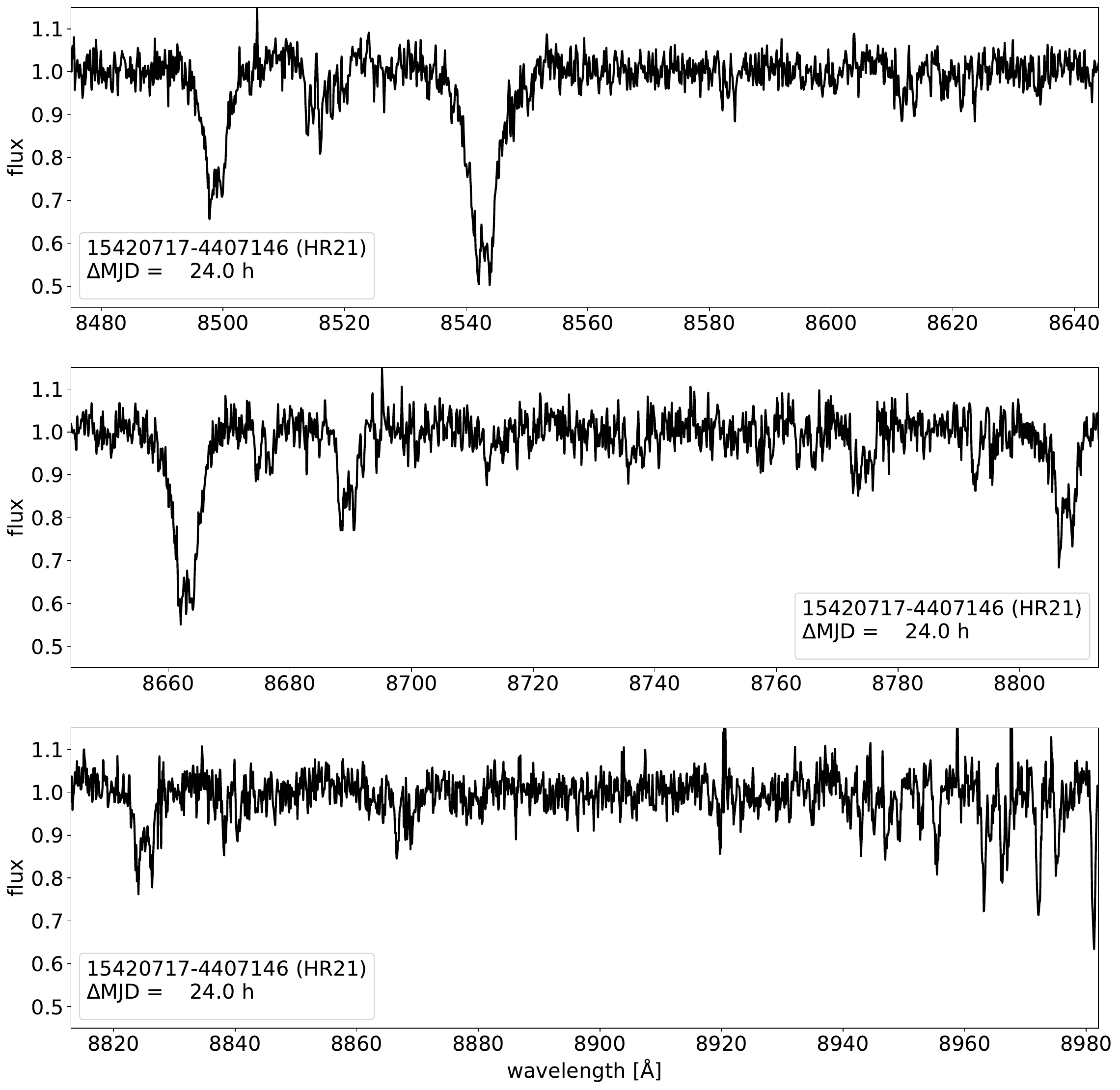}
  \captionof{figure}{\label{Fig:iDR5_SB3_atlas_15420717-4407146_4}(4/4) CNAME 15420717-4407146, at $\mathrm{MJD} = 56378.358684$, setup HR21.}
\end{minipage}
\cleardoublepage
\setlength\parindent{\defaultparindent}

\subsection{16003634-0745523}

\setlength\parindent{0cm}
\begin{minipage}{\textwidth}
  \centering
  \includegraphics[width=0.49\textwidth]{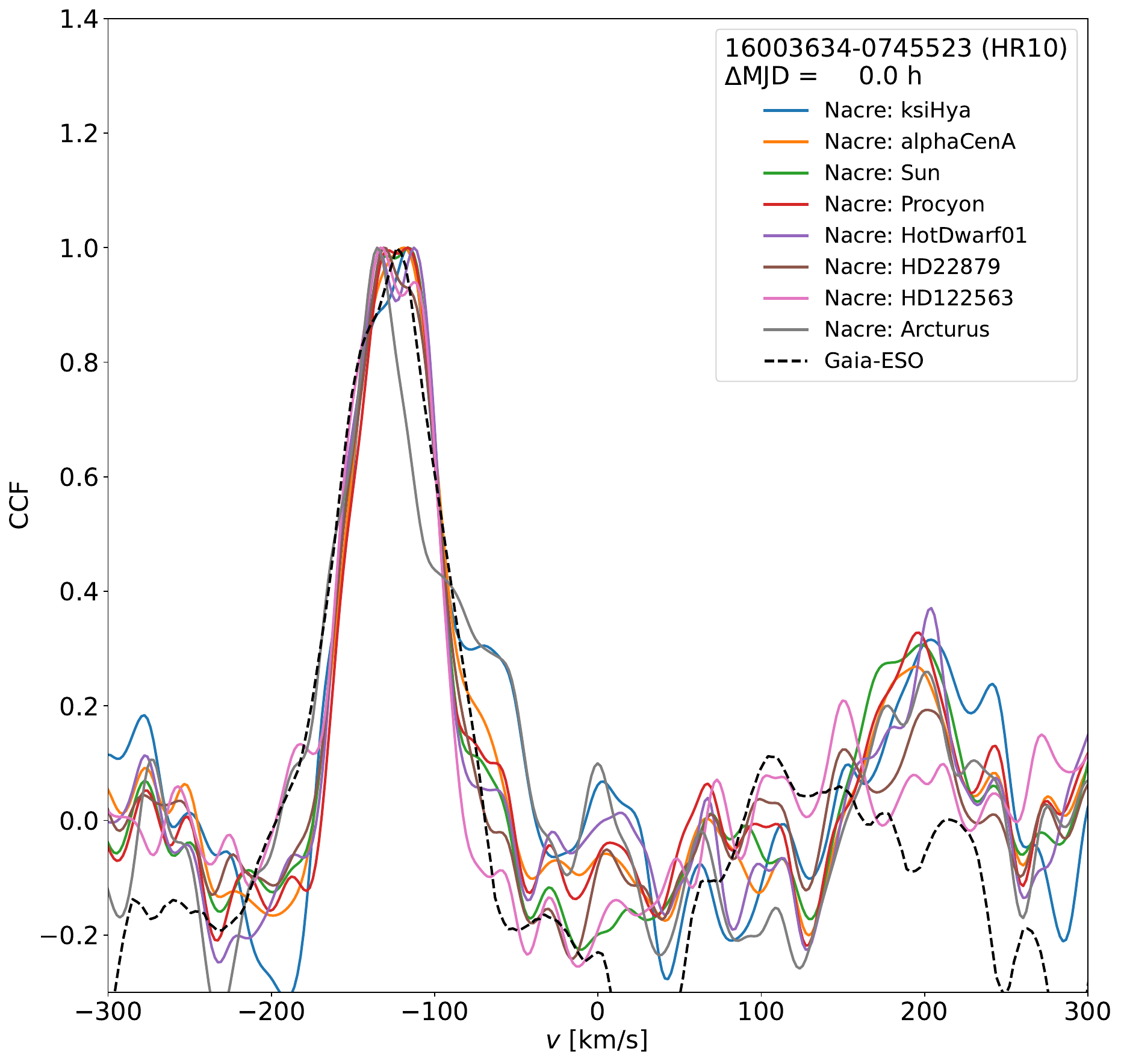}
  \includegraphics[width=0.49\textwidth]{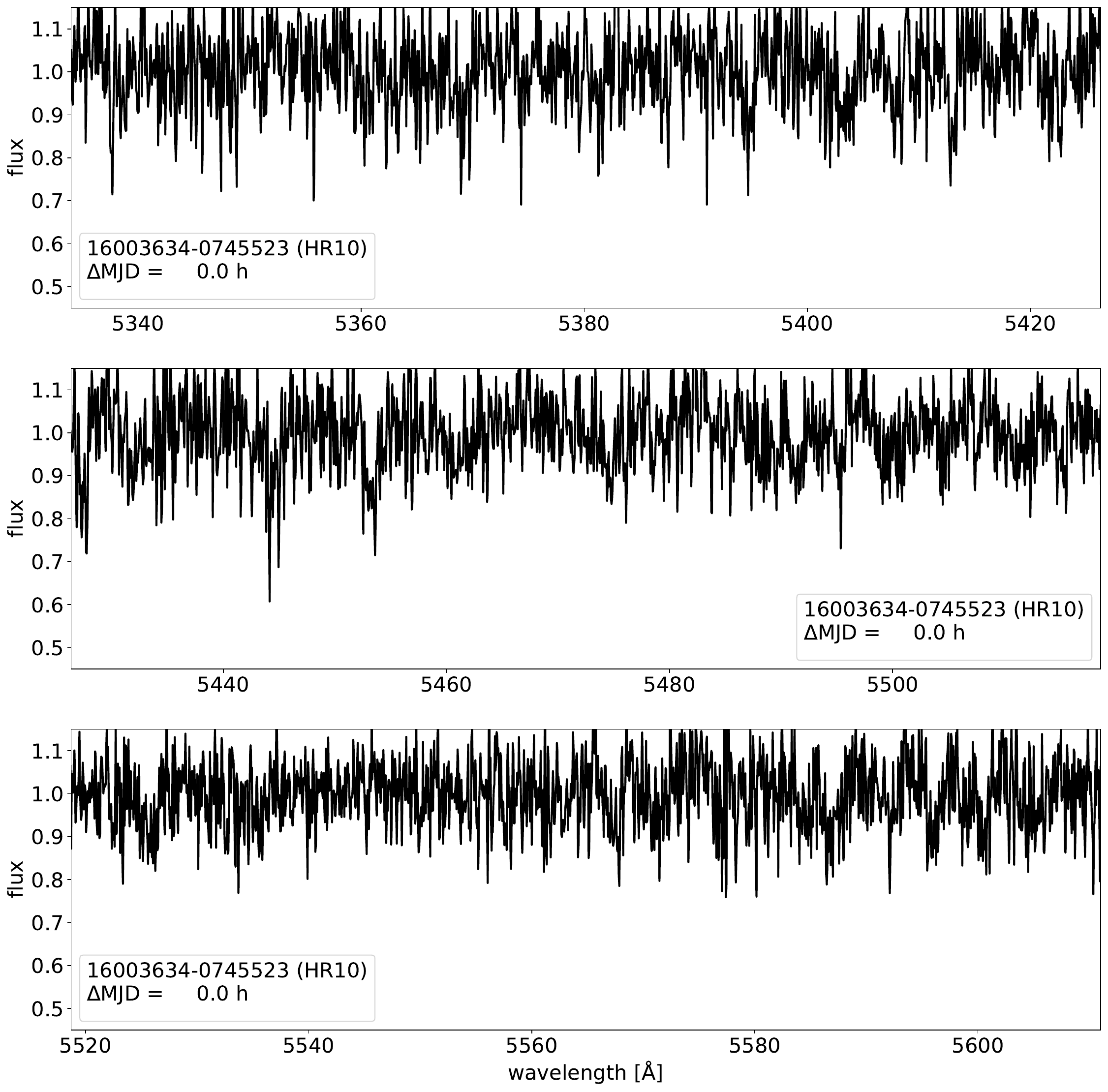}
  \captionof{figure}{\label{Fig:iDR5_SB3_atlas_16003634-0745523_1}(1/4) CNAME 16003634-0745523, at $\mathrm{MJD} = 56798.161286$, setup HR10.}
\end{minipage}
\begin{minipage}{\textwidth}
  \centering
  \includegraphics[width=0.49\textwidth]{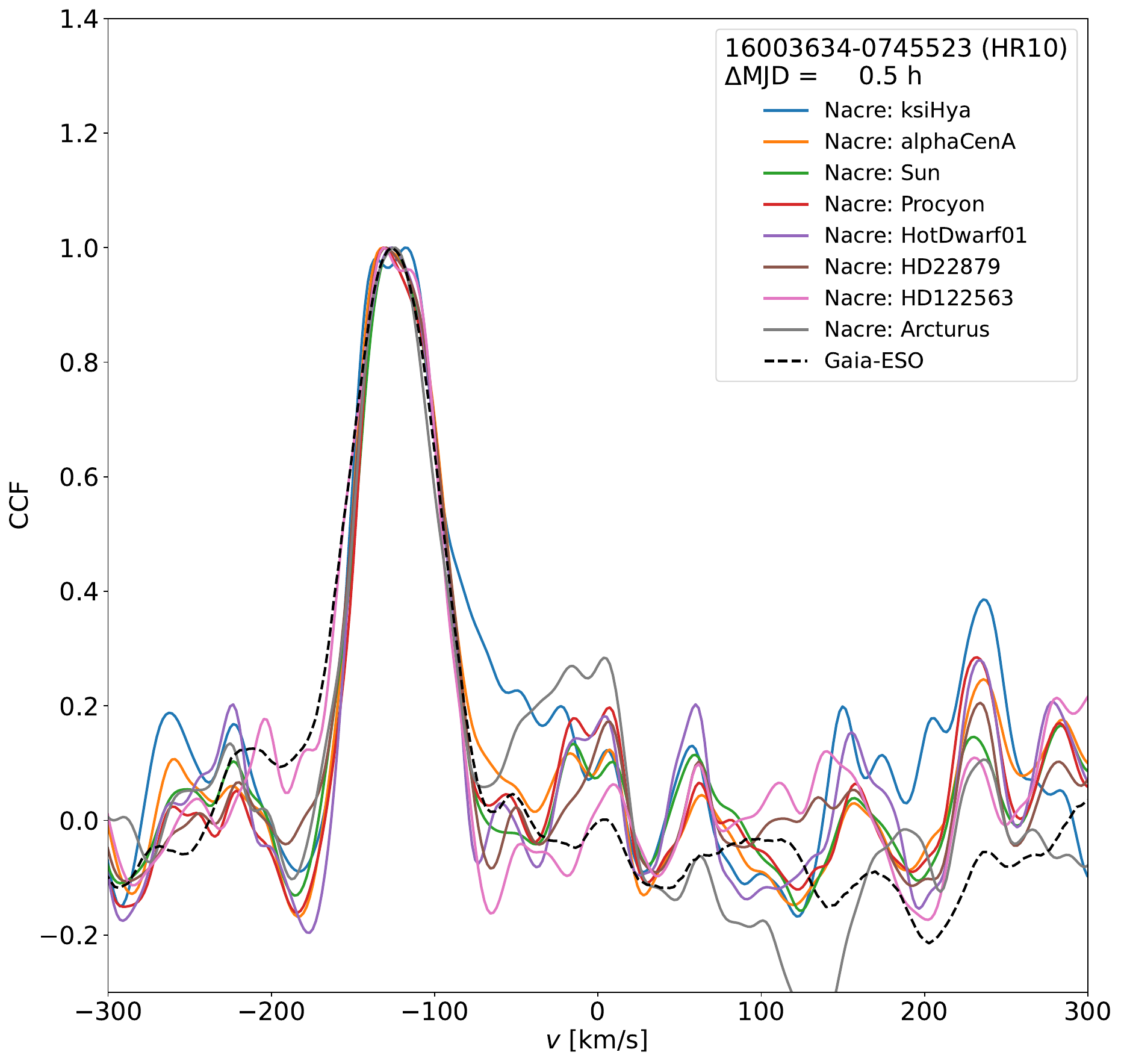}
  \includegraphics[width=0.49\textwidth]{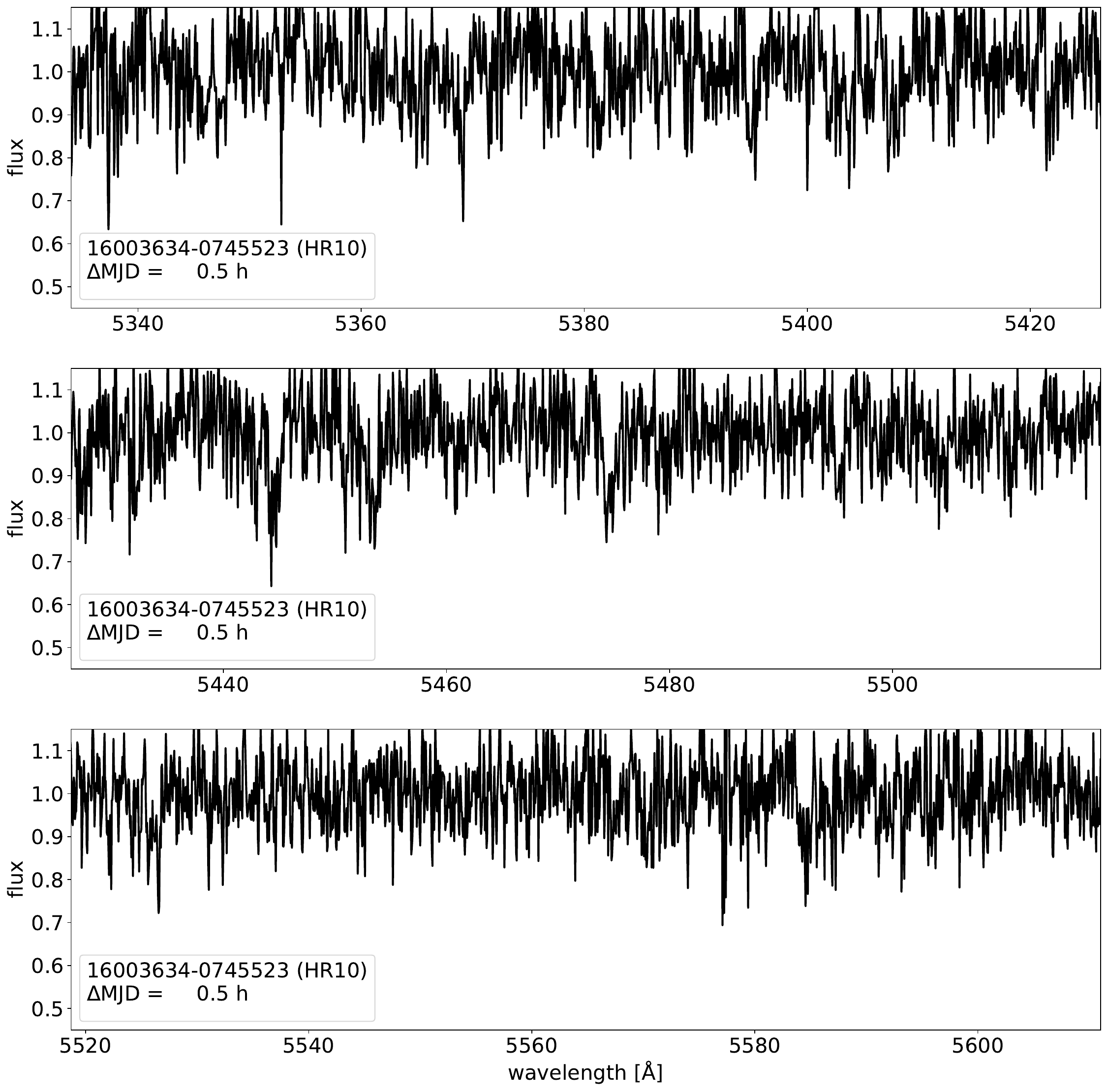}
  \captionof{figure}{\label{Fig:iDR5_SB3_atlas_16003634-0745523_2}(2/4) CNAME 16003634-0745523, at $\mathrm{MJD} = 56798.182565$, setup HR10.}
\end{minipage}
\clearpage
\begin{minipage}{\textwidth}
  \centering
  \includegraphics[width=0.49\textwidth]{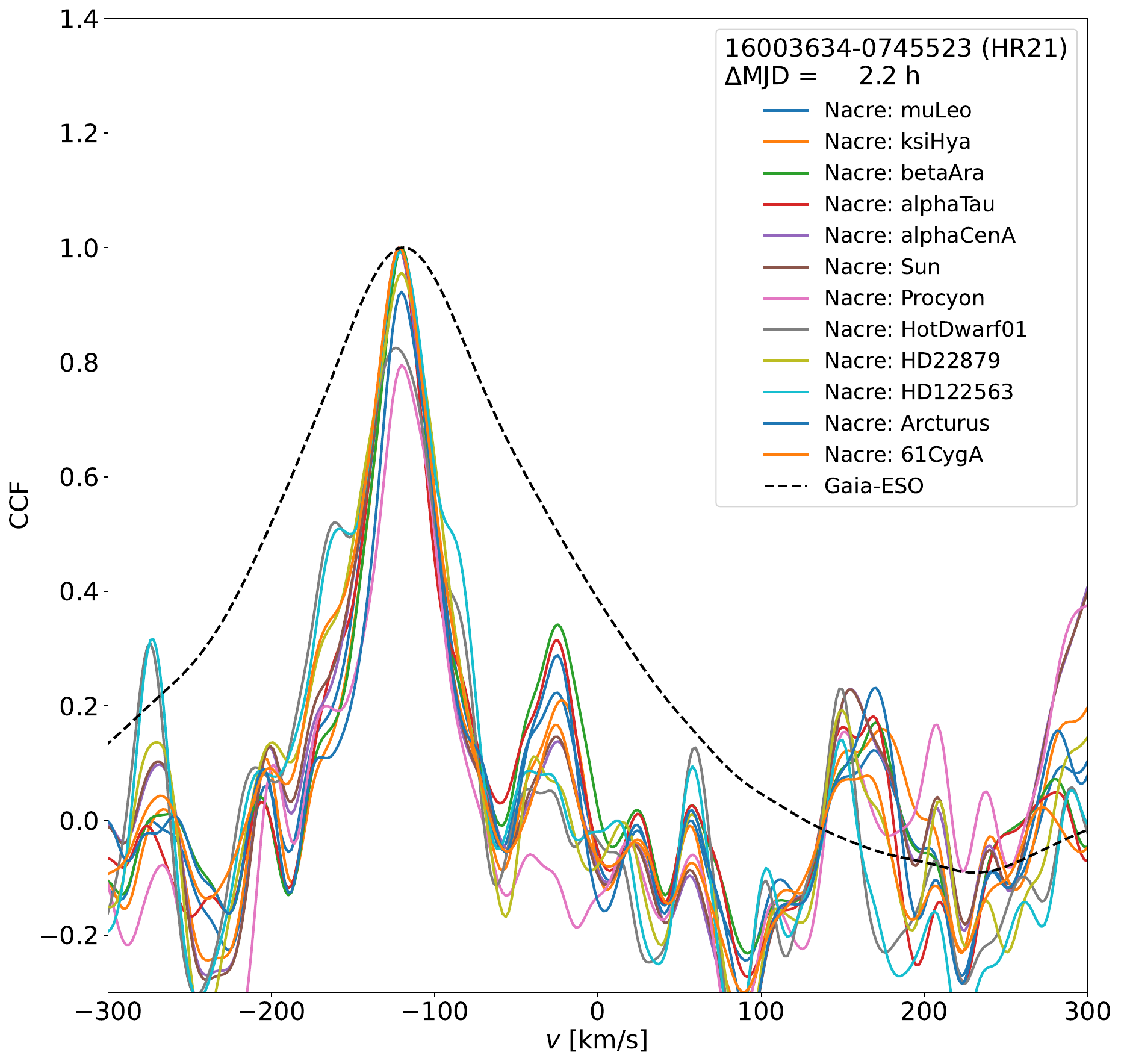}
  \includegraphics[width=0.49\textwidth]{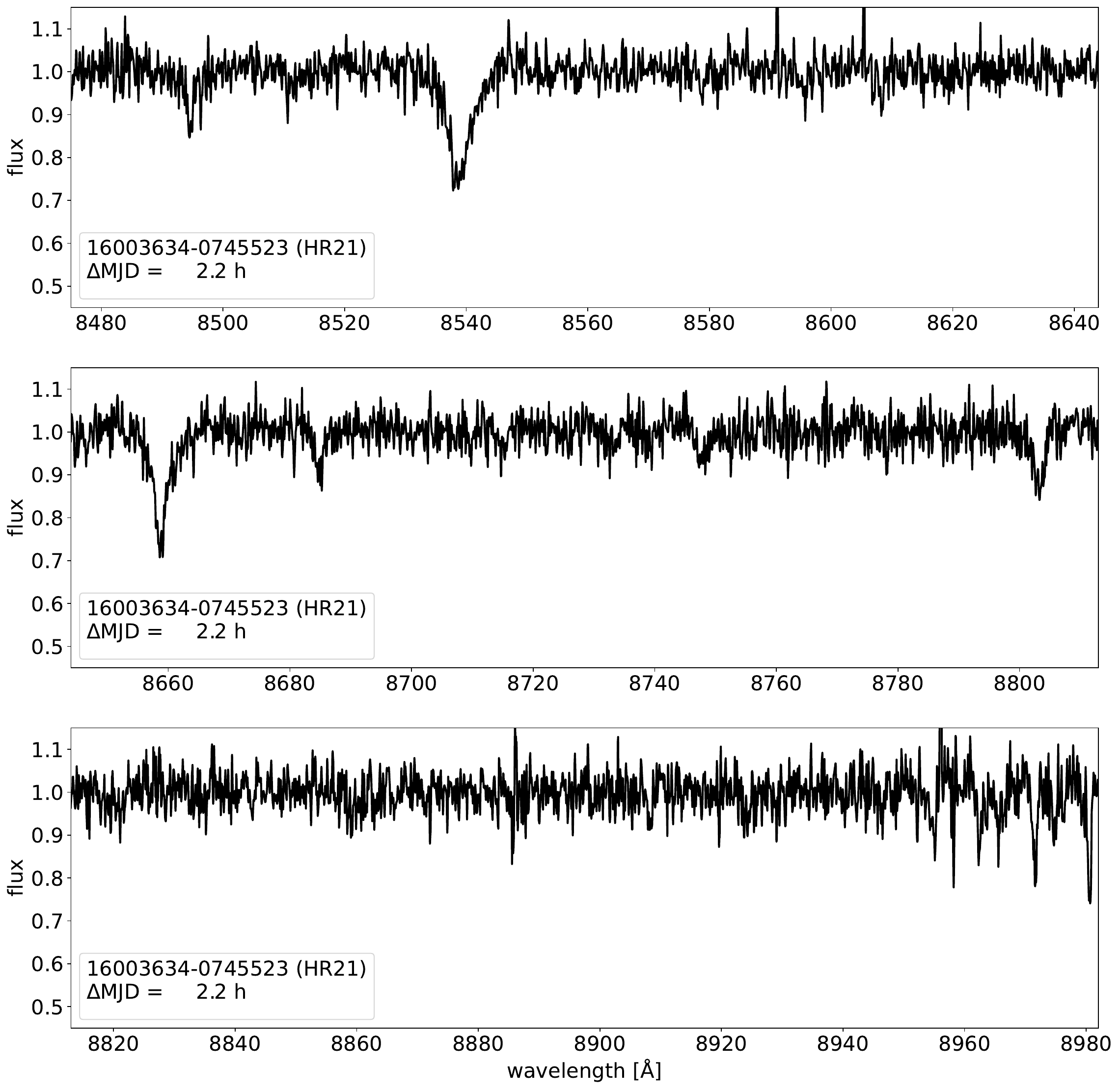}
  \captionof{figure}{\label{Fig:iDR5_SB3_atlas_16003634-0745523_3}(3/4) CNAME 16003634-0745523, at $\mathrm{MJD} = 56798.252817$, setup HR21.}
\end{minipage}
\begin{minipage}{\textwidth}
  \centering
  \includegraphics[width=0.49\textwidth]{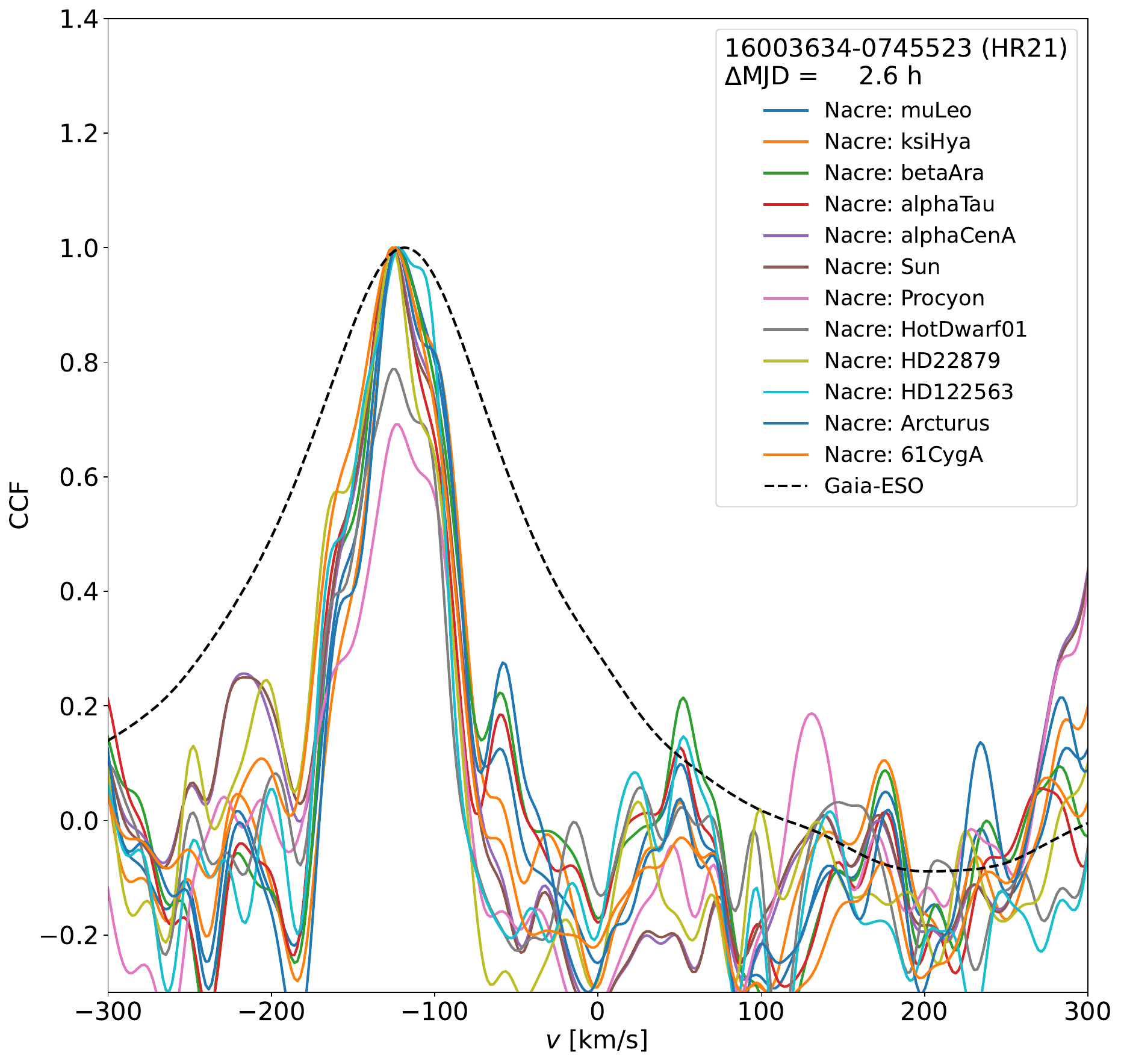}
  \includegraphics[width=0.49\textwidth]{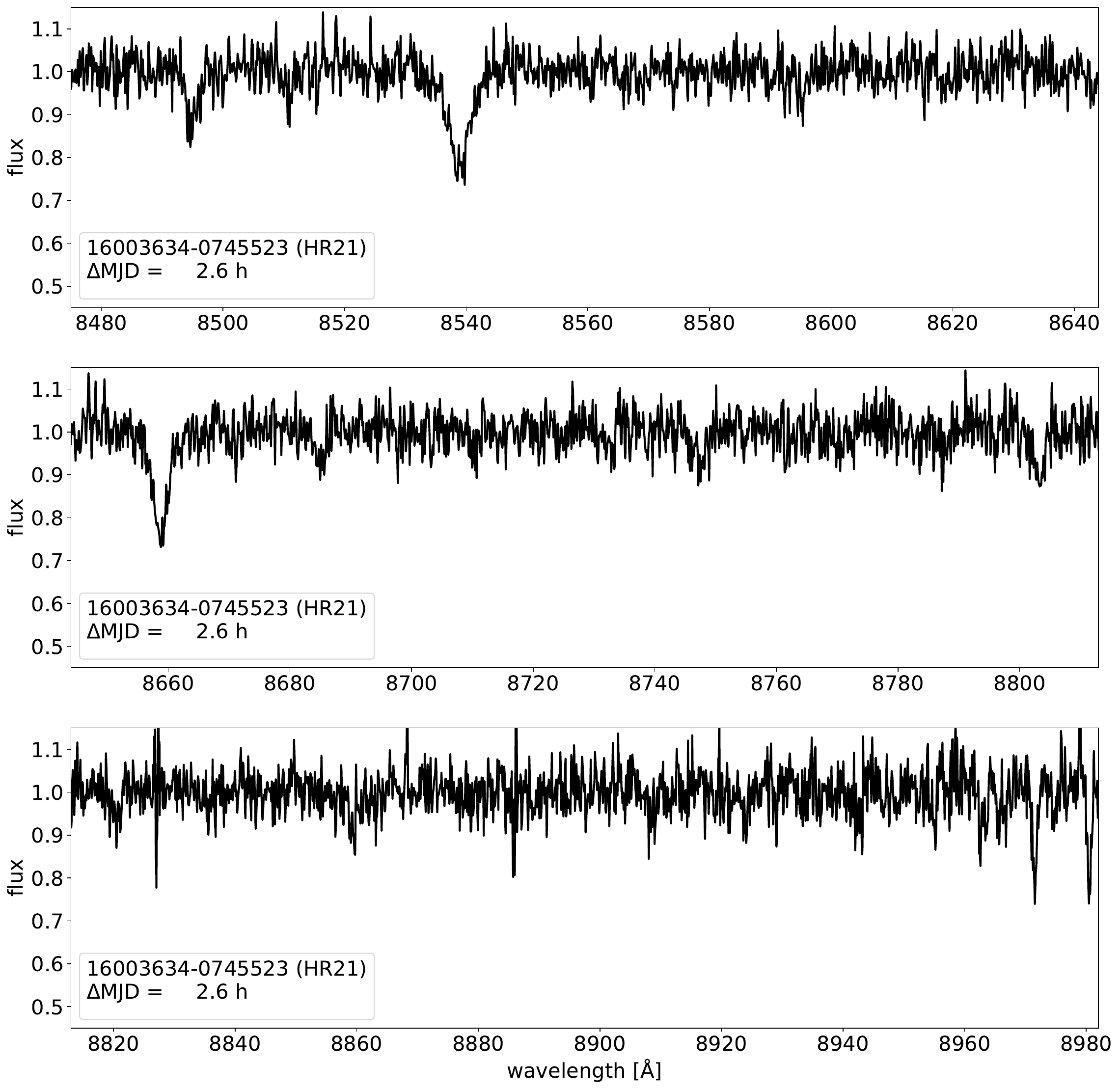}
  \captionof{figure}{\label{Fig:iDR5_SB3_atlas_16003634-0745523_4}(4/4) CNAME 16003634-0745523, at $\mathrm{MJD} = 56798.270793$, setup HR21.}
\end{minipage}
\cleardoublepage
\setlength\parindent{\defaultparindent}

\subsection{18170244-4227076}

\setlength\parindent{0cm}
\begin{minipage}{\textwidth}
  \centering
  \includegraphics[width=0.49\textwidth]{Figures/New_CCFs_-_SB3_CCFn_-_18170244-4227076_56821.258156_HR10.v202304.pdf}
  \includegraphics[width=0.49\textwidth]{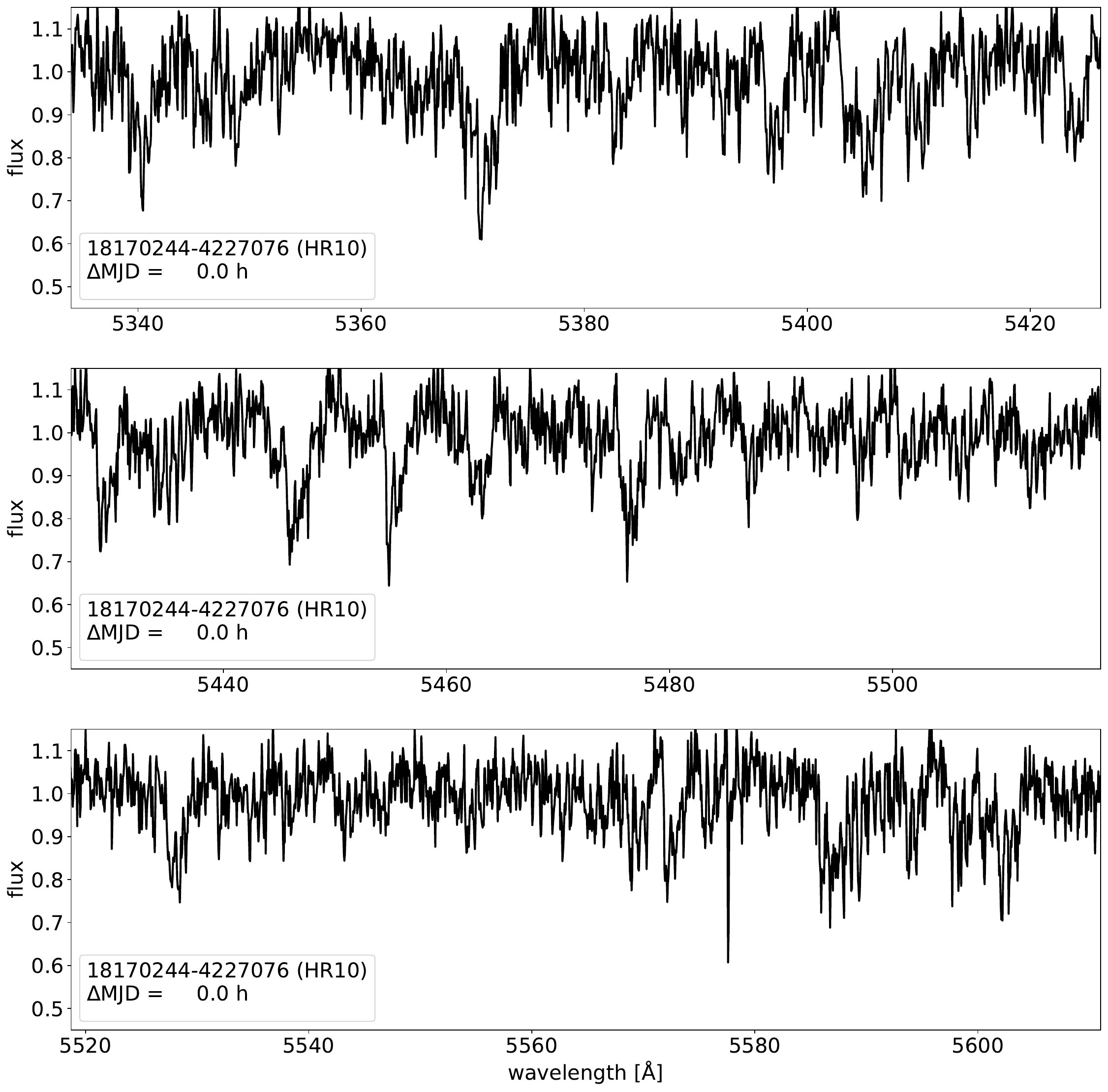}
  \captionof{figure}{\label{Fig:iDR5_SB3_atlas_18170244-4227076_1}(1/4) CNAME 18170244-4227076, at $\mathrm{MJD} = 56821.258156$, setup HR10.}
\end{minipage}
\begin{minipage}{\textwidth}
  \centering
  \includegraphics[width=0.49\textwidth]{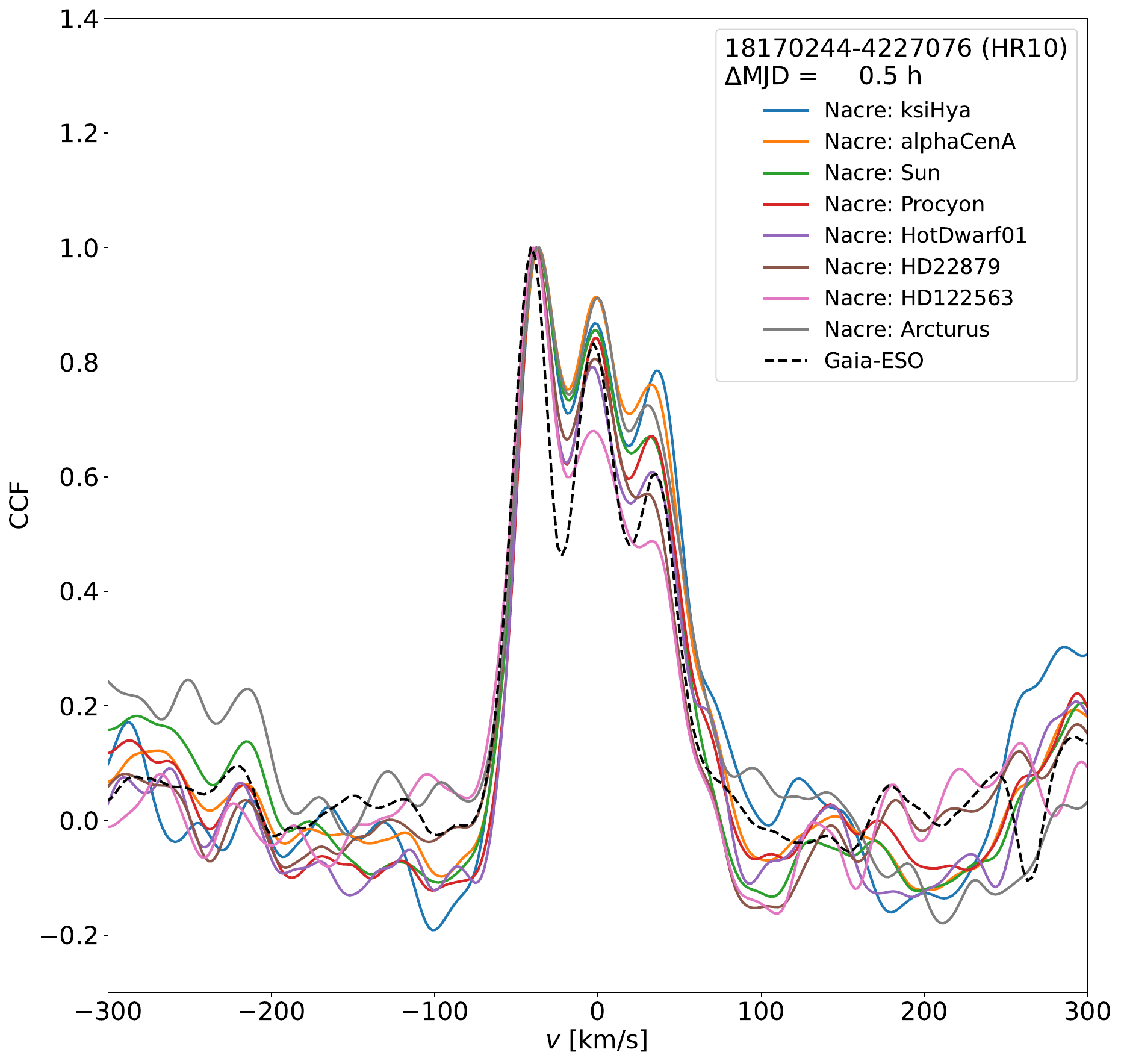}
  \includegraphics[width=0.49\textwidth]{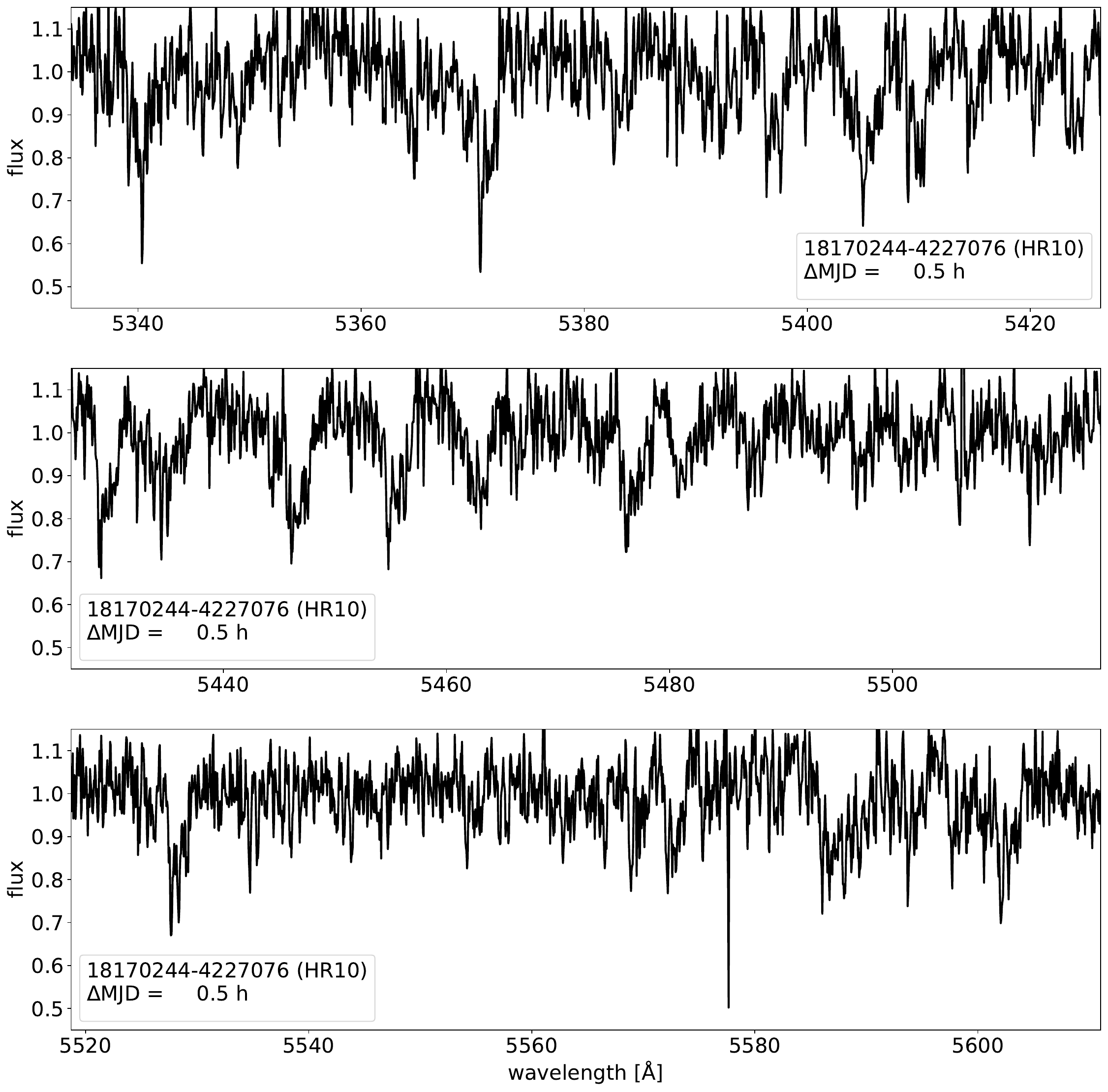}
  \captionof{figure}{\label{Fig:iDR5_SB3_atlas_18170244-4227076_2}(2/4) CNAME 18170244-4227076, at $\mathrm{MJD} = 56821.279464$, setup HR10.}
\end{minipage}
\clearpage
\begin{minipage}{\textwidth}
  \centering
  \includegraphics[width=0.49\textwidth]{Figures/New_CCFs_-_SB3_CCFn_-_18170244-4227076_57180.349352_HR21.v202304.pdf}
  \includegraphics[width=0.49\textwidth]{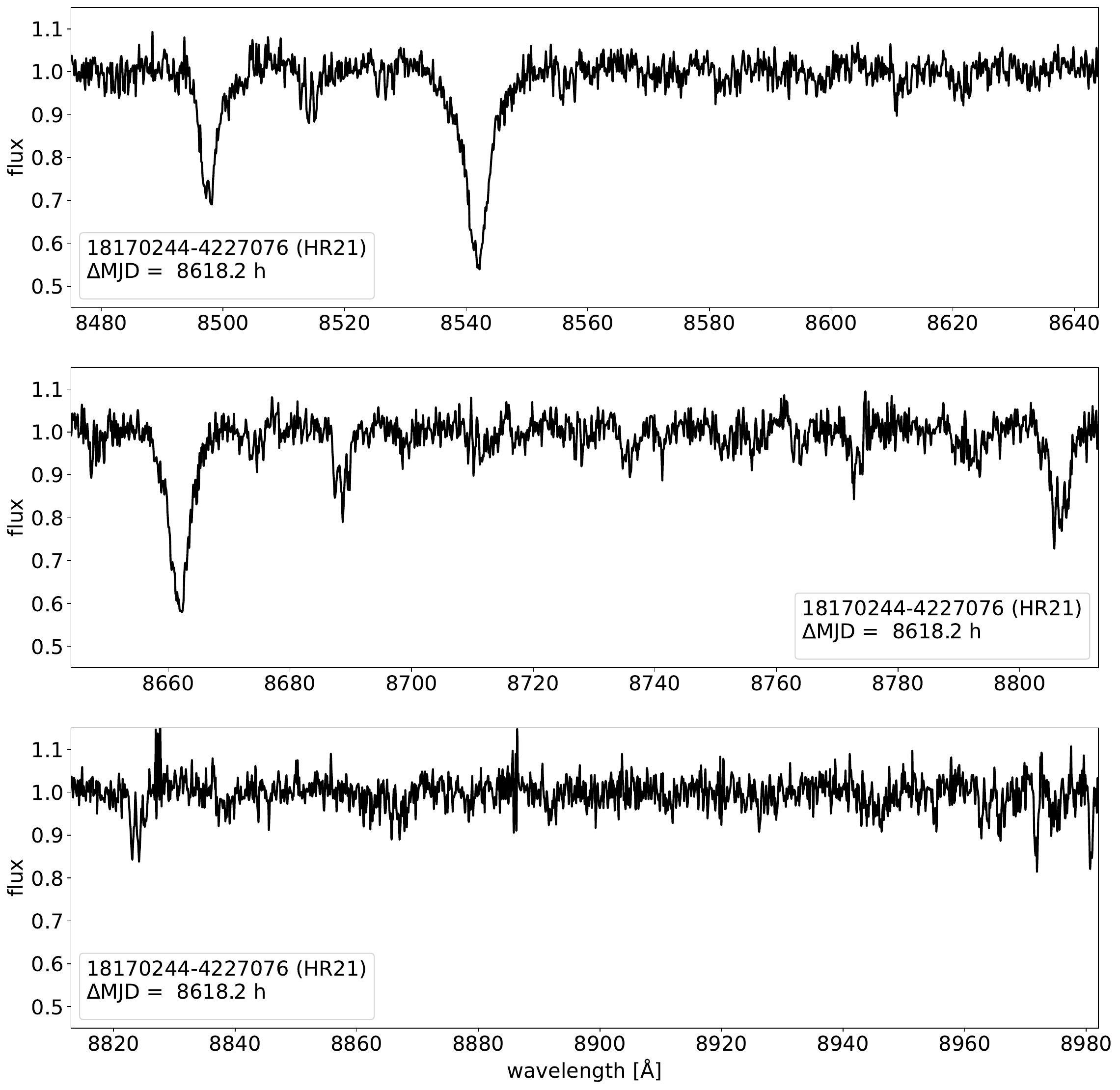}
  \captionof{figure}{\label{Fig:iDR5_SB3_atlas_18170244-4227076_3}(3/4) CNAME 18170244-4227076, at $\mathrm{MJD} = 57180.349352$, setup HR21.}
\end{minipage}
\begin{minipage}{\textwidth}
  \centering
  \includegraphics[width=0.49\textwidth]{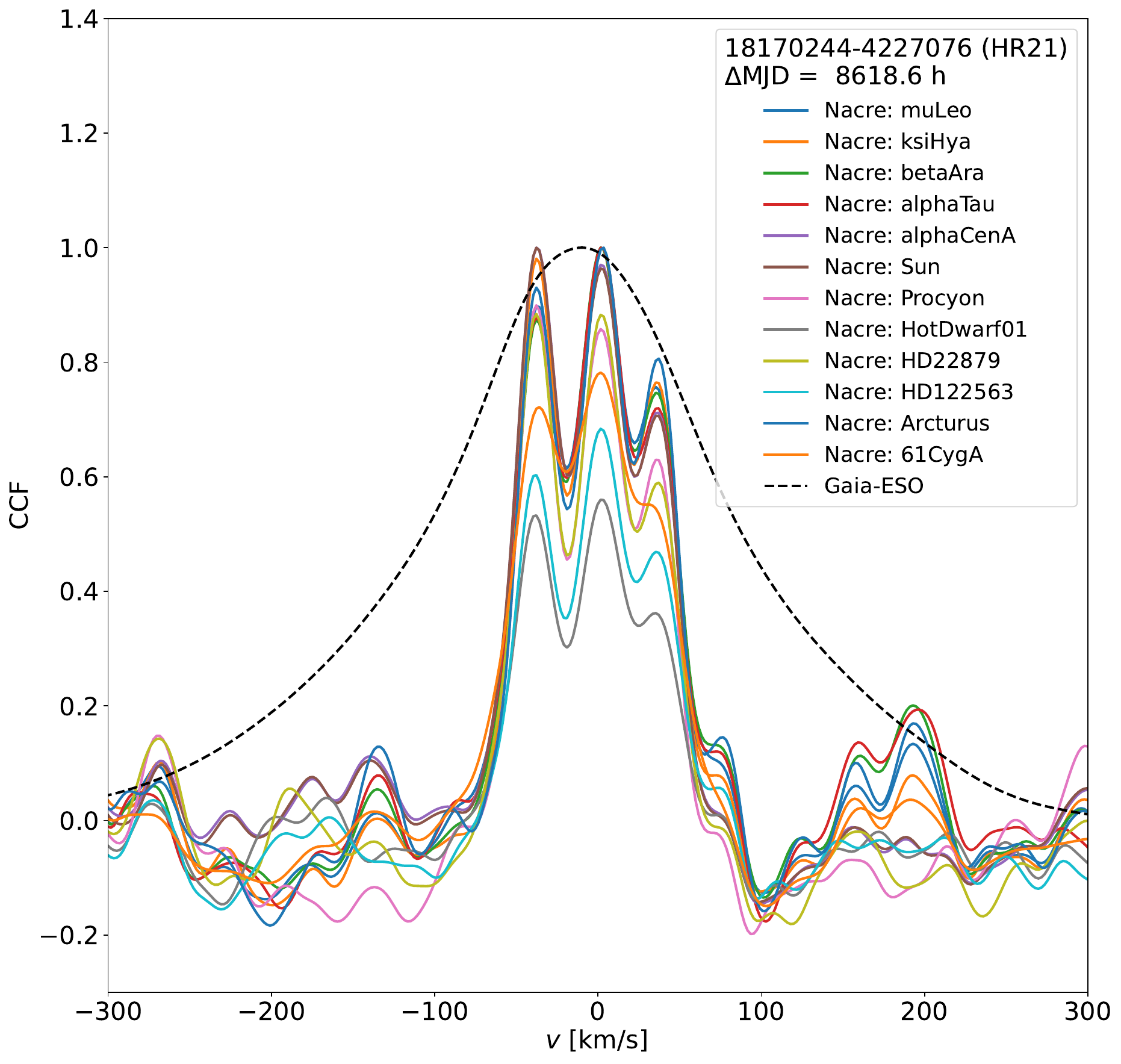}
  \includegraphics[width=0.49\textwidth]{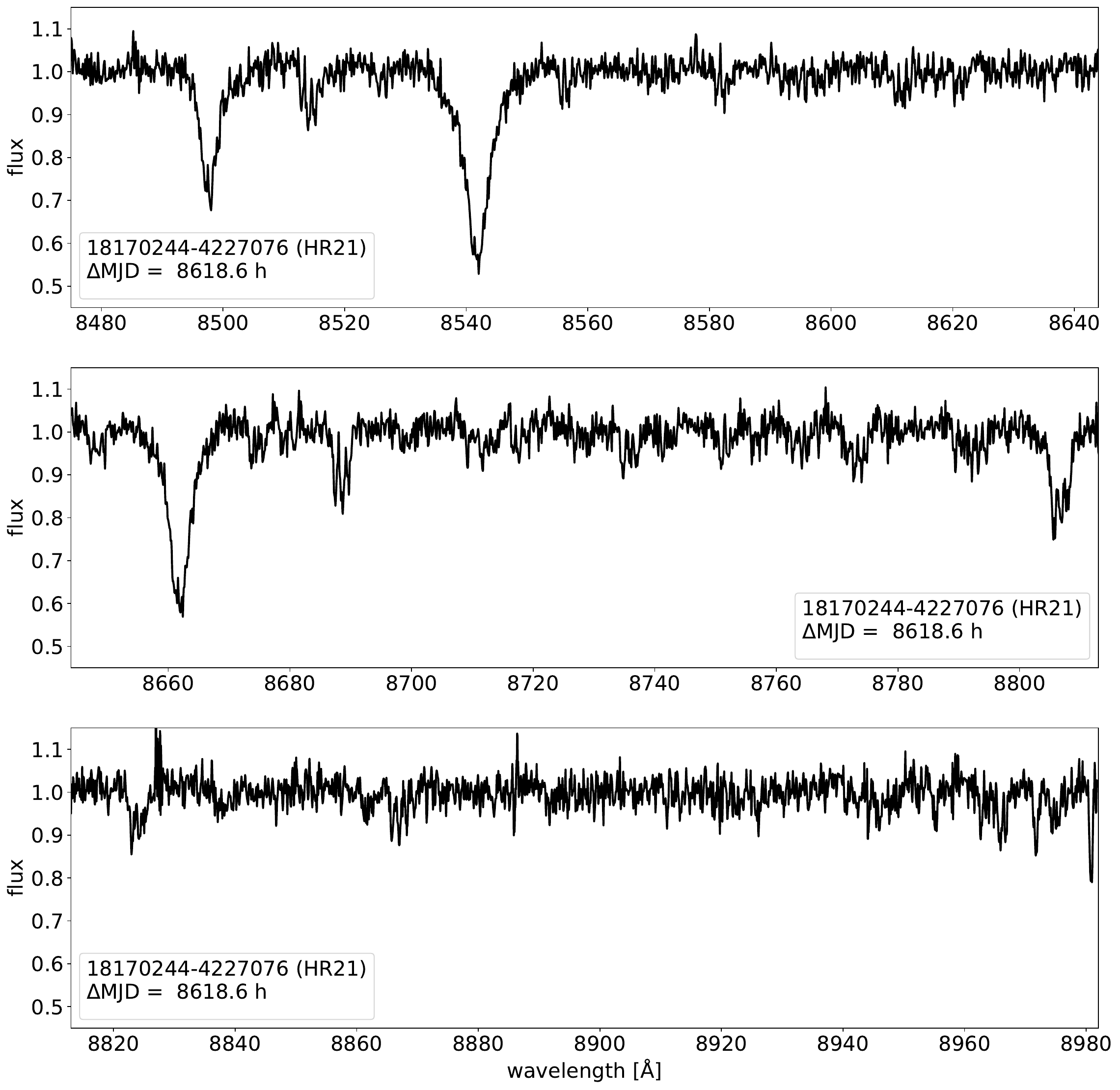}
  \captionof{figure}{\label{Fig:iDR5_SB3_atlas_18170244-4227076_4}(4/4) CNAME 18170244-4227076, at $\mathrm{MJD} = 57180.367370$, setup HR21.}
\end{minipage}
\cleardoublepage
\setlength\parindent{\defaultparindent}

\subsection{21393685-4659598}

\setlength\parindent{0cm}
\begin{minipage}{\textwidth}
  \centering
  \includegraphics[width=0.49\textwidth]{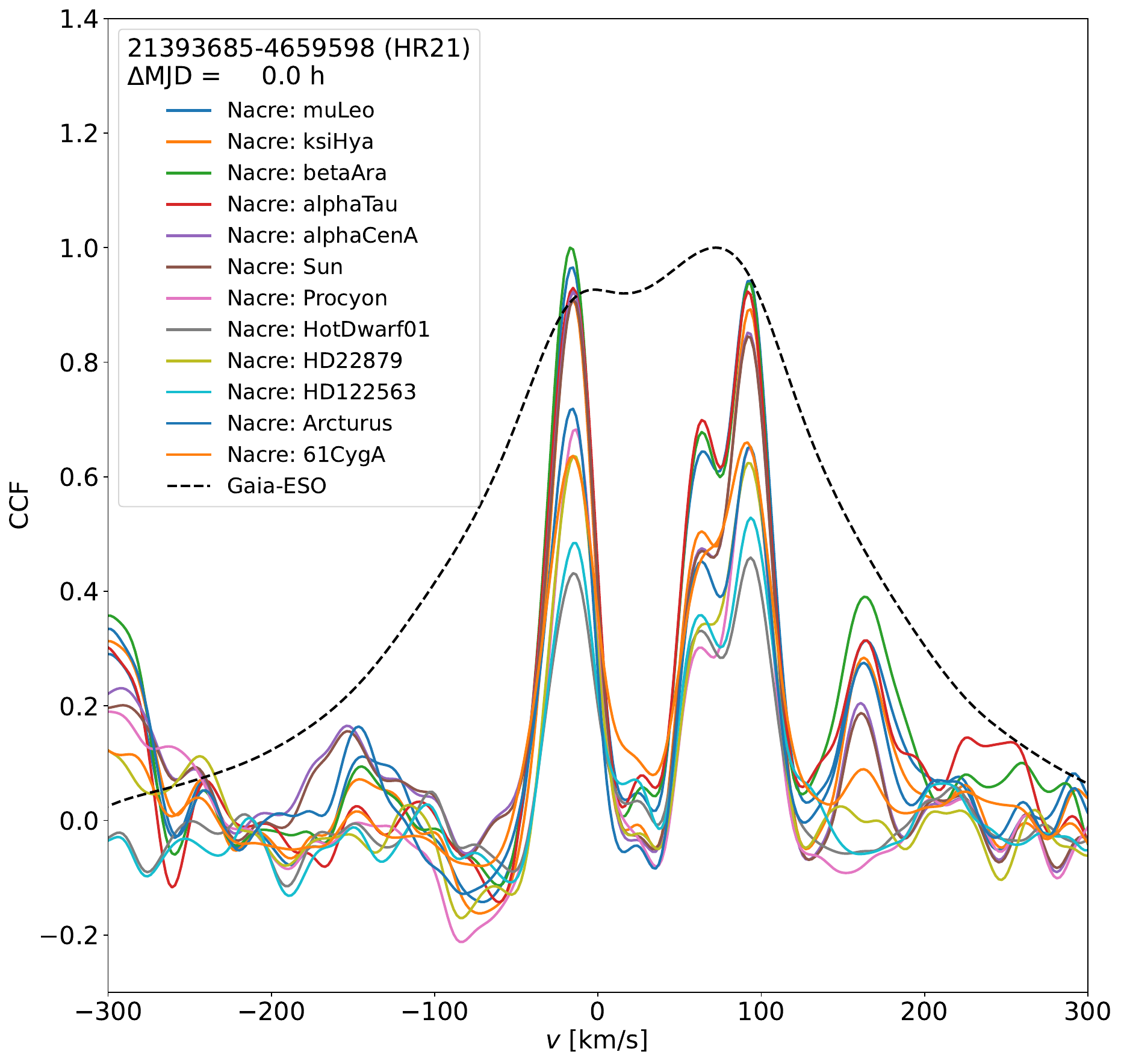}
  \includegraphics[width=0.49\textwidth]{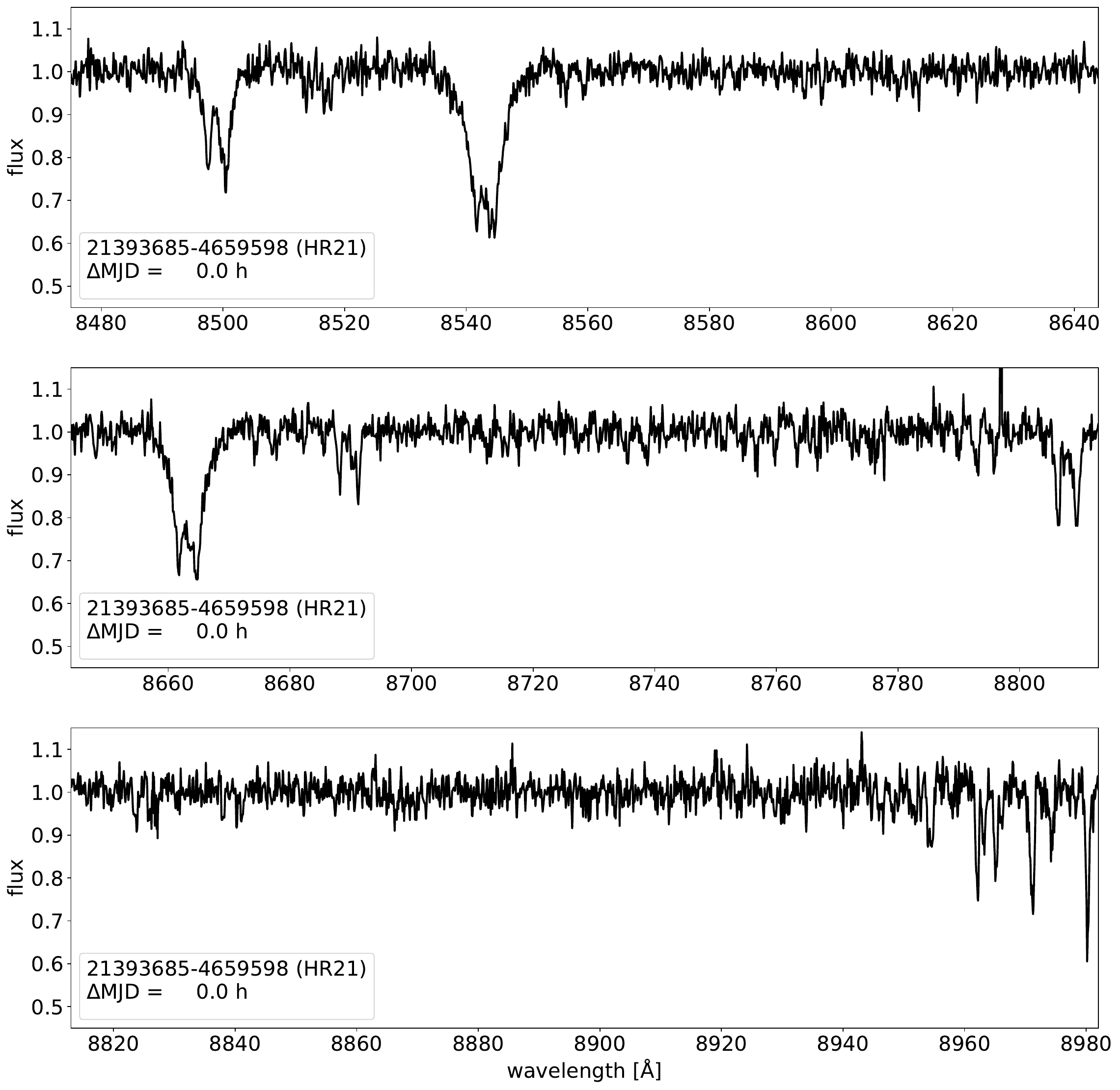}
  \captionof{figure}{\label{Fig:iDR5_SB3_atlas_21393685-4659598_1}(1/4) CNAME 21393685-4659598, at $\mathrm{MJD} = 57265.128815$, setup HR21.}
\end{minipage}
\begin{minipage}{\textwidth}
  \centering
  \includegraphics[width=0.49\textwidth]{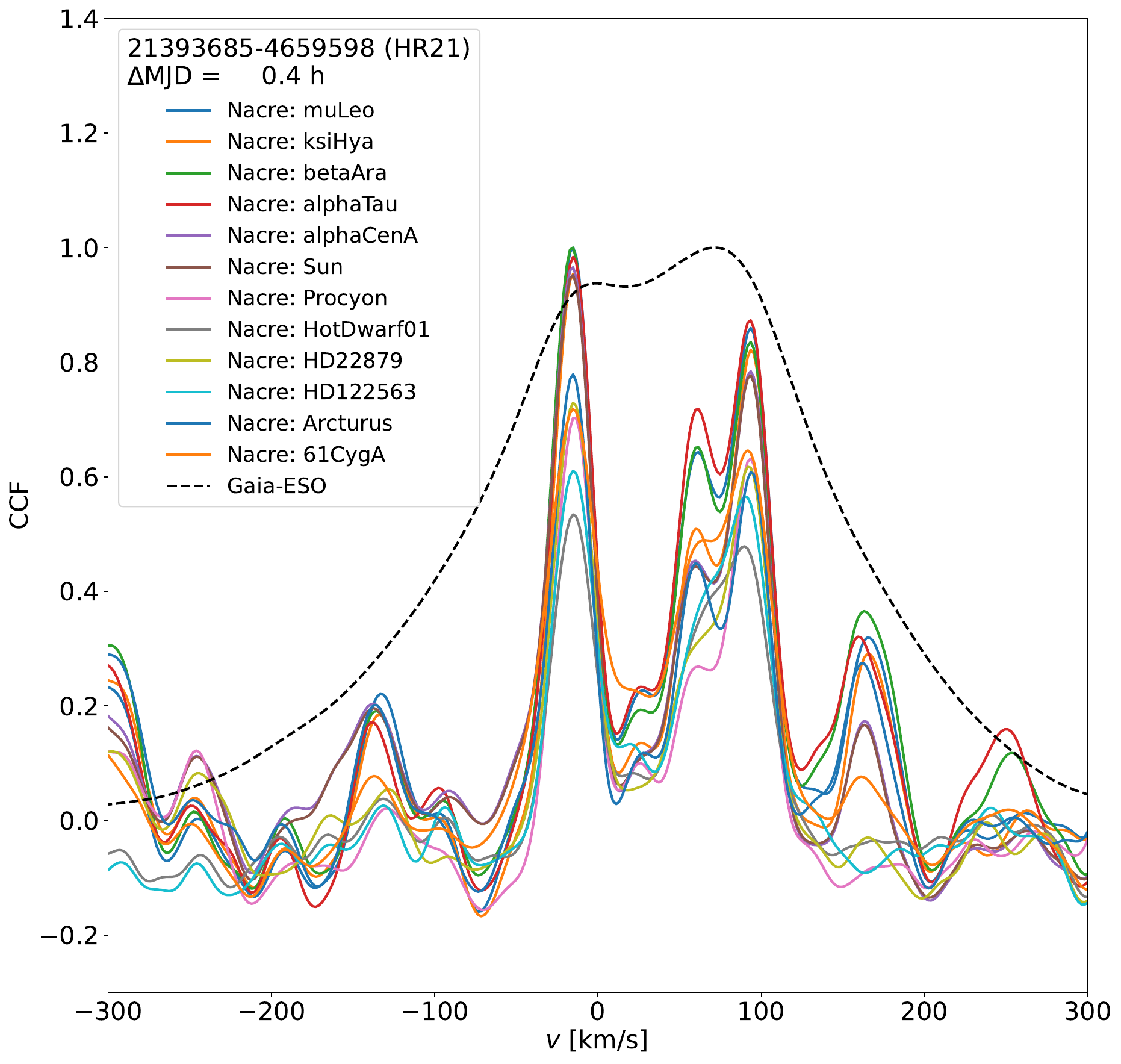}
  \includegraphics[width=0.49\textwidth]{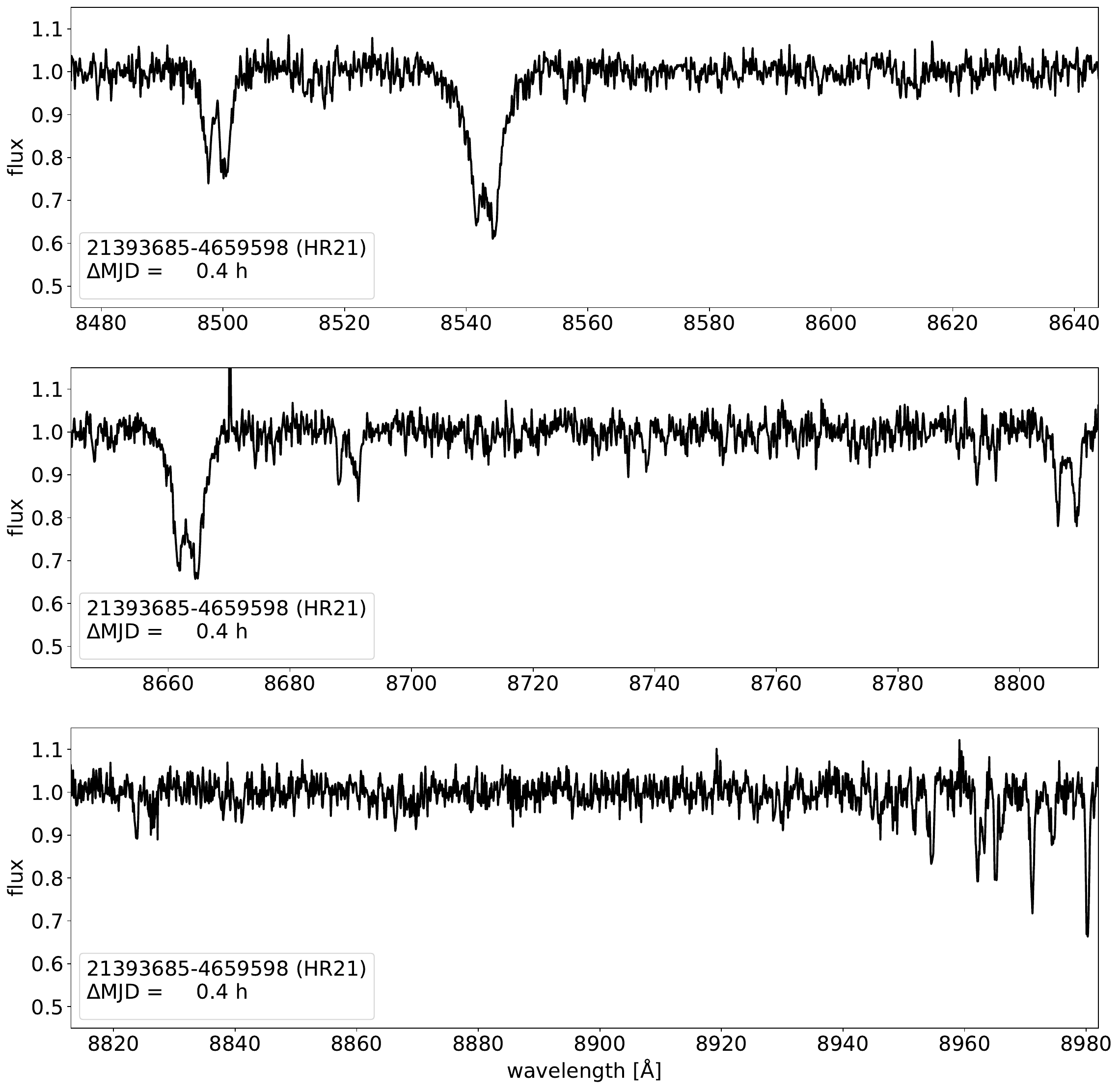}
  \captionof{figure}{\label{Fig:iDR5_SB3_atlas_21393685-4659598_2}(2/4) CNAME 21393685-4659598, at $\mathrm{MJD} = 57265.146791$, setup HR21.}
\end{minipage}
\clearpage
\begin{minipage}{\textwidth}
  \centering
  \includegraphics[width=0.49\textwidth]{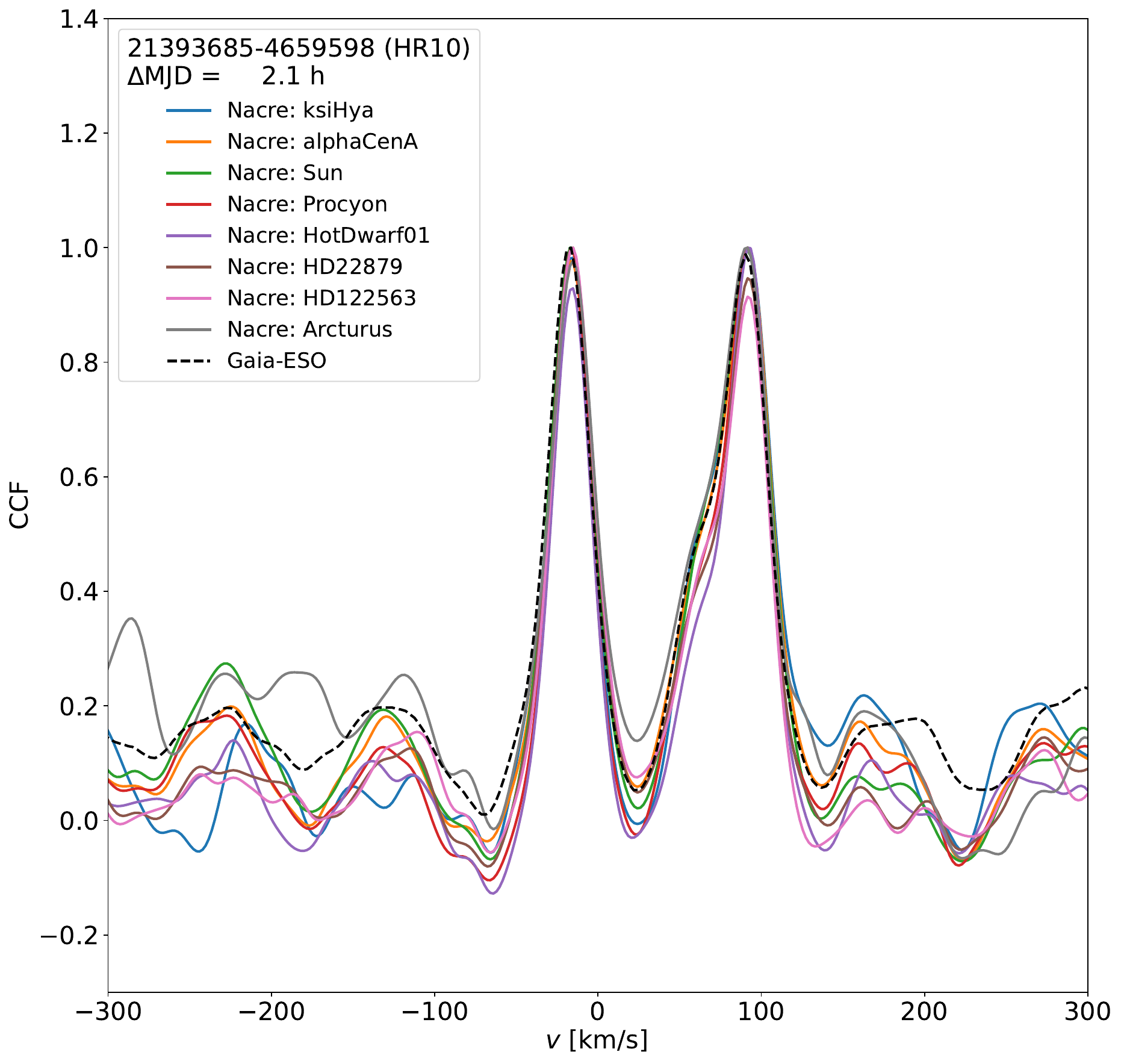}
  \includegraphics[width=0.49\textwidth]{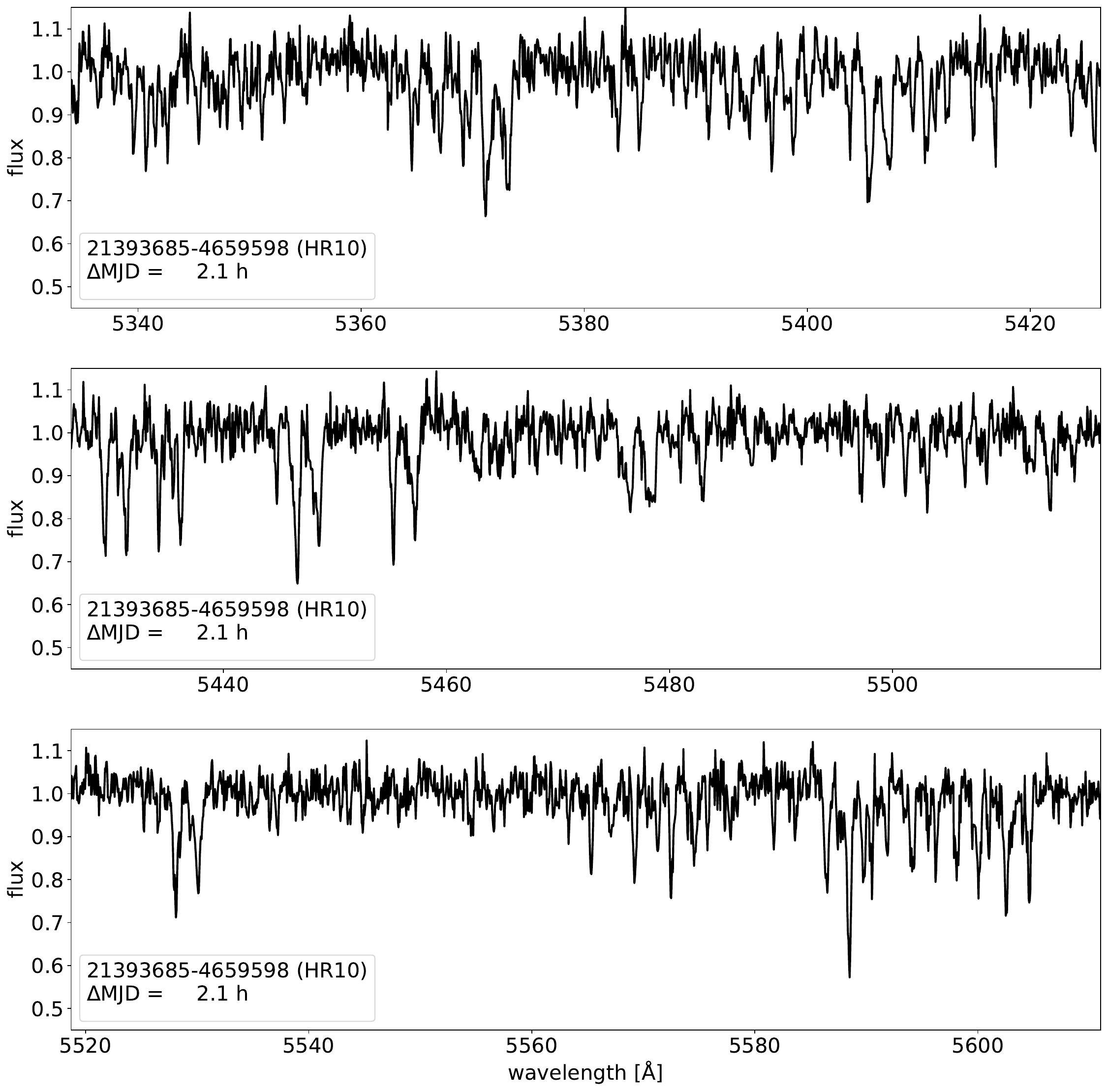}
  \captionof{figure}{\label{Fig:iDR5_SB3_atlas_21393685-4659598_3}(3/4) CNAME 21393685-4659598, at $\mathrm{MJD} = 57265.217895$, setup HR10.}
\end{minipage}
\begin{minipage}{\textwidth}
  \centering
  \includegraphics[width=0.49\textwidth]{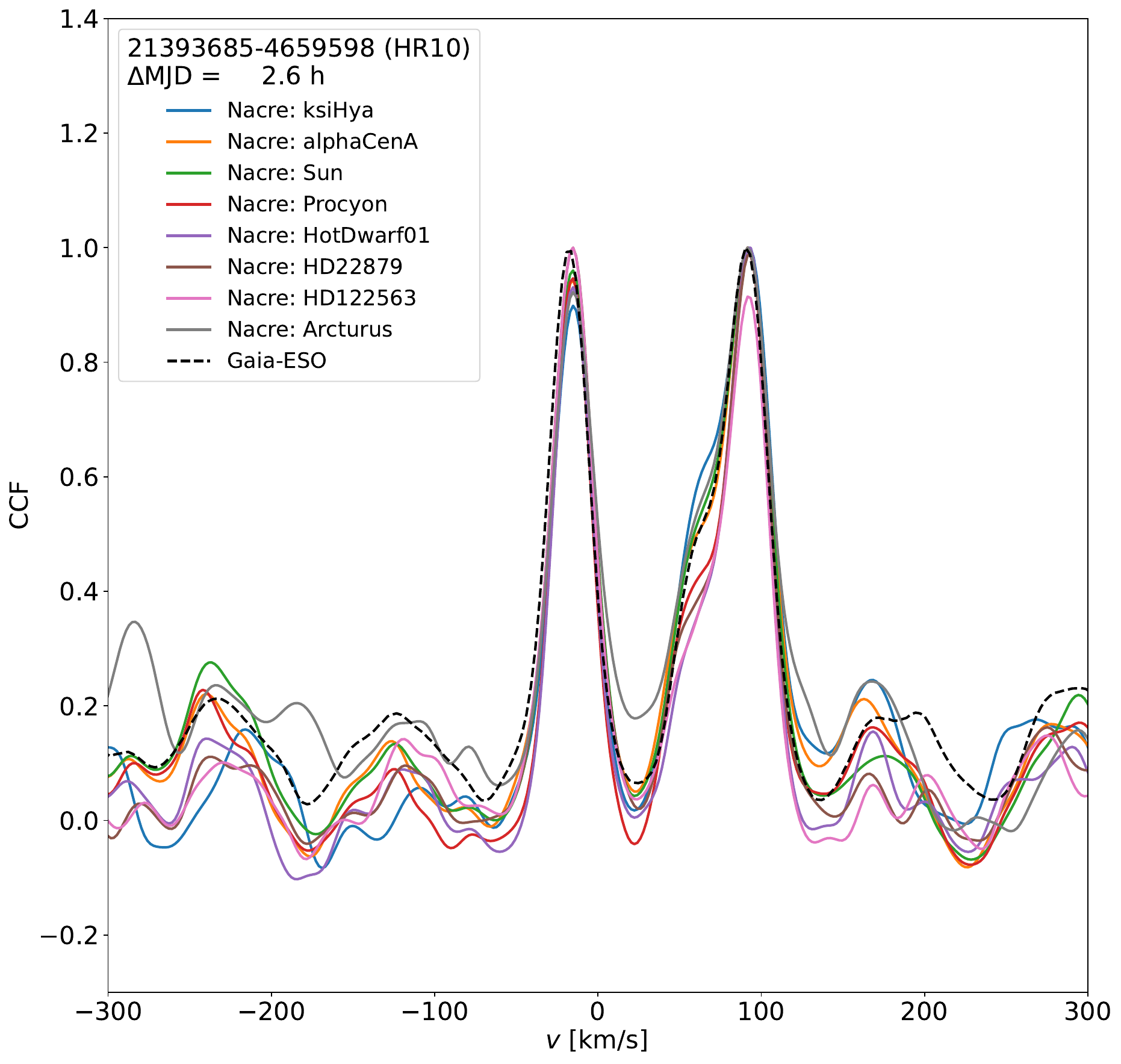}
  \includegraphics[width=0.49\textwidth]{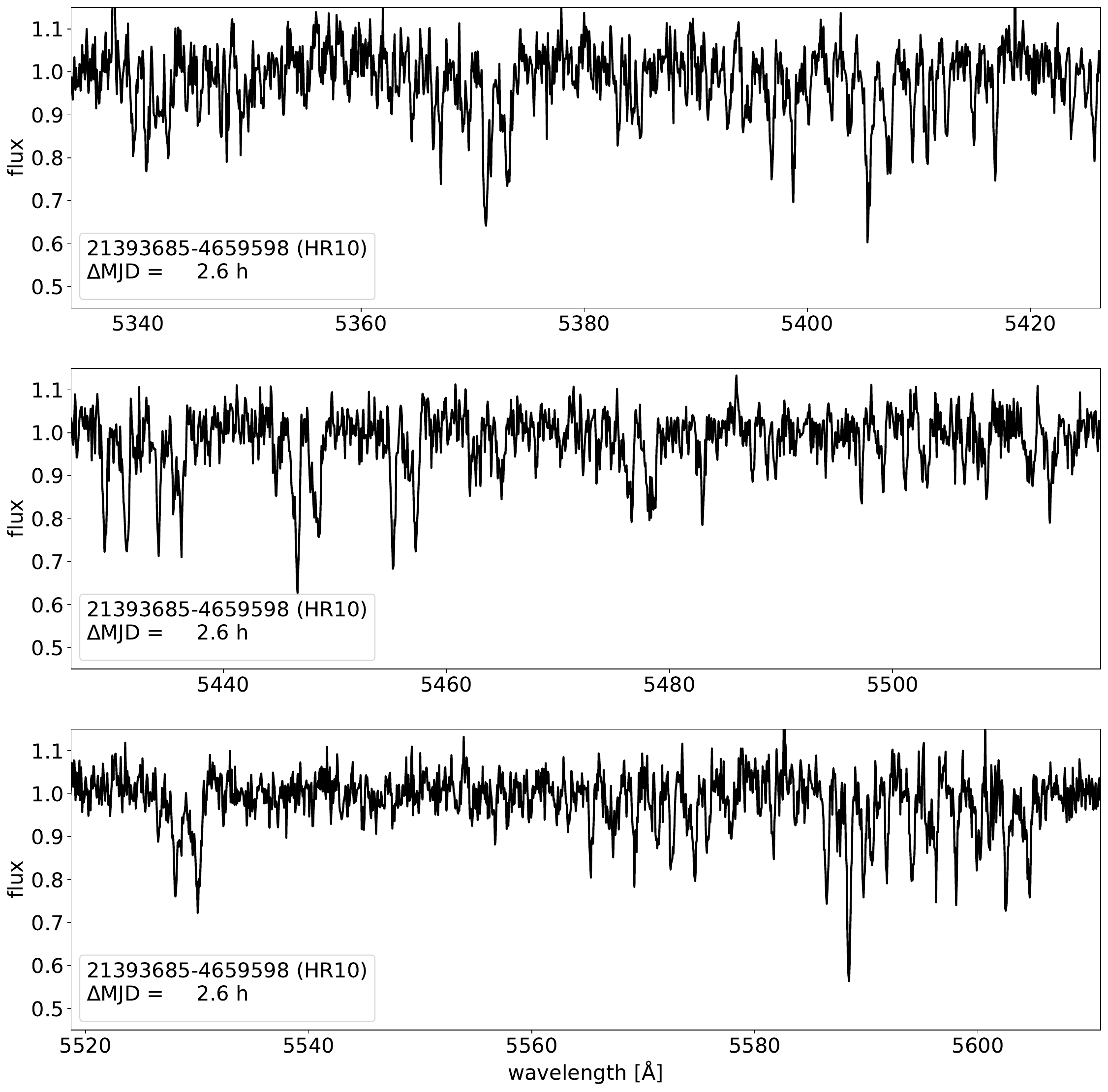}
  \captionof{figure}{\label{Fig:iDR5_SB3_atlas_21393685-4659598_4}(4/4) CNAME 21393685-4659598, at $\mathrm{MJD} = 57265.239211$, setup HR10.}
\end{minipage}
\cleardoublepage
\setlength\parindent{\defaultparindent}

\section{Atlas of \Gaia-ESO SB4}
\label{SecApp:_atlas_SB4}

{In this section, we show the CCFs and the associated spectrum at all epochs for each of the detected SB4. In each figure, the left panel shows the \nacre (coloured lines) and \Gaia-ESO (dashed black line) CCFs, while the right panel shows the corresponding spectrum (zoomed on the flux range $[0.45, 1.1]$ and with the setup's full wavelength range displayed over three sub-panels). The displayed $x$- and $y$-ranges are the same for each kind of plot (CCF or spectra), such that the reader can compare directly two epochs. All the CCFs (\nacre or \Gaia-ESO) are normalised such that the highest peak has its maximum at 1. The legend box reminds the CNAME of the system, the setup and the elapsed time $\Delta \mathrm{MJD}$ since the first available epoch. The caption indicates the CNAME, epoch (MJD) and GIRAFFE setup, plus a numbering to ease the reading of the time-series. The CCFs are ordered by ascending MJD. According to the stellar waltz occurring in the examined system, one can follow the motion of the different stellar components along time: depending on the orbital phase, one can see one, two, three or four components. These systems are highly likely at least SB3, while the existence of a fourth component is only tentative: for this reason, we briefly comment at the beginning of each section the interpretation of the CCFs pointing at an SB4.}
\cleardoublepage

\subsection{18263326-3146508}

{Figure~\ref{Fig:iDR5_SB4_atlas_18263326-3146508_1} shows two stellar components whose velocities are around \SI{120}{\kilo\metre\per\second} and \SI{180}{\kilo\metre\per\second}. Figure~\ref{Fig:iDR5_SB4_atlas_18263326-3146508_2} shows a third stellar component whose velocity is between the first two stellar components. A fourth tentative stellar component might exist at $v_{\mathrm{rad}} \approx \SI{20}{\kilo\metre\per\second}$. This local maximum exists at the two available epochs and in all CCFs, decreasing the probability for it to be correlation noise.}
\clearpage

\setlength\parindent{0cm}
\begin{minipage}{\textwidth}
  \centering
  \includegraphics[width=0.49\textwidth]{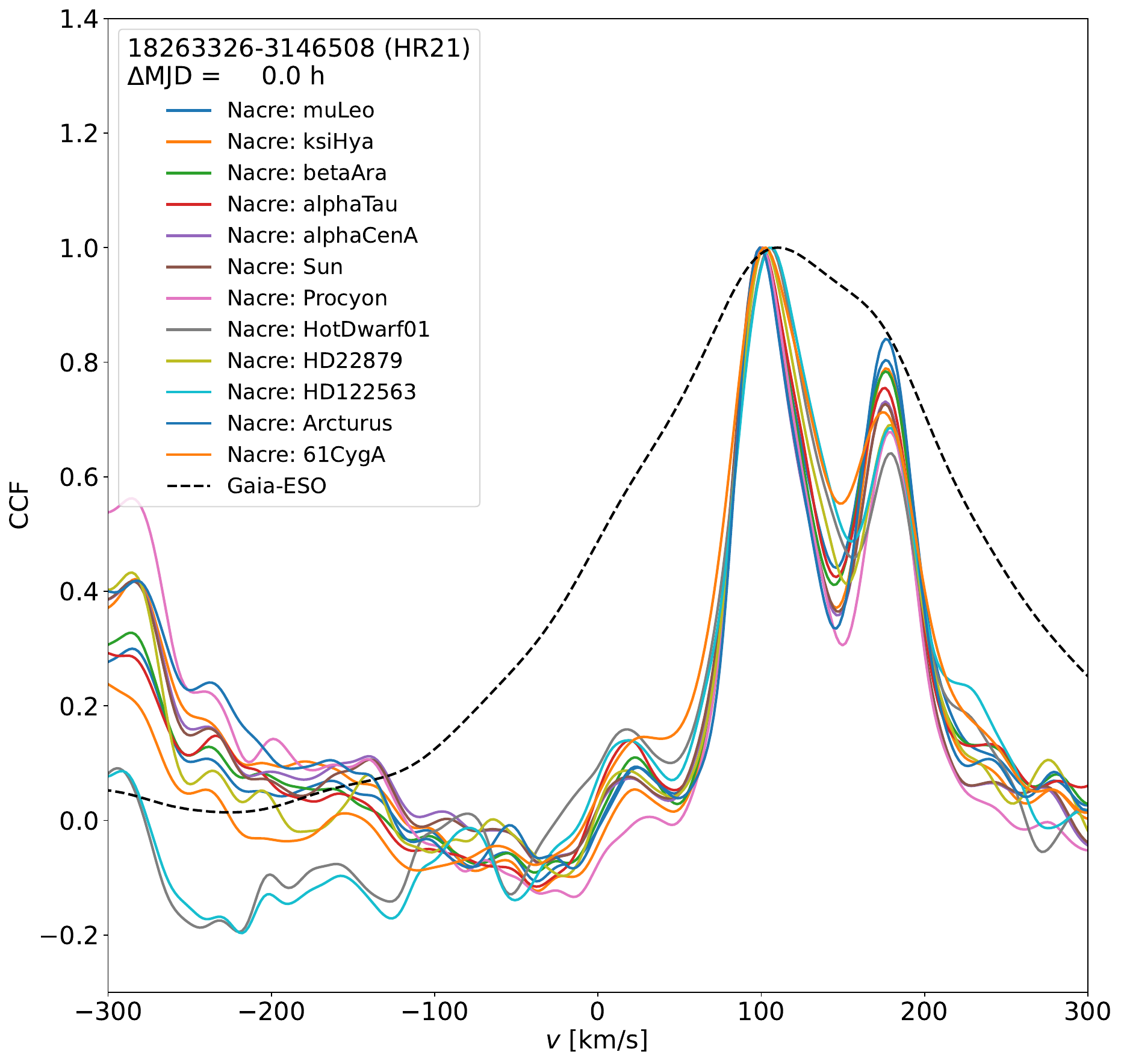}
  \includegraphics[width=0.49\textwidth]{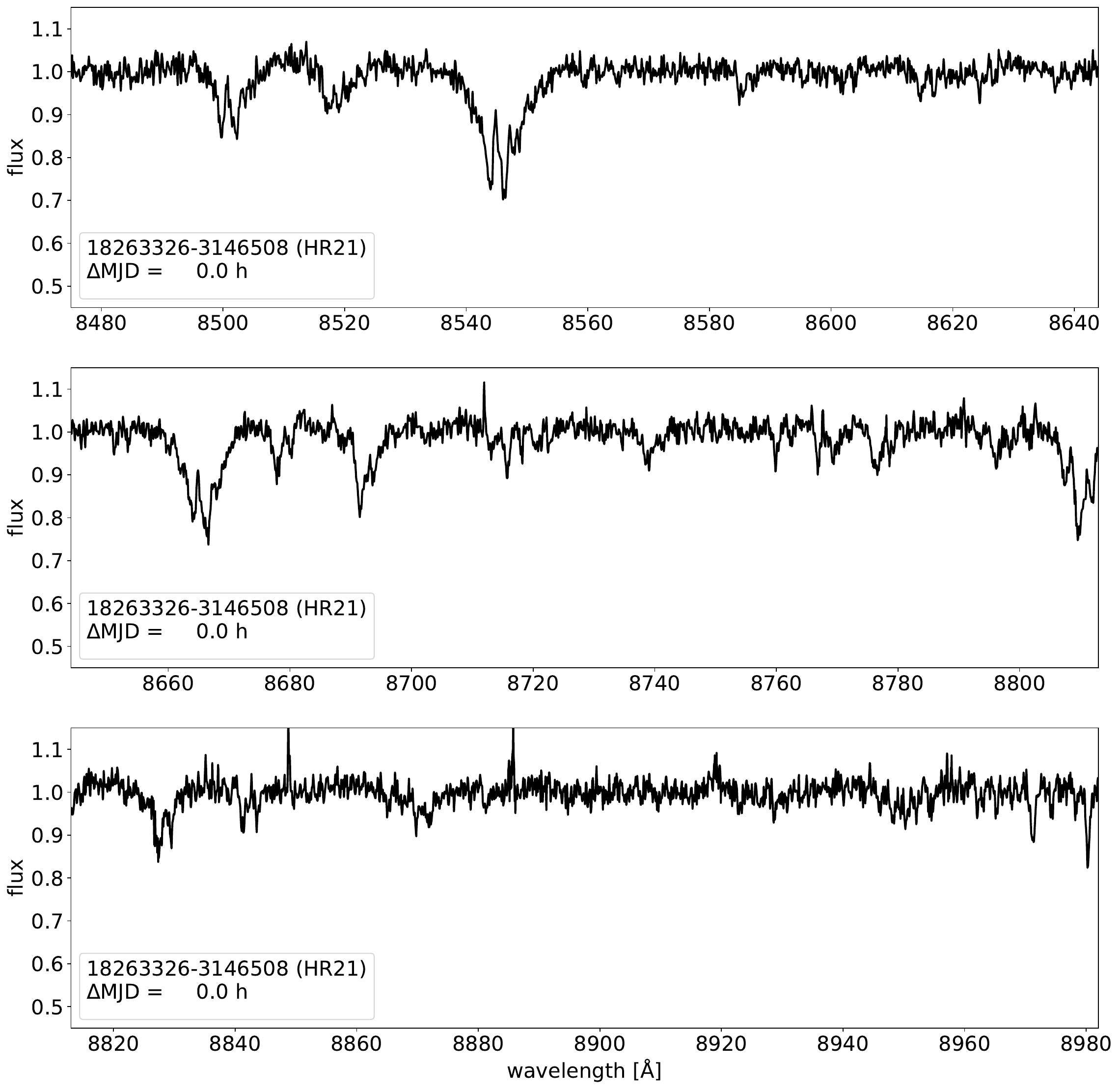}
  \captionof{figure}{\label{Fig:iDR5_SB4_atlas_18263326-3146508_1}(1/2) CNAME 18263326-3146508, at $\mathrm{MJD} = 56857.033274$, setup HR21.}
\end{minipage}
\begin{minipage}{\textwidth}
  \centering
  \includegraphics[width=0.49\textwidth]{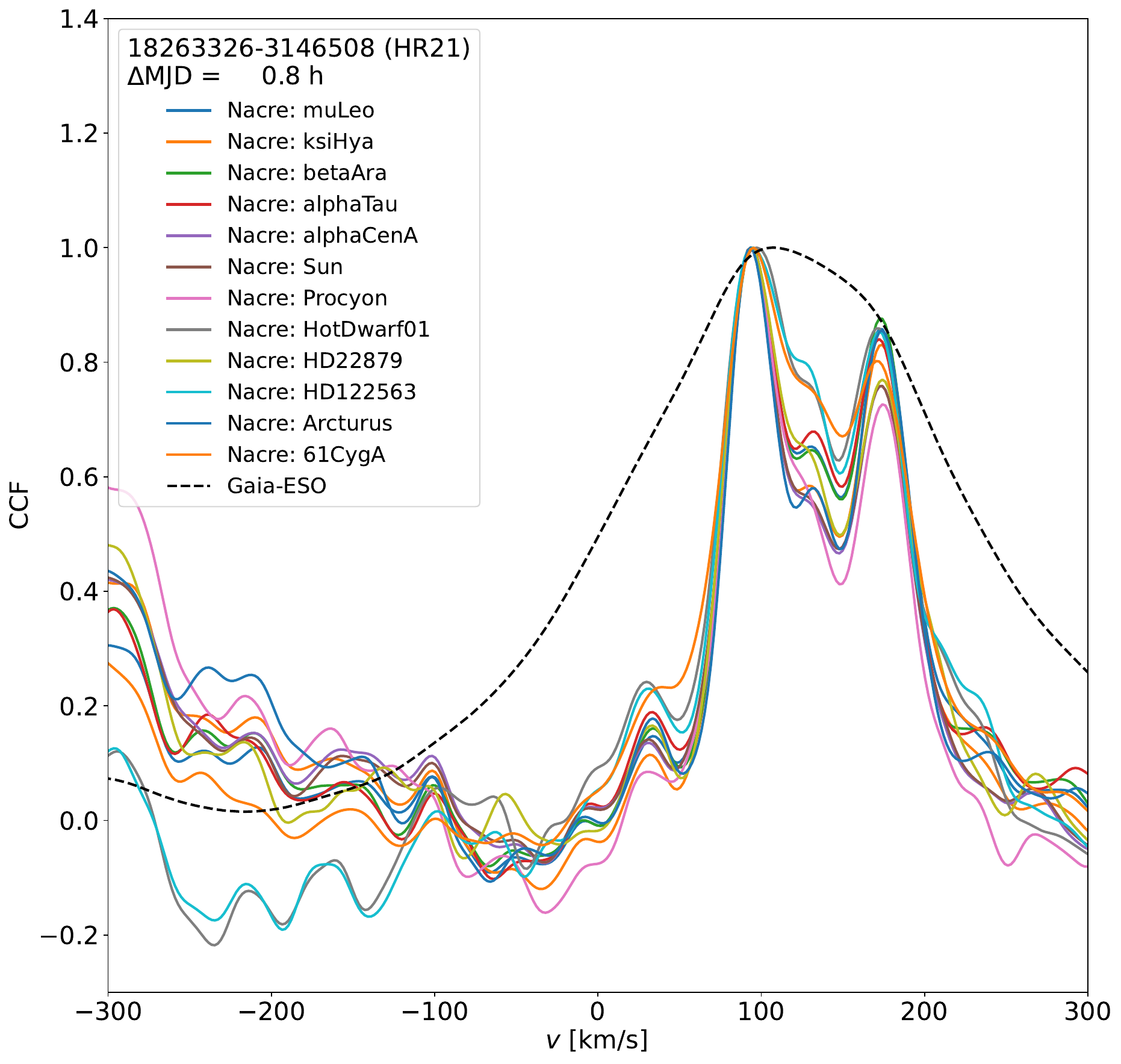}
  \includegraphics[width=0.49\textwidth]{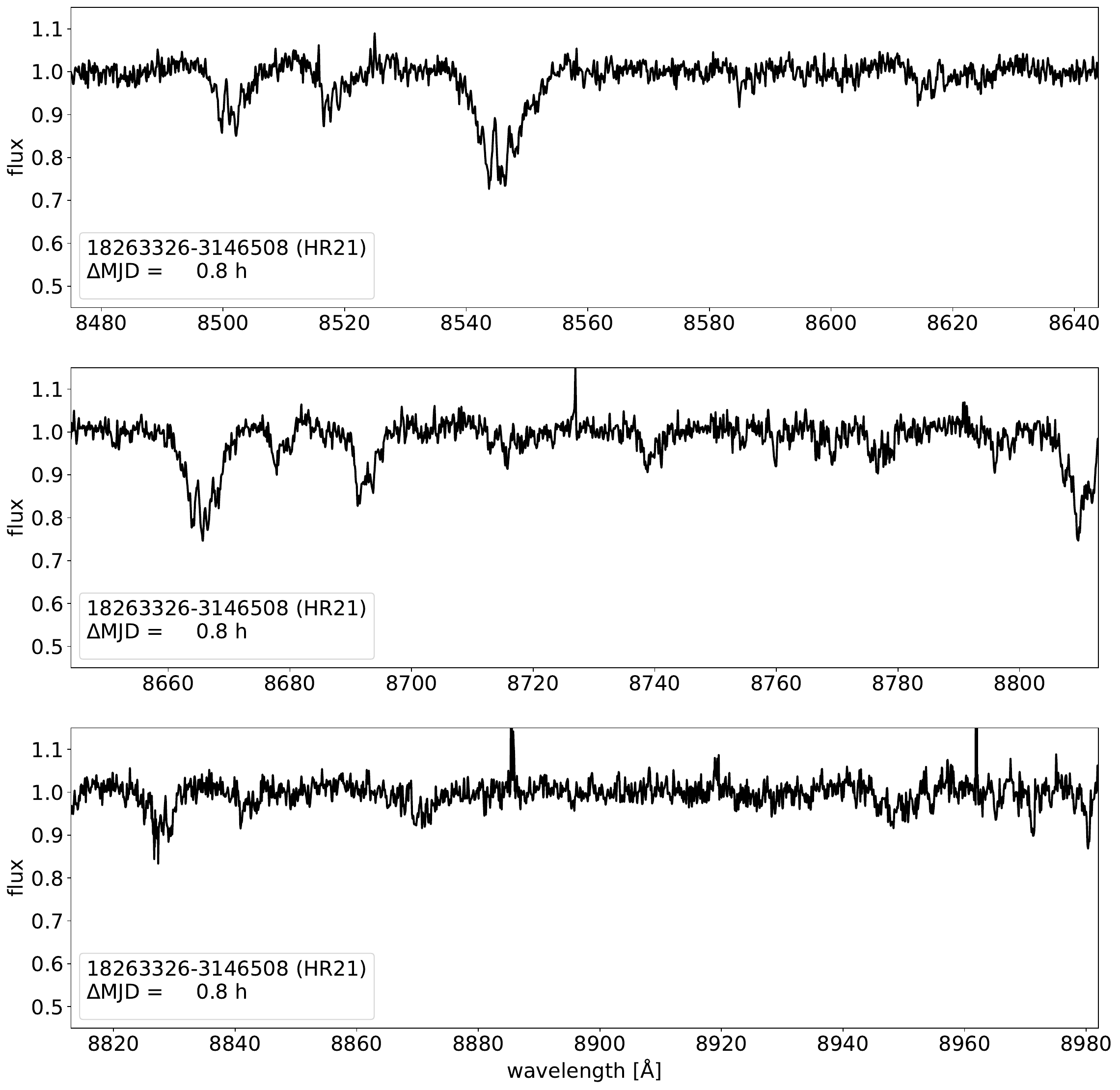}
  \captionof{figure}{\label{Fig:iDR5_SB4_atlas_18263326-3146508_2}(2/2) CNAME 18263326-3146508, at $\mathrm{MJD} = 56857.065142$, setup HR21.}
\end{minipage}
\cleardoublepage
\setlength\parindent{\defaultparindent}

\subsection{19243943+0048136}

{Figures~\ref{Fig:iDR5_SB4_atlas_19243943+0048136_1} and \ref{Fig:iDR5_SB4_atlas_19243943+0048136_2} show four stellar components whose velocities are between $\approx \SI{-100}{\kilo\metre\per\second}$ and $\approx \SI{180}{\kilo\metre\per\second}$. We note that they are all within the main peak of the very wide \Gaia-ESO CCF; an inflexion point can be seen in the \Gaia-ESO CCF where the component with the largest velocity lies. The local maximum at $v \approx \SI{200}{\kilo\metre\per\second}$ might be correlation noise since if it was a genuine stellar component, we would expect to see at least an inflexion point in the \Gaia-ESO CCF at this velocity (\Gaia-ESO CCF are less prone to correlation noise thanks to to their correlating masks). Figures~\ref{Fig:iDR5_SB4_atlas_19243943+0048136_7} and \ref{Fig:iDR5_SB4_atlas_19243943+0048136_8} might show only three components: the local maximum at $v \approx \SI{150}{\kilo\metre\per\second}$ might be correlation noise. The other epochs plotted here show either one, two or three stellar components.}
\clearpage

\setlength\parindent{0cm}
\begin{minipage}{\textwidth}
  \centering
  \includegraphics[width=0.49\textwidth]{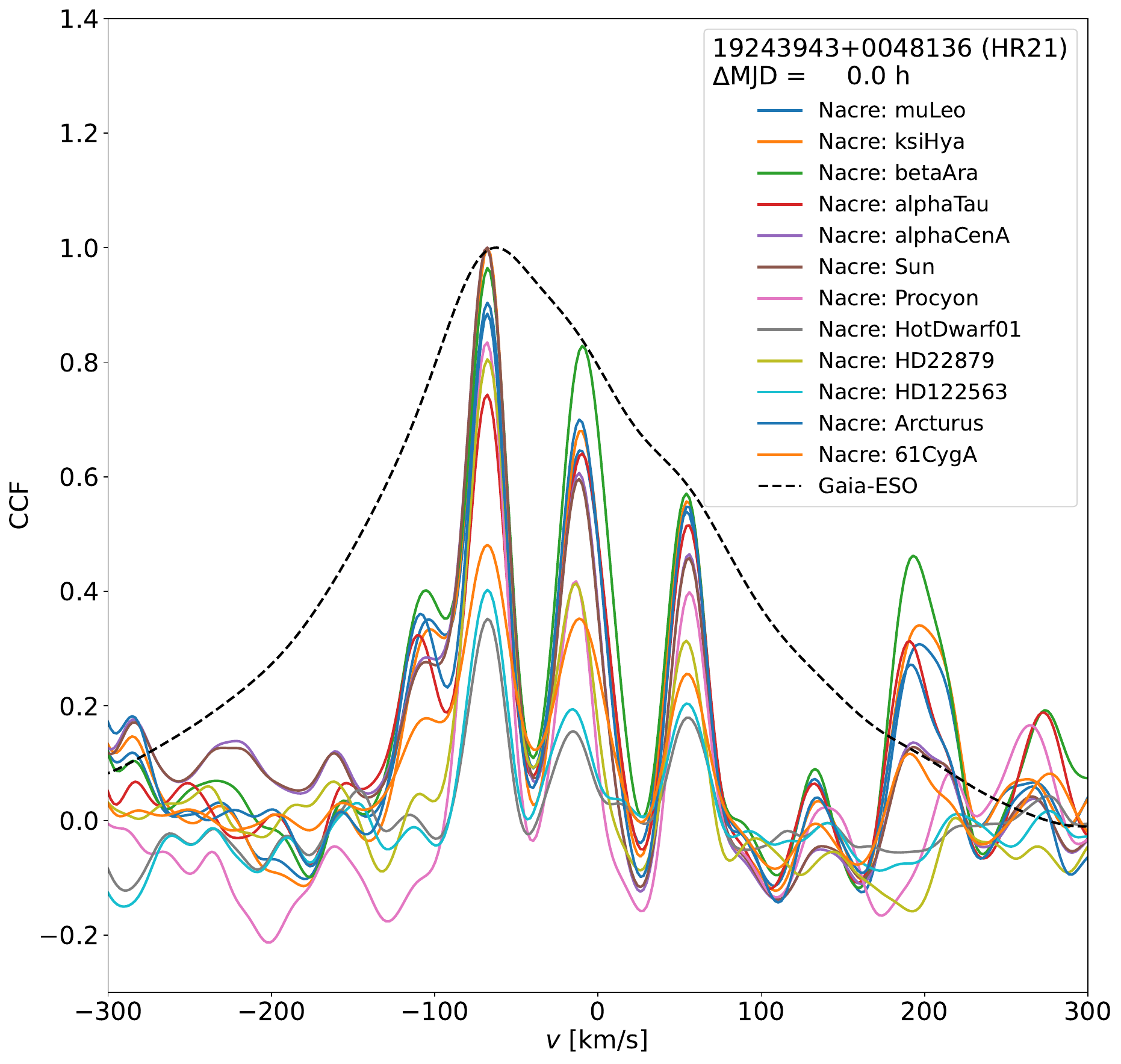}
  \includegraphics[width=0.49\textwidth]{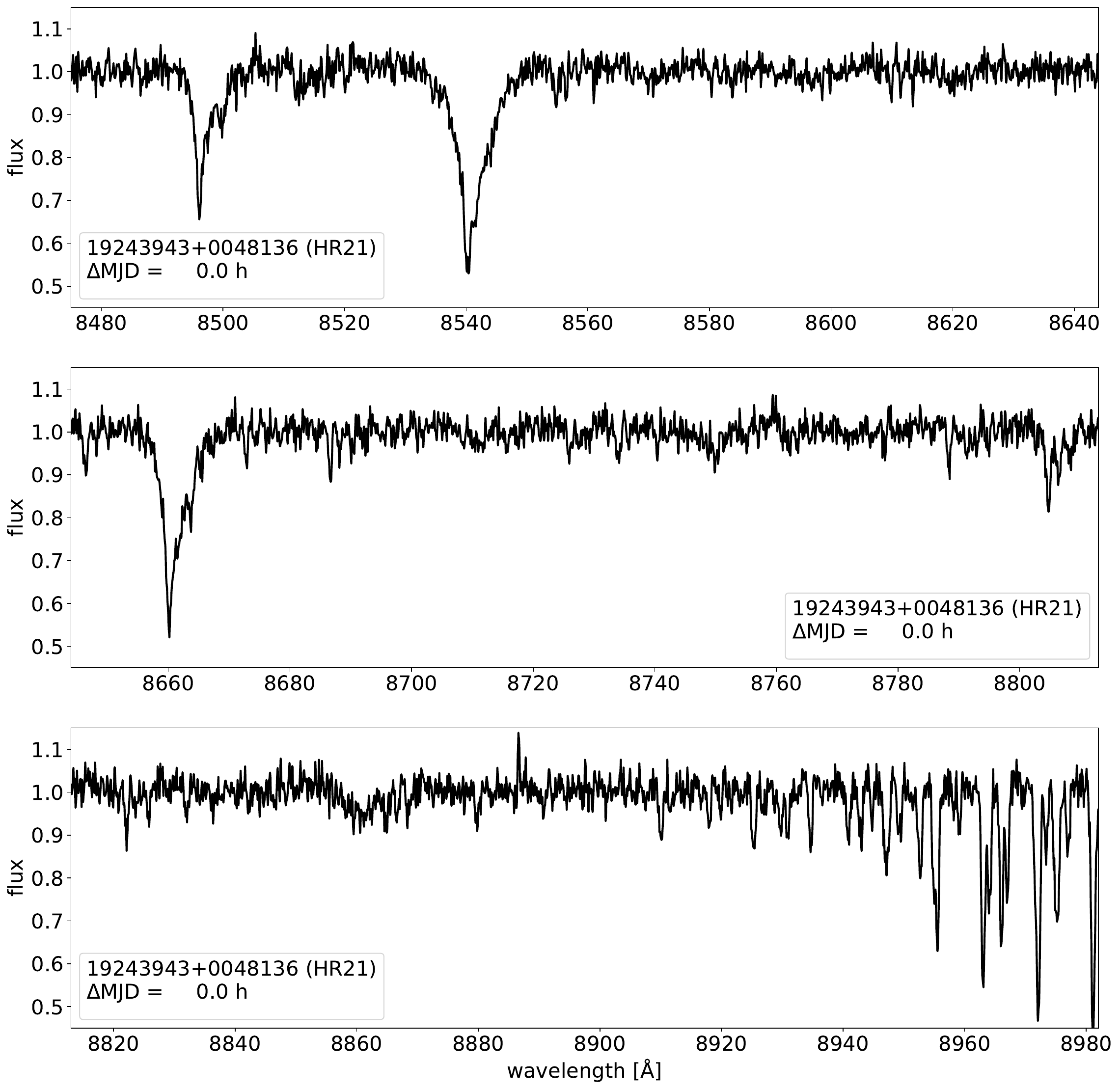}
  \captionof{figure}{\label{Fig:iDR5_SB4_atlas_19243943+0048136_1}(1/12) CNAME 19243943+0048136, at $\mathrm{MJD} = 56074.421068$, setup HR21.}
\end{minipage}
\begin{minipage}{\textwidth}
  \centering
  \includegraphics[width=0.49\textwidth]{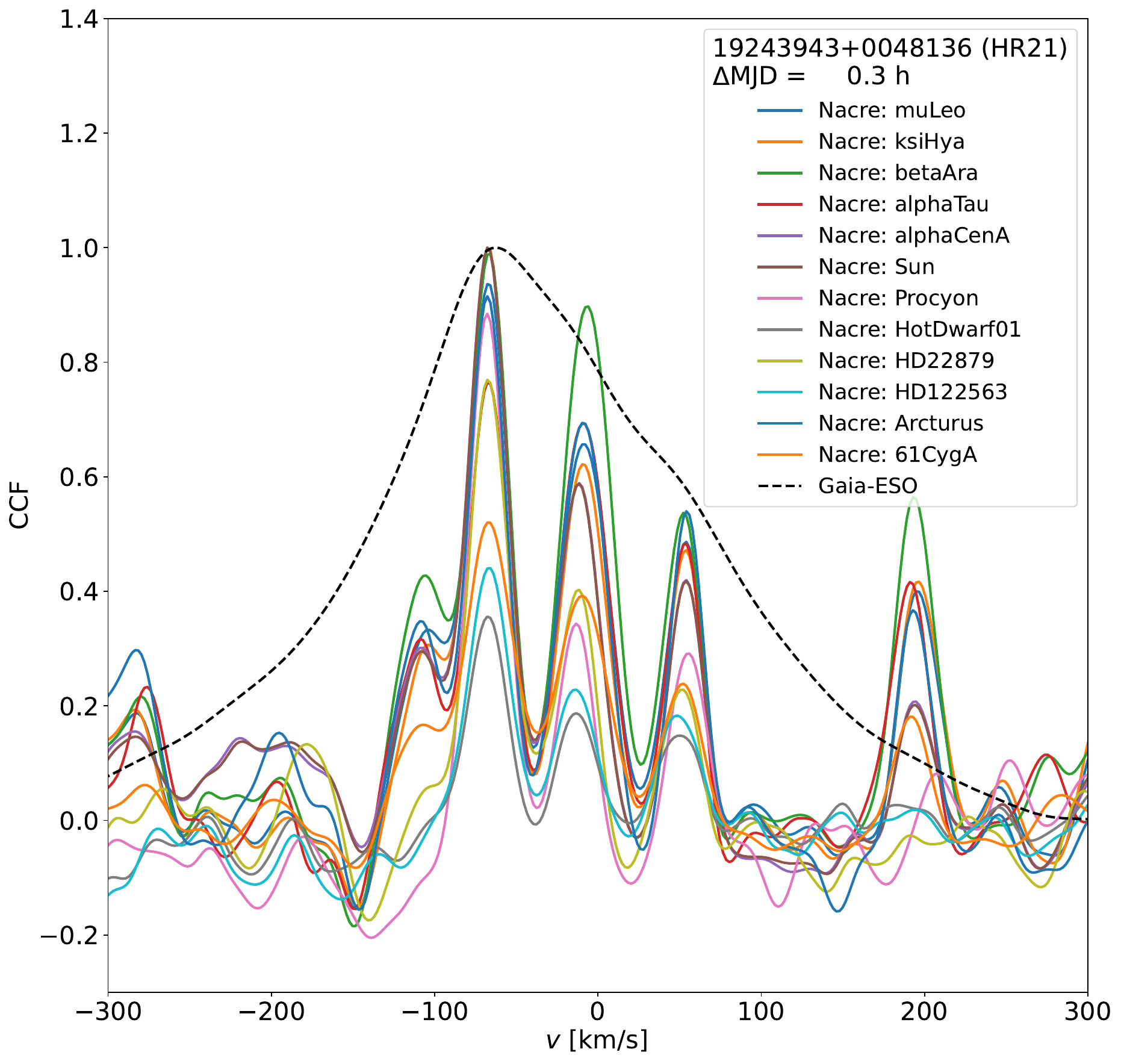}
  \includegraphics[width=0.49\textwidth]{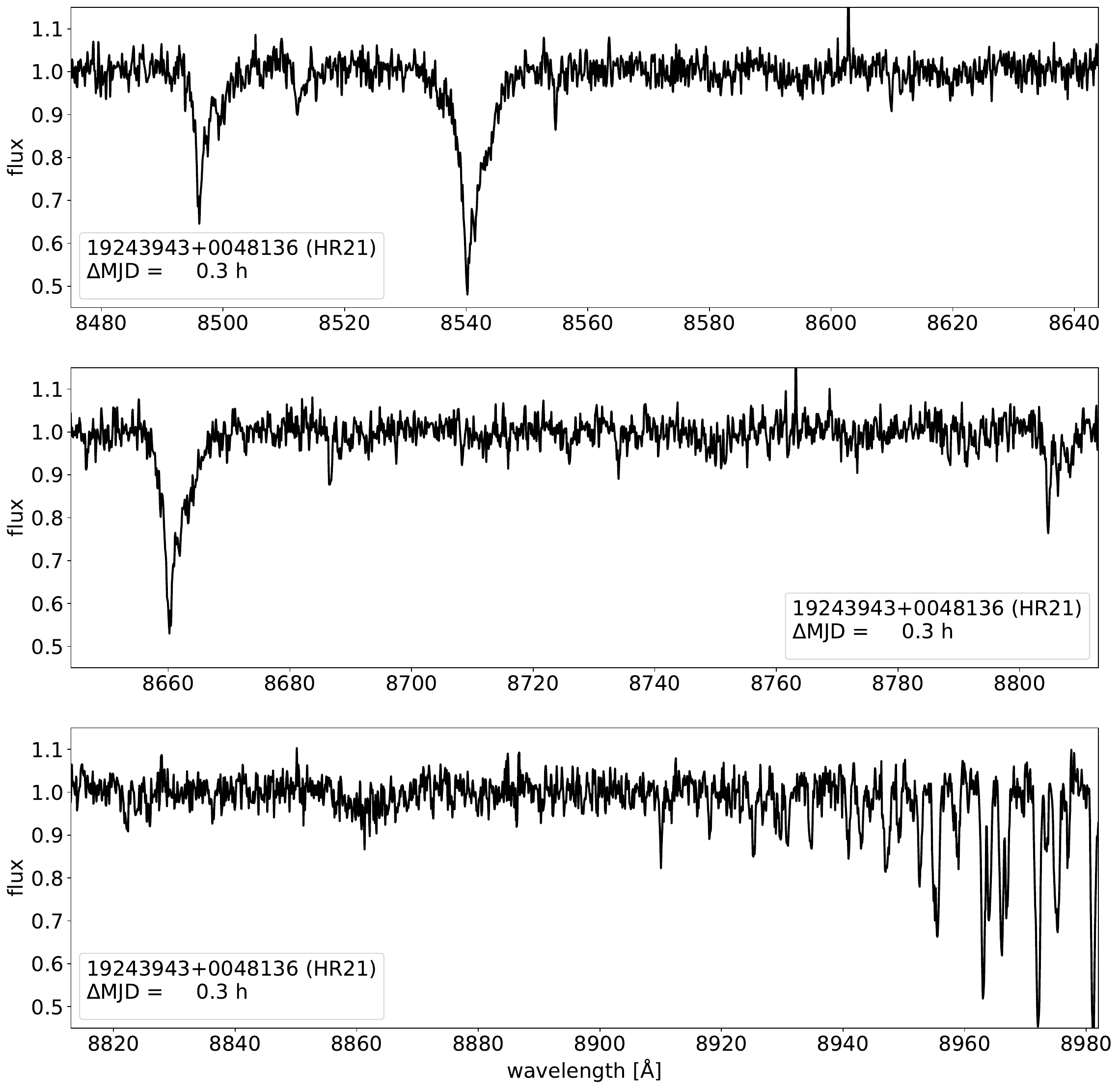}
  \captionof{figure}{\label{Fig:iDR5_SB4_atlas_19243943+0048136_2}(2/12) CNAME 19243943+0048136, at $\mathrm{MJD} = 56074.431550$, setup HR21.}
\end{minipage}
\clearpage
\begin{minipage}{\textwidth}
  \centering
  \includegraphics[width=0.49\textwidth]{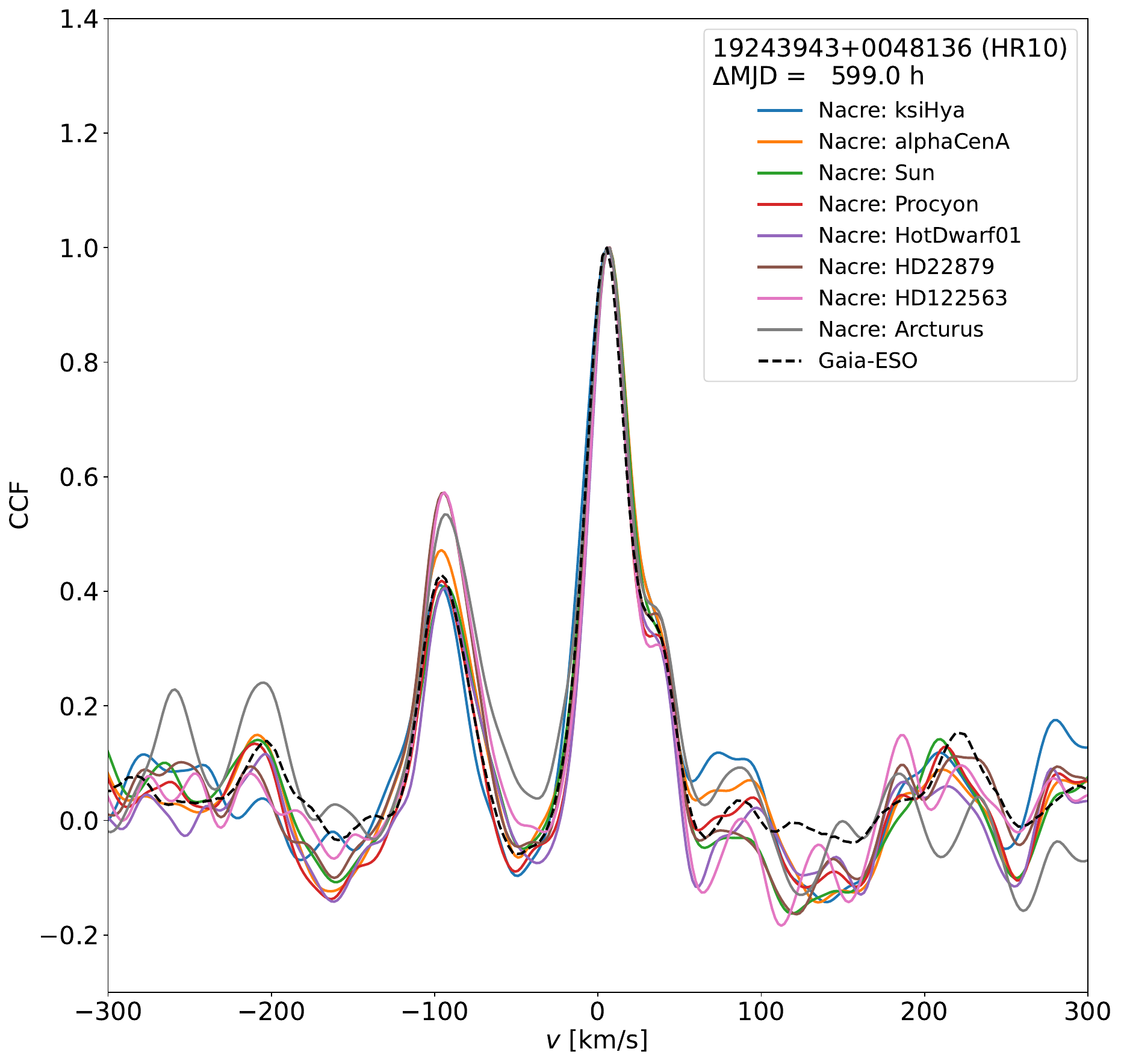}
  \includegraphics[width=0.49\textwidth]{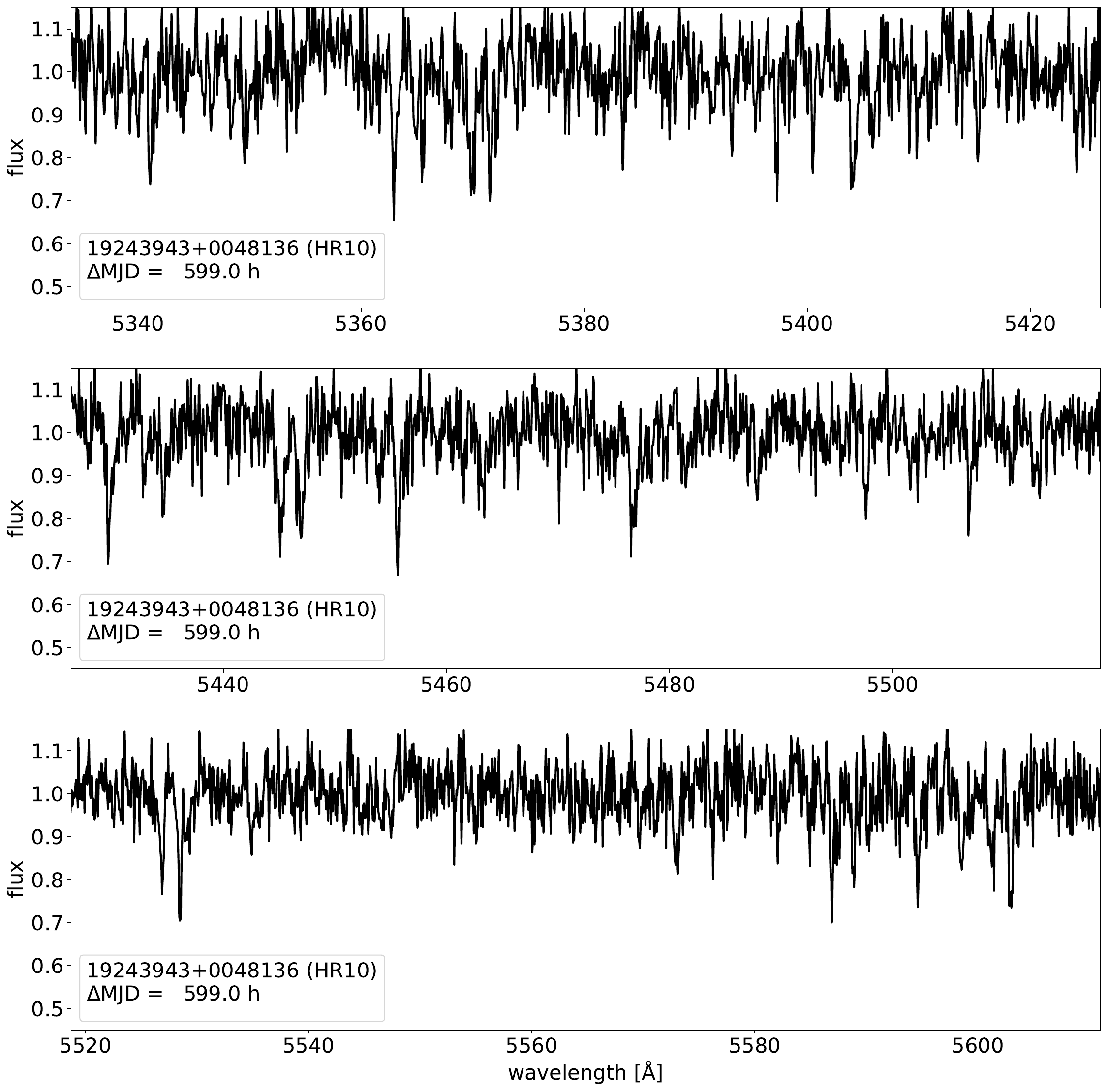}
  \captionof{figure}{\label{Fig:iDR5_SB4_atlas_19243943+0048136_3}(3/12) CNAME 19243943+0048136, at $\mathrm{MJD} = 56099.378781$, setup HR10.}
\end{minipage}
\begin{minipage}{\textwidth}
  \centering
  \includegraphics[width=0.49\textwidth]{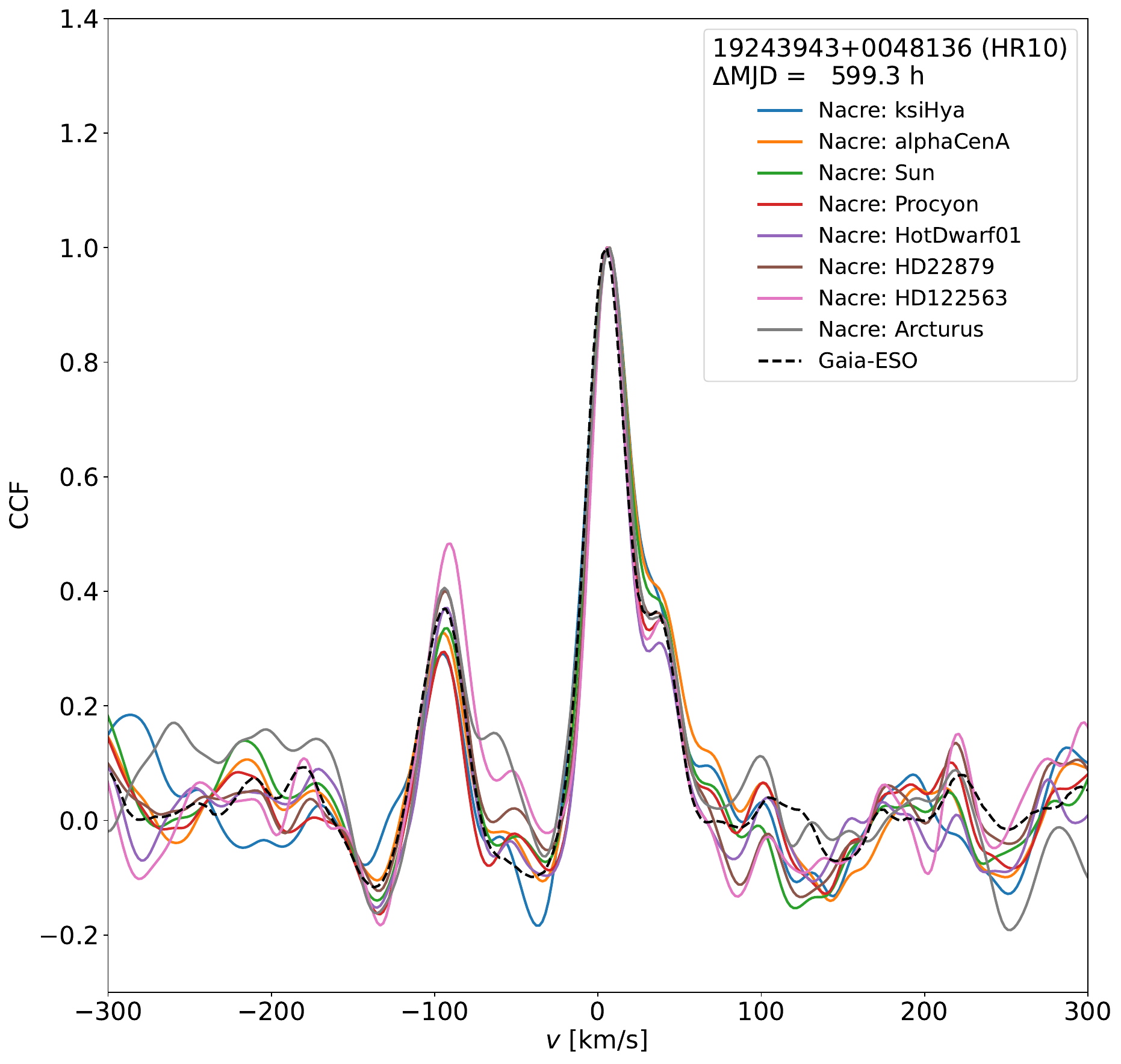}
  \includegraphics[width=0.49\textwidth]{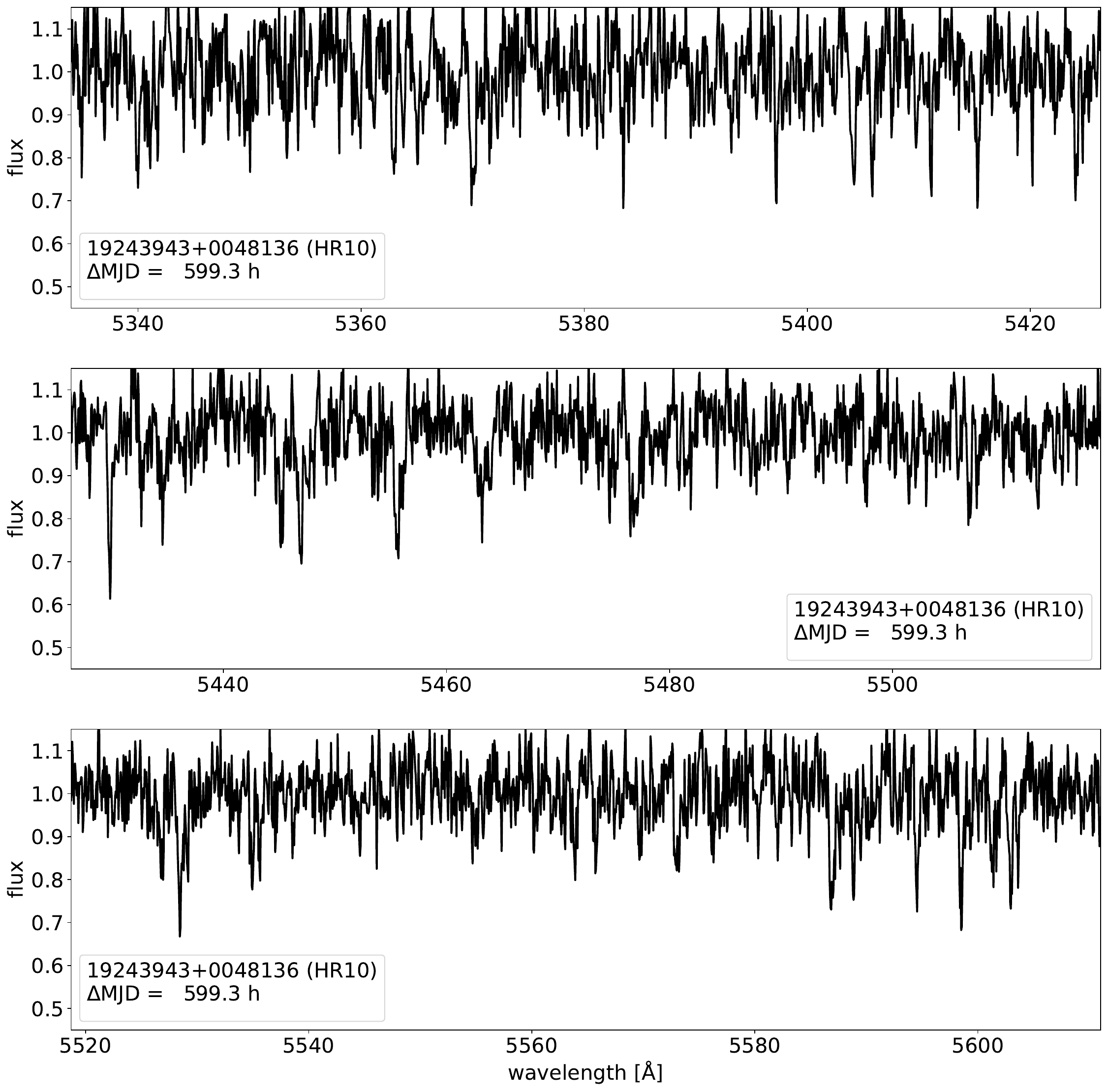}
  \captionof{figure}{\label{Fig:iDR5_SB4_atlas_19243943+0048136_4}(4/12) CNAME 19243943+0048136, at $\mathrm{MJD} = 56099.391848$, setup HR10.}
\end{minipage}
\clearpage
\begin{minipage}{\textwidth}
  \centering
  \includegraphics[width=0.49\textwidth]{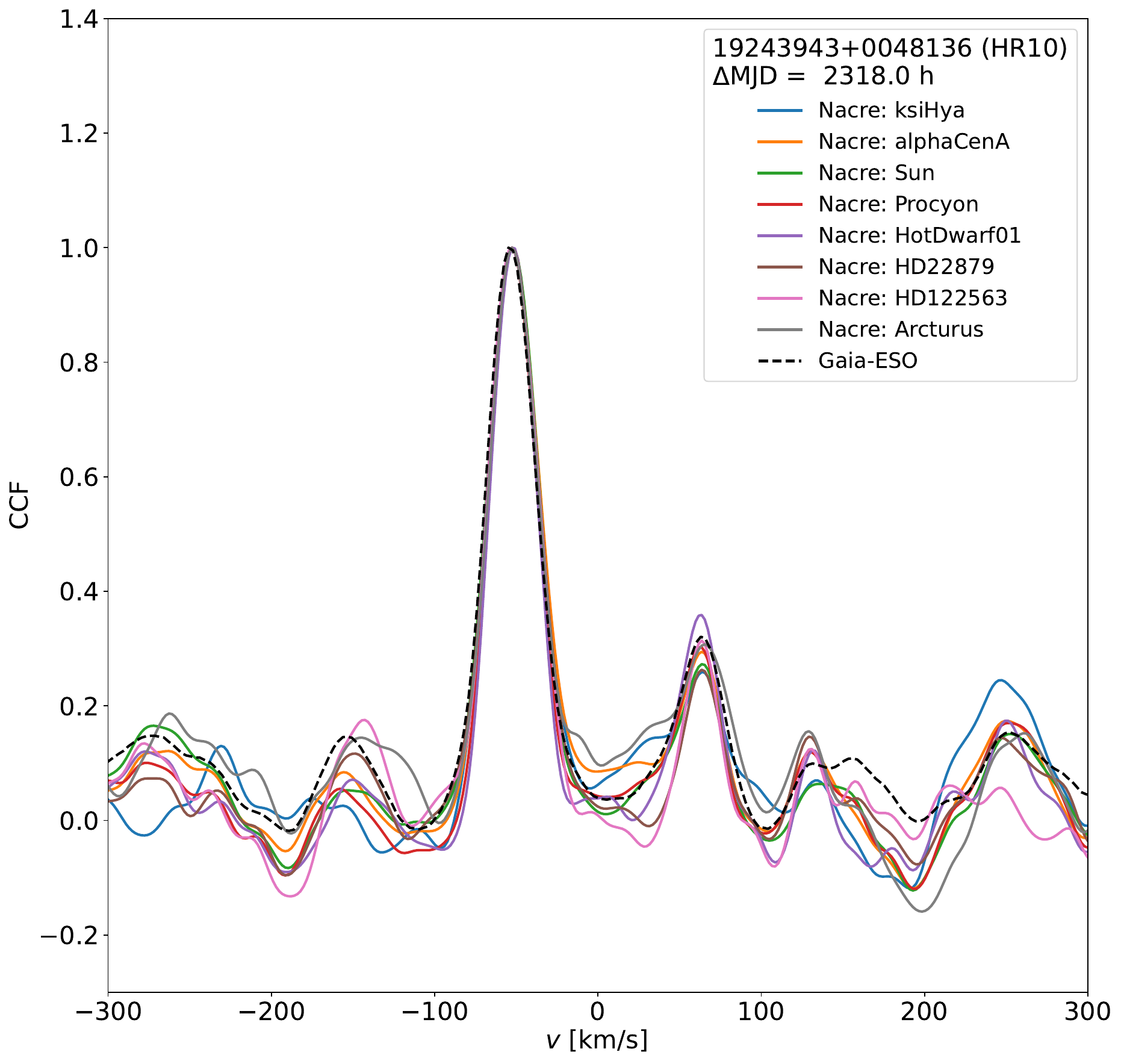}
  \includegraphics[width=0.49\textwidth]{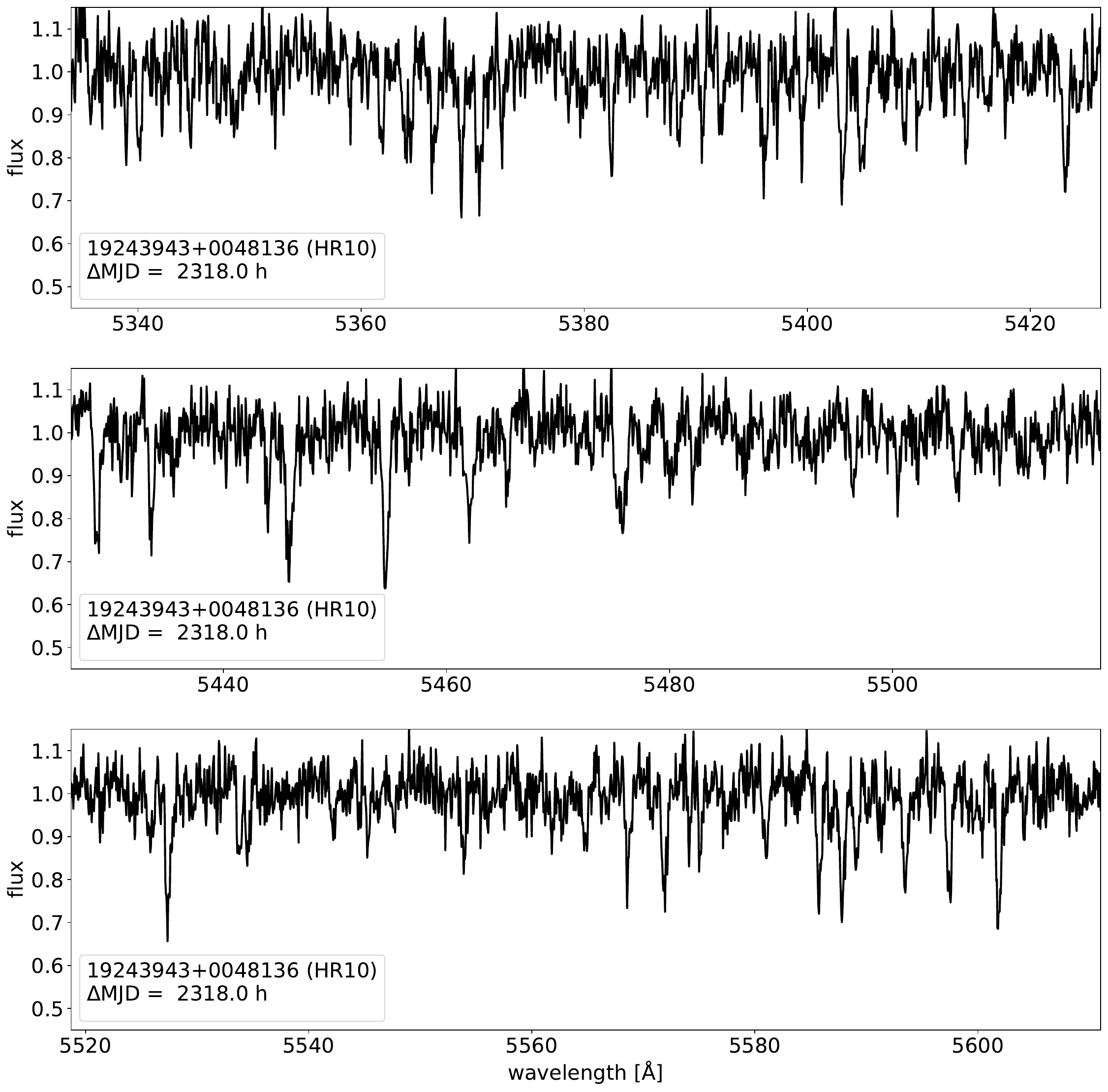}
  \captionof{figure}{\label{Fig:iDR5_SB4_atlas_19243943+0048136_5}(5/12) CNAME 19243943+0048136, at $\mathrm{MJD} = 56171.003662$, setup HR10.}
\end{minipage}
\begin{minipage}{\textwidth}
  \centering
  \includegraphics[width=0.49\textwidth]{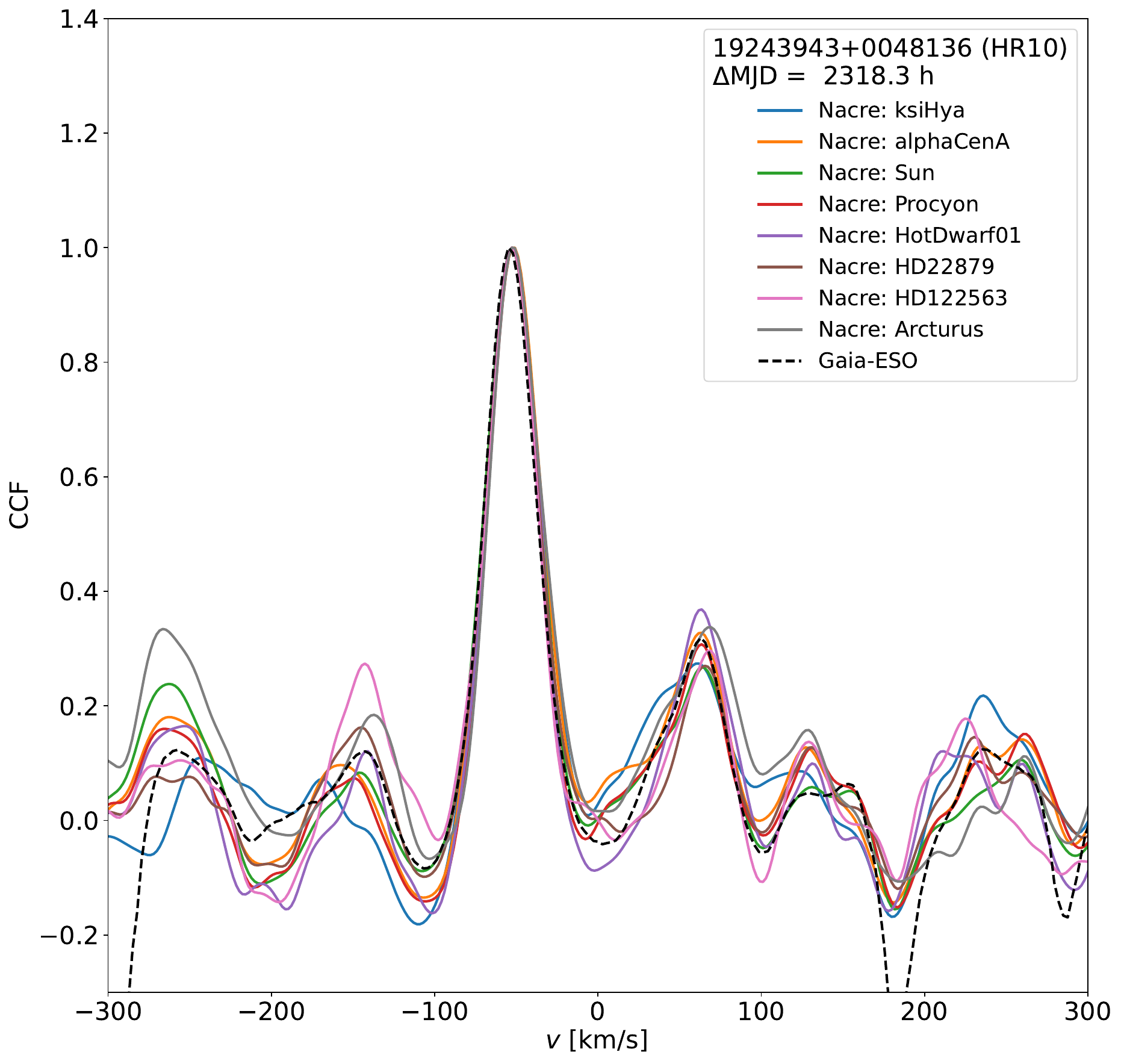}
  \includegraphics[width=0.49\textwidth]{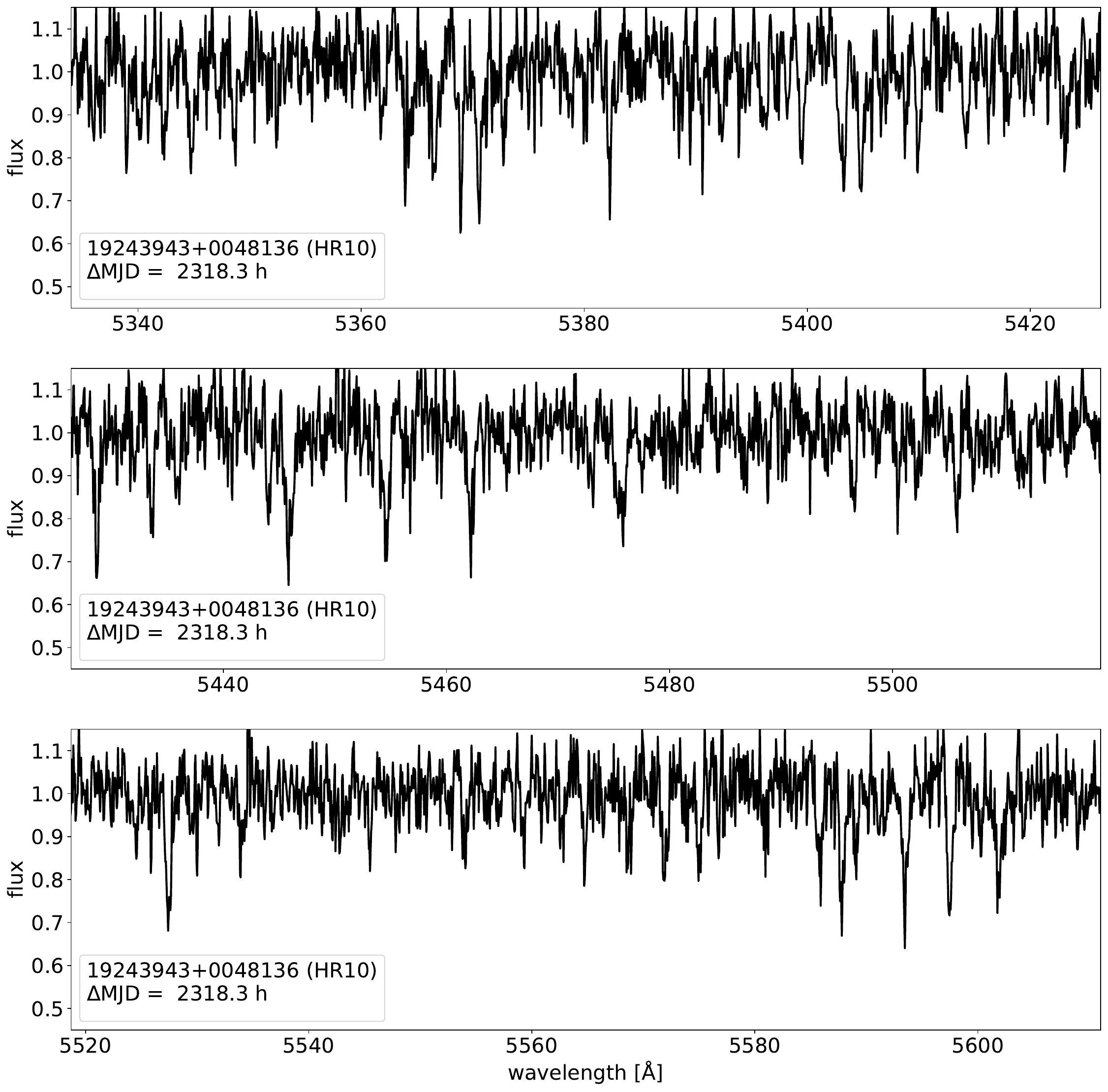}
  \captionof{figure}{\label{Fig:iDR5_SB4_atlas_19243943+0048136_6}(6/12) CNAME 19243943+0048136, at $\mathrm{MJD} = 56171.017543$, setup HR10.}
\end{minipage}
\clearpage
\begin{minipage}{\textwidth}
  \centering
  \includegraphics[width=0.49\textwidth]{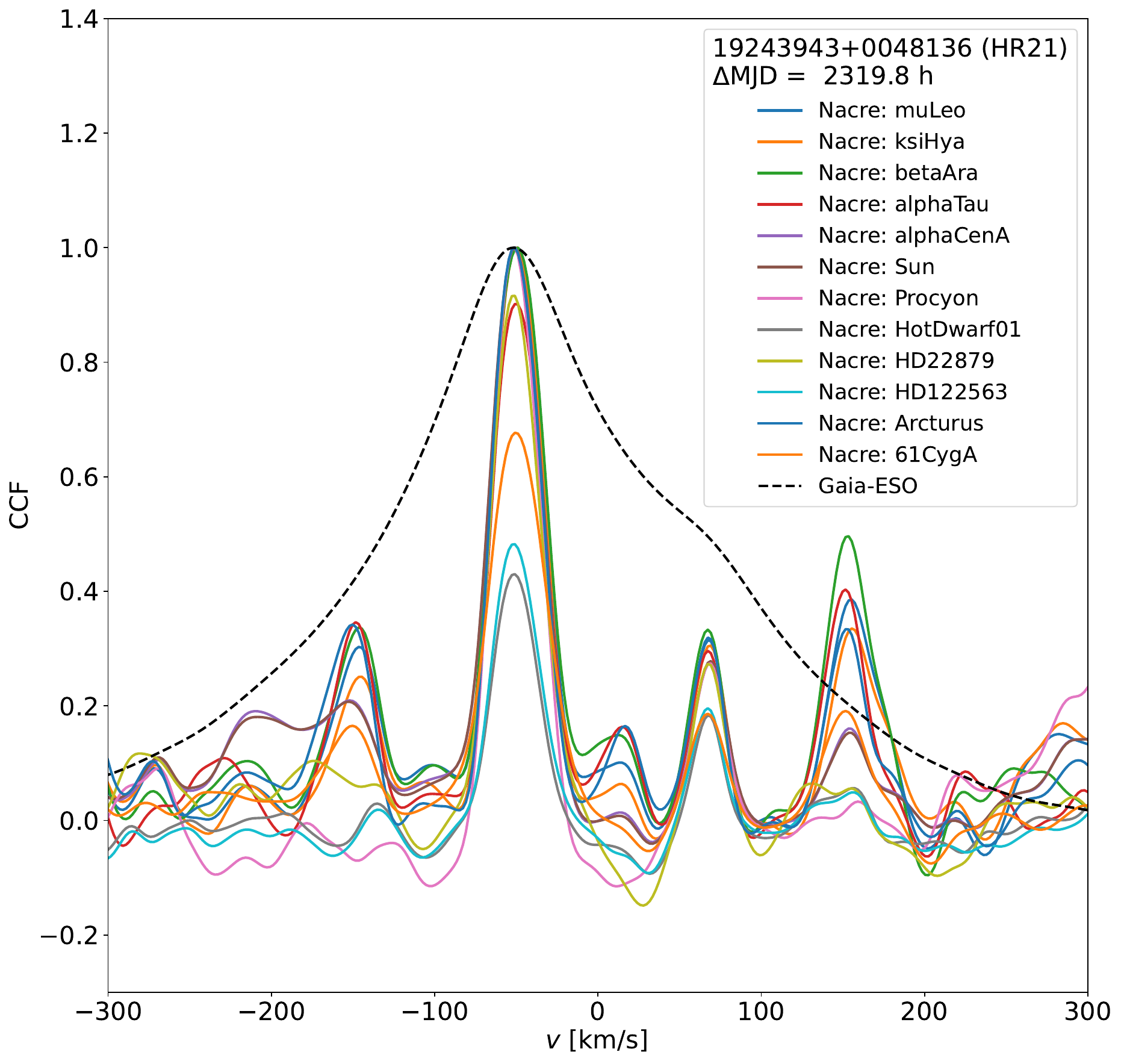}
  \includegraphics[width=0.49\textwidth]{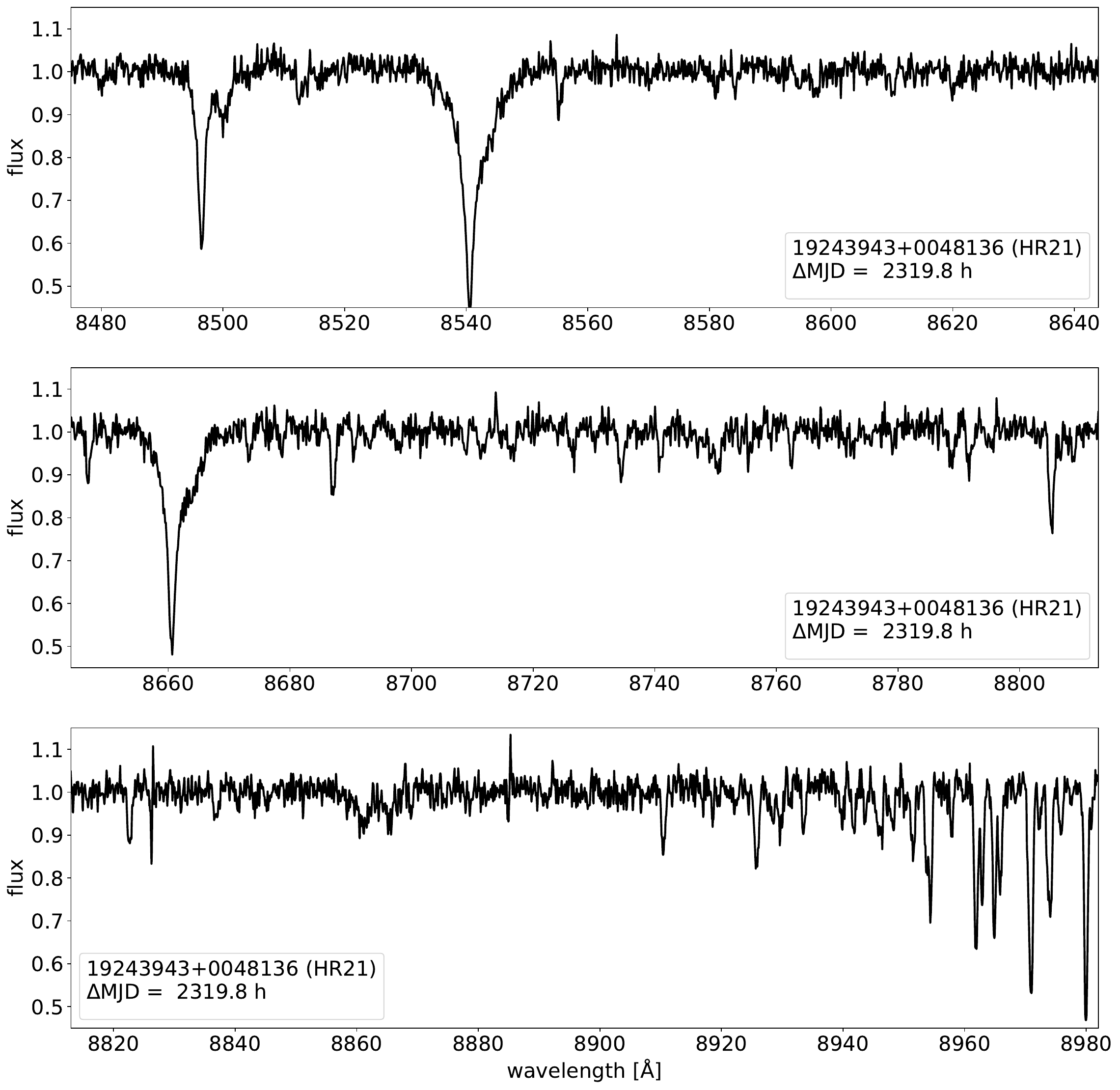}
  \captionof{figure}{\label{Fig:iDR5_SB4_atlas_19243943+0048136_7}(7/12) CNAME 19243943+0048136, at $\mathrm{MJD} = 56171.078342$, setup HR21.}
\end{minipage}
\begin{minipage}{\textwidth}
  \centering
  \includegraphics[width=0.49\textwidth]{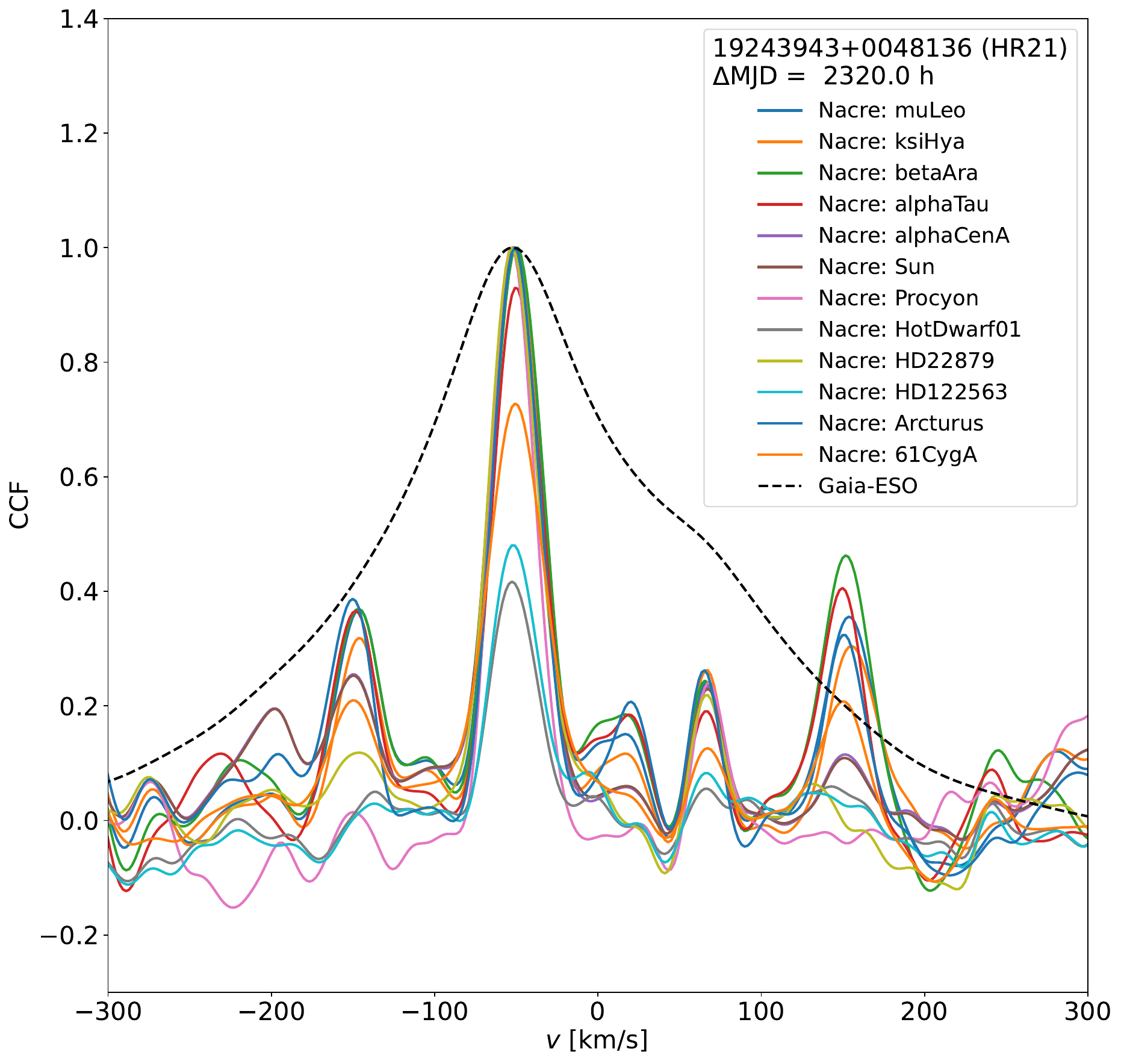}
  \includegraphics[width=0.49\textwidth]{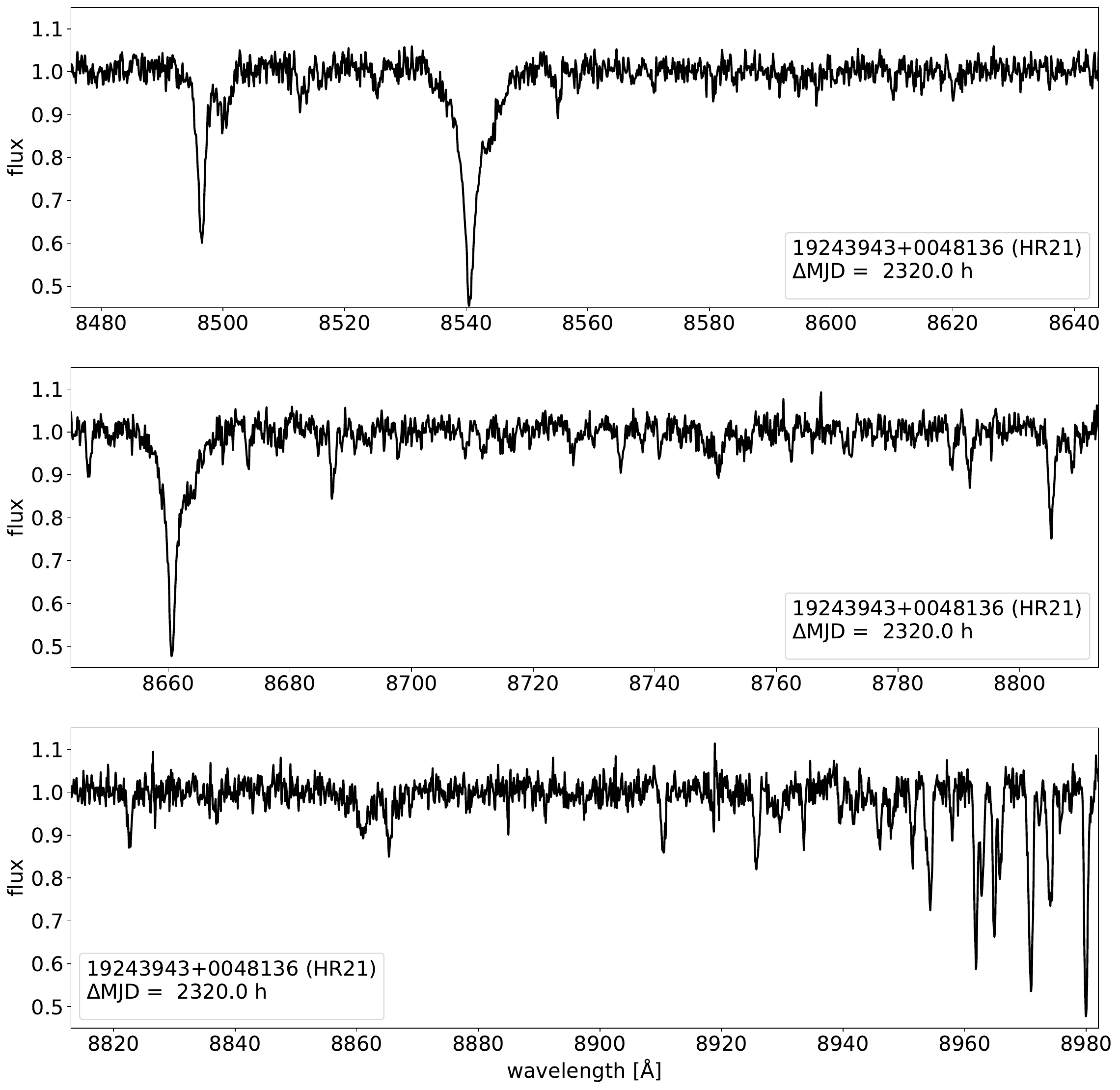}
  \captionof{figure}{\label{Fig:iDR5_SB4_atlas_19243943+0048136_8}(8/12) CNAME 19243943+0048136, at $\mathrm{MJD} = 56171.088809$, setup HR21.}
\end{minipage}
\clearpage
\begin{minipage}{\textwidth}
  \centering
  \includegraphics[width=0.49\textwidth]{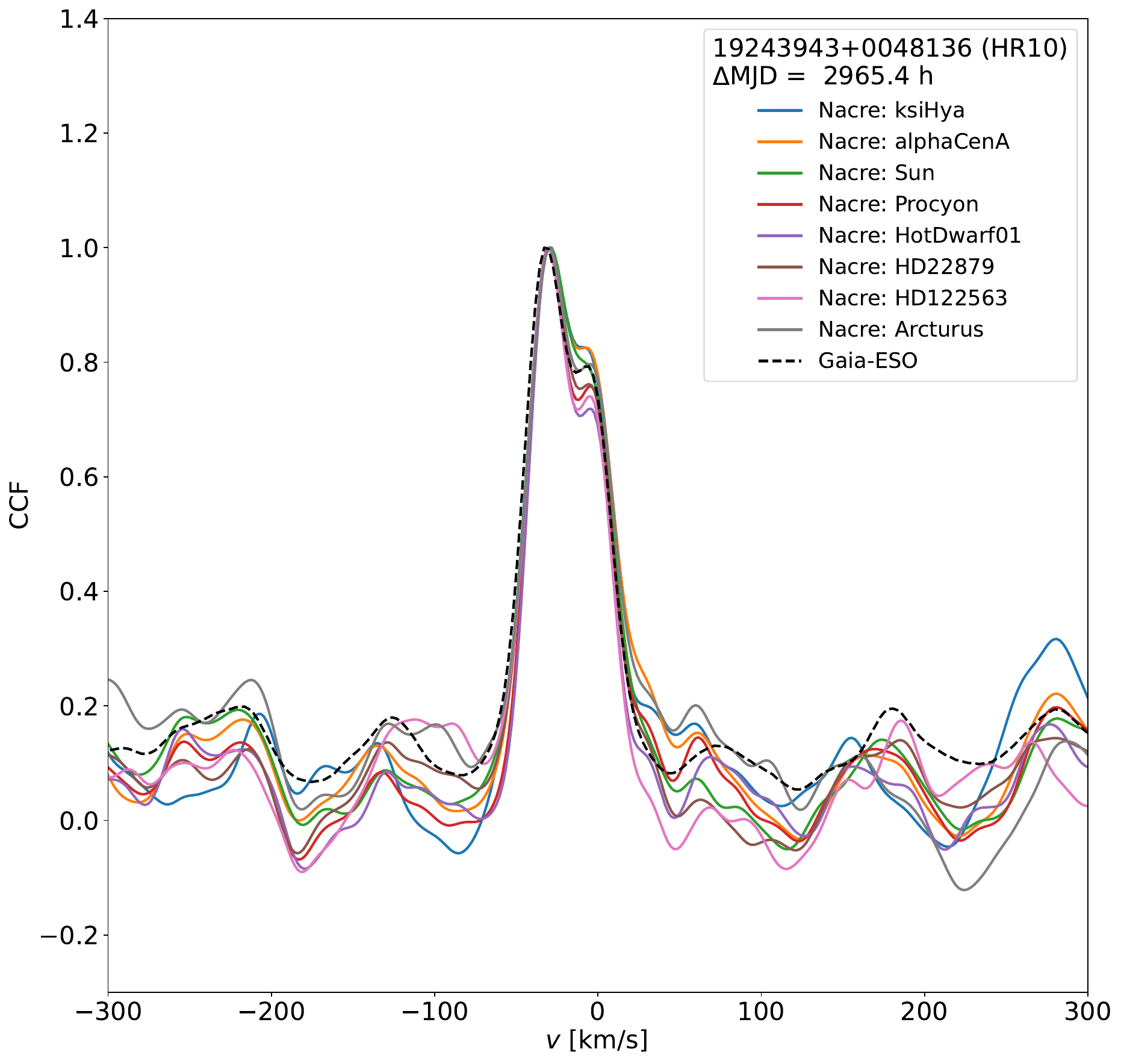}
  \includegraphics[width=0.49\textwidth]{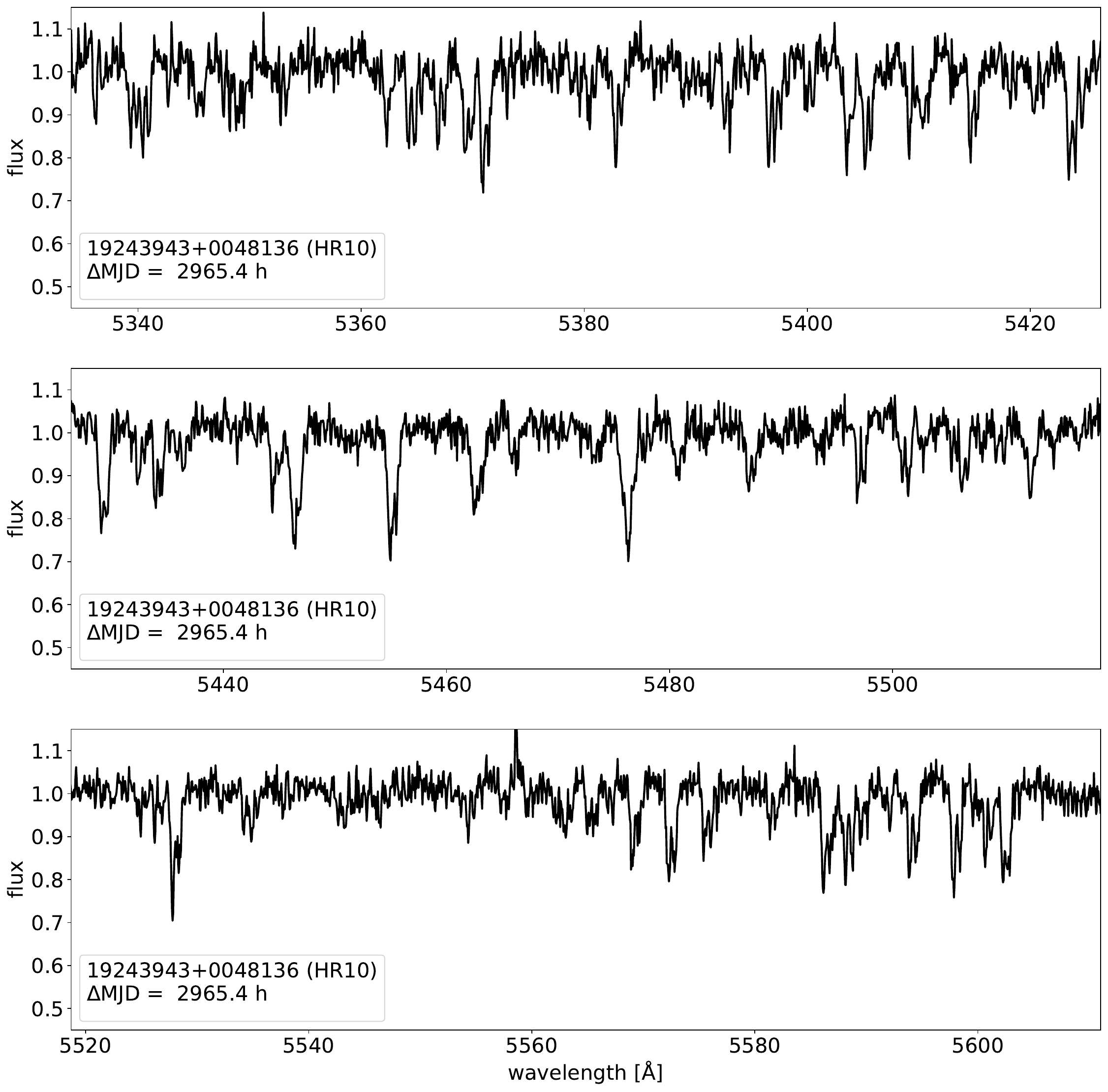}
  \captionof{figure}{\label{Fig:iDR5_SB4_atlas_19243943+0048136_9}(9/12) CNAME 19243943+0048136, at $\mathrm{MJD} = 56197.977533$, setup HR10.}
\end{minipage}
\begin{minipage}{\textwidth}
  \centering
  \includegraphics[width=0.49\textwidth]{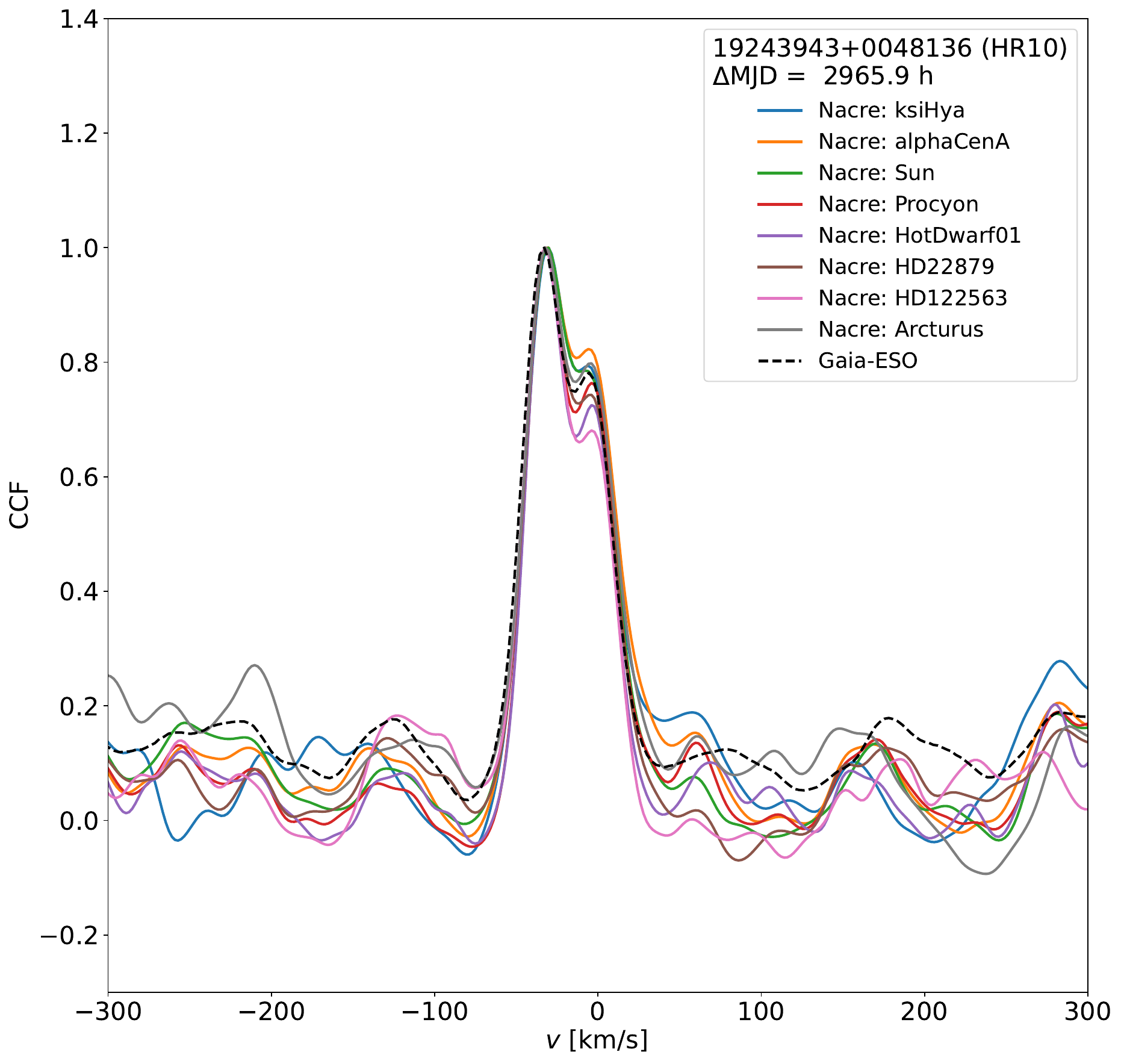}
  \includegraphics[width=0.49\textwidth]{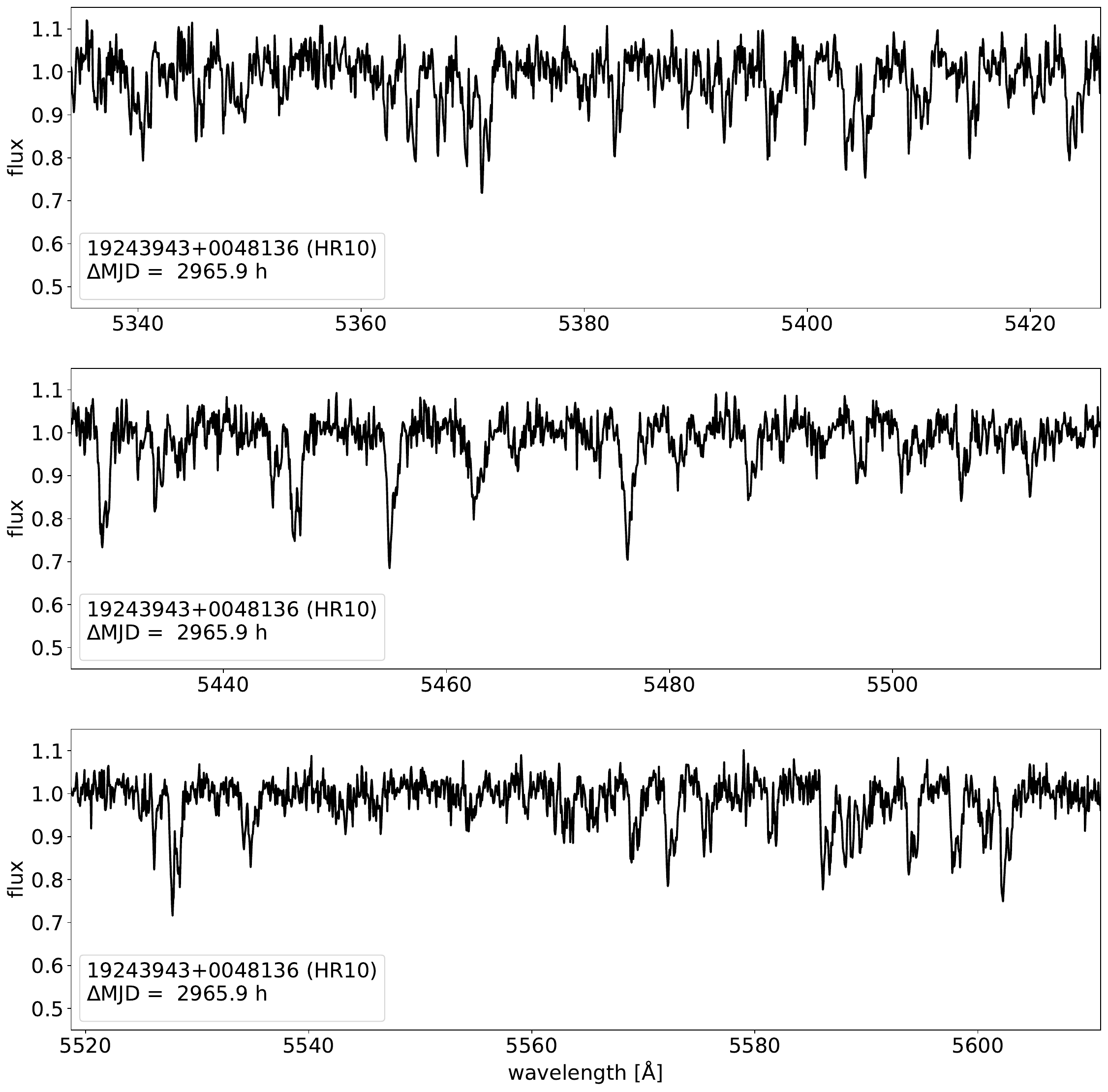}
  \captionof{figure}{\label{Fig:iDR5_SB4_atlas_19243943+0048136_10}(10/12) CNAME 19243943+0048136, at $\mathrm{MJD} = 56198.001631$, setup HR10.}
\end{minipage}
\clearpage
\begin{minipage}{\textwidth}
  \centering
  \includegraphics[width=0.49\textwidth]{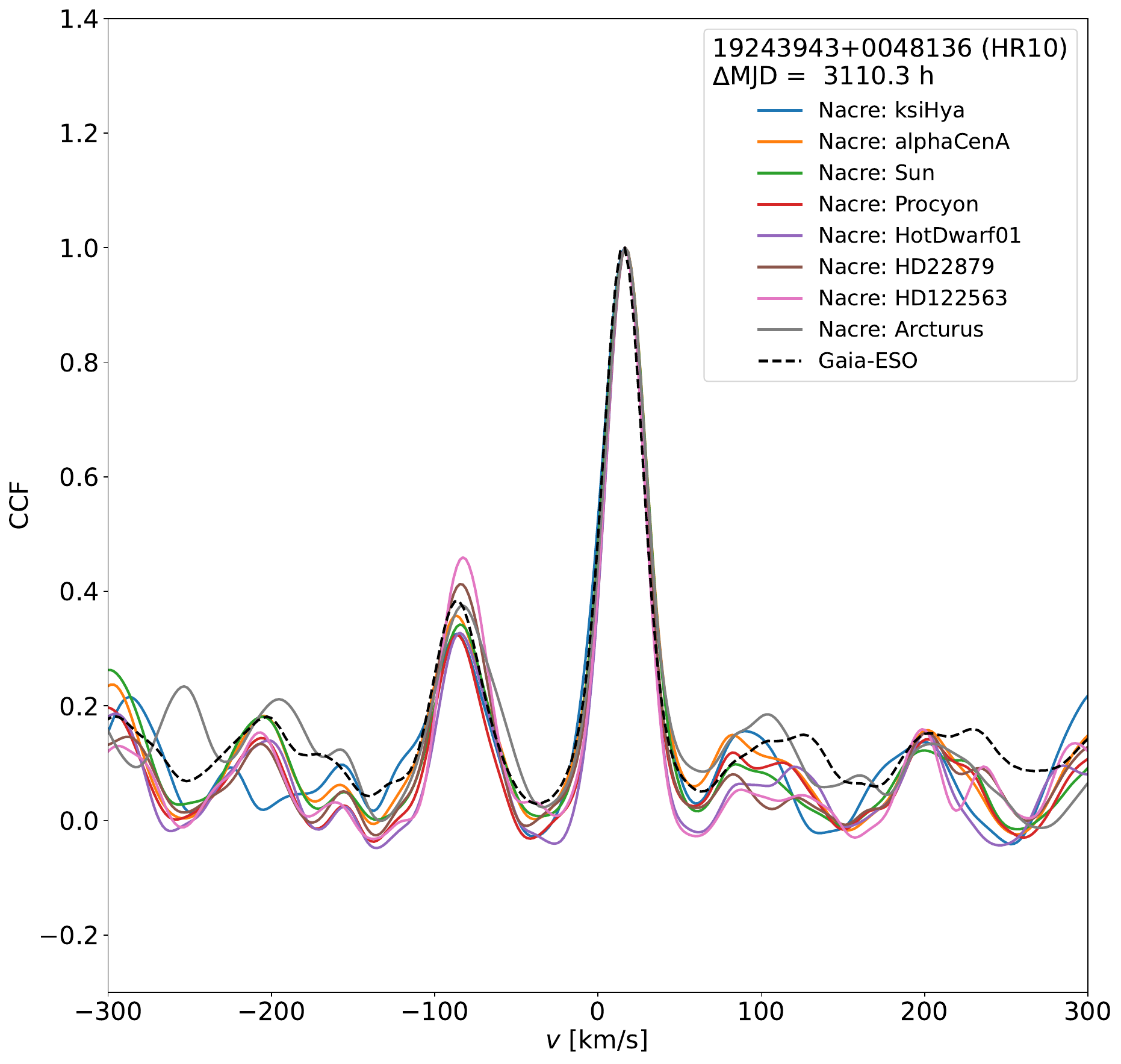}
  \includegraphics[width=0.49\textwidth]{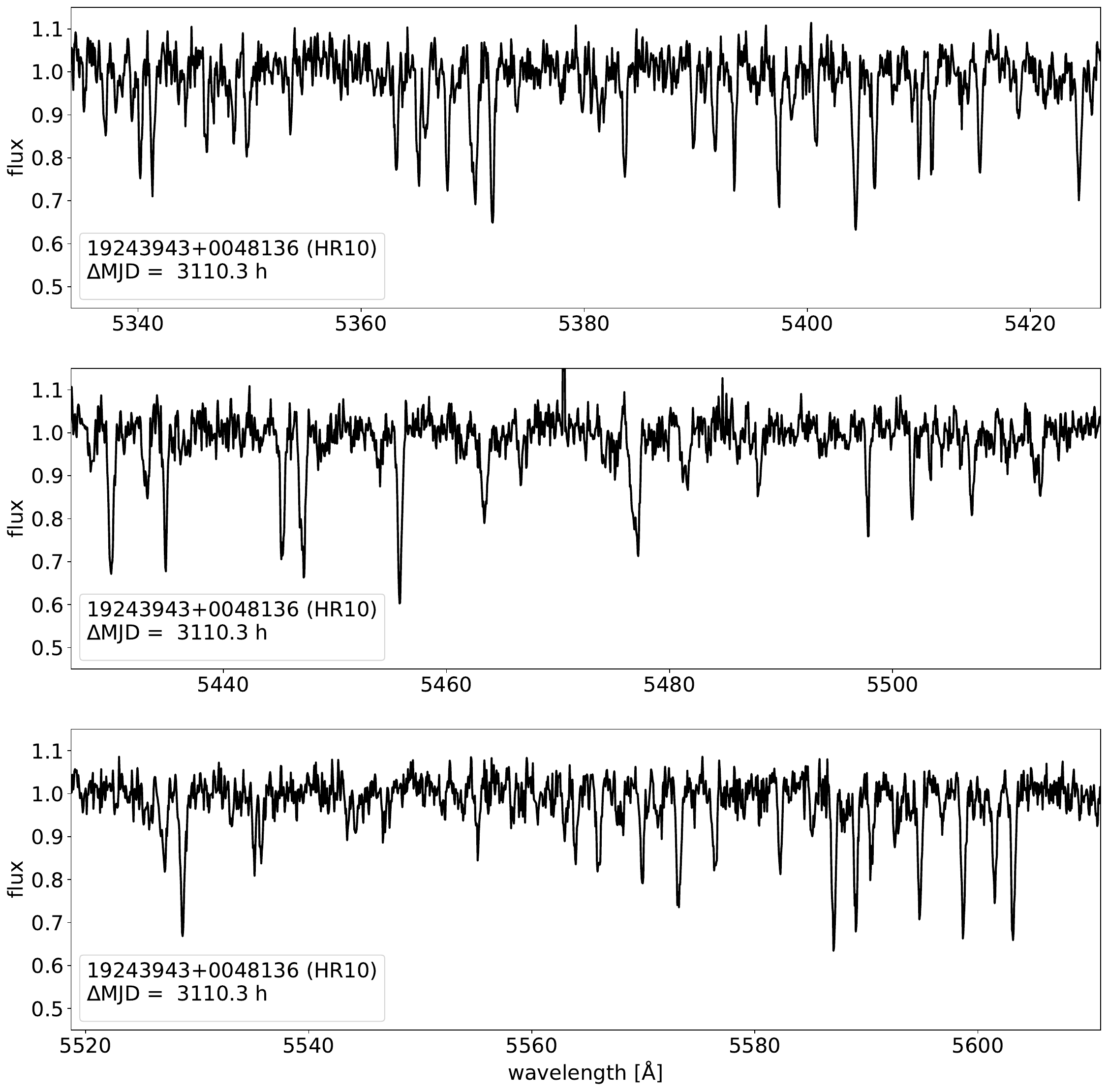}
  \captionof{figure}{\label{Fig:iDR5_SB4_atlas_19243943+0048136_11}(11/12) CNAME 19243943+0048136, at $\mathrm{MJD} = 56204.014924$, setup HR10.}
\end{minipage}
\begin{minipage}{\textwidth}
  \centering
  \includegraphics[width=0.49\textwidth]{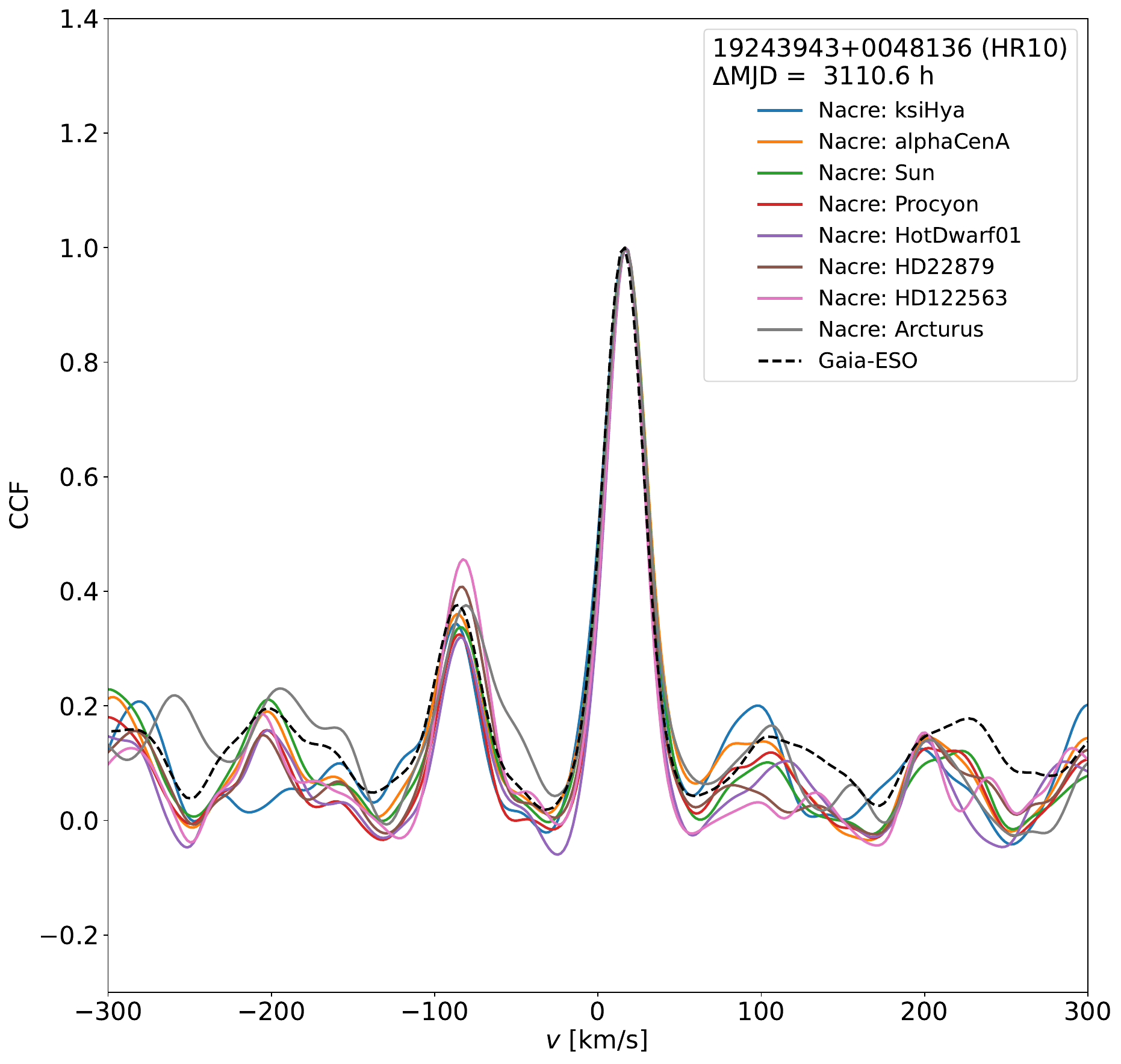}
  \includegraphics[width=0.49\textwidth]{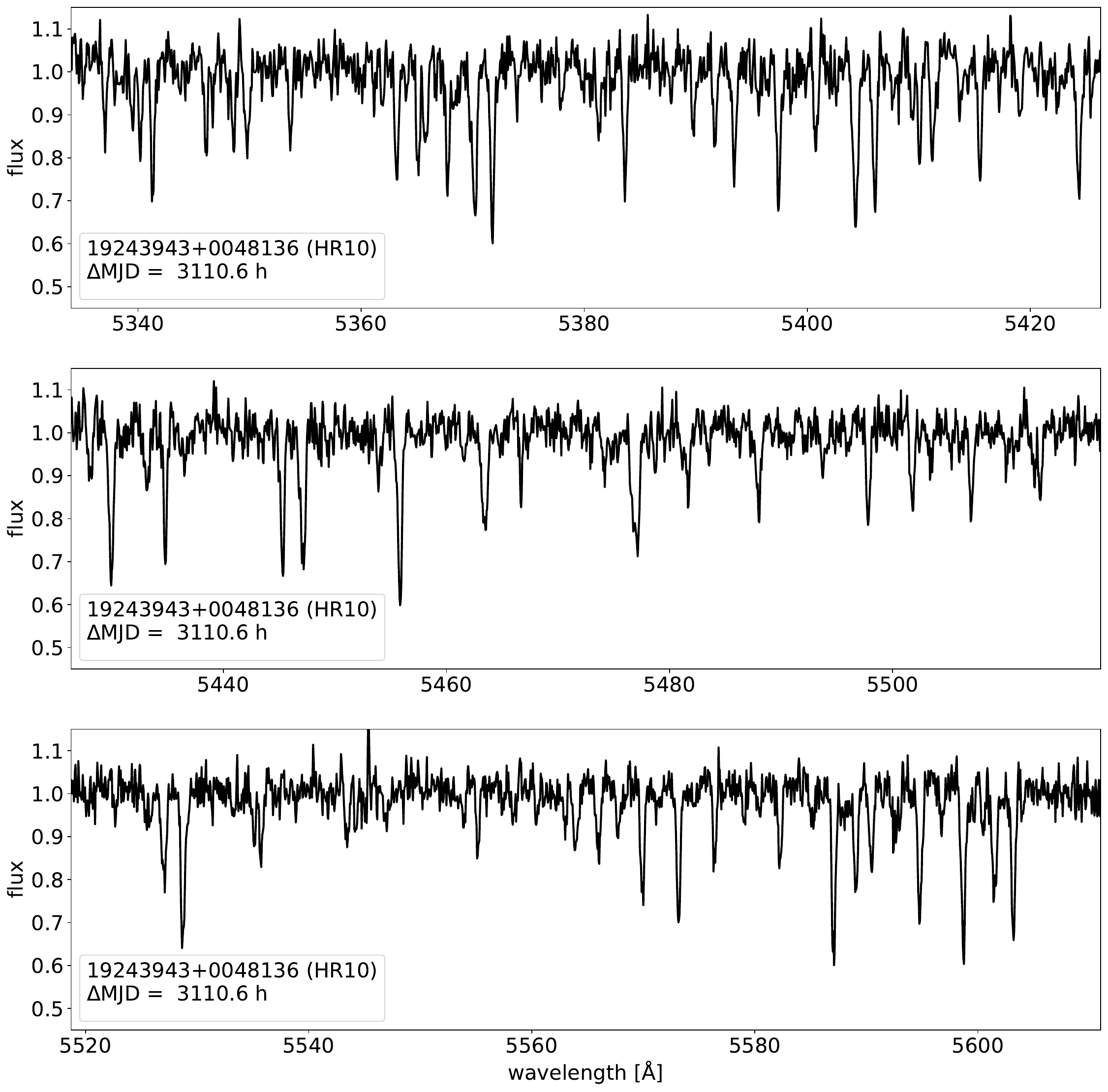}
  \captionof{figure}{\label{Fig:iDR5_SB4_atlas_19243943+0048136_12}(12/12) CNAME 19243943+0048136, at $\mathrm{MJD} = 56204.028780$, setup HR10.}
\end{minipage}
\cleardoublepage
\setlength\parindent{\defaultparindent}

\end{document}